\documentclass[11pt,letterpaper]{article}
\usepackage[T1]{fontenc}
\usepackage[utf8]{inputenc}
\usepackage{lmodern,microtype}
\usepackage[margin=1in]{geometry}
\usepackage{amsmath,amssymb,amsthm,mathtools,aliascnt}
\usepackage{booktabs,array,tabularx,enumitem,graphicx,xcolor,needspace}
\usepackage[all,defaultlines=3]{nowidow}
\usepackage[section]{placeins}
\usepackage{tikz}
\usetikzlibrary{arrows.meta,positioning,fit,backgrounds,calc,decorations.pathreplacing}
\usepackage[numbers,sort&compress]{natbib}
\usepackage{xurl}
\usepackage[bottom]{footmisc}
\usepackage[hidelinks,linktoc=all]{hyperref}
\usepackage[nameinlink,noabbrev]{cleveref}
\newcommand{\gcov}{\hyperref[sc:covering]{general covering relations}}
\newcommand{\gcovsec}{\hyperref[sc:covering]{the subsection on general covering relations}}
\newcommand{\Gcovsec}{\hyperref[sc:covering]{The subsection on general covering relations}}
\newtheorem{theorem}{Theorem}[section]
\newaliascnt{lemma}{theorem}
\newtheorem{lemma}[lemma]{Lemma}
\aliascntresetthe{lemma}
\newaliascnt{corollary}{theorem}
\newtheorem{corollary}[corollary]{Corollary}
\aliascntresetthe{corollary}
\newaliascnt{proposition}{theorem}
\newtheorem{proposition}[proposition]{Proposition}
\aliascntresetthe{proposition}
\newtheorem{mainthm}{Theorem}

\crefname{mainthm}{Theorem}{Theorems}
\theoremstyle{definition}
\newaliascnt{definition}{theorem}

\aliascntresetthe{definition}
\newaliascnt{remark}{theorem}

\aliascntresetthe{remark}
\newcommand{\gf}{\gamma_f}
\newcommand{\cost}{\mathcal C}

\newcommand{\BC}{\textnormal{\textsc{Congest-BC}}}
\newcommand{\LOCAL}{\textnormal{\textsc{Local}}}
\DeclareMathOperator{\OPT}{OPT}
\setlist{itemsep=3pt,topsep=5pt}
\definecolor{simpblue}{HTML}{24639A}
\definecolor{simporange}{HTML}{C65D25}
\definecolor{simpgreen}{HTML}{168477}
\tikzset{sv/.style={circle,draw=black!70,fill=white,inner sep=2.2pt},
 so/.style={sv,draw=simpblue,fill=simpblue!20},
 sb/.style={sv,draw=simporange,fill=simporange!25},
 se/.style={draw=black!60,line width=.6pt},
 arr/.style={-{Stealth[length=2mm]},draw=black!65}}

\hypersetup{pdftitle={Paying for Fractional Error: Tight Constant-Round Approximation of Covering Problems with Short Broadcasts},
 pdfauthor={Saeed Amiri, Sebastian Siebertz, Wojciech Wawrzyniak}}
\title{Paying for Fractional Error:\\
Tight Constant-Round Approximation of\\
Covering Problems with Short Broadcasts}
\author{Saeed Amiri\\Independent researcher\\\texttt{saeed.amiri.akh@gmail.com}
\and Sebastian Siebertz\\University of Bremen\\\texttt{siebertz@uni-bremen.de}
\and Wojciech Wawrzyniak\\Adam Mickiewicz University, Pozna\'n\\\texttt{wojciech.wawrzyniak@amu.edu.pl}}
\date{}
\usepackage{etoolbox}

\newlength{\stmtparindent}
\newlength{\statementsep}
\newcommand{\keepstatement}[1]{\BeforeBeginEnvironment{#1}{\par\addvspace{\statementsep}\noindent
  \begin{minipage}{\linewidth}\setlength{\parindent}{\stmtparindent}}\AfterEndEnvironment{#1}{\end{minipage}\par\addvspace{\statementsep}}}
\keepstatement{theorem}\keepstatement{lemma}\keepstatement{corollary}
\keepstatement{proposition}\keepstatement{mainthm}\keepstatement{definition}
\makeatletter
\AtBeginDocument{\let\savedtitle\@title\let\savedauthor\@author
 \let\saveddate\@date\let\savedmaketitle\@maketitle}
\newcommand{\candidatetitle}{\let\@title\savedtitle
 \let\@author\savedauthor\let\@date\saveddate\savedmaketitle}
\makeatother
\begin{document}
\pagenumbering{gobble}
\hypersetup{pageanchor=false}

\begin{titlepage}
\maketitle
\thispagestyle{empty}
\begin{abstract}\normalsize
We present a general framework for deterministic constant-round approximation of covering problems in sparse networks in the broadcast CONGEST model, where every vertex broadcasts the same~$\mathcal O(\log n)$-bit message to all its neighbors.
The framework combines two independent tools.
The first is a solver that computes a $(1+\delta)$-approximate solution of any local covering linear program with bounded positive integer costs on every class that excludes a fixed minor.
The second is a rounding principle: a rounding rule that saves a fixed fraction of its own output, for every feasible input, pays for a solver error fixed in advance, so that an approximate fractional solution yields an exact coefficient.

For ordinary domination, we achieve a $2a+1$ approximation against the fractional optimum~$\gamma_f(G)$ for every fixed rational $a\ge1$, whenever every minor has a ratio of edges to vertices of at most~$a$.
On planar graphs this is seven, matching the lower bound of Hilke, Lenzen, and Suomela.
The coefficients are also optimal at bounded treewidth and pathwidth, for~$K_t$-minor-free graphs with $3\le t\le9$, on triangle-free planar graphs (five), and on planar graphs of large girth, where it approaches the optimal three.
On every fixed surface the exact bound is~$7\gamma(G)$, and we show that a bound of~$7\gamma_f(G)$ is impossible on some fixed surfaces even sequentially.

For total domination, on graphs without isolated vertices and under the same minor-density assumption, the exact coefficient is~$2a$ against the fractional optimum, including the optimal six on planar graphs.
Red-blue domination, and hence set cover and hitting set on planar incidence graphs, has the optimal limiting coefficient four.
Our method also gives randomized constant-round and deterministic~$\mathcal O(\log^*n)$-round $(1+\varepsilon)$-approximation schemes, with the randomized cost guarantee holding in expectation.

\end{abstract}
\end{titlepage}

\clearpage
\setcounter{tocdepth}{2}
\begingroup
\normalsize
\tableofcontents
\endgroup
\clearpage
\pagenumbering{arabic}
\hypersetup{pageanchor=true}
\section{Introduction}\label{sc:introduction}

Many graph problems ask for a small set of vertices that meets a requirement near every vertex, including dominating set, total domination, and set cover on an incidence graph.
In a distributed network, the graph also carries the messages, so a fast algorithm must make good local choices with the information that the vertices can exchange.
Even a neighborhood of bounded radius may contain arbitrarily many vertices, which makes this a communication problem as well as an approximation problem.

We present a new, very general framework for local covering problems on sparse networks, with a fractional solver and a rounding principle that together achieve exact approximation coefficients.
The algorithms work in the \emph{broadcast CONGEST} model, where each vertex sends the same~$\mathcal O(\log n)$-bit message to all neighbors in a round, even when its degree is unbounded.
Vertices have distinct~$\mathcal O(\log n)$-bit identifiers, and local computation is unrestricted.
We obtain optimal coefficients against the fractional optimum for domination on several sparse classes and for total domination on planar graphs.
Our results settle the question of the optimal approximation factors after a long sequence of improvements in distributed algorithms, and they do so even in the restrictive broadcast CONGEST model.
For planar red-blue domination, the optimal limiting coefficient is four, with upper bound~$4+\varepsilon$ for every fixed $\varepsilon>0$.
On graphs of any fixed Euler genus, we also achieve~$7\gamma(G)$ without an embedding. However, we show that on some surfaces~$7\gamma_f(G)$ is impossible even for sequential algorithms.

\begingroup
\renewcommand{\thefootnote}{}
\makeatletter
\long\def\@makefntext#1{\noindent#1}
\let\@footnotetext\H@@footnotetext
\makeatother
\footnotetext{\begin{minipage}{\columnwidth}\footnotesize
We partly used generative AI tools in developing this work. The full disclosure appears in the \hyperref[stoc:ai-methodology]{AI methodology statement} at the end of the paper.
\end{minipage}}
\endgroup

\paragraph{The planar question.}
Lenzen, Oswald, and Wattenhofer~\cite{LOW2008} initiated constant-round planar domination, and the best coefficient improved from~$130$~\cite{LPW2013} through~$52$~\cite{Wawrzyniak2014}, $11+\varepsilon$~\cite{HSV2022}, and $8+\varepsilon$~\cite{Wawrzyniak2026} to $7+\varepsilon$ in the LOCAL model~\cite{AS2026}, where message sizes are unbounded.
Hilke, Lenzen, and Suomela~\cite{HLS2014} ruled out every coefficient below seven for deterministic constant-round algorithms, even with unbounded messages.
However, for the bounds $11+\varepsilon$, $8+\varepsilon$, and $7+\varepsilon$, the number of rounds depends on~$\varepsilon$.
Also, in the $(7+\varepsilon)$-approximation of Amiri and Siebertz~\cite{AS2026}, the treatment of cluster boundaries leaves a positive error for fixed parameters, so this approach does not give the exact coefficient seven and it seems unlikely that it can be adapted to achieve the optimum.
In this work we finally close this gap even in the more restrictive broadcast CONGEST model by computing a dominating set $D$ satisfying
\[
 |D|\le7\gf(G)\le7\gamma(G),
\]
where~$\gamma(G)$ is the minimum size of a dominating set and~$\gf(G)$ is the fractional optimum, assigning nonnegative weights so that each closed neighborhood receives at least one in total.

\subsection{A new framework for local covering problems}

We present a new framework with two separate parts: a distributed solver that computes an approximately optimal fractional cover with short messages, and a rounding method that turns this cover into an integral solution with an exact approximation coefficient.
The solver works on every class that excludes a fixed minor, and also gives integral approximation schemes. 
This separation lets the same communication procedures handle ordinary and total domination, restricted covering, and set cover and hitting set on sparse incidence graphs.

\paragraph{A solver for local covering programs.}
We consider the following setting.
Each demand~$r$ specifies a nonempty set~$C(r)$ of allowed candidates in its closed neighborhood. Each candidate~$v$ knows its positive integer cost $c_v\in\{1,\ldots,b\}$ for a fixed bound~$b$.
Every vertex knows whether it is a demand.
For each demand~$r$ and neighboring candidate~$v$, both vertices know whether~$v$ is allowed to cover~$r$, and each demand also knows whether it can cover itself.
This setting covers open neighborhoods, partial domination, forbidden candidates, and directed covering relations.
We write the minimum fractional cost as
\[
 \tau_c=\min\Bigl\{\sum_vc_vx_v:\ x\ge0,\ \sum_{v\in C(r)}x_v\ge1
 \text{ for every demand }r\Bigr\},
\]
and let~$\OPT_c$ be the minimum cost of an integral cover.
Distributed LP methods were developed in~\cite{KW2005,KMW2016}, but, to our knowledge, no constant-round $(1+\delta)$-solver for these programs was known with short messages and unbounded degrees.

For rational $a\ge1$, let~$\mathcal C_a$ be the class of graphs whose nonempty minors~$J$ satisfy $|E(J)|\le a|V(J)|$.
In the following, density or sparsity will always mean the ratio of edges to vertices. 
Every class excluding a fixed minor lies in some~$\mathcal C_a$ by a classical result of Mader~\cite{Mader1967}, and planar graphs lie in~$\mathcal C_3$.
Our first main theorem is the following.

\begingroup
\begin{mainthm}\label{sc:thmA}
For fixed rationals $\rho\ge1$ and $\delta>0$, and a fixed integer $b\ge1$, there is an algorithm that computes a feasible fractional cover of cost at most $(1+\delta)\tau_c$ on every $G\in\mathcal C_\rho$ in a constant number of deterministic broadcast CONGEST rounds.
The algorithm requires no auxiliary graph structure, degree bound, or knowledge of the graph order.
\end{mainthm}
\endgroup

We explain the main ideas in Section~\ref{simp:overview} and give full proofs in the main body of the paper. In a nutshell, the solver forms local covering tasks of bounded diameter that admit a cover of bounded cost.
It finds short descriptions of local solutions and combines them into a near-optimal fractional cover, using only short broadcasts.
The solver still introduces a small approximation error, and our second tool uses a saving in the rounding step to compensate for this error.

\begingroup
\paragraph{Paying for fractional error.}
The second tool uses the saving from rounding to absorb the error of the fractional solver.
We say that a local rounding rule producing a set~$D(x)$ from a feasible fractional cover~$x$ has \emph{slack} $\delta>0$ if
\[
 \cost(x)\ge(1+\delta)|D(x)|\qquad\text{for every feasible }x.
\]
Here $\cost(x)=\sum_vc_vx_v$ is the cost of the fractional cover.
The saving is proportional to $|D(x)|$ and holds for every feasible input, which allows us to use an approximately optimal solution.

\begin{mainthm}\label{sc:thmB}
Let a rounding rule have slack $\delta>0$ on a covering program with fractional optimum~$\tau$.
Then every feasible~$x$ with $\cost(x)\le(1+\delta)\tau$ satisfies $|D(x)|\le\tau$.
For dominating set on graphs of minor density at most a rational $a\ge1$, there is a transformation of the instance, computable in a constant number of broadcast rounds, after which threshold rounding has slack at least
\[
 \delta_a=\frac1{2a(2a+1)^2}\quad\text{if $2a$ is an integer},\qquad
 \delta_a=\min\Bigl\{\frac{1-\theta}{a(a+1)},\frac{\theta}{\lfloor2a\rfloor+1}\Bigr\}
 \quad\text{otherwise},
\]
where $\theta=2a-\lfloor2a\rfloor$. The transformed fractional optimum is at most $(2a+1)\gf(G)$ minus the number of vertices already selected.
If~$2a$ is not an integer, it suffices that all \emph{subgraphs} have a ratio of edges to vertices of at most~$a$.
\end{mainthm}
\endgroup

The first assertion follows from $(1+\delta)|D(x)|\le\cost(x)\le(1+\delta)\tau$.
The second shows how to obtain the required saving for domination: when~$2a$ is an integer, we first select the regions where the usual rounding bound could be tight, and assign the remaining candidates costs that record the budget left by this selection.
When~$2a$ is not an integer, the gaps between~$2a$ and integer degrees give the saving directly.
The solver accuracy can therefore be fixed independently of the input graph, and its error disappears from the final coefficient.
For each new application, we need to bound the transformed fractional optimum and prove the rounding saving.
The solver can then be reused.
The tools also work separately, since the solver gives approximation schemes.
Section~\ref{simp:endpoint:section} proves the general domination construction in Theorem~\ref{sc:thmB}.

\subsection{Exact coefficients for domination}

We combine the tools to obtain coefficients determined by minor density, or by a smaller subgraph-density bound when~$2a$ is noninteger.
The planar case gives seven, while larger girth and other sparse classes give the following guarantees.

\begingroup
\begin{mainthm}\label{sc:thmC}
There exist deterministic constant-round broadcast CONGEST algorithms that return a dominating set~$D$ with the following guarantees.
\begin{enumerate}[label=(\roman*),leftmargin=*]
\item\label{sc:thmC:minor} For fixed rational $a\ge1$ and $G\in\mathcal C_a$, the output satisfies $|D|\le(2a+1)\gamma_f(G)$.
\item\label{sc:thmC:sub} For fixed rationals $1\le a\le\rho$ with~$2a$ noninteger, the same bound holds for $G\in\mathcal C_\rho$ if every subgraph has a ratio of edges to vertices of at most~$a$.
\item\label{sc:thmC:girth} For planar~$G$ of girth at least a fixed integer $g\ge3$, we have $|D|\le(3+4/(g-2))\gamma_f(G)$.
\end{enumerate}
The algorithms require no embedding, orientation, initial cover, degree bound, or knowledge of~$n$.
\end{mainthm}
\endgroup

\begingroup
A planar graph of girth at least~$g$ has subgraph density at most $g/(g-2)$, so part~\ref{sc:thmC:sub} gives the girth bound except at $g\in\{3,4,6\}$.
The case $g=3$ follows from minor density, while girths four and six require a face-counting argument because contractions can create shorter cycles.
Section~\ref{new:planar} presents the planar construction, while Section~\ref{simp:endpoint:section} and Appendix~\ref{sc:girth} prove the general and high-girth bounds. As a corollary we obtain the optimal bounds in several important cases. 

\begin{corollary}\label{sc:optimal}
The coefficients~$7$ on planar graphs and on graphs of Euler genus at most two,~$5$ on triangle-free planar graphs, $2r+1$ on graphs of treewidth or pathwidth at most $r\ge1$, and $2t-3$ for~$K_t$-minor-free graphs, $3\le t\le9$, are optimal for deterministic constant-round algorithms.
The lower bounds hold even in the LOCAL model and with respect to~$\gamma(G)$.
\end{corollary}

For surfaces, this corollary gives~$7\gf(G)$ only up to Euler genus two, while Theorem~\ref{fv:thmE} below gives~$7\gamma(G)$ for every fixed Euler genus.

\endgroup

\begin{table}[ht]
\centering
\begingroup
\begin{tabular}{@{}lcc@{}}
\toprule
Graph class & Bound on $|D|$ & Optimal coefficient?\\
\midrule
Planar &~$7\gamma_f$ & Yes\\
Triangle-free planar &~$5\gamma_f$ & Yes\\
Treewidth or pathwidth at most~$r$ & $(2r+1)\gamma_f$ & Yes\\
$K_t$-minor-free, $3\le t\le9$ & $(2t-3)\gamma_f$ & Yes\\
Every minor has density at most~$a$ & $(2a+1)\gamma_f$ & Class-dependent\\
Euler genus at most fixed~$e$ &~$7\gamma$ & Yes\\
\bottomrule
\end{tabular}
\endgroup
\caption{Deterministic constant-round domination guarantees in broadcast CONGEST.
The lower bounds in the rows marked optimal hold in the LOCAL model and against the integral optimum.}
\label{sc:table}
\end{table}

\begingroup
The lower bounds come from~\cite{HLS2014} for planar graphs and~\cite{AS2026} for powers of paths.
For triangle-free planar graphs, Bonamy, Cook, Groenland, and Wesolek~\cite{BonamyCook2021} observed that large grids rule out every coefficient below five and conjectured that five could be achieved.
Our theorem proves this conjecture and achieves the coefficients previously approached within~$\varepsilon$ in LOCAL~\cite{AS2026}.
The proof of Corollary~\ref{sc:optimal}, including the tight lower bounds, is given in Appendix~\ref{sc:optimality}.

\medskip

To the best of our knowledge, constant-round constant-factor approximations in broadcast CONGEST were previously known for bounded-genus graphs, but not for general excluded-minor classes; see Section~\ref{sc:related-work}.
Our framework therefore extends these guarantees to a broader family of graph classes, while also giving exact approximation coefficients.
We must admit that this makes us quite proud, even though we were AI assisted on the final meters.
\endgroup

\subsection{Schemes and general covering}

\begingroup
The same construction also gives integral approximation schemes, using random bits in constant rounds or a deterministic partition in~$\mathcal O(\log^*n)$ rounds, which is of independent interest.

\begin{mainthm}\label{sc:thmD}
For fixed rational $\rho\ge1$, integer $b\ge1$, and $\varepsilon>0$, the following guarantees hold on every $G\in\mathcal C_\rho$:
\begin{enumerate}[label=(\roman*),leftmargin=*]
\item There exists a randomized algorithm that computes an integral cover that is always feasible and has expected cost at most $(1+\varepsilon)\OPT_c$ in a constant number of broadcast CONGEST rounds.
\item If vertices know a common polynomial bound on the identifiers, there exists a deterministic algorithm that computes a feasible integral cover of cost at most $(1+\varepsilon)\OPT_c$ in~$\mathcal O(\log^*n)$ broadcast CONGEST rounds.
\end{enumerate}
The constants in the round bounds depend only on~$\rho,b,\varepsilon$.
\end{mainthm}

For domination, both schemes were known in LOCAL~\cite{AS2026,CHWW2018} while our algorithms use short broadcasts and apply to all the stated covering problems with bounded costs.
A direct adaptation of the argument of Lenzen and Wattenhofer~\cite{LW2008} shows that the deterministic round bound is necessary already on cycles: for every fixed $0<\varepsilon<1/4$, approximation within~$1+\varepsilon$ requires~$\Omega(\log^*n)$ rounds even in LOCAL.
We present this observation in Appendix~\ref{app:rounds-lower} (Proposition~\ref{sc:rounds-lower}).

\paragraph{Other covering problems.}
Appendix~\ref{sc:applications} gives a general rounding rule with coefficient $2a+1+\varepsilon$, or $2a+\varepsilon$ when no demand covers itself, against~$\tau_c$, where every subgraph of the covering-incidence graph has a ratio of edges to vertices of at most~$a$.
This includes weighted, partial, restricted, directed, total, and red-blue domination (Corollary~\ref{sc:rounding-all}).
For set cover and hitting set on planar incidence graphs the coefficient is $4+\varepsilon$, with no bound on set sizes or element frequencies, and Theorem~\ref{sc:thmD} also gives approximation schemes.
The stronger total-domination results appear in Appendix~\ref{fv:total}, the red-blue results in Appendix~\ref{fv:rb}, and further high-girth bounds in Appendices~\ref{fv:imp} and~\ref{fv:hg}.
\par\endgroup

\begingroup
\subsection{Fixed surfaces}

On every fixed surface, the framework achieves the same exact coefficient as in the planar case, with the integral optimum as benchmark.

\begin{mainthm}\label{fv:thmE}\label{new:surface-main}
For every fixed $e\ge0$, there is a deterministic constant-round broadcast CONGEST algorithm that returns a dominating set~$D$ with $|D|\le7\gamma(G)$ on every graph~$G$ of Euler genus at most~$e$.
The algorithm works on orientable and nonorientable surfaces and does not need an embedding.
\end{mainthm}

The proof is given in Section~\ref{new:surface}.
Seven is optimal for deterministic constant-round algorithms, and on some fixed surfaces even a sequential algorithm cannot guarantee~$7\gf(G)$, so we cannot replace~$\gamma(G)$ by~$\gf(G)$ in this theorem (Proposition~\ref{sc:gap} in Appendix~\ref{simp:closing}).

\paragraph{Organization.}
Section~\ref{simp:overview} explains the main ideas of the framework and its planar application. Section~\ref{simp:preliminaries} introduces the notation, and Section~\ref{prop:framework} presents the framework and its planar application. Detailed proofs, further applications, and lower bounds appear in the subsequent sections and appendices.
\par\endgroup

\begingroup
\section{An intuitive overview of the framework}\label{simp:overview}

In this section we explain how the fractional solver and the rounding principle work together.
The solver is the same in all applications, while each application chooses its first selection, its costs, and its rounding rule.
Let us begin with this connection and the planar example, before explaining how short messages are enough to find the fractional solution.

\subsection{Transferring the budget before rounding}

We first select a set~$F$ of vertices by a local rule and then define a new covering problem, whose demands, candidates, and costs depend on this selection.
For ordinary domination, the demands are the vertices not covered by~$F$, and every vertex outside~$F$ remains a candidate, even if it is already covered.
Other applications may keep additional demands or restrict the candidates.
Write $\cost(x)=\sum_vc(v)x_v$ for the cost of a feasible fractional cover of this new problem, and let~$\tau'$ be its minimum fractional cost, with $|F|$ counted separately.

The rounding must compensate for the error of the fractional solver.
Suppose rounding~$x$ gives a set~$D(x)$ such that $F\cup D(x)$ covers the original instance.
To obtain a target coefficient~$\alpha$ against a benchmark~$B(G)$ for the original problem, we need to show that
\[
 |F|+\tau'\le\alpha B(G),\qquad
 (1+\delta)|D(x)|\le\cost(x)
 \quad\text{for every feasible }x,
\]
where $\delta>0$ is fixed for the graph class.
The solver returns a feasible vector with $\cost(x)\le(1+\delta)\tau'$, so the second inequality gives $|D(x)|\le\tau'$, and the first gives
\[
 |F\cup D(x)|\le |F|+|D(x)|\le\alpha B(G).
\]
Thus the solver may make a fixed positive relative error, because the rounding saves the same fraction of its output size.
The selection, the costs, and the rounding must all be computed in a constant number of broadcast CONGEST rounds.

We show that the saving is large enough by using the same quantities to bound the saving from below and the size of the rounded solution from above.
For nonnegative quantities~$S_i(x)$, suppose fixed positive constants $\alpha_i,\beta_i$ satisfy
\[
 \cost(x)-|D(x)|\ge\sum_i\alpha_iS_i(x),\qquad
 |D(x)|\le\sum_i\beta_iS_i(x).
\]
Then $\delta=\min_i\alpha_i/\beta_i$ gives $\cost(x)-|D(x)|\ge\delta|D(x)|$, which is the required inequality.
Only this fixed accuracy is given to the solver; the global quantities~$S_i(x)$ are used in the analysis and need not be computed.
Proposition~\ref{sc:principle} states the principle formally.
\par\endgroup

\subsection{The planar case: an exact factor seven}

The coefficient seven on planar graphs comes from a degree count.
A degree-six vertex belongs to seven closed neighborhoods, so if all degrees were at most six, summing the fractional constraints would give $|V(G)|\le7\gamma_f(G)$.
To extend this observation to planar graphs with vertices of arbitrarily large degree, we need a rounding argument, and this argument must also give a saving that compensates for the fractional solver's error.
The same degree count explains where this saving can fail: inside large regions of degree-six vertices, our rounding bound can be tight locally, leaving only a~$7(1+\delta)$ guarantee.
Our first selection is designed to handle these regions: we select the vertices well inside them and account for them directly, while adjusting the costs of the remaining candidates so that rounding the remaining problem gives the required saving.

\begingroup
This first selection and the adjusted costs can be determined locally from the degrees.
Call a vertex a \emph{defect} if its degree differs from six, let~$B$ be the set of defects, and put
\[
 F=V(G)\setminus N_G[B],\qquad
 R_0=V(G)\setminus N_G[F],\qquad
 c(v)=7-|N_G(v)\cap F|\quad(v\notin F).
\]

Every vertex of~$F$ and all its neighbors have degree six.
The new demands are~$R_0$, while all vertices outside~$F$ remain candidates, with costs between one and seven.
For example, a candidate with two neighbors in~$F$ receives cost five, since those two incidences already contribute to paying for~$F$.
\par\endgroup

These costs give the budget inequality $|F|+\tau'\le7\gamma_f(G)$.
To see this, restrict an optimal fractional dominating vector~$y$ to the candidates outside~$F$, which still covers~$R_0$.
Sum the original constraints over~$F$ and add the cost of this restricted vector.
A candidate $v\notin F$ contributes $|N_G(v)\cap F|y_v+c(v)y_v=7y_v$, while a vertex of~$F$ contributes at most~$7y_v$.
The total is therefore at most~$7\gamma_f(G)$.

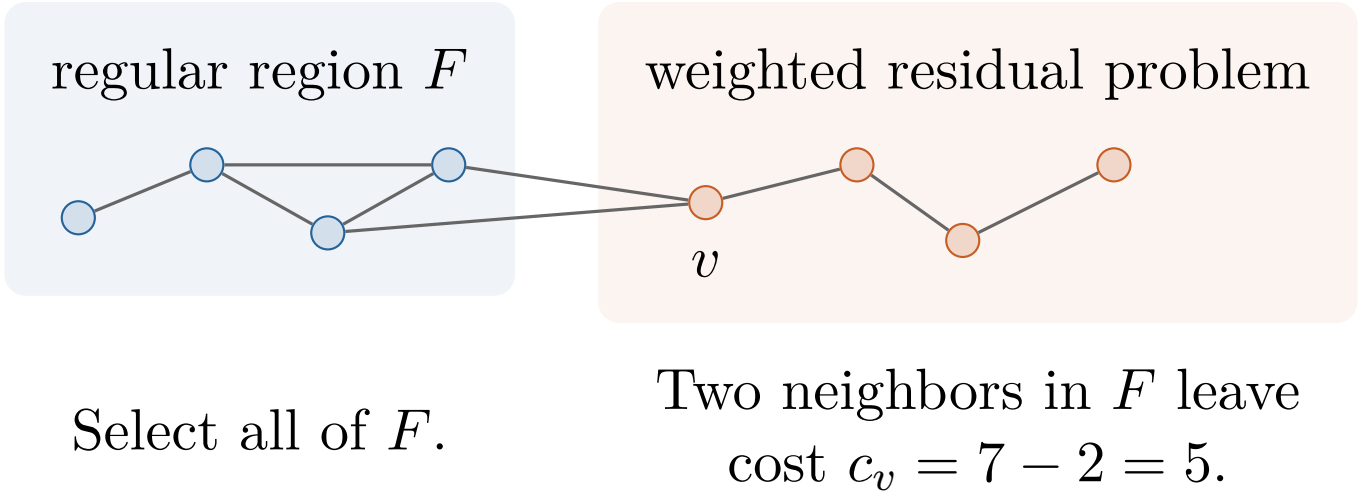
\begin{figure}[ht]
\centering
\begin{tikzpicture}[x=1cm,y=1cm,font=\normalsize]
 \node[align=center,inner sep=0pt] (regular-title) at (1.85,1.45) {regular region $F$};
 \node[align=center,inner sep=0pt] (residual-title) at (6.6,1.45) {weighted residual problem};
 \foreach \x/\y/\n in {.65/.5/a,1.5/.85/b,2.3/.4/c,3.1/.85/d}
    \node[so] (\n) at (\x,\y) {};
 \foreach \x/\y/\n in {4.8/.6/v,5.8/.85/e,6.5/.35/f,7.5/.85/g}
    \node[sb] (\n) at (\x,\y) {};
 \draw[se] (a)--(b)--(c)--(d);
 \draw[se] (b)--(d);
 \draw[se] (c)--(v)--(d);
 \draw[se] (v)--(e)--(f)--(g);
 \node[below=2mm of v,inner sep=0pt] (v-label) {$v$};
 \begin{scope}[on background layer]
  \node[fill=simpblue!7,rounded corners,inner sep=3mm,
        fit=(regular-title)(a)(b)(c)(d)] {};
  \node[fill=simporange!7,rounded corners,inner sep=3mm,
        fit=(residual-title)(v)(e)(f)(g)(v-label)] {};
 \end{scope}
 \node[align=center] at (1.85,-.9) {Select all of $F$.};
 \node[align=center] at (6.6,-.9) {Two neighbors in $F$ leave\\cost $c_v=7-2=5$.};
\end{tikzpicture}
\caption{The costs are the remaining fractional budget.
The drawing shows only selected incidences. 
Vertices in~$F$ have degree six in the full graph.
All candidates outside~$F$ are kept, including candidates that no longer require coverage themselves.}
\label{simp:fig:budget}
\end{figure}

The solver's fractional cover is now rounded by three local choices.
For a feasible~$x$ with coordinates at most one, select the cost-seven candidates with $x_v>1/4$, forming~$W$.
Next select every remaining candidate covering at least seven demands still uncovered by~$W$, forming~$H$.
Finally select all demands still uncovered, forming~$U$, and return $D=W\cup H\cup U$.
These choices use only local counts and selection bits.

\begingroup
We now compare the saving with the size of the output.
Let $w=|W|$, and let~$\ell$ count the demands that are still uncovered after the first stage and then covered by~$H$ without belonging to~$H$ themselves.
These demands need not be selected in the last stage.
Let~$P$ be the fractional cost outside $W\cup H$ that remains after subtracting its contribution to covering the demands left after~$W$.
These quantities are nonnegative, and the proof gives
\[
 \cost(x)-|D|\ge(w+\ell)/4+P,\qquad
 |D|\le49(w+\ell)+294P.
\]

The first inequality follows by summing the covering constraints and using a planar edge count to bound the contributions of~$H$.
For the second inequality, we simplify certain already covered parts of the graph before counting degrees.
Let~$R$ be the demands still uncovered after choosing~$W$.
In the proof, we contract each connected component of~$G-R$ that contains a vertex of~$W$ to a single vertex, keeping all other vertices separate and removing loops and repeated edges.
This replaces possibly large components by at most~$w$ vertices, and the resulting graph is still planar.
In this graph, the selected defects in~$H\cap R$ have degree at least seven, while every other vertex has degree six except for the contracted vertices, the~$\ell$ demands covered without being selected, and~$U$.
Since a planar graph has average degree at most six, these high-degree defects must be balanced by the exceptional vertices, which bounds their number in terms of~$w+\ell+|U|$.
Every other vertex of~$H\cap R$ has degree six and a defect neighbor, so counting edges at the selected defects and the~$\ell$ demands also bounds the number of these remaining selections.
Together with~$|H\setminus R|\le\ell$ from the first edge count and~$|U|\le6P$ from the covering constraints, this gives the second inequality.
The algorithm does not need to find or contract these components, they are used only in the proof.

Dividing the second inequality by~$294$ and comparing it with the first gives $(1+1/294)|D|\le\cost(x)$.
Thus, with solver accuracy~$1/294$, the rounding compensates for the solver's error and gives the exact factor seven.
In Section~\ref{new:planar} we present the full planar proof and in 
Section~\ref{simp:endpoint:section} we give the general construction for integer and noninteger~$2a$.
\par\endgroup

\subsection{Finding local fractional covers with short messages}

The solver must now find the fractional cover used by the rounding.
For this, we form covering subproblems, which we call \emph{local tasks}, each requiring coverage of a specified group of demands.
A task contains every candidate that can help cover one of these demands, including candidates that are not themselves demands of the task.
Each candidate keeps the cost assigned to it in the covering problem given to the solver.
Below we explain how we choose the groups and handle their boundaries, so that the task solutions can be combined at a small additional cost.
After solving the tasks with a fixed accuracy, we add the vertices selected at their boundaries.
An exact average of the resulting covers over different choices of the groups gives the fractional solution for the full problem.
A near-optimal cover can be represented by a short list of identifiers and found using short messages, even when a task contains arbitrarily many vertices.

\paragraph{Local tasks.}
A task uses a connected subgraph~$H$ of the physical communication graph, with a specified demand set~$R\subseteq V(H)$ and integer candidate costs~$1\le c_v\le B_0$.
For the closed-neighborhood case, every vertex of~$H$ is a candidate, and each demand can use itself and its neighbors in~$H$.
The task is to assign nonnegative weights~$x_v$ to the candidates so that every demand receives total weight at least one, while keeping the cost~$\sum_vc_vx_v$ as small as possible.
Only the vertices of~$R$ must be covered, the other vertices of~$H$ remain available as candidates and for communication.
Write~$\tau$ for the minimum fractional cost of this task.

Each task is local because~$H$ has bounded diameter, so the whole task lies within a fixed distance of one vertex in the physical communication graph.
The construction also ensures that its demands can be covered by a bounded number of candidates, which together with the cost bound gives a known bound~$\tau\le K_0$.
The number of available candidates and demands may still be arbitrarily large.
These bounds depend only on the graph class, cost bound, and requested accuracy, while the actual task optimum remains unknown.
For example, two large stars joined at their centers have diameter three and can be covered by their two centers, while collecting all identifiers at one center would require those of the other star to cross their connecting edge.
Instead of collecting the graph, the network searches for a short description of a cover.

\paragraph{A short description.}
We describe a fractional cover by a short list of candidate identifiers, where an identifier may occur more than once.
Fix an integer $q>0$ and let~$\mu_v$ be the number of occurrences of~$v$ in a list, representing the fractional value $x_v=\mu_v/q$.
For example, $(u,u,v)$ with $q=3$ represents weights $2/3$ and $1/3$, so a demand that can use both vertices receives total weight one.
The list is feasible when the candidates of each demand occur at least~$q$ times in total.
Its number of entries is $\sum_v\mu_v$, its cost is $\sum_vc_v\mu_v$, and the fractional cost is the latter divided by~$q$.
If $c_u=2$ and $c_v=1$, the example list has three entries and cost five, giving fractional cost $5/3$.

\begingroup
Sparsity guarantees a sufficiently accurate list of bounded length.
The fixed graph class gives a minor-density bound~$a$, which also bounds the density of every subgraph of~$H$.
Putting $p=\lceil a\rceil$, Lemma~\ref{simp:certificate} gives a list satisfying
\[
 \sum_{v\in N_H[r]}\mu_v\ge q\quad(r\in R),\qquad
 \sum_vc_v\mu_v\le(q+2p)\tau.
\]
\endgroup
For a fixed accuracy $\xi>0$, choosing $q\ge2p/\xi$ gives fractional cost at most $(1+\xi)\tau$.
Since each occurrence costs at least one, the list has at most $M=\lceil(q+2p)K_0\rceil$ entries, a known bound independent of the task size.
Each entry still needs~$\mathcal O(\log n)$ bits.

To show that such a list exists, we gradually replace fractional values with integer multiplicities. The extra $2p$ in the scaling factor $q+2p$ compensates for the coverage lost in this process.
Scale an optimal vector by $q+2p$ and orient the graph with at most~$p$ outgoing edges per vertex.
Keep the vector's integer parts fixed and preserve each incoming demand sum while it contains more than~$p$ noninteger values.
Each movable value occurs in at most~$p$ such sums, so fewer equations than movable values remain, and some values can move until one becomes integral without increasing cost.
Once a sum is no longer preserved, its loss is less than~$p$, while the demand itself and its outgoing neighbors together lose less than~$p+1$.
The final coverage is therefore an integer greater than $q-1$, hence at least~$q$.
The orientation and the optimal vector are needed only for this proof.

\paragraph{Finding the list.}
The algorithm finds the list by a search with boundedly many choices.
If~$\ell$ demands still need coverage and $m>0$ entries remain available, every feasible completion of the list has a candidate covering at least~$\ell/m$ of these demands.
When $\ell\ge2(a+1)m$, density bounds the number of such candidates by~$2am$, and the search tries each one, allowing repetitions.
When fewer demands remain, collect their identifiers and remaining coverage requirements and keep a cheapest candidate for each covered subset, breaking ties by identifier.
Candidates covering the same subset are interchangeable for finishing the list, so all completions can then be tested locally.
Demands outside this remaining list are already covered by the earlier entries, which are kept throughout the search.
Complete lists are compared by cost, and branches without a possible completion are discarded.

Short messages suffice for every step of this search.
A rooted tree of bounded depth collects counts, minimum identifiers, and minimum cost-identifier pairs, while bounded lists are sent one entry at a time.
The depth of the search is at most~$M$, and its number of choices depends only on the fixed parameters.
Dividing the cheapest list's multiplicities by~$q$ gives the required fractional cover, and values above one can be reduced to one.
Lemma~\ref{simp:local-solver} gives the details; with unit costs and~$q=1$, the same search finds an exact integral cover within a fixed size bound, as used for surfaces in Section~\ref{new:surface}.

\subsection{Owners and interactions between tasks}

The local tasks must have bounded diameter and a cover of bounded cost. They must also be chosen so that shared candidates are not paid for twice.
If one candidate covers demands in two groups, solving the groups separately may count its cost twice although a global cover pays once.
The construction records these interactions before dividing the demands.
We continue with closed neighborhoods and write~$R$ for the demand set and~$\tau_c$ for the fractional optimum of the full problem given to the solver.

\paragraph{Choosing owners.}
An initial cover gives the demands a small set of group centers.
A pruning algorithm finds~$D_0$ with $|D_0|=\mathcal O(\tau_c)$ in a constant number of broadcast CONGEST rounds, without using a fractional solution.
We explain this step below in Section~\ref{sc:initialization-overview}.
Every demand outside~$D_0$ chooses a covering neighbor in~$D_0$ as its owner, and each owner owns itself.
The resulting disjoint stars~$V_i$ contain exactly $R\cup D_0$, while other vertices remain candidates.
The set~$D_0$ organizes the work and is not added in full to the answer.

\paragraph{Shared candidates and the owner graph.}
The support of a candidate records which groups it can help.
Let~$S_0(v)$ be the set of owners whose stars meet~$N_G[v]$.
If $|S_0(v)|>L$, call~$v$ heavy and select it, forming a set~$P$.
In the analysis, contracting suitable parts of the stars gives $|P|\le2a|D_0|/L$, so bounded costs and a large fixed~$L$ make this selection inexpensive relative to~$\tau_c$.
Every candidate that can cover a remaining demand in $R'=R\setminus N_G[P]$ is now light, with at most~$L$ support labels.
The original stars and supports stay fixed, and covered vertices continue to communicate.

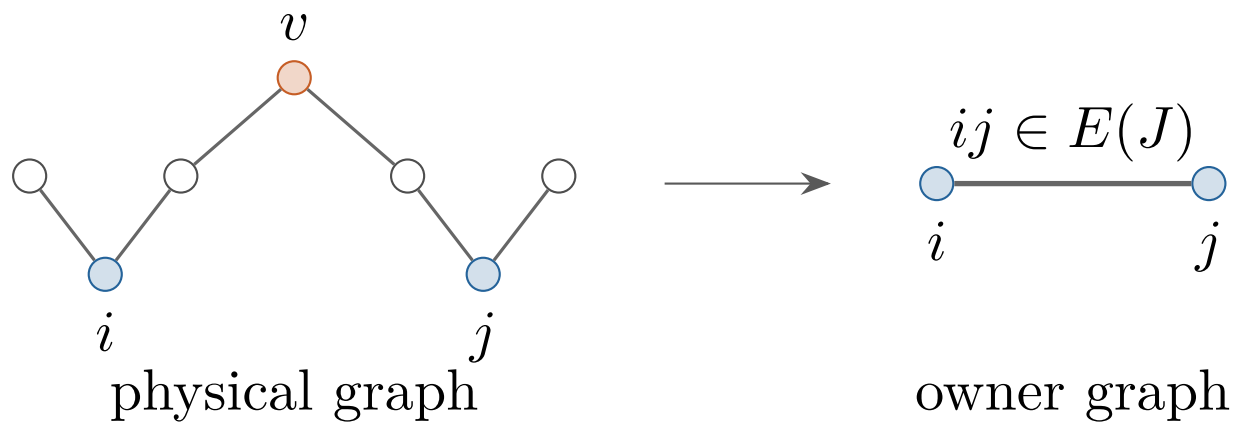
\begin{figure}[ht]
\centering
\begin{tikzpicture}[x=1cm,y=1cm,font=\normalsize]
 \node[so,label=below:$i$] (i) at (0,0) {};
 \node[sv] (u) at (-.5,.65) {};
 \node[sv] (w) at (.5,.65) {};
 \node[so,label=below:$j$] (j) at (2.5,0) {};
 \node[sv] (z) at (2,.65) {};
 \node[sv] (t) at (3,.65) {};
 \node[sb,label=above:$v$] (v) at (1.25,1.3) {};
 \draw[se] (i)--(u) (i)--(w) (j)--(z) (j)--(t);
 \draw[se] (w)--(v)--(z);
 \node at (1.25,-.8) {physical graph};
 \draw[arr] (3.7,.6)--(4.8,.6);
 \node[so,label=below:$i$] (ii) at (5.5,.6) {};
 \node[so,label=below:$j$] (jj) at (7.3,.6) {};
 \draw[se,line width=1pt] (ii)--node[above] {$ij\in E(J)$}(jj);
 \node at (6.4,-.8) {owner graph};
\end{tikzpicture}
\caption{A shared light candidate~$v$ creates an edge between owners~$i$ and~$j$, so the partition must account for their interaction.}
\label{simp:fig:owners}
\end{figure}

Shared light candidates define a graph used for partitioning.
The owner graph~$J$ has vertex set~$D_0$, and each light candidate makes its support pairwise adjacent, as in Figure~\ref{simp:fig:owners}.
Section~\ref{simp:owners} proves that every induced subgraph has a ratio of edges to vertices of at most $C_L=2aL$ and has small separators.
Random contractions establish a density bound in the proof, but the algorithm uses deterministic exchanges of owner identifiers.
Select the owners~$Q$ whose degree in~$J$ exceeds a fixed~$\lambda$, covering their stars, and remove them from~$J$ to obtain~$J_0$.
An owner needs at most $\lambda+1$ distinct neighbor identifiers to detect that its degree is too large; an owner below this threshold learns its complete neighbor list.
The bound $|Q|\le2C_L|D_0|/\lambda$ makes their cost small when~$\lambda$ is chosen large enough.
The graph~$J_0$ has bounded degree, and its broadcasts can be simulated through stars and light candidates in a constant number of rounds.

\paragraph{Forming local tasks.}
Each small connected part of~$J_0$ gives one local task after its boundary owners are selected.
Suppose a part~$C$ has at most~$k$ owners, and call an owner boundary if it has a neighbor in~$J_0$ outside~$C$. The other owners of~$C$ are interior.
Selecting these owners covers their demands, leaving the demands in~$R'$ assigned to interior owners as the task for~$C$.
Its communication graph~$H_C$ contains the full stars of~$C$ and all light candidates whose supports meet~$C$, with all physical edges between these vertices.
This retains every useful candidate, even one belonging to the star of an owner already removed in~$Q$.

These tasks have the bounds needed by the local solver.
Every owner edge within~$C$ is represented inside~$H_C$ by a walk of at most four edges, so~$H_C$ has diameter at most~$4k$, while the owners of~$C$ cover its demands at cost at most $B_0k$.

\paragraph{Adding local costs.}
If a candidate could cover interior demands in two different parts, its support would make their owners adjacent, forcing them to be boundary owners instead.
Thus useful candidates of different tasks are disjoint, although communication regions may overlap.
To compare costs, restrict one global optimal fractional cover to the useful candidates of each task.
The restricted vector still covers the task's demands, since every possible covering candidate is present in~$H_C$, and no candidate weight is charged twice.
Consequently, the local fractional optima~$\tau_C$ satisfy $\sum_C\tau_C\le\tau_c$.
The next step controls the boundary cost and combines the local solutions. Section~\ref{simp:averaging} also bounds participation in overlapping tasks, and \gcovsec{} explains the additional covering partners needed for general allowed incidences.

\subsection{Exact averaging and integral schemes}

\begingroup
We combine the local solutions while keeping the expected cost of selecting boundary owners small.
For one partition, let~$B$ be its boundary owners.
Let~$X$ be the combined fractional cover of the full problem.
Its coordinate~$X_v$ is the sum of the weights assigned to~$v$ by the local solutions, plus one for each of~$P$,~$Q$, and~$B$ containing~$v$.
Every demand is covered in every partition, and in the closed-neighborhood case the cost satisfies
\[
 \cost(X)\le(1+\xi)\tau_c+B_0\bigl(|P|+|Q|+|B|\bigr).
\]
The sets~$P$ and~$Q$ are already fixed, so only the boundary term depends on the partition.
Since $|D_0|=\mathcal O(\tau_c)$, it is enough to make the expected boundary a small fraction of $|D_0|$, and then use the exact average of these feasible covers.
\par\endgroup

\paragraph{Choosing parts locally.}
The partition rule lets small connected sets compete for their owners.
For a sufficiently large constant~$k$, consider every nonempty connected set of at most~$k$ owners in~$J_0$ as a possible part, named by its sorted identifier list.
The degree of~$J_0$ is bounded, so every owner belongs to boundedly many possible parts, and each part intersects boundedly many competitors.
In each stage, a part is eligible if its owners are unassigned and it has few edges to other unassigned owners, relative to its size.
Give every possible part a fresh independent fair bit. Accept an eligible part if its bit is one and the bits of all other intersecting possible parts are zero, including those that are not eligible.
Accepted parts are disjoint, and their owners become assigned.
After a fixed number of stages, each owner still unassigned becomes a singleton part.

The separator property ensures that this competition makes steady progress.
Among the unassigned owners, disjoint eligible sets can cover at least half, and each has a positive constant probability of winning against its bounded number of competitors.
Thus the expected number of unassigned owners decreases by a fixed factor in each stage.
To count crossing edges, charge each edge touching an accepted part to the part accepted first, whose eligibility test already counted this edge.
The other crossing edges join final singletons, and bounded degree controls their number.
For any fixed $\eta>0$, suitable fixed parameters therefore give $\mathbb E|B|\le2\eta|D_0|$.
The separators are used only in this proof, and the algorithm does not compute them; Section~\ref{simp:partitions} gives the details.

\paragraph{Taking the average without random choices.}
For a fractional solution, we can average the coordinate at each vertex over all outcomes.
Every demand receives total weight at least one in every outcome, so it receives at least one in the average, and linearity of cost gives exactly the expected cost.
Choosing the accuracy~$\xi$, the selection thresholds~$L$ and~$\lambda$, and the boundary allowance~$\eta$ makes the total cost at most $(1+\delta)\tau_c$.
These choices depend only on the fixed graph class, cost bound, and accuracy, and require no knowledge of~$\tau_c$ or the number of vertices.

The average can be computed locally because every possible part depends on only boundedly many random bits.
Indeed, a fixed number of stages in the bounded-degree graph~$J_0$ gives a bounded neighborhood containing every bit that can affect whether a particular part~$C$ appears.
The owners collect this neighborhood and try all assignments to these bits, which gives the exact probability~$p_C$ that~$C$ is a final part.
They do not enumerate partitions of the whole network.
For each possible part, fix one local answer~$z^C$ using identifier-based choices, and multiply this solution and its boundary contributions by~$p_C$.
Both depend only on~$C$ and the input, so adding these contributions, together with those from~$P$ and~$Q$, gives the required coordinate average.
Independence between the events for different parts is not needed, because expectations always add.

The bounded number of tasks at each physical vertex keeps this calculation within the communication limits.
Their searches run in a fixed schedule, and each bit has the same name in every local view, determined by its possible part and stage.
The probabilities have bounded power-of-two denominators, while the local solutions also have a common fixed denominator, so the sums can be computed exactly with short messages.
Finally, reducing any coordinate above one to one preserves coverage and decreases cost.
The fractional algorithm is therefore deterministic; random bits describe the average but are never sampled, as detailed in Section~\ref{simp:averaging}.

\paragraph{Obtaining integral covers.}
For an integral cover, use one actual partition and solve its tasks exactly with integral choices.
The owners of a part, with partners when needed, give a cover of cost at most $2B_0k$, so an optimum uses at most $2B_0k$ candidates.
The same short-list search finds it with multiplicity $q=1$, and the disjoint useful candidate sets ensure that these local optima add up to at most~$\OPT_c$.
A random run of the partition rule then gives an always feasible cover with expected cost at most $(1+\varepsilon)\OPT_c$ in constant rounds.

For a deterministic cover, a coloring of a fixed power of~$J_0$, combined with local indices, gives colors to possible parts so that intersecting parts have different colors.
Processing colors in order replaces random competition and makes the boundary small deterministically.
With a known common polynomial bound on identifiers, the coloring takes~$\mathcal O(\log^*n)$ rounds, and the remaining steps take constant rounds.
The cover then has cost at most $(1+\varepsilon)\OPT_c$.
Section~\ref{sc:schemes} proves both guarantees for all covering problems of~Theorem~\ref{sc:thmA}.

\subsection{Finding the initial cover}
\label{sc:initialization-overview}

It remains to find the cover~$D_0$ that organizes the solver, with $|D_0|=\mathcal O(\tau_c)$ and without knowing an optimal solution.
This constant-factor bound is enough because~$D_0$ only provides the owners and is not added in full to the final answer.
Any deterministic constant-round broadcast CONGEST algorithm with this guarantee suffices.
The algorithms of~\cite{CHWW2019} find a dominating set on graphs of bounded orientable genus. Here we need to cover arbitrary demand sets on every class that excludes a fixed minor.
We consider first closed neighborhoods, where every vertex is a candidate and a demand may cover itself.
The algorithm makes a fixed number of passes, covering demands with few useful choices, and then selects all remaining demands.
Coordinating the shared choices keeps each pass small, while a sparse-minor argument bounds the final selection.

\paragraph{Pruning demands with few choices.}
Fix a positive integer~$h$, and call a candidate useful if it covers more than~$h$ currently uncovered demands.
A demand with no useful candidate selects itself, while demands with between one and~$h$ useful candidates submit their lists to the procedure below.
Demands with longer lists may wait, although choices made for other demands can already cover them.
The list contains only useful choices, so even a demand with a short list may have arbitrarily many neighbors, and a useful candidate may cover arbitrarily many demands.
Let~$S_i$ be the set selected in pass~$i$. We remove the demands it covers and recompute usefulness; all vertices remain available as candidates and for communication.
If every minor has density at most~$a$, we let $p=\lceil a\rceil$, take $h=3p$, and make $T=p+1$ passes.
Thus planar graphs need four passes with lists of at most nine useful candidates.
Let $S=\bigcup_iS_i$ and let~$X$ now be the set of remaining uncovered demands. The output $D_0=S\cup X$ covers every original demand.

\paragraph{Making shared choices.}
The submitted lists must be handled together, since independent choices can waste many vertices.
We call a set meeting all lists a list cover.
For example, one vertex~$v$ meets all lists $\{v,u_1\},\ldots,\{v,u_m\}$, whereas choosing the distinct~$u_i$ uses~$m$ vertices.
We look for shared candidate sets that are expensive to avoid.
Fix a sufficiently large constant~$t$ depending only on~$a$ and~$h$, and call a nonempty subset~$C$ of a list essential if every set disjoint from~$C$ that meets all lists containing~$C$ has at least~$t$ vertices.
One member of~$C$ meets all those lists, while avoiding~$C$ needs many alternatives.
Every original list is essential, because a disjoint set cannot meet that list itself.
Within a pass, the original list family stays fixed through all phases.
We process essential sets by increasing size, selecting the smallest-identifier member of each set that has no previously selected member, with all choices of one size made together.
After~$h$ phases every list is met.

Our density bound on minors bounds the number of choices made by this rule.
An essential set still unmet at the start of its phase contains no smaller essential set, which limits how many such sets can share a vertex of an optimum list cover.
An essential set avoiding that optimum forces many pairs between its selected member and vertices of the optimum, through the demands carrying the lists.
Too many such choices would give too many edges after contracting suitable paths through demands.
Lemma~\ref{simp:bounded-lists} shows how to turn these observations into a constant-factor bound.
For communication, each demand broadcasts its short list.
A candidate can then test every subset containing it, since it receives all lists containing that subset, and selection bits are relayed through the same lists.
This takes boundedly many broadcasts.

\paragraph{Bounding a pass and its survivors.}
Let~$O$ be a minimum-cardinality cover of the original demands.
Its useful vertices meet all short lists except at most $h|O|$, since each nonuseful vertex covers at most~$h$ current demands.
Adding one candidate for each missed list gives a list cover of size~$\mathcal O(|O|)$; the same count bounds demands with empty lists.
Thus the constant number of passes gives $|S|=\mathcal O(|O|)$.

To bound the survivors, their many useful choices give paths with disjoint interiors, using Hall's theorem.
Contracting these paths produces a minor with many edges to the selected vertices.
The minor-density bound then gives $|X|\le T|S|$.
These paths are used only in the proof; the algorithm simply selects~$X$.
Section~\ref{simp:initialization-section} gives the full construction and count.

Together, the two bounds give $|D_0|=\mathcal O(|O|)$, while Corollary~\ref{simp:integrality-gap} gives $|O|\le(2p+1)\tau_c$, since every candidate costs at least one.
The initializer does not know~$\tau_c$, does not use the costs, and never calls the solver.
For general allowed covering sets, useful candidates and lists use only allowed incidences, and a demand that would select itself instead requests an allowed candidate.
\Gcovsec{} proves that the same bounds remain valid.

\label{new:main-end}

\section{Related work}\label{sc:related-work}

\paragraph{Local domination and sparse graph classes.}

Lenzen, Oswald, and Wattenhofer~\cite{LOW2008} initiated the study of constant-round planar domination, and Lenzen, Pignolet, and Wattenhofer~\cite{LPW2013} later presented the full analysis.
The later improvements cited above brought the coefficient down to $7+\varepsilon$ in LOCAL~\cite{AS2026}, while Heydt, Kublenz, Ossona de Mendez, Siebertz, and Vigny~\cite{HeydtEtAl2025} developed a broader treatment of domination on sparse graph classes.
With short messages, Wawrzyniak~\cite{Wawrzyniak2015} gave a planar algorithm even in anonymous networks with port numbering, and Czygrinow, Han\'ckowiak, Wawrzyniak, and Witkowski~\cite{CHWW2019} gave constant-factor broadcast CONGEST approximation on graphs of bounded orientable genus.
To the best of our knowledge, no corresponding constant-round guarantee was known for all graphs of treewidth at most three or for all~$K_{3,3}$-minor-free graphs.

Constant-round LOCAL approximation was also obtained beyond planarity~\cite{ASS2019}, for excluded topological cliques~\cite{CHWW2018}, and for bounded-expansion classes~\cite{KublenzSiebertzVigny2020}.
Bonamy, Cook, Groenland, and Wesolek~\cite{BonamyCook2021} achieved the optimal coefficient five on outerplanar graphs.
Bonamy, Gavoille, Picavet, and Wesolek~\cite{BGPWLarge2025} obtained factor~$50$ on~$K_{2,t}$-minor-free graphs in the LOCAL model, with a round bound depending on~$t$ and a factor independent of it.
Work on locally nice graphs~\cite{BonamyGavoillePicavetWesolek2025} and on transfer between graph classes~\cite{BGDSPW2026} gave further refinements.
Picavet~\cite{Picavet2026} obtained $d+1+\varepsilon$ in LOCAL under maximum average degree~$d$ and bounded depth-one shallow-minor density, including $7+\varepsilon$ on fixed surfaces.

For sparse graphs of high girth, Amiri and Wiederhake~\cite{AmiriWiederhake2022} gave constant-round CONGEST algorithms for distance-$r$~covering, and Lenzen and Wenning~\cite{LenzenWenning2024} improved the bounds and showed that they were tight up to constant factors on high-girth graphs of low expansion.

\paragraph{Domination through set cover.}

Set cover also led to approximation algorithms for domination on graphs of bounded arboricity.
For graphs of arboricity~$\alpha$, Lenzen and Wattenhofer~\cite{LW2010} gave a randomized~$\mathcal O(\alpha^2)$-approximation in~$\mathcal O(\log n)$ CONGEST rounds.
Amiri~\cite{Amiri2021} obtained the same asymptotic bounds deterministically through a simple reduction to set cover with bounded element frequency.
The deterministic set-cover algorithm of Even, Ghaffari, and Medina~\cite{EvenGhaffariMedina2018} could then be applied.
Dory, Ghaffari, and Ilchi~\cite{DGI2024} also proved a LOCAL lower bound on arboricity-two graphs, which showed why bounded arboricity alone was not enough for the constant round bounds obtained here from excluded minors.
Assuming that the vertices knew~$\Delta$, Alipour and Farokhnejad~\cite{AlipourFarokhnejad2026} obtained a deterministic~$\mathcal O(\alpha\log\Delta/\log\log\Delta)$-approximation in~$\mathcal O(\log\Delta/\log\log\Delta)$ rounds with one-bit messages, so this faster round bound came with a degree-dependent factor.

The reduction also explains why set cover is a useful language for our framework: a demand is an element and a candidate represents the set of demands it serves, while the physical graph determines which messages can be exchanged.
The guarantee of Even, Ghaffari, and Medina depended on element frequency, and the round bound also depended on the largest set size.
Here the assumption concerns the sparse communication graph, so both quantities may be unbounded.

\paragraph{Covering problems beyond domination.}

For unrestricted set cover, Johnson~\cite{Johnson1974} and Chv\'atal~\cite{Chvatal1979} proved logarithmic approximation guarantees for the greedy method, and Feige~\cite{Feige1998} proved a matching asymptotic barrier unless NP had algorithms running in time~$n^{\mathcal O(\log\log n)}$.
When every element belonged to at most~$f$ sets, Hochbaum~\cite{Hochbaum1982} obtained factor~$f$ by LP rounding.
Rajagopalan and Vazirani~\cite{RajagopalanVazirani1998} developed parallel primal-dual algorithms for set cover and covering integer programs.
Our incidence-graph assumptions give another source of structure, which permits a constant factor without bounding the set sizes or frequencies.

Several other covering problems were studied in sequential and parameterized algorithms.
On planar incidence graphs, set cover is red-blue domination, for which Garnero, Sau, and Thilikos~\cite{GarneroSauThilikos2017} gave a linear kernel.
For several geometric hitting-set problems, Mustafa and Ray~\cite{MustafaRay2010} obtained approximation schemes using local search with a planar analysis.
Total domination was introduced by Cockayne, Dawes, and Hedetniemi~\cite{CockayneDawesHedetniemi1980} and was treated in the monograph of Henning and Yeo~\cite{HenningYeo2013}.
Garnero and Sau~\cite{GarneroSau2018} also gave a linear kernel for total domination on planar graphs.
This range of problems is why the solver is stated for general covering relations and bounded costs, while each exact coefficient has its own rounding argument.

\paragraph{Fractional optimization and rounding.}

Kuhn and Wattenhofer~\cite{KW2005} gave distributed LP methods for domination, and later work extended this approach to covering and packing programs~\cite{KMW2006,KMW2016}.
Sparse LP rounding played a central role in the work of Bansal and Umboh~\cite{BansalUmboh2017}, Dvo\v{r}\'ak~\cite{Dvorak2019}, and Amiri and Siebertz~\cite{AS2026}.

\paragraph{Decomposition and short certificates.}

Solving local problems and paying for their boundaries was already a common idea in approximation, from the geometric shifting method of Hochbaum and Maass~\cite{HochbaumMaass1985} and Baker's planar scheme~\cite{Baker1994} to distributed algorithms based on sparse partitions~\cite{CH2006,CHW2008,CHWW2018}.
An initial dominating set was already used in~\cite{CH2006} to organize stars and partial-domination tasks.
Deterministic~$\mathcal O(\log^*n)$-round LOCAL schemes for domination were given on bounded-genus graphs by Amiri and Schmid~\cite{AmiriSchmid2016} and on excluded-minor classes in~\cite{CHWW2018}, while Amiri and Siebertz~\cite{AS2026} gave a randomized constant-round LOCAL scheme.

Local fractional solutions were averaged over decompositions in~\cite{KMW2005}, and partition oracles gave local descriptions of partitions in the bounded-degree query model~\cite{HassidimKelnerNguyenOnak2009,LeviRon2015}.
The new ingredients make this route work with short broadcasts when degrees are unbounded.
The owner graph records shared candidates, short fractional certificates avoid collecting the tasks, and a finite partition rule permits an exact average with bounded denominators.
On surfaces, the small-optimum approach of~\cite[full version, Proposition~D.2]{BGDSPW2026} is applied to the remaining covering problem, where the two budgets give the exact coefficient.

\section{Notation and elementary proof tools}\label{simp:preliminaries}

In this section we fix the notation used in the detailed proofs.
The graph serves both as the instance to be dominated and as the communication network, so it matters which vertices still require coverage and which remain available to cover or relay messages.

All graphs are finite, simple, and undirected.
Write~$N_G(v)$ and~$N_G[v]$ for open and closed neighborhoods, $N_G[X]=\bigcup_{v\in X}N_G[v]$, and $N_G^2[X]=N_G[N_G[X]]$, so these last two sets contain the vertices at distance at most one and two from~$X$, respectively.
The subscript~$G$ is omitted when it is clear.
A set~$D$ dominates a demand set~$R$ if $R\subseteq N[D]$, and candidates may lie anywhere in the graph.
The minimum size of a set dominating all vertices is~$\gamma(G)$.
In a fractional solution, each vertex~$v$ contributes an amount $x_v\ge0$, and the total contribution from the closed neighborhood of each demand must be at least one.
Thus the fractional domination number~$\gf(G)$ is the minimum of $\sum_v x_v$ subject to $\sum_{v\in N[r]}x_v\ge1$ for every vertex~$r$.

For a specified demand set~$R$ and integer candidate costs $1\le c_v\le B_0$, let
\begin{equation}\label{simp:eq:weighted-lp}
 \tau_{R,c}(G)=\min\left\{\sum_{v\in V(G)}c_vx_v:
 x_v\ge0,\quad \sum_{v\in N_G[r]}x_v\ge1\quad(r\in R)\right\}.
\end{equation}
Every fractional weight greater than one may be replaced by one: that weight alone still covers every demand using it, and the cost can only decrease.
Every vertex knows whether it is a demand and knows its own cost.
Throughout the paper~$B_0$ is fixed, so the weighted solver is a statement about bounded positive costs.
The same notation is used with positive real costs when the proof only shows that a short list exists, without yet computing it.

A minor of a graph is obtained by deleting vertices and edges and contracting disjoint connected sets, after which loops are removed and just one edge is retained from each group of parallel edges.
For fixed rational $a\ge1$, the class~$\mathcal C_a$ consists of graphs for which every nonempty minor~$J$ satisfies $|E(J)|\le a|V(J)|$.
The density of a graph is its ratio of edges to vertices, with density zero for the empty graph.
The same bound applies to subgraphs without further comment.
Put $p=\lceil a\rceil$.
Every subgraph or minor of a graph in this class is still in the class, and no graph in the class has a~$K_s$ minor when $s>2a+1$.

In broadcast CONGEST, vertices have distinct identifiers of~$\mathcal O(\log(n+1))$ bits, and in each synchronous round a vertex sends the same message of this size to every neighbor.
Local computation and memory are unrestricted.
Every bound called constant may depend on~$a$,~$B_0$, and the fixed accuracy, but is independent of~$n$, degrees, and the optimum.
The vertices need not know~$n$.
The schedule is fixed, and empty messages are sent whenever a step has already finished, so all vertices can stop after the prescribed number of rounds.
If the graph is disconnected, the same procedure runs in each component.

Direct the edges so that only boundedly many leave any one vertex; this bounds how often a fractional weight occurs in certain sums.
The orientation is needed only for the proof, and a short matching argument shows that it exists.
This is the relevant special case of Hakimi's orientation theorem~\cite{Hakimi1965}.

\begin{lemma}
\label{simp:orientation}
If every subgraph~$H$ of~$G$ satisfies $|E(H)|\le p|V(H)|$ for an integer $p\ge1$, then~$G$ has an orientation of maximum outdegree at most~$p$.
\end{lemma}

Separators are sets of vertices whose removal breaks the graph into smaller parts.
The excluded-minor separator theorem of Alon, Seymour, and Thomas~\cite{AlonSeymourThomas1990} implies that, for fixed~$a$, there is a constant~$s_a$ such that every~$m$-vertex graph in~$\mathcal C_a$, with arbitrary nonnegative vertex weights, has a separator of size at most $s_a\sqrt m$ leaving components of weight at most two thirds of the total.
Here a component's weight is the sum of its vertex weights; the bound concerns this sum, not necessarily the number of vertices.
The theorem serves only to prove that certain partitions exist; the distributed algorithm does not receive or construct a separator.

\begingroup
\section{The framework and its planar application}
\label{prop:framework}

The framework combines a fractional solver with a rounding rule whose saving pays for the solver's approximation error.
Each application has three steps: we \emph{prepare the remaining covering problem}, use the \emph{fractional solver} to find a near-optimal fractional cover, and apply \emph{rounding with slack} to obtain an integral cover.
In this section, we show how the solver and rounding work together to obtain an exact approximation coefficient for planar domination.
Section~\ref{prop:local-tools} gives the construction of the solver; Section~\ref{simp:endpoint:section} develops the budget transfer and rounding for general minor-density bounds.

\subsection{Preparing the remaining covering problem}\label{prop:budget-transfer}

To obtain an exact coefficient, rounding must save enough to pay for the fractional solver's approximation error.
Such a saving need not be available on the original instance: for domination on graphs in~$\mathcal C_a$ with integer~$2a$, our rounding rule can select every vertex deep inside a region of degree $2a$ where the fractional weights are $1/(2a+1)$.
With cost $2a+1$ per candidate, each such vertex contributes one to both the fractional cost and the output size, leaving no saving there.
For planar graphs, we take $a=3$, giving degree six, weight $1/7$, and cost seven.
We therefore allow a first selection~$F$ that is included in the final answer and whose size we count directly.
The remaining problem and its costs are chosen so that the rounding gives the required saving.
When the saving is already available without a first selection, we may take~$F=\varnothing$.

We choose the remaining demands, their allowed candidates, and the candidate costs so that every integral cover~$D$ of the remaining problem gives a cover~$F\cup D$ of the original instance.
We use these costs to account for both the first selection and the remaining cover. For a target coefficient~$\alpha$ and a reference value~$B(G)$, we require
\[
 |F|+\tau'\le\alpha B(G).
\]
Here~$\tau'$ is the fractional optimum of the remaining problem.
We call this inequality the \emph{budget transfer}. After selecting~$F$, the remaining budget is at least~$\tau'$.
The value~$\tau'$ is used only in the proof and need not be computed.

The output of this step is~$F$ and the locally specified covering instance for the solver below.
In our applications, both are computable in a constant number of short broadcasts, and the instance satisfies the solver's assumptions.

\subsection{The fractional solver}\label{sc:solver}

We apply the solver of Theorem~\ref{sc:thmA} to the remaining instance.
Each demand~$r$ has a nonempty set $C(r)\subseteq N_G[r]$ of allowed candidates with integer costs $1\le c_v\le B_0$.
The demands, costs, and allowed incidences are locally known as specified there.
Write $\cost(x)=\sum_vc_vx_v$, summing over allowed candidates, and let~$\tau'$ be the minimum cost of a nonnegative vector satisfying $\sum_{v\in C(r)}x_v\ge1$ for every demand~$r$.
Closed and open neighborhoods give ordinary and total domination, respectively.

For any fixed rational accuracy~$\delta>0$, the solver returns a feasible fractional cover~$x$ satisfying
\[
 \cost(x)\le(1+\delta)\tau'.
\]
The algorithm requires a constant number of broadcast CONGEST rounds, depending only on the minor-density bound,~$B_0$, and~$\delta$.
The rounding analysis fixes~$\delta$ before the solver runs. The resulting vector~$x$ is then rounded to an integral cover of the remaining instance.

\begingroup
\subsection{Rounding with slack}
\label{new:rounding-framework}

We require the rounding saving to be at least a fixed positive multiple of the output size. The following proposition shows how bounds on the rounding saving and the output size determine a suitable~$\delta$.

Consider any covering program with nonnegative costs and fractional optimum~$\tau$, and a rounding rule that maps every feasible fractional solution~$x$ to a set~$D(x)$.
As in the introduction, the rule has slack $\delta>0$ if $\cost(x)\ge(1+\delta)|D(x)|$ for every feasible~$x$.
\begin{proposition}[Rounding with slack]
\label{sc:principle}
If a rounding rule has slack $\delta>0$, then every feasible~$x$ with $\cost(x)\le(1+\delta)\tau$ satisfies $|D(x)|\le\tau$.
Let $S_1(x),\ldots,S_k(x)$ be nonnegative quantities, and let $\alpha_i,\beta_i>0$ be fixed constants.
If
\[
 \cost(x)-|D(x)|\ge\sum_i\alpha_iS_i(x)
 \quad\text{and}\quad
 |D(x)|\le\sum_i\beta_iS_i(x)
\]
for every feasible~$x$, then the rounding rule has slack
\[
 \delta=\min_i\frac{\alpha_i}{\beta_i}.
\]
\end{proposition}
\begin{proof}
Here $(1+\delta)|D(x)|\le\cost(x)\le(1+\delta)\tau$.
For the certificate, with $\delta=\min_i\alpha_i/\beta_i$, $\delta|D(x)|\le\sum_i\delta\beta_iS_i(x)\le\sum_i\alpha_iS_i(x) \le\cost(x)-|D(x)|$.
\end{proof}
\par\endgroup

To apply this proposition, we must prove that the rounding rule has a fixed positive slack for every feasible input.
Choose a fixed positive rational~$\delta$ no larger than this slack and use it as the solver's accuracy. This choice depends only on the fixed constants in the rounding analysis, so no knowledge of~$\tau'$ is needed.

With~$\tau=\tau'$, the proposition gives
\[
  (1+\delta)|D(x)|\le\cost(x)\le(1+\delta)\tau',
\]
so $|D(x)|\le\tau'$. The rounding rule must produce a cover of the remaining demands in a constant number of short broadcasts.
Then~$F\cup D(x)$ covers the original instance, and the budget bound gives
\[
 |F\cup D(x)|\le|F|+\tau'\le\alpha B(G).
\]
For each new problem, we need to prove the budget bound and the required rounding slack. The same fractional solver and Proposition~\ref{sc:principle} can be used whenever the solver's assumptions hold.

\subsection{Planar domination}
\label{new:planar}

We apply the framework with $\alpha=7$ and $B(G)=\gamma_f(G)$ to prove Theorem~\ref{sc:thmC} for planar graphs.
Large regions where every vertex and all its neighbors have degree six can prevent a proportional rounding saving on the original instance.
With fractional weights $1/7$, our rounding rule would select every vertex deep inside such a region, using the full factor-seven budget and leaving no saving there. Boundary savings may be too small to absorb a fixed relative solver error.
The algorithm therefore selects these vertices first and adjusts the costs of the remaining candidates so that the budget includes the first selection and the remaining cover. This accounts for the selected vertices directly, so only the remaining problem incurs the solver's error.
Planar edge bounds then give the required proportional rounding saving.

\paragraph{The first selection and the remaining problem.}
\begingroup
The first step selects every degree-six vertex whose neighbors also have degree six.
Let~$G$ be planar, call a vertex a defect if its degree differs from six, and put
\[
 B=\{v:\deg_G(v)\ne6\},\qquad
 F=V(G)\setminus N_G[B],\qquad
 R_0=V(G)\setminus N_G[F].
\]
Select every vertex of~$F$, so the demands still requiring coverage are exactly~$R_0$.
Every vertex outside~$F$ remains a candidate, including a vertex that is already dominated, since it may still cover a demand in~$R_0$.
Write $C=V(G)\setminus F$ and give each candidate the cost
\[
 c(v)=7-|N_G(v)\cap F|.
\]
The number seven is the approximation factor we aim to obtain. Think of it as seven units of budget per unit of fractional weight. Selecting~$F$ uses part of this budget: for each candidate, we deduct one unit for every neighbor in~$F$. Its cost records what remains available for covering~$R_0$. For example, two neighbors in~$F$ leave cost $7-2=5$. The budget-transfer argument below shows that this accounting pays for~$F$ and the remaining fractional cover within a total budget of~$7\gamma_f(G)$.

\par\endgroup

\begingroup
By definition of~$F$, no defect lies in~$F$ or has a neighbor in~$F$. Thus every defect remains a demand: $B\subseteq R_0$.
A candidate has cost seven exactly when it has no neighbor in~$F$, that is, when it belongs to~$R_0$.
If $c(v)<7$, then~$v$ has a neighbor in~$F$ and therefore has degree six. Both~$v$ and its $7-c(v)$ neighbors in~$F$ are already dominated by~$F$. Only its other neighbors can remain demands, so~$v$ covers at most $c(v)-1$ of them.
Consequently
\begin{equation}\label{new:planar-cost-properties}
 1\le c(v)\le7,\qquad
 c(v)<7\ \Longrightarrow\ |N_G[v]\cap R_0|\le c(v)-1.
\end{equation}
\par\endgroup
A demand has no neighbor in~$F$, so its covering constraint is the same in~$G$ and in $G-F$.
Let~$\tau'$ be the minimum fractional cost of covering~$R_0$ with candidates in~$C$.
It remains to cover~$R_0$, since~$F$ already dominates every other vertex. The chosen costs serve two purposes. First, they let us bound the combined cost of~$F$ and a fractional cover of~$R_0$ by~$7\gamma_f(G)$. Second, inequality~\eqref{new:planar-cost-properties} says that, for $c(v)<7$, the cost~$c(v)$ exceeds the number of remaining demands that~$v$ can cover by at least one. Therefore, a fractional weight~$x_v$ contributes $c(v)x_v$ to the cost but at most $(c(v)-1)x_v$ to the sum of the remaining covering constraints. The difference is at least~$x_v$ and contributes to the rounding saving. We prove the budget bound next and then give the rounding rule.

\paragraph{Budget transfer.}
\begingroup
The aim is to show that the first selection and the remaining fractional cover together cost at most~$7\gamma_f(G)$.
Let~$y$ be an optimal fractional dominating vector of~$G$.
Deleting its weights on~$F$ leaves a fractional cover of~$R_0$, because no demand has a neighbor in~$F$, and therefore $\tau'\le\sum_{v\in C}c(v)y_v$.
To bound~$|F|$, sum the original covering constraints over~$F$:
\[
 |F|\le\sum_{v\in V(G)}|N_G[v]\cap F|y_v.
\]
For $v\in F$, the coefficient is at most seven, while for $v\in C$ it is exactly $7-c(v)$.
After adding the cost of the restricted vector, each~$y_v$ has coefficient at most seven. Hence
\begin{equation}\label{new:planar-budget}
 |F|+\tau'\le7\sum_vy_v=7\gamma_f(G).
\end{equation}
It therefore suffices to cover~$R_0$ with at most~$\tau'$ additional vertices. Together with~$F$, they form a dominating set of size at most~$7\gamma_f(G)$.
These degree properties and the budget calculation do not use planarity, so they also apply to the same first selection on a surface.
\par\endgroup

\paragraph{Rounding with slack.}
\label{prop:planar-rounding-section}
\begingroup
Consider any feasible fractional cover $x\in[0,1]^C$ of~$R_0$, and write $\cost(x)=\sum_{v\in C}c(v)x_v$.
First, form~$W$ by selecting the cost-seven candidates with fractional value greater than one quarter. Next, form~$H$ by selecting every remaining candidate that covers at least seven demands still uncovered by~$W$.
Finally, select all demands still uncovered, forming~$U$. In formulas,
\begin{align*}
W&=\{v\in C:c(v)=7,\ x_v>1/4\},
 &R&=R_0\setminus N_G[W],\\
h(v)&=|N_G[v]\cap R|\quad(v\in C\setminus W),
 &H&=\{v\in C\setminus W:h(v)\ge7\},\\
U&=R\setminus N_G[H],
 &D&=W\cup H\cup U.
\end{align*}
Every demand of~$R_0$ is covered by one of these choices, and all selected vertices lie in~$C$.
In particular, $F\cup D$ dominates~$G$.
Figure~\ref{fig:planar-rounding-sets} shows the first selection and the sets used in the rounding rule.

\begin{figure}[!htbp]
\begingroup
\centering
\begingroup
\definecolor{roundingF}{HTML}{168477}
\definecolor{roundingW}{HTML}{C65D25}
\definecolor{roundingH}{HTML}{24639A}
\definecolor{roundingU}{HTML}{AD3434}
\begin{tikzpicture}[
  x=1cm,y=1cm,
  font=\fontsize{9}{11}\selectfont,
  every node/.style={inner sep=2pt,align=center,text=black},
  region/.style={draw=black!65,line width=.45pt,rounded corners=2pt},
  selected/.style={draw=black,line width=1pt,rounded corners=2pt},
  heading/.style={anchor=north west,align=left},
  region heading/.style={heading,font=\fontsize{9}{11}\selectfont\bfseries\boldmath},
  setlabel/.style={font=\fontsize{11}{13}\selectfont\bfseries\boldmath}
]

\draw[selected,fill=roundingF!25] (.18,6.07) rectangle (2.55,7.10);
\node[setlabel] at (1.365,6.83) {$F$};
\node at (1.365,6.37) {Selected first};
\node[font=\fontsize{12}{14}\selectfont\bfseries\boldmath] at (9.535,6.585)
  {Dominating set: $F\cup D=F\cup W\cup H\cup U$};

\draw[region,fill=black!3] (0,-.35) rectangle (16.3,5.50);
\node[region heading] at (2.91,5.37) {Candidates $C=V(G)\setminus F$};
\draw[region,fill=black!10] (.18,-.12) rectangle (2.55,4.87);
\node[setlabel] at (1.365,2.675) {$C\setminus R_0$};
\node at (1.365,2.075) {$c(v)<7$};

\draw[draw=black,line width=.65pt,-{Stealth[length=1.8mm]}]
  (1.365,6.05) -- (1.365,4.91);
\node[anchor=west] at (1.51,5.78) {dominates};

\draw[region,fill=roundingW!5] (2.77,-.12) rectangle (16.12,4.87);
\node[region heading] at (2.91,4.73)
  {Demands $R_0$ \quad (Price $7$)};

\draw[region,fill=violet!7] (2.98,.10) rectangle (7.68,4.25);
\node[region heading] at (3.12,4.11) {Dominated by $W$\\$R_0\cap N_G[W]$};
\draw[region,fill=white] (7.90,.10) rectangle (15.91,4.25);
\node[region heading] at (8.04,4.11) {Residual demands $R=R_0\setminus N_G[W]$};

\draw[selected,fill=roundingW!38] (3.23,.43) rectangle (5.31,2.35);
\node[setlabel,anchor=base] at (4.27,1.78) {$W$};
\node at (4.27,1.41) {$x_v>1/4$};
\node at (4.27,.85) {Step 1};

\draw[selected,fill=roundingH!28] (5.69,.43) rectangle (7.43,2.35);
\node[setlabel,anchor=base] at (6.56,1.78) {$H\setminus R$};
\draw[selected,fill=roundingH!28] (8.12,.43) rectangle (10.64,2.35);
\node[setlabel,anchor=base] at (9.38,1.78) {$Q=H\cap R$};
\draw[draw=black,line width=1pt,dash pattern=on 3pt off 2pt,
      rounded corners=3pt] (5.52,.27) rectangle (10.81,3.31);
\node[draw=black!45,line width=.35pt,rounded corners=1.5pt,
      fill=roundingH!12,inner sep=3pt,minimum width=4.3cm,minimum height=.55cm]
  at (8.165,2.85)
  {{\boldmath$H$}: $h(v)\ge7$\quad Step 2};

\draw[region,fill=roundingH!8] (11.08,.43) rectangle (13.20,2.35);
\node[setlabel,anchor=base] at (12.14,1.78) {$L$};
\node at (12.14,.85) {not selected};
\draw[selected,fill=roundingU!22] (13.48,.43) rectangle (15.66,2.35);
\node[setlabel,anchor=base] at (14.57,1.78) {$U$};
\node at (14.57,.85) {Step 3};

\draw[draw=black,line width=.65pt,-{Stealth[length=1.8mm]}]
  (10.82,2.85) to[out=12,in=115] (12.14,2.43);
\node[anchor=south] at (12.30,3.00) {dominates};

\end{tikzpicture}
\endgroup
\caption{The first selection~$F$ and the rounding sets. Thick outlines mark the selected regions, whose union is $F\cup D$; the dashed outline is~$H$. The sets $Q=H\cap R$, $L=R\setminus(H\cup U)$, and~$U$ partition~$R$.}
\label{fig:planar-rounding-sets}
\par\endgroup
\end{figure}
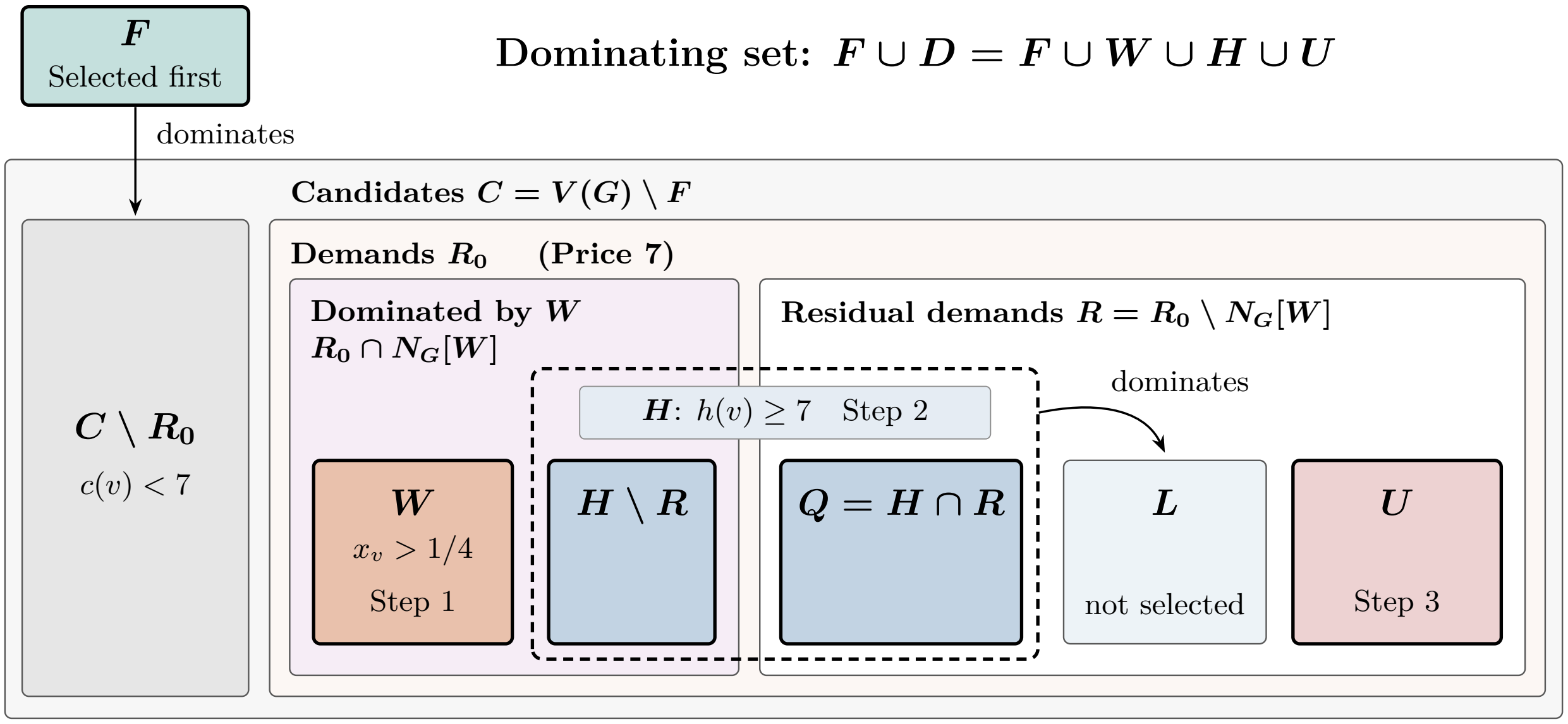

By \eqref{new:planar-cost-properties}, a candidate of cost less than seven has $h(v)\le c(v)-1$, so it cannot belong to~$H$.
Thus every vertex of $W\cup H$ has cost seven, and every candidate in $O=C\setminus(W\cup H)$ satisfies $h(v)\le6$ and $c(v)-h(v)\ge1$.
To measure the saving, put
\[
L=R\setminus(H\cup U),\qquad
w=|W|,\qquad \ell=|L|,\qquad
P=\sum_{v\in O}(c(v)-h(v))x_v.
\]
\par\endgroup
The set~$L$ consists of demands of~$R$ covered by~$H$ without being selected into~$H$. These demands do not need to be selected in the last step.
Each~$v\in O$ contributes~$h(v)x_v$ to the sum of the covering constraints for~$R$. Its cost is~$c(v)x_v$, so~$P$ is the total cost left after subtracting these contributions.
All three quantities $w,\ell,P$ are nonnegative.

The next lemma bounds the rounding saving and the output size in terms of the same quantities~$w,\ell,P$. Comparing these bounds will give a saving proportional to~$|D|$.

\begingroup
\begin{lemma}\label{prop:planar-rounding}
For every feasible vector~$x$ as above,
\[
 \cost(x)-|D|\ge\frac{w+\ell}{4}+P,
 \qquad
 |D|\le49(w+\ell)+294P.
\]
\end{lemma}
\par\endgroup

\begingroup
\begin{proof}
\textbf{The rounding saving.}
Let $Q=H\cap R$ be the demands selected into~$H$. The sets~$Q$,~$L$, and~$U$ partition~$R$.
Each vertex of~$Q$ covers itself, contributing one to~$\sum_{v\in H}h(v)$.
All other contributions come from edges of~$G[H\cup L]$. An edge is counted twice only when both ends lie in~$Q$.
Write $e(X)=|E(G[X])|$, and~$e(X,Y)$ for the number of edges between disjoint sets~$X,Y$. There is no edge from~$H$ to~$U$.
Applying the planar edge bound to~$G[H\cup L]$ and~$G[Q]$ gives
\[
\begin{aligned}
 \sum_{v\in H}h(v)
 &=2e(Q)+e(Q,L)+e(H\setminus R,Q\cup L)+|Q|\\
 &\le e(H\cup L)+e(Q)+|Q|\\
 &\le3(|H|+\ell)+4|Q|.
\end{aligned}
\]
Since $|H|=|Q|+|H\setminus R|$, it follows that
\begin{equation}
 \label{prop:planar-excess}
 \sum_{v\in H}(h(v)-7)+4|H\setminus R|\le3\ell.
\end{equation}
In particular, we have $|H\setminus R|\le\ell$, because $h(v)\ge7$ for every~$v\in H$.

From the constraints of the demands in~$R$, we have $\sum_{v\in C\setminus W}h(v)x_v\ge|R|$.
Let \mbox{$E=\sum_{v\in H}(h(v)-7)$} denote the excess.
On~$H$, we have $x_v\le1/4$ because these vertices have cost seven and lie outside~$W$. Hence the contribution of~$H$ to the constraint sum exceeds its cost by at most $\sum_{v\in H}(h(v)-7)/4=E/4$. On~$O$, the cost equals its contribution to the constraint sum plus~$P$. Finally, each vertex of~$W$ contributes more than~$7/4$ to the cost. Combining these observations gives
\[
 \cost(x)\ge\frac74w+|R|+P-\frac14E.
\]
The partition $R=Q\cup L\cup U$ gives $|D|=w+|R|-\ell+|H\setminus R|$.
Subtracting this value and using~\eqref{prop:planar-excess} gives the slightly stronger estimate
\[
 \cost(x)-|D|\ge\frac34w+\frac14\ell+P,
\]
which proves the first inequality.

\textbf{The output size.}
Since no candidate in $W\cup H$ covers a demand of~$U$, their constraints use only candidates in~$O$. Summing these constraints and using $h(v)\le6$ and $c(v)-h(v)\ge1$ on~$O$ gives
\begin{equation}
 \label{prop:planar-U}
 |U|\le\sum_{v\in O}h(v)x_v\le6P.
\end{equation}
Only~$|Q|$ remains to be bounded. We first count the defects in~$Q$ and then the other vertices of~$Q$ through their defect neighbors.

The contraction below collects all defects outside~$R$ into at most~$w$ vertices.
Contract every component of $G-R$ that meets~$W$ to one vertex, keep all other vertices unchanged, remove loops, and retain one edge from each group of parallel edges.
This contraction is used only in the proof; the algorithm does not compute the components of~$G-R$.
We call the resulting minor~$K$, write~$A$ for its contracted vertices, and put
\[
B_Q=B\cap Q,\qquad T=|A|+\ell+|U|,\qquad
S=A\cup L\cup U\cup B_Q.
\]
We have $|A|\le w$, and all edges of~$G[R]$ remain in~$K$.
Here~$B_Q$ is the set of defects in~$Q$, and $T\le w+\ell+|U|$.

We first observe that every uncontracted vertex outside~$R$ has degree six in~$K$.
Indeed, $B\subseteq R_0$, so every defect outside~$R$ belongs to $R_0\setminus R\subseteq N_G[W]$, and hence lies in~$W$ or has a neighbor in~$W$. Its component in $G-R$ is therefore contracted.
Since an uncontracted component has no edge to a contracted component, all its vertex degrees stay the same.

Every $q\in Q$ has at least six neighbors in~$R$ because $h(q)\ge7$ and this count includes~$q$ itself. These neighbors belong to $Q\cup L$, since no vertex of~$H$ is adjacent to~$U$.
If $q\in B_Q$, its original degree is at least seven, so either seven neighbors already lie in~$R$, or at least one additional neighbor outside~$R$ survives as a vertex or as a contracted component.
In both cases, we have $\deg_K(q)\ge7$.
If $q\in Q\setminus B_Q$, its six neighbors in $Q\cup L$ are all its neighbors, so its degree stays six.
We conclude that every vertex outside~$S$ has degree exactly six.

Since~$K$ is planar, its total degree is at most~$6|V(K)|$. Subtracting the contribution of the vertices outside~$S$ gives
\[
 7|B_Q|\le\sum_{v\in S}\deg_K(v)
 =2|E(K)|-6|V(K)\setminus S|
 \le6|S|.
\]
Since $|S|=T+|B_Q|$, we obtain
\begin{equation}
 \label{prop:planar-defect-count}
 |B_Q|\le6T,
 \qquad |S|\le7T.
\end{equation}
It remains to count the vertices of~$Q$ that are not defects.
Every $q\in Q\setminus B_Q$ lies outside~$F\cup B$, so it has a defect neighbor by the definition $F=V(G)\setminus N_G[B]$.
Since all its neighbors lie in $Q\cup L$, this neighbor belongs to $B_Q\cup L\subseteq S$.
We count one such edge for each~$q$ and use the same degree sum to obtain
\[
 |Q\setminus B_Q|
 \le\sum_{v\in S}\deg_K(v)
 \le6|S|\le42T.
\]
Together with \eqref{prop:planar-defect-count}, this gives us
\[
 |Q|\le48T
       \le48(w+\ell+|U|).
\]
Since $|D|=w+|Q|+|H\setminus R|+|U|$, the bounds $|H\setminus R|\le\ell$ and \eqref{prop:planar-U} give
\[
 |D|\le49(w+\ell+|U|)
       \le49(w+\ell)+294P,
\]
as required.

\end{proof}

\paragraph{Combining the three ingredients.}
\label{prop:planar-algorithm}

The two bounds show that rounding saves at least~$|D|/294$, regardless of the feasible vector~$x$.
Indeed, dividing the second bound by~$294$ gives
\[
 \frac{|D|}{294}\le\frac{w+\ell}{6}+P
 \le\cost(x)-|D|,
\]
and hence
\begin{equation}\label{new:planar-slack}
 \left(1+\frac1{294}\right)|D|\le\cost(x).
\end{equation}

Run the fractional solver of Theorem~\ref{sc:thmA} on the planar graph $G-F$, with demands~$R_0$, cost bound seven, class parameter three, and accuracy $\delta=1/294$.
It returns a feasible vector $x\in[0,1]^C$ satisfying $\cost(x)\le(1+1/294)\tau'$.
By \eqref{new:planar-slack}, or equivalently by Proposition~\ref{sc:principle}, the factor~$1+1/294$ cancels, giving $|D|\le\tau'$.
Together with \eqref{new:planar-budget}, this proves
\[
 |F\cup D|=|F|+|D|\le|F|+\tau'\le7\gamma_f(G).
\]

\paragraph{Communication.}
Every algorithmic step uses a constant number of broadcast CONGEST rounds.
After learning their degrees, vertices broadcast whether they are defects and then whether they belong to~$F$. These messages determine~$R_0$ and the candidate costs. The fractional solver then uses a fixed number of rounds for the parameters above.
When the solver runs on $G-F$, vertices ignore messages across edges leaving that graph.
Three further broadcasts announce membership in~$W$,~$R$, and~$H$, after which each remaining demand decides whether to join~$U$.
All messages have~$\mathcal O(\log n)$ bits, and the algorithm needs no embedding, orientation, initial cover, degree bound, or knowledge of~$n$.
\par\endgroup

\subsection{Reusing the framework}\label{prop:reuse}

The covering relation, costs, and rounding argument determine the coefficient, while the distributed solver can be reused.
The table below summarizes applications of the framework and the changes required in each case.

\begin{center}
\renewcommand{\arraystretch}{1.15}
\setlength{\tabcolsep}{8pt}
\begin{tabularx}{\linewidth}{@{}>{\raggedright\arraybackslash}p{0.25\linewidth}>{\centering\arraybackslash}p{0.13\linewidth}>{\raggedright\arraybackslash}X@{}}
\toprule
\textbf{Application} & \textbf{Coefficient} & \textbf{What changes}\\
\midrule
\textbf{Domination on other sparse classes} & $2a+1$ & Budget transfer and rounding count for minor density~$a$ (Section~\ref{simp:endpoint:section}).\\
\midrule
\textbf{Total domination} & $2a$ & Open neighborhoods and a different rounding argument at minor density~$a$; coefficient six for planar graphs (Appendix~\ref{fv:total}).\\
\midrule
\textbf{Red-blue domination,\newline set cover,\newline and hitting set} & $4+\varepsilon$ & Bipartite density on planar incidence graphs. No saving eliminating~$\varepsilon$ is proved here (Appendix~\ref{fv:rb}).\\
\bottomrule
\end{tabularx}
\end{center}

The solver and rounding may use different density bounds: planar red-blue instances use minor density three for the solver and subgraph density two for rounding.
\par\endgroup

{
\section{Constructing the fractional solver}
\label{prop:local-tools}

In this section, we construct the fractional solver and prove Theorem~\ref{sc:thmA}. We first consider the case where every vertex is a candidate and can cover itself and its neighbors. We first outline the five steps of the construction, then give the details and proof.
\begin{enumerate}[label={\textbf{Step \arabic*.}},leftmargin=*]
\item \textbf{Find an initial cover.} We first find a cover whose size is at most a constant times the fractional optimum. Its vertices, called owners, serve as centers for communication and organize the demands into groups. The interactions between these groups are used to form partitions and define local covering tasks used to obtain the final solution. The constant-factor bound controls the cost of later selections.
\item \textbf{Build the owner graph.} The owners from the previous step form the vertices of the owner graph. A candidate can connect two groups if it can cover a vertex in each. We select candidates that can cover vertices in many groups and owners whose groups are connected to many others. These selections cover some demands at small cost compared with the fractional optimum. The remaining owner graph has bounded degree, so the local communication needed in the next steps takes a bounded number of rounds.
\item \textbf{Choose parts locally.} The owner graph from the previous step has bounded degree, but its components can still have large diameter. We use a local rule to divide the owner graph into small connected parts, each containing a bounded number of owners and therefore having bounded diameter. This bound concerns the owners; their groups in the original graph may still contain arbitrarily many vertices. The rule uses finitely many fair bits. On average, few owners have a neighbor in another part; we call them boundary owners.
\item \textbf{Solve the local tasks.} For each possible part, we define a local covering task. Its diameter and fractional optimum are bounded, although it may contain arbitrarily many vertices. A local solver finds a fractional cover of cost close to this optimum and gives a short description of it.
\item \textbf{Compute the exact average.} For each partition, the local covers, boundary owners, and vertices selected in Step~2 together cover all demands. We average these solutions using the exact probabilities of the parts. The resulting fractional vector still covers every demand, and its cost equals the average cost. Each vertex computes the probabilities by enumerating the possible local bit assignments; the algorithm does not sample random bits.
\end{enumerate}

All parameters are fixed before execution.
They are chosen so that the local approximation error and the cost of the extra selections sum to at most $\delta$ times the fractional optimum.

\subsection[Finding the initial cover]{Finding the initial cover}\label{prop:solver-preparation}
\label{prop:decomposition}
\label{simp:initialization-section}

Our goal here is to find a set~\(D_0\) that dominates the prescribed demands \(R\subseteq V(G)\). Every vertex remains available as a candidate, including vertices outside~\(R\), and every minor~\(J\) of~\(G\) satisfies \(|E(J)|\le a|V(J)|\), for fixed \(a\ge1\).

First consider the case where each demand has a short list of allowed choices.
The procedure identifies shared subsets that would be expensive to avoid and selects from them in order of increasing size.
This procedure then runs in a fixed number of passes on the original graph, covering every demand that has only a few useful choices.
To see why few demands can remain, paths are built from the remaining demands, using distinct choices of the next vertex.
For each pass, every successful path reaches some demand covered in that pass, and contracting the paths would give a dense minor if there were too many remaining demands.
The paths and their contractions are only used in the proof.

\Needspace{11\baselineskip}
\begin{theorem}
\label{simp:initializer}
For every fixed \(a\ge1\), there exist a constant~\(A_a\) and a deterministic broadcast CONGEST algorithm with the following property.
Given a graph~\(G\) of minor density at most~\(a\) and a locally specified set \(R\subseteq V(G)\), the algorithm returns \(D_0\subseteq V(G)\) such that
\[
 R\subseteq N_G[D_0],
 \qquad |D_0|\le A_a\,\tau_{R,c}(G)
\]
for every choice of costs \(c_v\ge1\). The algorithm requires~\(\mathcal O_a(1)\) rounds, does not use the costs, and does not need an orientation, embedding, decomposition, initial solution, or the number of vertices as input.
\end{theorem}

The size of~\(D_0\) is first compared with~\(\gamma_R(G)\), the minimum number of vertices dominating~\(R\). We later relate this integral optimum to~\(\tau_{R,c}(G)\), using an argument independent of the initial-cover construction.

\subsubsection{Coordinating bounded lists}
\label{simp:bounded-lists-section}

Suppose each demand~\(r\) has a nonempty list \(\Lambda(r)\subseteq N_G[r]\) of at most~\(f\) candidates, where \(f\ge1\) is a fixed integer.
Let~\(\mathcal H\) be the family of distinct lists and~\(\tau(\mathcal H)\) the minimum size of a set meeting every list.
Repeated lists ask for the same thing and are counted only once.
If the family is empty, its optimum is zero.

Choosing separately from every list can be costly, so the aim is first to find choices that help many lists at once.
For example, if the lists are \(\{v,u_1\},\ldots,\{v,u_q\}\), with distinct \(u_i\ne v\), choosing~\(v\) covers them all, whereas avoiding~\(v\) requires all~\(q\) other vertices.
The same idea applies to larger shared sets, with the fixed constants
\[
 \beta=f(a+f),\qquad q=\lceil2\beta\rceil.
\]
A nonempty subset~\(C\) of a list is \emph{essential} if every set disjoint from~\(C\) that intersects all lists containing~\(C\) has at least~\(q\) vertices. Thus an essential set is expensive to avoid, even though any vertex of~\(C\) belongs to every list containing~\(C\). Every original list is essential, because no set disjoint from that list can intersect it.

We construct a set~\(D\) of candidates that intersects every list. Start with \(D=\varnothing\) and process the essential sets in~\(f\) phases, in order of increasing size. In phase~\(j\), add to~\(D\) the smallest-identifier vertex of every essential set of size~\(j\) that is disjoint from~\(D\) at the start of the phase. This vertex is the set's representative, and all choices in a phase are made together. The list family stays fixed, so essential sets do not change; only their intersections with~\(D\) change. Since every original list is essential, the final set~\(D\) intersects every list.

To bound the size of~\(D\), we first count the pairs of vertices that appear together in a list. Each pair can be represented by a path through a demand whose list contains both endpoints. When that demand is outside the pair's endpoint sets, a fixed fraction of these paths can be kept as different edges of a minor. This is the only density argument that the list procedure needs.

\begin{lemma}
\label{simp:list-pairs}
{
For disjoint sets \(X,Y\subseteq V(G)\), the number of pairs \((x,y)\in X\times Y\) that occur together in a list is at most
\[
 \beta(|X|+|Y|).
\]
}
\end{lemma}
\begin{proof}
{
Put \(U=X\cup Y\), and assign each counted pair to one demand~\(r\) whose list contains both endpoints.
Demands inside~\(U\) receive at most~\(f^2\) pairs each, since their lists have at most~\(f\) entries, so these account for at most \(f^2|U|\) pairs.

For every demand \(r\notin U\) receiving pairs, at most~\(f\) different endpoints in~\(Y\) occur among them.
Choose an endpoint~\(y_r\) that occurs most often and keep the pairs assigned to~\(r\) with that endpoint, retaining at least a \(1/f\) fraction.
Every vertex is then deleted except those in~\(U\) and the demands receiving these pairs, and each edge~\(ry_r\) is contracted.
Several demands may be merged with the same \(y\in Y\), but these stars are disjoint and no vertex of~\(X\) is merged.
A retained pair \((x,y_r)\) gives the path \(x,r,y_r\), and hence the edge~\(xy_r\) after contraction.
Different retained pairs have different endpoints, so their edges remain different when parallel edges are suppressed.
The resulting minor has \(|U|\) vertices and at most \(a|U|\) edges.
There were therefore at most \(fa|U|\) pairs assigned outside~\(U\), and adding the first group gives \(f^2|U|+fa|U|=\beta(|X|+|Y|)\).
}
\end{proof}

Processing smaller essential sets first keeps the answer small. A set that still needs a choice cannot contain a smaller essential set, since that subset would already contain a selected vertex. This limits how many such sets can share one vertex, and the pair bound controls the choices that avoid an optimum solution entirely.

\begin{lemma}
\label{simp:bounded-lists}
{
For fixed \(a\ge1\) and integer \(f\ge1\), the selection rule above covers every list and selects at most \(c_{a,f}\tau(\mathcal H)\) candidates, where
\[
 c_{a,f}=(q-1)^{f-1}+1.
\]
It uses~\(\mathcal O(f^2)\) broadcast CONGEST rounds.
The graph may have arbitrary degree, and each vertex may be both a demand and a candidate.
}
\end{lemma}
\begin{proof}
An essential set is \emph{active} if it is disjoint from the selected vertices at the start of its phase. By the end of phase~\(j\), every essential set of size~\(j\) contains a selected vertex, so an active set has no smaller essential subset. Since every original list is essential, every list eventually contains a selected vertex.

{
\paragraph{How many active sets can share a vertex.}
Every nonempty set~\(C\) of size at most~\(f\) lies in at most \((q-1)^{f-|C|}\) active sets, counted over all phases.
The proof is by induction on \(f-|C|\).
When \(|C|=f\), only~\(C\) itself can be such a set, which proves the base case.
The bound is also immediate whenever at most one active set contains~\(C\), because \(q\ge2\).

Otherwise, some active set properly contains~\(C\), so~\(C\) is a subset of a list but is not essential.
Hence there is a set~\(B\) of at most \(q-1\) vertices outside~\(C\) intersecting every list containing~\(C\).
Every active set \(A\supseteq C\) must intersect~\(B\): the set~\(B\) also intersects every list containing~\(A\), and if it avoided~\(A\), it would contradict the definition of an essential set.
Each such~\(A\) therefore contains \(C\cup\{b\}\) for some \(b\in B\).
Applying induction to these larger sets gives at most
\[
 |B|(q-1)^{f-|C|-1}\le(q-1)^{f-|C|}
\]
active sets containing~\(C\).
In particular, every vertex lies in at most \((q-1)^{f-1}\) active sets.

\paragraph{Comparison with an optimum.}
Fix a minimum set~\(O\) intersecting the original lists.
For every selected vertex~\(v\), choose one active set~\(C_v\) for which it was the representative when first selected.
These sets are different for different selected vertices, because each set has only one representative.
The vertices whose~\(C_v\) intersects~\(O\) number at most \((q-1)^{f-1}|O|\), by the preceding bound.
}

{
Let~\(X\) be the set of other selected vertices.
For each \(v\in X\), the vertices of~\(O\) appearing in lists containing~\(C_v\) intersect all those lists and lie outside~\(C_v\).
There are at least~\(q\) of them because~\(C_v\) is essential, and each appears together with~\(v\) in a list.
Moreover, \(X\cap O=\varnothing\), since \(v\in C_v\).
Lemma~\ref{simp:list-pairs} now gives
\[
 2\beta|X|\le q|X|\le\beta(|X|+|O|),
\]
so \(|X|\le|O|\).
The two groups together have size at most \(((q-1)^{f-1}+1)|O|\), as required.
}

{
\paragraph{Finding essential sets from their original lists.}
Each demand first broadcasts its list, one identifier per round, padded to~\(f\) rounds.
A candidate~\(v\) now knows every distinct list containing it, with a possible self-incidence handled locally.
For any subset~\(C\) of one of these lists with \(v\in C\), all lists containing~\(C\) are among the known lists.
To determine whether~\(C\) is essential, it suffices to consider sets of fewer than~\(q\) vertices from their union outside~\(C\), because vertices outside this union belong to none of these lists.
Thus each candidate can find all essential sets containing it by local computation, with no bound needed on its degree or memory.

At the start of each phase, all vertices broadcast their selection bits.
Each original demand then relays the identifier and bit of each member of its list in~\(f\) padded rounds.
Every candidate therefore knows which essential sets containing it contained no selected vertex at the start of the phase, and selects itself when it has the smallest identifier in at least one such set of the current size.
There are~\(f\) phases, each using \(f+1\) rounds, after the initial list broadcasts and a possible identifier exchange.
Each message contains one identifier and a constant number of flags.
The empty family selects nothing.
}
\end{proof}

{
\subsubsection{Pruning arbitrary neighborhoods}
\label{simp:pruning-section}

The procedure of Lemma~\ref{simp:bounded-lists} is applied to the original graph, with one fixed threshold~\(h\): a candidate is useful when it covers more than~\(h\) remaining demands, and a demand gets a list when it has at most~\(h\) useful choices.
Put
\[
 p=\lceil a\rceil,\qquad T=p+1,\qquad h=3p.
\]
For planar graphs we may take \(a=3\), giving \(p=3\), \(T=4\), and \(h=9\).
Thus there are four passes, each covering the lists with at most nine entries.
Let \(R_1=R\), and in pass \(i=1,\ldots,T\) define
\[
 C_i=\{v:|N_G[v]\cap R_i|>h\},\qquad
 \Lambda_i(r)=N_G[r]\cap C_i\quad(r\in R_i).
\]
The lists of sizes \(1,\ldots,h\) are covered using Lemma~\ref{simp:bounded-lists}, and each demand with an empty list selects itself.
Write~\(S_i\) for these selections and update \(R_{i+1}=R_i\setminus N_G[S_i]\).
After all passes, every vertex of~\(R_{T+1}\) is selected as well.
Membership in the sets and lists is obtained by a constant number of extra broadcasts per pass.

Before bounding the remaining demands, we bound the number of selections in each pass.
Fix a minimum set~\(O\) dominating~\(R\).
Its eligible members \(O\cap C_i\) intersect every nonempty short list except possibly the lists of demands covered only by \(O\setminus C_i\).
There are at most \(h|O\setminus C_i|\) such demands, because each ineligible member covers at most~\(h\) residual demands.
Keeping \(O\cap C_i\) and choosing one member from every short list still missed therefore gives a list cover of size at most
\[
 |O\cap C_i|+h|O\setminus C_i|\le h|O|.
\]
Demands with empty lists are also covered only by ineligible members of~\(O\), so at most \(h|O|\) of them select themselves.
Thus
\begin{equation}
 \label{simp:pass-cost}
 |S_i|\le h(c_{a,h}+1)\gamma_R(G).
\end{equation}
}

The remaining difficulty is to bound the demands surviving all passes.
Their many useful choices allow two paths to start from each remaining demand, without using any other vertex twice.
A path either stops before adding any vertex of~\(R_{T+1}\) (including its own start) or selected vertex, or reaches outside all remaining useful choices.
There cannot be too many paths of the first kind, because their stopping vertices are different.
A path of the second kind must leave every earlier set of useful choices, so it contains a demand covered in each pass.
These paths give the minor that completes the proof.

\begin{lemma}
\label{simp:residual-bound}
{
The pruning procedure satisfies
\[
 |R_{T+1}|\le T\left|\bigcup_{i=1}^T S_i\right|.
\]
}
\end{lemma}
\begin{proof}
{
Let \(X=R_{T+1}\) and \(S=\bigcup_iS_i\), and define
\[
 U_i=R_{i+1}\cup C_i\qquad(1\le i\le T).
\]
The sets~\(R_i\),~\(C_i\), and~\(U_i\) decrease with~\(i\), and \(X\subseteq U_T\subseteq\cdots\subseteq U_1\).
Every vertex of~\(U_i\) has at least~\(h\) distinct neighbors in \(R_i\cup C_i\).
Indeed, a surviving demand has more than~\(h\) candidates, since every short list was covered, and an eligible candidate covers more than~\(h\) residual demands.
Removing a possible self-incidence leaves at least~\(h\) neighbors in either case.
}

{
\paragraph{One set of neighbor choices for all passes.}
For \(v\in U_1\), let~\(i\) be the last index for which~\(v\in U_i\), and call its neighbors in \(R_i\cup C_i\) eligible for~\(v\).
It has at least~\(h\) such neighbors.
This choice has a useful property: if \(v\in U_j\) and~\(w\) is eligible for~\(v\), then \(w\in R_j\cup C_j\), since \(j\le i\).
In particular,
\begin{equation}
 \label{simp:path-exit}
 v\in U_j,\quad w\notin U_j
 \quad\Longrightarrow\quad
 w\in R_j\setminus R_{j+1}\subseteq N_G[S_j].
\end{equation}
This implication applies whenever~\(w\) is an eligible neighbor of~\(v\).

Two eligible neighbors are chosen for every \(x\in X\) and one for every vertex of \(U_1\setminus X\), with no physical vertex chosen twice.
Each required choice is one request, and every request at~\(v\) may use any eligible neighbor of~\(v\).
For a set~\(Q\) of requests, let~\(V\) be the set of vertices making these requests and~\(W\) the set of all their eligible neighbors.
Since each vertex makes at most two requests, \(|Q|\le2|V|\).
Each vertex of~\(V\) has at least \(h=3p\) choices, and counting these incidences gives
}
{
\[
\begin{aligned}
 3p|V|&\le |E(G[V\cup W])|+|E(G[V\cap W])|\\
       &\le p(|V|+|W|).
\end{aligned}
\]
An edge can be counted twice only when both endpoints belong to~\(V\) and to~\(W\), which explains the extra term for \(V\cap W\).
Thus \(|W|\ge2|V|\ge|Q|\), even when the two physical vertex sets overlap.
This is Hall's condition for the bipartite graph of requests and possible choices, so a matching gives the required distinct neighbors.
}

{
\paragraph{Following the choices until a path must stop.}
Two paths start at every \(x\in X\), one through each chosen neighbor.
Whenever the chosen next vertex lies in \(X\cup S\), the path stops before adding it and is called \emph{interrupted}.
Otherwise that vertex is added. If it lies outside~\(U_1\), the path stops and is called \emph{successful}; if the vertex is still in \(U_1\setminus(X\cup S)\), its one chosen neighbor gives the next step.
}

{
No path can revisit a vertex outside~\(X\), because a first repeat would require two different predecessors to have chosen that vertex.
A return to a starting vertex is stopped before it is added again.
The same reasoning shows that different paths cannot first meet outside~\(X\); the two paths from the same start also have different first neighbors.
Thus the paths share only their possible common start, and finiteness forces every path to stop.

If~\(m\) paths succeed, the other paths stop before different vertices of \(X\cup S\), again because no vertex was chosen twice.
There were \(2|X|\) paths at the start, so
\begin{equation}
 \label{simp:path-count}
 2|X|\le m+|X|+|S|,
 \qquad |X|\le m+|S|.
\end{equation}
}

\begin{figure}[ht]
\centering
\begin{tikzpicture}[x=1cm,y=1cm,font={\normalsize}]
 \node[so,label=left:$x\in X$] (x) at (0,1) {};
 \node[sv] (a) at (1,1.8) {};
 \node[sv] (b) at (2,1.8) {};
 \node[sv,draw=simpgreen,fill=simpgreen!25,
       label=above:{outside $U_1$}] (c) at (3,1.8) {};
 \draw[se] (x)--(a) (b)--(c);
 \draw[se,densely dotted] (a)--(b);
 \node[sv] (d) at (1,.2) {};
 \node[sv] (e) at (2,.2) {};
 \node[sb,label=below:{$X\cup S$}] (f) at (3,.2) {};
 \draw[se] (x)--(d);
 \draw[se,densely dotted] (d)--(e);
 \draw[arr,dashed] (e)--(f);
 \node[align=center] at (1.5,-1.3) {One path succeeds.
 The other stops\\before adding its next vertex.};
 \draw[arr] (4.3,1)--(5.5,1);
 \node[align=center] at (4.9,3.2) {Remove the start,\\contract the path.};
 \node[sv,draw=simpgreen,fill=simpgreen!25,
       label=above:{one path}] (P) at (7.5,1.4) {};
 \node[sb,label=below:{$S_1$}] (s1) at (6.3,0) {};
 \node[sb,label=below:{$S_T$}] (sT) at (8.7,0) {};
 \draw[se] (P)--(s1) (P)--(sT);
 \node at (7.5,-.15) {$\cdots$};
 \node[align=center] at (7.5,-1.3) {The minor vertex has a neighbor\\in every selection set $S_i$.};
\end{tikzpicture}
\caption{Each chosen next vertex is used at most once.
Paths may share only their start in~\(X\).
Dotted segments may contain further vertices, while the dashed last edge of an interrupted path is not included.
Every successful path leaves each~\(U_i\) at a vertex covered by~\(S_i\).
After removing its start, it therefore still touches a vertex of each~\(S_i\), as shown on the right.}
\label{simp:fig:paths}
\end{figure}
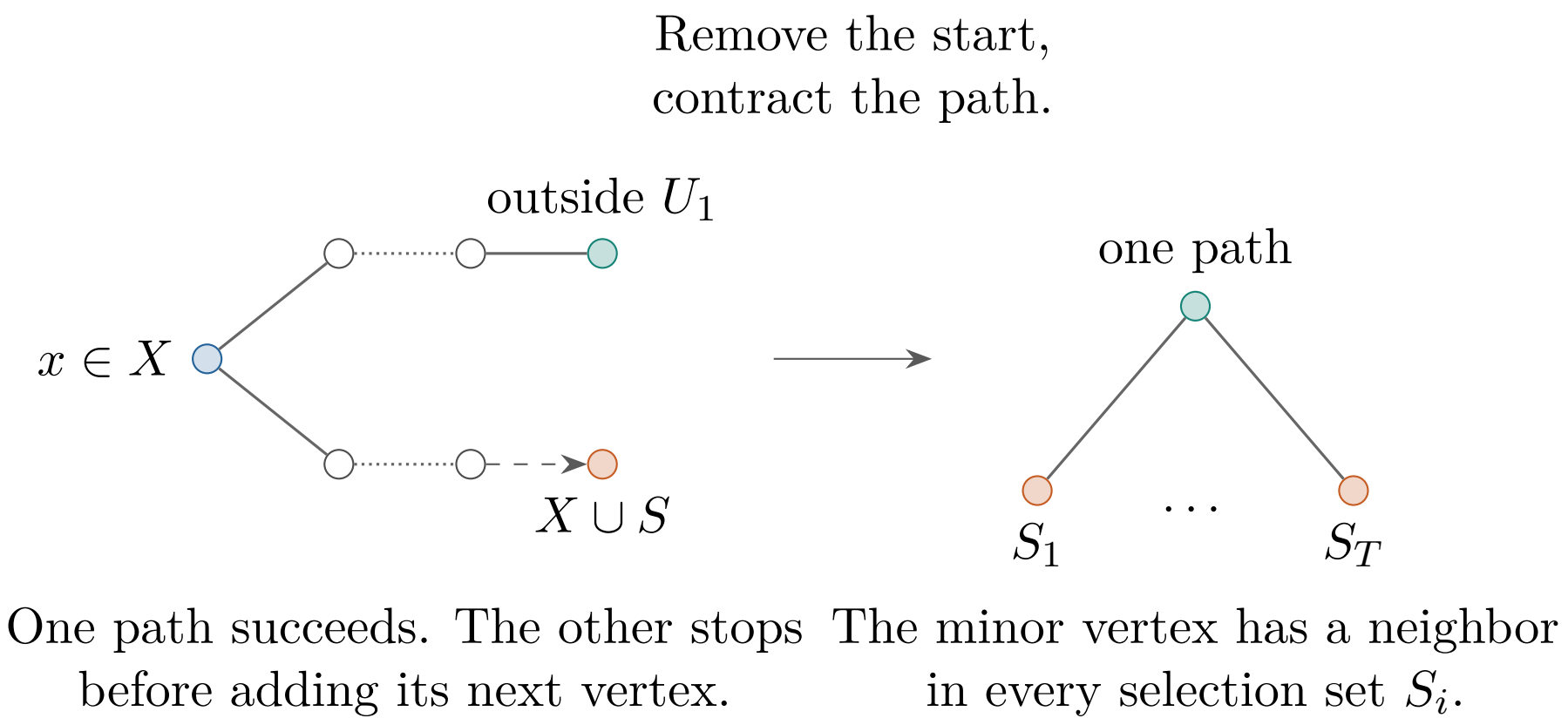

{
\paragraph{From successful paths to many edges of a minor.}
A successful path begins in \(X\subseteq U_i\) and ends outside~\(U_1\), so for every pass~\(i\) it has a first step leaving~\(U_i\).
By \eqref{simp:path-exit}, the vertex reached in this step belongs to \(R_i\setminus R_{i+1}\).
It is adjacent to a vertex of~\(S_i\), since it is covered by~\(S_i\) but the path avoids~\(S\).
Next, the starting vertex is deleted from every successful path.
Each path remains nonempty, and the remaining paths are pairwise disjoint and avoid~\(S\).
Their exit vertices also remain, because these lie outside \(U_i\supseteq X\).

\Needspace{12\baselineskip}
Contract the~\(m\) remaining paths, keep the vertices of~\(S\) separate, and delete all other vertices.
Each contracted path has a neighbor in every~\(S_i\).
These selection sets are pairwise disjoint: after a vertex is selected, its closed neighborhood has no later residual demand, so it cannot be a later eligible candidate or a later demand selecting itself.
The minor consequently has \(m+|S|\) vertices and at least~\(Tm\) different edges.
Its density is at most \(a\le p\), which gives
\[
 Tm\le p(m+|S|),\qquad m\le p|S|,
\]
where \(T=p+1\) was used.
Combined with \eqref{simp:path-count}, this gives \(|X|\le(p+1)|S|=T|S|\), as required.
The same inequalities also cover empty sets and an empty minor.
}
\end{proof}

\begin{proof}[Proof of Theorem~\ref{simp:initializer}]
{
The algorithm returns \(D_0=S\cup R_{T+1}\), which dominates every original demand; \eqref{simp:pass-cost} and Lemma~\ref{simp:residual-bound} give
\[
 |D_0|\le(T+1)|S|
 \le K_a\gamma_R(G),\qquad
 K_a=T(T+1)h(c_{a,h}+1).
\]
Corollary~\ref{simp:integrality-gap} gives \(\gamma_R(G)\le(2p+1)\tau_{R,c}(G)\) for costs at least one, so we may take \(A_a=(2p+1)K_a\).
For an empty demand set, the algorithm returns the empty solution.
Each pass uses~\(\mathcal O(h^2)\) broadcast rounds, and~\(T,h\) depend only on~\(a\), so the total is~\(\mathcal O_a(1)\).
The optimum set, neighbor matching, paths, and minor are only used in the analysis and are not computed by the algorithm.
}
\end{proof}

\subsection[Building the owner graph]{Building the owner graph}
\label{simp:owners}

{
The initial cover now organizes the remaining work.
It gives a small number of named groups, although each group may still contain arbitrarily many vertices.
The aim is to record which groups can use the same candidate, since these are precisely the interactions that a partition will have to respect.
Selecting first the few candidates that can cover vertices in many groups, and then the few groups that have many remaining interactions, leaves a graph on the group names with bounded degree.

Start with the set~$D_0$ from Theorem~\ref{simp:initializer} and put $N=|D_0|$.
Every vertex of~$D_0$ owns itself, and each vertex of $R\setminus D_0$ chooses the neighbor in~$D_0$ with the smallest identifier as its owner.
Write~$m(u)$ for the owner of $u\in O:=R\cup D_0$.
The classes $V_i=m^{-1}(i)$ are disjoint stars centered at the owners $i\in D_0$, and together they contain exactly the vertices of~$O$.
To describe how a physical vertex~$v$ can help these stars, define its support
\[
 S_0(v)=\{m(u):u\in N_G[v]\cap O\}.
\]
Thus the support lists the stars whose members~$v$ can dominate.
The set~$O$, the stars, and the supports stay fixed throughout the algorithm, and vertices continue to pass on messages after they have been dominated.

Choose an integer $L\ge\max\{2,4a\}$ and let $P=\{v:|S_0(v)|>L\}$ be the set of heavy vertices; all other vertices are light.
The value of~$L$ and the later thresholds are chosen at the end of the proof, when we bound the total cost.
For now, one broadcast of owner names and membership in~$O$ lets every vertex determine its support, even when its degree is unbounded.

The first count explains why selecting heavy vertices is affordable.
Each heavy vertex is kept separate while the other members of the stars are contracted, so that its many different support labels still give many different neighbors in a minor.
}

{
\begin{lemma}\label{simp:heavy}
The set~$P$ of heavy vertices satisfies $|P|\le 2aN/L$.
\end{lemma}
\begin{proof}
Keep the owners and the heavy nonowners, contract every other owned vertex to its owner, and delete the remaining vertices.
The resulting minor~$K$ has $N+h$ vertices, where $h=|P\setminus D_0|$; let $z=|P\cap D_0|$.
Each heavy vertex that is not an owner has at least $L+1$ different neighbors in~$K$, one for each support label.
If the vertex itself has an owner, its edge to that owner accounts for its own label.
For every other label, an adjacent representative either remains as a heavy vertex or is contracted to its owner, and representatives of different labels remain different.
A heavy owner has at least~$L$ neighbors by the same argument, after its own label is omitted.

In the sum of degrees at the heavy vertices, an edge is counted twice only if both endpoints are heavy, so this sum is at most $e(K)+e(K[P])$.
The minor-density bound therefore gives
\[
 (L+1)h+Lz\le a(N+h)+a(h+z).
\]
Since $a\ge1$, this gives $(L+1-2a)(h+z)\le aN$; using $L+1-2a\ge L/2$ then gives the claimed bound on $|P|=h+z$.
\end{proof}

Put $R'=R\setminus N_G[P]$, so that every candidate that can dominate a remaining demand is light.
The owner graph~$J$ on~$D_0$ is defined by making~$S_0(v)$ a clique for each light vertex~$v$.
In this way, an edge records that one candidate can serve vertices in both owner classes.
Two facts about this graph are needed: it has few edges, and its induced subgraphs have small separators.

The edge bound uses a random contraction, but only in the proof; the distributed algorithm never performs it.
Each owner edge comes from a light candidate whose support contains both endpoints; call this candidate its witness.
The contraction is arranged so that the witness preserves the edge with a fixed positive probability.
Since every outcome is a sparse minor of~$G$, the owner graph must be sparse as well.

\begin{lemma}\label{simp:owner-density}
Every induced subgraph of~$J$ has at most~$C_L$ times as many edges as vertices, where $C_L=2aL$.
\end{lemma}
\begin{proof}
Fix an owner set~$I$ and a light witness~$v$ for each edge $ij\in E(J[I])$.
Each owner of~$I$ is sampled independently with probability $1/2$.
The full star of every sampled owner is contracted, light vertices outside these stars are retained, and the other vertices are deleted.
Each light vertex that remains has at most~$L$ neighbors among the sampled owners.
If it has any such neighbors, it is assigned to one of them with equal probability; if it has none, it is deleted.
Each assigned vertex is then contracted into the sampled owner chosen for it.
The contracted sets are connected and disjoint, so the result is a minor on the sampled owners.

Consider the witness of~$ij$.
If it belongs to the star of~$i$ or~$j$, sampling both endpoints creates the edge, with probability $1/4$.
Otherwise, both endpoints must be sampled and the witness must remain outside the sampled stars.
This has probability at least $1/8$:~$i$ and~$j$ are sampled and, if the witness has a third owner in~$I$, that owner is not.
On this event the witness is assigned to~$i$ or~$j$ with probability at least $2/L$, and either choice creates the edge.
Both sampled owners are indeed neighbors of the witness, because their representatives have been contracted into their stars.
Thus every edge of~$J[I]$ appears with probability at least $1/(4L)$.
Every outcome has at most~$a$ times as many edges as sampled owners, so
\[
 \frac{|E(J[I])|}{4L}\le \mathbb E e(K)
 \le a\,\mathbb E|V(K)|=\frac{a|I|}{2}.
\]
\end{proof}
}

{
For the separator bound, every owner edge is represented by a short physical walk, and the weighted separator theorem is applied to the union of these walks.
Only the owners get weight, since it is their number that must be split, rather than the potentially much larger number of physical vertices.
A physical separator may contain a witness or one of its representatives, but in either case that vertex can be replaced by at most~$L$ owner labels.

\begin{lemma}\label{simp:owner-separator}
Every induced~$m$-vertex subgraph of~$J$ has a separator of size at most $C\sqrt m$ leaving components of size at most $2m/3$, where
\[
 C=s_a L\sqrt{1+3C_L}.
\]
\end{lemma}
\begin{proof}
Let $I\subseteq D_0$; all owners in~$I$ are retained, including isolated ones.
For every edge~$ij$ of~$J[I]$, a light witness~$v$ is chosen, together with vertices $u_i,u_j\in O$ in its closed neighborhood, with owners~$i$ and~$j$, respectively.
The walk $i,u_i,v,u_j,j$ is kept, merging consecutive equal vertices when they occur.
The union of these walks is a subgraph~$H$ of~$G$ with at most $(1+3C_L)|I|$ vertices.
Owners in~$I$ get weight one and all other vertices weight zero.
The weighted separator theorem gives a separator~$Z$ of size at most $s_a\sqrt{(1+3C_L)|I|}$, with each remaining component containing at most $2|I|/3$ owners of~$I$.

Every light vertex $z\in Z$ is replaced by the labels $S_0(z)\cap I$, and a heavy owned vertex by its owner, if that owner lies in~$I$.
A heavy unowned vertex cannot occur in a retained walk.
This adds at most~$L$ labels per vertex of~$Z$.
If one of the walks representing an owner edge intersects~$Z$, at least one of that edge's endpoints has been added: a light witness adds both endpoints, and an endpoint or a vertex representing its star adds its own owner.
Thus every path in~$J[I]$ that avoids the added labels gives a walk in $H-Z$.
All owners of its component lie in one component of $H-Z$, which contains at most $2|I|/3$ owners, as required.
\end{proof}

\subsubsection{Bounded degree and broadcast simulation}
\label{simp:owner-simulation}

Choose an integer $\lambda\ge1$, let $Q=\{i\in D_0:\deg_J(i)>\lambda\}$, and put $J_0=J[D_0\setminus Q]$.
Selecting~$Q$ dominates the original stars of its owners, and the degree sum bound from Lemma~\ref{simp:owner-density} gives
\begin{equation}\label{simp:high-owner-bound}
 |Q|\le\frac{2C_LN}{\lambda}.
\end{equation}
The graph~$J_0$ has maximum degree at most~$\lambda$ and inherits the separators of Lemma~\ref{simp:owner-separator}.

Bounded neighborhoods of~$J_0$ can now be used, provided that owners can find their neighbors and exchange messages through their physical stars.
This is possible because a light candidate has at most~$L$ owner labels to forward, while an owner outside~$Q$ needs messages from at most $\lambda+1$ different sources, including itself.

\begin{lemma}\label{simp:simulation}
Owners determine their membership in~$Q$ and, outside~$Q$, their complete neighbor lists in~$J_0$ in a parameter-bounded number of broadcast CONGEST rounds.
A broadcast round on~$J_0$ can then be simulated in a parameter-bounded number of physical rounds.
\end{lemma}
\begin{proof}
First, each owner determines whether its degree exceeds the threshold and, if not, finds all its neighbors.
Each light vertex gets~$L$ rounds to broadcast its support, sending one label per round and empty entries after its list ends.
Each owned vertex~$u$ then takes the union of the lists received in~$N_G[u]$, removes~$m(u)$, and broadcasts the smallest $\lambda+1$ remaining labels for its owner.
Each owner takes the union of these lists, including its own, and again keeps the smallest $\lambda+1$ labels.
If all lists were kept in full, the union at~$i$ would be exactly~$N_J(i)$, since a label $j\ne i$ arrives exactly when a light candidate has both labels in its support.
Keeping only the smallest $\lambda+1$ labels at each step preserves the smallest $\lambda+1$ in the final union: a discarded label already has that many smaller labels in a single input list.
Thus an owner either learns its full neighbor list or detects that its degree is greater than~$\lambda$.

Second, the owners exchange one round of messages.
Each owner broadcasts its message, with its identifier as the source, and each member of its star repeats this message.
A light candidate then knows the messages for its support labels and forwards them in~$L$ rounds, again filling unused rounds with empty entries.
An owned vertex with owner~$i$ can receive sources only from $\{i\}\cup N_J(i)$ through these light candidates, so if $i\notin Q$, it forwards at most $\lambda+1$ messages for its owner in that many slots.
Every owner edge has a witness that provides this route, so all required messages arrive.
This procedure is first used to exchange bits saying whether each owner belongs to~$Q$, which lets the low-degree owners identify~$J_0$, and afterwards only messages from owners in~$J_0$ are delivered.

A vertex's owned and candidate roles use separate phases.
In particular, heavy owned vertices still repeat and relay messages for their owners, while only light vertices forward messages for support labels.
When a step uses the vertex itself as a member of its closed neighborhood, its own information is used locally.
Repeated copies of a message with the same source are also discarded, since those copies are identical.
Every transmission is a single broadcast to all neighbors; bounded lists are sent one entry at a time, with a source identifier and one short message in each entry.
\end{proof}

In particular, each owner can learn its neighborhood of any fixed radius~$r$ in~$J_0$.
This neighborhood has at most $\sum_{j=0}^{r}\lambda^j$ vertices, so the required lists of vertices and edges can be sent one entry at a time.
This bound concerns the owner graph; the physical stars may still have arbitrarily many vertices, and they are not gathered.

}

\subsection[Choosing parts locally]{Choosing parts locally}
\label{simp:partitions}

{
 We now have an owner graph of bounded degree and want to divide it into small connected parts while cutting few edges.
Separators show first that such parts  exist.  A local random rule then produces small parts with few crossing edges in expectation.
The rule uses only a fixed number of fair bits, each equally likely to be zero or one, in a fixed neighborhood.
 Enumerating the finitely many bit assignments gives the exact probabilities needed for averaging.
The constants may be large, but none of them depends on the number of vertices.

\begin{lemma}\label{simp:partition}
Fix a positive integer~$\lambda$, a constant $C>0$, and rational $0<\eta\le1$, and let~$H$ be a graph of maximum degree at most~$\lambda$.
Assume that every induced~$m$-vertex subgraph has a separator of size at most $C\sqrt m$ leaving components of size at most $2m/3$.
There are constants $k,T,b,r_*$ depending only on these parameters and a randomized local algorithm that partitions~$V(H)$ into connected sets of size at most~$k$, with at most $\eta|V(H)|$ crossing edges in expectation.
Every vertex outputs the complete identifier list of its part, using only its radius-$r_*$~neighborhood and~$b$ independent fair bits at each vertex of that neighborhood.
On a bounded-degree graph this algorithm uses a parameter-bounded number of broadcast CONGEST rounds, without knowing $|V(H)|$.
\end{lemma}
}
\begin{proof}
{
Put $\varepsilon=\eta/8$ and choose an integer $k\ge1$ such that
\begin{equation}\label{simp:partition-size}
 \frac{C\lambda}
 {\sqrt{k}\bigl(1-\sqrt{2/3}\bigr)}
 \le\varepsilon.
\end{equation}

\emph{A good partition exists.}
 We repeatedly remove a separator from each component larger than~$k$, make every removed vertex a part by itself, and apply the same step to the components that remain.
A component of size at most~$k$ is kept as a part, so every part is connected.
}
A separator in a component of size~$m$ is paid for by charging $C/\sqrt m$ to each vertex of that component, so the total charge is enough for its at most $C\sqrt m$ removed vertices.

{
For a fixed vertex, the components containing it shrink by a factor of at most $2/3$ along the recursion, until the vertex is removed or its component has size at most~$k$.
Read backwards, its charges satisfy: the last charge is at most $C/\sqrt k$ and each earlier charge is at most~$\sqrt{2/3}$ times the next one.
Its total charge is therefore at most
\[
 \frac{C}{\sqrt k}\sum_{j\ge0}(2/3)^{j/2}
 =\frac{C}{\sqrt k\bigl(1-\sqrt{2/3}\bigr)}.
\]
}
{
Summing over vertices bounds the number of removed vertices by this quantity times $|V(H)|$.
Every edge between different parts has a removed vertex as an endpoint, so the maximum-degree bound and \eqref{simp:partition-size} give at most $\varepsilon|V(H)|$ crossing edges.
The same argument works in every induced subgraph.
These partitions only serve to prove that suitable small sets exist. The distributed algorithm does not compute separators.

\emph{All small connected sets can be listed once.}
Put
\[
 A=\sum_{j=0}^{2k-2}\lambda^j,\qquad M=kA.
\]
At most~$A$ connected sets of size at most~$k$ contain a specified vertex.
To see this, choose a spanning tree of each such set by a fixed rule, root it at the specified vertex, and walk through the tree, crossing each edge once in each direction.
A fixed identifier order settles every choice.
The resulting walk has length at most $2k-2$ and determines the original set, since it visits exactly the vertices of that set.
There are at most~$A$ walks of these lengths.
It follows that any one set meets at most~$M$ of the candidate sets, counting itself.
}

{
We use the sorted list of vertex identifiers as the name of each nonempty connected set of size at most~$k$ in the original graph, and assign the set to its smallest-identifier vertex.
Each vertex receives at most~$A$ sets, which we order by their lists.
Each assigned set gets one independent fair bit per stage, and we add unused slots so that every vertex has~$A$ slots per stage.
These names and bit positions stay fixed throughout the procedure, including for sets whose vertices have already been removed.

\emph{Fresh bits remove some eligible sets.}
At the start of a stage, let~$U$ be the set of vertices that remain.
Call a candidate set~$S$ eligible if $S\subseteq U$ and
\[
 |E_H(S,U\setminus S)|\le4\varepsilon|S|.
\]
An eligible set is accepted when its bit is one and the bit of every other candidate set that meets it is zero.
This test uses all original candidate sets, whether or not they are currently eligible, so a competing set's eligibility never has to be checked.
Accepted sets are disjoint, because two intersecting sets cannot both pass the bit test.
Their vertices are removed, and each accepted set becomes a part.

Once all choices from earlier stages are fixed, the eligible sets are known, while the new bits remain independent.
Since a candidate meets at most~$M$ candidates including itself, an eligible set is accepted with probability at least~$2^{-M}$.
The existence argument above applies to~$H[U]$.
Summing the numbers of edges leaving its resulting parts counts every crossing edge twice, so this sum is at most $2\varepsilon|U|$.
The parts with boundary greater than~$4\varepsilon$ times their size contain less than half of~$U$, so eligible parts contain at least half of its vertices.
Each vertex in those parts is removed whenever its part is accepted, and hence, writing~$U'$ for the next remaining set, we obtain
\[
 \mathbb E\bigl[|U'|\mid U,\text{ all earlier choices}\bigr]
 \le\bigl(1-2^{-(M+1)}\bigr)|U|.
\]
Choose an integer $T\ge0$ with
\[
 \bigl(1-2^{-(M+1)}\bigr)^T\le\eta/\lambda,
\]
run~$T$ stages, and make each remaining vertex a singleton part.
The expected number of remaining vertices is at most $(\eta/\lambda)|V(H)|$.

\emph{The edges cut by this procedure.}
}
If a crossing edge has an endpoint in an accepted part, we charge it to an endpoint part removed first.
At that time its other endpoint is still present, including when both parts are removed in the same stage, so the edge belongs to the boundary used in the eligibility test.
Summing these boundaries over the disjoint accepted parts costs at most $4\varepsilon|V(H)|=\eta|V(H)|/2$.
The only edges left to count join vertices that remain as individual parts at the end. There are at most $\lambda|U|/2$ of these, with expectation at most $\eta|V(H)|/2$.
This proves the required expected cut bound.

\emph{The information needed for a local decision.}
To decide whether a vertex belongs to an accepted set, one needs to know which candidate sets containing it are eligible, and the bits of all sets that meet those candidates.
The eligibility tests use current membership in~$U$ only within distance $k+1$.
All intersecting candidate sets lie within distance~$2k$, and their bits are stored at the vertices with the smallest identifiers in those sets.
The position of a bit among the sets assigned to its vertex is determined within distance~$k$ of that vertex.
Thus all the original names and bit positions needed for a stage are determined within distance $3k+1$.

If each vertex's current state depends only on information within radius~$\rho$, its state after the next stage depends on radius at most $\max\{\rho+k+1,3k+1\}$, using the fresh bits at their already fixed positions.
The same information identifies the complete list of any accepted part containing the vertex.
Hence it suffices to take the following bounds
\begin{equation}\label{simp:partition-radius}
 r_*=(4k+4)(T+2),\qquad b=AT.
\end{equation}
For $T=0$ the output is the singleton partition, and no bits are used.

Because the graph has bounded degree, every neighborhood of this radius has bounded size.
Its vertices, edges, identifiers, and bits can be exchanged by a fixed number of broadcasts, sending bounded lists one entry at a time, and the local rule is then evaluated.
Vertices continue to relay this information and provide their reserved bits after they have been assigned to a part.
Every message uses~$\mathcal O(\log(n+1))$ bits, and the schedule depends only on the fixed parameters, so no knowledge of the number of vertices is needed.
\end{proof}

Let~$B$ be the set of vertices that have a neighbor in another part of the resulting partition.
Each such edge has two endpoints, so the lemma also gives
\begin{equation}\label{simp:boundary-expectation}
 \mathbb E|B|\le2\eta|V(H)|.
\end{equation}
The fractional solver will use this bound on the number of boundary owners.
To make that solver deterministic, we go through all possible choices of the finitely many relevant bits and count the resulting outcomes exactly.

\subsection{Solving the local tasks}\label{simp:certificates}

\subsubsection{Short descriptions and local search}

Step~4 requires a procedure for local tasks of bounded diameter and optimum, although each task may contain arbitrarily many vertices.
 We solve each task by communicating a short list of selected identifiers, with repetitions allowed.
We first prove that such a list can describe a sufficiently good fractional solution when the number of times each identifier appears is divided by a fixed integer.
The list itself is then found by a search with a bounded number of choices, using only counts, minima, and bounded lists.
An orientation and an optimal fractional vector prove that a suitable list exists, but the search needs neither of them as input.

\begin{lemma}\label{simp:certificate}
Assume $a\ge1$ and that every subgraph of~$H$ has a ratio of edges to vertices of at most~$a$.
Put $p=\lceil a\rceil$ and consider a demand set $R\subseteq V(H)$ with positive vertex costs~$c_v$.
Write $\tau=\tau_{R,c}(H)$.
For every positive integer~$q$, there are nonnegative integers~$\mu_v$ giving the number of copies of each candidate, such that
\[
 \sum_{v\in N_H[r]}\mu_v\ge q\quad(r\in R),
 \qquad
 \sum_vc_v\mu_v\le(q+2p)\tau.
\]
\end{lemma}
\begin{proof}
Scale an optimal fractional vector by $h=q+2p$, which leaves room for rounding.
Its integer parts are kept as copies already chosen, and $y\in[0,1)^{V(H)}$ denotes the fractional parts, which are now rounded to zero or one without increasing their cost.
The analysis uses the orientation of outdegree at most~$p$ from Lemma~\ref{simp:orientation}.

A coordinate is \emph{floating} while it lies strictly between zero and one.
At a stage with~$m$ floating coordinates, the incoming sum is preserved at every demand whose incoming neighborhood contains more than~$p$ of them.
Each floating coordinate occurs in at most~$p$ incoming neighborhoods, because its vertex has at most~$p$ outgoing edges, so there are at most $pm/(p+1)<m$ sums to preserve.
There are fewer sums to preserve than floating coordinates, so the coordinates can be varied along a line while all these sums stay fixed.
Choose one of the two directions along this line so that the cost does not increase, and move until a floating coordinate reaches zero or one, keeping all coordinates within $[0,1]$.
Integral coordinates stay fixed, so this process ends after finitely many steps.

Consider what a demand can lose.
Its incoming sum stays fixed until at most~$p$ incoming coordinates remain floating, after which their total possible loss is strictly less than~$p$.
The rest of its closed neighborhood consists of the demand itself and at most~$p$ outgoing neighbors, whose total possible loss is strictly less than $p+1$.
After the integer parts are added back, its coverage is therefore an integer strictly greater than $h-(2p+1)=q-1$, and hence is at least~$q$.
Cost has only decreased, as required.
\end{proof}

The case $q=1$ compares the initializer's integral benchmark with the fractional optimum.

\begin{corollary}
\label{simp:integrality-gap}
Under the hypotheses of Lemma~\ref{simp:certificate}, let~$\gamma_R(H)$ be the minimum cardinality of a set dominating~$R$.
If every cost is at least one, then
\[
 \gamma_R(H)\le(2\lceil a\rceil+1)\tau_{R,c}(H).
\]
There also exists a set dominating~$R$ whose cost is at most this same multiple of~$\tau_{R,c}(H)$.
\end{corollary}
\begin{proof}
Apply Lemma~\ref{simp:certificate} with $q=1$ and retain each vertex whose multiplicity is positive just once.
This preserves unit coverage and can only decrease cost.
Cardinality is at most cost when~$c_v\ge1$.
\end{proof}

We now show how to find such a certificate algorithmically.
At any point in the search, a bounded number of copies is still available.
If many demands remain, one of those copies must cover many of them, and density ensures that only a bounded number of candidates can do so.
If few demands remain, they can be listed and one cheapest candidate kept for each subset of these demands that a candidate can cover.
In either case the next choice reduces to a bounded list, even when the physical graph has large degree.

\Needspace{10\baselineskip}
\begin{lemma}\label{simp:local-solver}
Fix $a\ge1$, $B_0,K_0,D$ and a positive rational~$\xi$.
Assume that~$H$ is connected, has diameter at most~$D$, and satisfies the subgraph-density bound~$a$, that $R\subseteq V(H)$ is known by local membership bits, and that locally known integer costs satisfy $1\le c_v\le B_0$.
If $\tau_{R,c}(H)\le K_0$, a deterministic broadcast CONGEST algorithm returns a feasible vector of cost at most $(1+\xi)\tau_{R,c}(H)$.
Its round bound and a common denominator for all coordinates depend only on the fixed parameters.
\end{lemma}
\begin{proof}
Put $p=\lceil a\rceil$, choose a positive integer $q\ge2p/\xi$, and set $M=\lceil(q+2p)K_0\rceil$.
A~$q$-fold cover gives each demand at least~$q$ copies among its candidates.
By the certificate lemma, there is a~$q$-fold cover of cost at most $(q+2p)\tau_{R,c}(H)$, and the lower bound of one on costs means that this cover uses at most~$M$ copies.
The search finds the cheapest~$q$-fold cover using at most~$M$ copies, and its multiplicities are divided by~$q$.
Thus the search uses the known bound~$K_0$, without needing to know the optimum.

\paragraph{Branching when many demands remain.}
A search state is described by the multiset~$S$ of copies already selected, where $|S|\le M$ counts copies, including repetitions.
Each demand uses the selected identifiers to compute how many further copies it needs, called its \emph{deficit}.
Let~$U$ be the set of demands with positive deficit, $\ell=|U|$ their number, and $m=M-|S|$ the number of copies still available.
If $\ell=0$, the state is recorded as a completed cover, while if $m=0<\ell$, it is discarded because no completion is possible.

When $\ell\ge2(a+1)m$ and $m>0$, consider
\[
 C=\{v:|N_H[v]\cap U|\ge\ell/m\}.
\]
Every completion by at most~$m$ copies contains a member of~$C$, because these copies must cover every one of the~$\ell$ demands at least once.
The number of choices is bounded by counting how many of these demands each candidate covers, including itself when it is a demand:
\[
 \sum_{v\in C}|N_H[v]\cap U|
 \le e(H[U\cup C])+e(H[U\cap C])+|U\cap C|
 \le a\ell+(a+1)|C|.
\]
An edge can be counted twice only when both endpoints lie in $U\cap C$, and the last term allows each vertex in the intersection to cover itself.
Since every member of~$C$ contributes at least $\ell/m$, this gives
\[
 \left(\frac{\ell}{m}-(a+1)\right)|C|\le a\ell.
\]
The assumption on~$\ell$ makes the factor in parentheses at least $\ell/(2m)$, so $|C|\le2am$.
There is one branch for each member of~$C$, adding one copy of that vertex to ~$S$. Identifiers already in~$S$ are allowed, since repeated copies may still be needed.

\paragraph{Finishing the search when few demands remain.}
When $\ell<2(a+1)m$, the vertices collect the bounded list of identifiers and deficits in~$U$.
A candidate is now described by the subset of~$U$ that it covers, its \emph{mask}.
 For each nonempty mask, we keep a cheapest candidate, breaking ties by identifier. This leaves at most~$2^\ell$ representatives.
The algorithm tests every choice of at most~$m$ copies of these representatives, with repetitions allowed, to see whether it meets the deficits.

This last step finds a cheapest completion of the current state.
Indeed, each copy in any completion may be replaced by the representative of its mask without changing what it covers in~$U$ or increasing  cost. A copy with an empty mask is discarded.
The previously selected multiset~$S$ already covers the demands outside~$U$, so they remain covered throughout this replacement.

At every branching state, one branch therefore remains consistent with any fixed optimum completion, and at every terminal state a cheapest completion is found directly.
At most~$M$ copies are added along any branch, there are at most $\lceil2aM\rceil$ choices at a branching step, and the terminal lists have bounded size, so an optimum certificate is found in a number of stages depending only on the fixed parameters, with deterministic tie breaking.
After division by~$q$, its cost is at most $(1+2p/q)\tau_{R,c}(H)\le(1+\xi)\tau_{R,c}(H)$.
Capping every coordinate at one preserves each unit-demand constraint and can only decrease cost.

\paragraph{Communicating only bounded records.}
Vertices first exchange identifiers, elect the minimum identifier in~$D$ flooding rounds, and construct a breadth-first tree of depth at most~$D$.
Each vertex announces its parent identifier, so counts and minima can be aggregated in decreasing depth order: a vertex broadcasts its aggregate once, and only its parent uses it.
Results are sent back down the tree in increasing depth order.

These operations provide every piece of information used in the search.
Each demand computes its own deficit from the announced selected identifiers and their costs, and the positive-deficit bits are broadcast before candidates count how many remaining demands they cover.
Successive minimum queries, excluding identifiers already reported, collect bounded candidate or demand lists.
In the small-demand case, all masks are iterated over and the minimum pair $(c_v,\mathrm{ID}(v))$ is aggregated for each nonempty mask that occurs.

Counts are at most a fixed multiple of the ambient graph order, while the number of identifiers in a record and the sizes of costs, masks, and multiplicities depend only on the fixed parameters.
Each record therefore fits in~$\mathcal O(\log(n+1))$ bits, and the search and all lists are padded to fixed parameter bounds.
No topology collection, knowledge of~$n$, or distributed orientation is used, and every output coordinate has the same denominator~$q$, also after capping.
\end{proof}

\subsubsection{Defining and solving the tasks}\label{simp:task-definition}

Fix a positive rational~$\xi$ and take the partition distribution of Lemma~\ref{simp:partition} on~$J_0$.
For each nonempty connected set $C\subseteq V(J_0)$ of size at most~$k$, put
\[
 F(C)=\{i\in C:N_{J_0}(i)\not\subseteq C\},
 \qquad I(C)=C\setminus F(C).
\]
The set~$F(C)$ contains the owners in~$C$ that have a neighbor outside~$C$, while~$I(C)$ contains the remaining owners, called its interior.
Thus~$C$ itself determines which boundary owners are selected and which owners remain for the local  task. The other parts of the partition have no effect on this choice.

Let~$H_C$ be the subgraph of~$G$ induced by the full original stars~$V_i$ for $i\in C$, together with all light vertices~$v$ satisfying $S_0(v)\cap C\ne\varnothing$, including every edge of~$G$ between these vertices.
The demands of the task are
\[
 R_C=\{r\in R':m(r)\in I(C)\}.
\]
Every vertex of~$H_C$ may serve as a candidate, with its original cost.
Some candidates belong to other stars, including stars of selected high-degree  owners. They are kept because they may still help cover an interior demand.
The demands themselves have owners in~$I(C)$.
A demand may also remain in the task after some other selection has covered it, since covering it again can only make the local requirement stronger.

The task has the two bounds required by our local solver.
First, each edge~$ij$ of~$J_0[C]$ has a light witness and representatives giving a walk $i,u_i,v,u_j,j$ of length at most four inside~$H_C$, and every vertex of~$H_C$ is within distance two of an owner in~$C$.
A spanning tree of~$J_0[C]$ therefore shows that~$H_C$ is connected and has diameter at most $2+4(|C|-1)+2\le4k$.
Second, selecting the owners in~$C$ covers~$R_C$ at cost at most $B_0k$.
Lemma~\ref{simp:local-solver} therefore applies with $D=4k$ and $K_0=B_0k$ and gives a deterministic local vector~$z^C$ of cost at most $(1+\xi)\tau_C$, where $\tau_C=\tau_{R_C,c}(H_C)$.
Fixed deterministic rules settle every choice of the solver, so~$z^C$ depends on~$C$ and the input alone.
This is why the same local answer can be used whenever~$C$ appears in a partition.
}

{
\subsection{Computing the exact average}\label{prop:solver-global}

Combining the local solver with the prepared structure gives the following guarantee for closed neighborhoods.

\begin{theorem}
\label{simp:weighted-solver}
For every fixed rational $a\ge1$, integer $B_0\ge1$, and rational $\delta>0$, there is a deterministic broadcast CONGEST algorithm with the following guarantee.
On a graph $G\in\mathcal C_a$, with demands~$R$ and integer costs $1\le c_v\le B_0$, it computes a feasible vector $x\in[0,1]^{V(G)}$ such that
\[
 \sum_vc_vx_v\le(1+\delta)\tau_{R,c}(G).
\]
The number of rounds and a common denominator for all output coordinates depend only on $a,B_0,\delta$.
The algorithm requires no embedding, orientation, decomposition, initial solution, graph order, or maximum-degree bound.
\end{theorem}

\subsubsection{Combining local solutions}\label{prop:solver-construction}
\label{simp:averaging}

We combine the local solver and the partition to obtain the global fractional cover, then choose the parameters to obtain the required cost bound.
For any one partition, the algorithm selects the boundary owners and solves the demands in the interior of each part.
These tasks can be charged to different coordinates of a global fractional optimum, so their total cost is small.
Averaging over the finite partition distribution then preserves coverage and leaves only the expected boundary cost to be paid.
The last part of this section explains how each physical vertex computes its coordinate of this average using short broadcasts.

To add these local costs, consider the candidates that can actually cover an interior demand.
If one candidate served two different parts, its support would create an edge between their interior owners, contrary to the meaning of interior.
This gives the separation we need, even though the larger graphs used to relay messages may overlap.

\begin{lemma}\label{simp:tasks}
For every partition outcome, $\sum_{C\text{ a part}}\tau_C \le\tau_{R,c}(G)$.
\end{lemma}
\begin{proof}
Let~$T_C$ be the set of all physical vertices that dominate at least one vertex of~$R_C$.
If $v\in T_C$ serves~$r$, then $r\in R'$ makes~$v$ light and $m(r)\in S_0(v)\cap C$ places~$v$ in~$H_C$.
If~$v$ also served a task in a different part, its support would contain an interior owner in each part.
These owners would be adjacent in~$J_0$, contradicting their being interior.
The sets~$T_C$ are therefore pairwise disjoint.

Take a fractional solution of minimum cost for the whole graph, keep its coordinates on~$T_C$, and put zero on the other vertices of~$H_C$.
This vector still covers~$R_C$: every candidate that can cover any one of its demands belongs to~$T_C$.
Its cost bounds~$\tau_C$.
Summing over the disjoint sets~$T_C$ proves the assertion.
\end{proof}

For one partition outcome, extend each~$z^C$ by zero outside~$H_C$ and set
\[
 X=\mathbf1_P+\mathbf1_Q+
 \sum_{C\text{ a part}}\bigl(z^C+\mathbf1_{F(C)}\bigr).
\]
This nonnegative vector satisfies every demand.
A vertex in~$N_G[P]$ is covered by~$P$.
Any other demand has either an owner in~$Q$, an owner in the boundary of its part, or an interior owner whose local task covers it.
Writing~$B$ for all boundary owners, the cost is at most
\[
 (1+\xi)\tau_{R,c}(G)+B_0(|P|+|Q|+|B|).
\]
 We now average these vectors over the partition distribution.
Each demand stays covered because its total coverage is a sum of coordinates, and for every $r\in R$ we have
\[
 \sum_{v\in N_G[r]}\mathbb E X_v
 =\mathbb E\!\left[\sum_{v\in N_G[r]}X_v\right]\ge1.
\]
The bounds on~$P$,~$Q$, and the expected boundary give
\begin{equation}\label{simp:compiler-cost}
 \mathbb E\sum_vc_vX_v
 \le (1+\xi)\tau_{R,c}(G)
 +B_0\left(\frac{2a}{L}+\frac{2C_L}{\lambda}+2\eta\right)N.
\end{equation}
 We compute the average exactly in two steps.
First, each owner computes the probability of each possible part containing it. It then sends these probabilities to the physical vertices together with the descriptions of their local tasks.
}

{
\subsubsection{Exact probabilities from a fixed number of fair bits}

For each possible part~$C$, put
\[
 p_C=\Pr(C\text{ is a part}).
\]
This event occurs exactly when the owner in~$C$ with the smallest identifier outputs the complete list~$C$ as its part.
By Lemma~\ref{simp:partition}, that output is determined by a fixed neighborhood and the fair bits assigned to its vertices.
The fixed names and bit positions of the original candidate sets ensure that we refer to the same independent bit whenever a set appears in several local views.

The radius is $R_*=r_*+1$, where~$r_*$ is given in \eqref{simp:partition-radius}.
From the radius-$R_*$~neighborhood of the owner with the smallest identifier, we try every possible choice of the relevant bits and count the choices for which its part output is~$C$.
Only the decisions needed for that owner's output are computed, following the local rule proved in the partition lemma, so all required information lies inside the gathered neighborhood.
Dividing the successful count by the total number of choices gives~$p_C$ exactly.
There may be many choices to try, but this computation requires no further messages.

Every owner in~$C$ must know this probability before it passes the task to physical vertices.
Each owner gathers its radius-$(R_*+k)$ neighborhood in~$J_0$, using Lemma~\ref{simp:simulation}.
If it belongs to~$C$, the owner of smallest identifier in~$C$ is at distance at most $k-1$, so this larger neighborhood contains the radius-$R_*$~neighborhood needed for the count.
Each owner can therefore compute~$p_C$  for every possible part containing it.
 These computations all give the same value~$p_C$, which the owners send to the physical vertices together with the task description.

Every vertex has $b=AT$ fair bits, so a common upper bound on the number used by one event is
\begin{equation}\label{simp:probability-bits}
 B_*=b\sum_{j=0}^{R_*}\lambda^j.
\end{equation}
Unused bits can pad the count for an event, so every probability can be written with the same denominator~$2^{B_*}$.
Gathering a larger neighborhood to handle several events does not change this bound, since each event still uses only the radius-$R_*$~neighborhood of its owner with the smallest identifier.

{
\subsubsection{A bounded number of tasks at each physical vertex}

Consider all nonempty connected sets~$C$ of at most~$k$ vertices in~$J_0$.
Each owner belongs to at most $A=\sum_{j=0}^{2k-2}\lambda^j$ such sets, by the walk count in the partition proof.
Their sorted identifier lists serve as task names.
The bounded owner neighborhoods already gathered also determine the boundary sets~$F(C)$ and the probabilities~$p_C$.

 Each physical vertex participates in only boundedly many tasks.
A light vertex joins~$H_C$ whenever~ $S_0(v)$ intersects~$C$, giving at most~$LA$ tasks.
If it also has an owner, that owner is already in~$S_0(v)$, so belonging to its own star adds no tasks to this count.
A heavy vertex can join only through its own star, if it has one, and hence belongs to at most~$A$ tasks.
Thus every physical vertex has at most
\begin{equation}\label{simp:role-bound}
 K_{\rm roles}=LA
\end{equation}
roles, where a role means its participation in one task~$H_C$.

To tell vertices their roles, each owner broadcasts its at most~$A$ task records to its full star, and each owned vertex repeats these records with its owner's identifier.
A record contains the task name and~$p_C$, together with one bit for each owner in~$C$ saying whether that owner belongs to~$F(C)$.
A light vertex keeps the tasks  whose sets intersect its support, while a heavy owned vertex keeps the tasks containing its own owner.
Every support label has a representative in the vertex's closed neighborhood, so all required records arrive, and repeated copies are discarded.
The boundary bits also tell each owned demand whether it belongs to~$R_C$.

Each vertex then broadcasts its bounded list of task names.
Along with the physical neighbor identifiers, these lists tell the endpoints exactly which physical edges belong to each induced graph~$H_C$.
All local solvers run in blocks of~$K_{\rm roles}$ physical rounds: in one block each vertex broadcasts one message for each of its roles, including the task name, and sends empty entries in the unused rounds.
Its neighbors collect messages by task name and deliver them to the corresponding local solver at the end of the block.
The order of roles can differ between vertices, because the task name identifies every message.
Thus each solver receives exactly the messages from its neighbors in~$H_C$, and all solvers stay in the schedule until their common fixed round bound, with empty messages sent for any solver that has already finished.
The size and degrees of~$H_C$ may be  unbounded. The bounded role count and the certificate solver are what keep the schedule fixed.

Each vertex now knows its coordinate in every relevant~$z^C$ and the corresponding probability, so it computes
\begin{equation}\label{simp:exact-average}
 \widehat x_v=\mathbf1_P(v)+\mathbf1_Q(v)
 +\sum_{C:v\in H_C}
   p_C\bigl(z^C_v+\mathbf1_{F(C)}(v)\bigr).
\end{equation}
This is exactly~$\mathbb E X_v$, because the local solution and boundary set depend only on~$C$, and each contribution is weighted by the probability that~$C$ occurs.
The output is $x_v=\min\{1,\widehat x_v\}$.
Reducing coordinates to at most one preserves coverage: if a constraint contains a coordinate that was reduced, that coordinate is now equal to one and therefore  satisfies the constraint by itself, and otherwise the constraint has not changed.

All local vectors have the same denominator~$q$ from Lemma~\ref{simp:local-solver}, so $q2^{B_*}$ is a common denominator for \eqref{simp:exact-average} and the capped output.
Each vertex adds only boundedly many terms.
Costs, lists of bits, numbers of copies, and representations of fractions have lengths bounded by the fixed parameters, while identifiers and counts collected from vertices use~$\mathcal O(\log(n+1))$ bits.
Bounded records are sent one entry at a time, which preserves the broadcast CONGEST round bound.
}

\subsubsection{Choosing the parameters}\label{prop:parameters}

It remains to make the four sources of error fit the requested accuracy.
Half of the error budget goes to the local solvers and one sixth to each of the three kinds of selected vertices: heavy candidates, high-degree owners, and boundary owners.
The initializer gives $N\le A_a\tau_{R,c}(G)$, so these choices can be made without knowing either~$N$ or the optimum.
For the target rational $\delta>0$, take rational $0<\xi\le\min\{1,\delta/2\}$, then an integer
\[
 L\ge\max\{2,4a,12B_0aA_a/\delta\}.
\]
With $C_L=2aL$, take an integer $\lambda\ge\max\{1,12B_0C_LA_a/\delta\}$ and rational
\[
 0<\eta\le\min\{1,\delta/(12B_0A_a)\}.
\]
With these choices, each selection term in \eqref{simp:compiler-cost} is at most $\delta\tau_{R,c}(G)/6$, while $\xi\le\delta/2$ pays for the local solver error.
The total cost is therefore at most $(1+\delta)\tau_{R,c}(G)$.
The partition, search, and denominator parameters are then fixed in this order of dependence, so every constant is determined by $a,B_0,\delta$ alone.

In a connected component without demands, the initializer selects no owners, so there are no tasks or extra selections.
The fixed schedule is still followed and the output is zero, without any need to test whether the whole component is empty of demands.
The construction proves Theorem~\ref{simp:weighted-solver}.

\subsection{General covering relations}\label{sc:covering}

This section completes the proof of Theorem~\ref{sc:thmA}.
The proof of Theorem~\ref{simp:weighted-solver} uses the closed neighborhood of a demand in two different roles.
It is the set of candidates that may cover the demand, and it is also the set of physical neighbors that carry messages and appear in edge counts.
For general covering sets, the second role stays unchanged and only the first is replaced.
The single new difficulty is that an owner may be unable to cover  itself. Such an owner is given a \emph{partner} that covers it.

\paragraph{The input.}
The input is a graph $G\in\mathcal C_a$ together with a demand set~$R$ and, for each $r\in R$, a nonempty covering set $C(r)\subseteq N_G[r]$.
For every edge~$rv$, both endpoints know whether $v\in C(r)$, and each demand knows whether $r\in C(r)$.
A vertex is \emph{admissible} if it belongs to  some covering set. Every vertex knows whether it is admissible after one round.
Admissible vertices have costs $c_v\in\{1,\ldots,B_0\}$, and
\[
 \tau_c=\min\Bigl\{\sum_vc_vx_v:\ x\ge0,\ \sum_{v\in C(r)}x_v\ge1\
 (r\in R)\Bigr\},
\]
where~$x$ ranges over vectors on the admissible vertices.
Total domination is the case $C(r)=N_G(r)$, red-blue domination the case in which~$C(r)$ consists of the blue vertices of~$N_G[r]$, and Theorem~\ref{simp:weighted-solver} the case $C(r)=N_G[r]$.
The covering sets must be nonempty, so total domination requires that no demand is isolated, and red-blue domination that every red vertex has a blue neighbor or is blue itself.
A vertex set~$X$ \emph{covers} a demand~$r$ if $X\cap C(r)\ne\varnothing$.

We adapt the five steps of the solver in turn.
In each step, the covering relation $v\in N_G[r]$ is replaced by $v\in C(r)$, while every physical edge is kept for communication and for counting.
Since $C(r)\subseteq N_G[r]$, a covering incidence is always a physical edge or the self-incidence of a demand.

\paragraph{Certificates.}
In Lemma~\ref{simp:certificate}, the coverage of a demand~$r$ becomes $\sum_{v\in C(r)}\mu_v$, and vertices that are not admissible receive no copies.
The floating argument is unchanged: the coordinates of~$C(r)$ that enter~$r$ along the orientation belong to incoming neighbors, and the remaining coordinates of~$C(r)$ belong to~$r$ and to at most~$p$ outgoing neighbors, so the two possible losses are again less than~$p$ and $p+1$.
Corollary~\ref{simp:integrality-gap} follows as before.
In the search of Lemma~\ref{simp:local-solver}, the branching set becomes $\{v:|\{r\in U:v\in C(r)\}|\ge\ell/m\}$, and its size bound remains valid because  $|\{r\in U:v\in C(r)\}|\le|N_H[v]\cap U|$. Masks record the sets $\{r\in U:v\in C(r)\}$.
The search uses only that every covering set of a demand of~$H$ lies in~$H$, which holds for~$G$ and, as shown below, for  every task graph. Vertices that are not admissible receive the dummy cost one and are never chosen.
A demand computes its deficit from the announced identifiers and its own incidence bits, and a candidate counts the remaining demands that it covers from its incidence bits and their deficit bits.
The communication is otherwise unchanged.

\paragraph{Initialization.}
In pass~$i$ of the pruning procedure, the eligible candidates,  previously denoted by~$C_i$, and the lists become
\[
 E_i=\{v:|\{r\in R_i:v\in C(r)\}|>h\},\qquad
 \Lambda_i(r)=C(r)\cap E_i.
\]
A demand with an empty list, and after the last pass  every remaining demand, no longer selects itself. Instead, it \emph{requests} the member of~$C(r)$ with the smallest identifier, which then selects itself.
This costs two extra rounds: one for the requests and one for the new selection bits.
The pass bound \eqref{simp:pass-cost} is proved as before, with a minimum cover~$O$ and its eligible members $O\cap E_i$: a demand whose list is  empty or still uncovered is covered only by ineligible members of~$O$, each of which covers at most~$h$ residual demands.
The lists are subsets of closed neighborhoods, so Lemmas~\ref{simp:list-pairs} and~\ref{simp:bounded-lists} apply.

In the proof of Lemma~\ref{simp:residual-bound}, put $U_i=R_{i+1}\cup E_i$ and call the physical neighbors in $R_i\cup E_i$ eligible.
 A remaining demand has more than~$h$ eligible candidates in $C(r)\subseteq N_G[r]$, and an eligible candidate covers more than~$h$ residual demands, all in its closed neighborhood, so every vertex of~$U_i$ still has at least~$h$ eligible neighbors.
In \eqref{simp:path-exit}, the vertex~$w$ is covered by~$S_j$, so some $s\in S_j$ lies in  $C(w)\subseteq N_G[w]$. As the path avoids~$S$, we have $s\ne w$, so~$s$ is a physical neighbor of~$w$, which is all that the minor argument uses.
The selection sets remain disjoint: after~$s$ is selected, it covers no later residual demand, so it is neither eligible nor requested later.
 Every demand remaining after the last pass requests one vertex, so the set~$D_0$ of all selected and requested vertices covers~$R$, consists of admissible vertices, and satisfies the bound of Theorem~\ref{simp:initializer}.

\paragraph{Owners and partners.}
Put $O=R\cup D_0$.
Every vertex $i\in D_0$ owns itself, and every $r\in R\setminus D_0$ chooses as its owner~$m(r)$ the member of $D_0\cap C(r)$ with the smallest identifier, which exists because~$D_0$ covers~$r$.
The classes~$m^{-1}(i)$ are again disjoint stars.
 We keep the supports $S_0(v)=\{m(u):u\in N_G[v]\cap O\}$ defined by physical neighborhoods.
The algorithm selects the admissible heavy vertices $P=\{v\text{ admissible}:|S_0(v)|>L\}$.
Put
\[
 R'=\{r\in R:C(r)\cap P=\varnothing\},
\]
and build the owner graph~$J$ from the supports of the light admissible vertices.
Lemmas~\ref{simp:heavy} and~\ref{simp:owner-density} count only admissible heavy vertices and  admissible witnesses. The proofs do not change.
In Lemma~\ref{simp:owner-separator}, a vertex on a retained walk  that is not both light and admissible occurs only as an owner or as a representative~$u_i$, and is replaced  by its owner. A light admissible vertex is replaced by its at most~$L$ labels as before.
The selection of high-degree owners and the simulation of Lemma~\ref{simp:simulation} use only physical edges and do not change, provided that only light admissible vertices broadcast and forward support labels, and that a vertex that is not admissible keeps only the tasks of its own owner.
 The partitions of Lemma~\ref{simp:partition} concern only the owner graph.

An owner $i\in R'$ with $i\notin C(i)$ is not covered by itself.
It receives the \emph{partner}~$\pi(i)$, the member of $D_0\cap C(i)$ with the  smallest identifier. For all other owners~$\pi(i)$ is undefined.
Selecting an owner together with its partner, if it has one, covers every demand of~$R'$ in its class, at cost at most~$2B_0$.
One broadcast of the bits for membership in~$P$ and in~$D_0$ lets every demand decide whether it lies in~$R'$, and lets every owner in~$R'$ choose its partner.

\paragraph{Tasks and averaging.}
The task graph~$H_C$ consists of the stars of the owners in~$C$ and the light admissible vertices  whose supports intersect~$C$, and its demands are the vertices $r\in R'$ with $m(r)\in I(C)$.
If $v\in C(r)$ for such a demand, then~$v$ is admissible and not heavy, because $r\in R'$, and $m(r)\in S_0(v)$, because $r\in N_G[v]\cap O$.
Thus every covering set of a task demand lies in~$H_C$, and the set~$T_C$ of candidates that cover some demand of the task consists of light admissible vertices whose supports contain an interior owner of~$C$.
The disjointness and restriction argument in the proof of Lemma~\ref{simp:tasks} therefore applies and gives $\sum_C\tau_C\le\tau_c$.
The owners in~$C$ and their partners, which lie in~$H_C$ by the same argument, cover the task at cost at most $2B_0k$, so Lemma~\ref{simp:local-solver} applies with $K_0=2B_0k$.

For one partition outcome, select every high-degree owner and every boundary owner together with its partner, counting partners with multiplicity:
\[
 X=\mathbf1_P+\sum_{i\in Q}\bigl(\mathbf1_i+\mathbf1_{\pi(i)}\bigr)
 +\sum_{C\text{ a part}}\Bigl(z^C+\sum_{i\in F(C)}
   \bigl(\mathbf1_i+\mathbf1_{\pi(i)}\bigr)\Bigr),
\]
where $\mathbf1_{\pi(i)}=0$ when~$\pi(i)$ is undefined.
A demand outside~$R'$ is covered by~$P$, a demand of~$R'$ whose owner is selected is covered by that owner or its partner, and every other demand of~$R'$ lies in a task.
The cost bound \eqref{simp:compiler-cost} holds with its last two terms doubled, so the parameters must satisfy $\lambda\ge24B_0C_LA_a/\delta$ and $\eta\le\delta/(24B_0A_a)$, where~$A_a$ is the constant of the  extended initializer. The partition parameters are chosen afterwards, as before.

To compute~$\mathbb EX_v$, every owner~$i$ computes the probability $\beta_i=\sum_{C\ni i,\ i\in F(C)}p_C$ that it is a boundary owner, from the neighborhood that it has already gathered, and broadcasts the identifier of~$\pi(i)$, the bit $[i\in Q]$, and~$\beta_i$.
A partner $v=\pi(i)$ is a physical neighbor of~$i$, since $\pi(i)\in C(i)\subseteq N_G[i]$.
Moreover $v\in C(i)$ with $i\in R'$ shows that~$v$ is light and $i\in S_0(v)$, so every vertex is the partner of at most~$L$ owners, and it adds at most~$L$ terms $[i\in Q]+\beta_i$ to its coordinate in \eqref{simp:exact-average}.
All other steps of the averaging are unchanged.
The output is a feasible vector on the admissible vertices of cost at most $(1+\delta)\tau_c$, with a common bounded denominator, which proves Theorem~\ref{sc:thmA}.

}

\section{General domination on excluded-minor classes}\label{simp:endpoint:section}

We prove the general domination guarantees of Theorems~\ref{sc:thmB} and~\ref{sc:thmC}, using the rounding principle from Section~\ref{new:rounding-framework} and the fractional solver of Section~\ref{prop:local-tools}.
The planar case is already presented in Section~\ref{new:planar}; here the density parameter may be any fixed rational satisfying the stated hypotheses.
The fractional solver stays unchanged; the two ingredients to establish are the budget transfer from Section~\ref{prop:budget-transfer} and rounding with slack from Section~\ref{new:rounding-framework}.

Two features of Proposition~\ref{sc:principle} matter in applications.
The accuracy~$\delta$ depends only on the rounding rule, so it can be fixed before the solver runs, and no estimate of~$\tau$ is needed.
The hypothesis holds for every feasible~$x$, so applying the proposition to the solver's output requires only its feasibility and cost bound.
In the rest of this section, we prove the required slack bound for dominating set.
Lemma~\ref{simp:endpoint:rounding} expresses the rounding saving in terms of quantities~$S_i$, and Lemma~\ref{simp:endpoint:count} and the non-integer case then bound the output by the same quantities.

In the proof of part~\ref{sc:thmC:minor} of Theorem~\ref{sc:thmC}, we distinguish between the cases where~$2a$ is an integer and where it is not.
The rounding rule is the same in both cases.
When~$2a$ is an integer, which includes planar graphs, a large region of degree~$2a$ can make the rounding exact, with no saving at all.
A vertex is a \emph{defect} if its degree differs from~$2a$.
The rule first selects the vertices that are neither defects nor adjacent to defects; the remaining candidates get costs that record how much of their budget is still available.
When~$2a$ is not an integer, the gaps between~$2a$ and the integer degrees provide the saving directly.

\paragraph{The fractional solver used in this section.}
We run the solver from Theorem~\ref{simp:weighted-solver} on a graph $J\in\mathcal C_\rho$, with locally specified demands~$R_0$ and integer costs $1\le c(v)\le B_0$.
The solver returns a feasible vector $x\in[0,1]^{V(J)}$ with
\[
 \cost(x):=\sum_{v\in V(J)}c(v)x_v\le(1+\delta)\cost^*,
\]
where~$\cost^*$ is the minimum fractional cost of covering~$R_0$.
All vertices of~$J$ are candidates, including those outside~$R_0$. The number of broadcast rounds and a common denominator for all coordinates of~$x$ depend only on the fixed parameters $\rho,B_0,\delta$.
In this section $\rho=a$, except in the non-integer case, where $\rho\ge a$ may be larger.
If all costs are equal, the solver may run with unit costs instead, since multiplying all costs by one number changes neither the feasible vectors nor the ratio to the optimum.
The solver theorem is proved in the main text; it is the only algorithmic result needed here.

On graphs whose minors have edge density at most a fixed rational $a\ge1$, the solver and rounding procedure together produce a dominating set of size at most $(2a+1)\gf(G)$.
All planar minors have edge density at most three, so on planar graphs the coefficient is seven.

\Needspace{8\baselineskip}
\subsection{The rounding calculation}

\paragraph{The rounding rule.}
Fix $a\ge1$, and put $b=2a+1$ and $t=1/(a+1)$.
Let $C\subseteq V(G)$ be a candidate set, with demands $R_0\subseteq C$, costs $0<c(v)\le b$, and a feasible fractional cover $x\in[0,1]^C$ of~$R_0$. Write $\cost(x)=\sum_{v\in C}c(v)x_v$.
\begin{enumerate}[label=\textbf{Step \arabic*.},leftmargin=*,itemsep=2pt,parsep=0pt,topsep=4pt]
\setlength{\abovedisplayskip}{6pt}
\setlength{\belowdisplayskip}{4pt}
\setlength{\abovedisplayshortskip}{4pt}
\setlength{\belowdisplayshortskip}{4pt}
\item Select candidates of cost~$b$ with fractional weight greater than~$t$, and let~$R$ be the demands still uncovered:
\[
 W=\{v\in C:c(v)=b,\ x_v>t\},
 \qquad R=R_0\setminus N_G[W].
\]
\item For each remaining candidate, count the demands in~$R$ that it covers, and select those covering at least~$b$:
\[
 h(v)=|N_G[v]\cap R|\quad(v\in C\setminus W),
 \qquad H=\{v\in C\setminus W:h(v)\ge b\}.
\]
\item Select every demand still uncovered, and return~$D$:
\[
 U=R\setminus N_G[H],\qquad D=W\cup H\cup U.
\]
\end{enumerate}

\paragraph{Notation for the analysis.}
Let $O=C\setminus(W\cup H)$ be the candidates not selected in the first two steps, and let $L=R\setminus(H\cup U)$ be the demands covered by~$H$ but not selected into~$H$.
Write $w=|W|$, $\ell=|L|$, and
\[
 P=\sum_{v\in O}(c(v)-h(v))x_v.
\]
Here~$P$ is the difference between the fractional cost of candidates in~$O$ and their total fractional contribution to covering~$R$.

To compare $|D|$ with~$\cost(x)$, we must account for the candidates selected in Step~2 that lie outside the remaining demand set, that is, the vertices in $H\setminus R$.
Each such vertex adds one to the output size, but the density bound will give a saving that compensates for this increase.

\begin{lemma}
\label{simp:endpoint:rounding}
In the setting above, suppose that every subgraph of~$G$ has edge density at most~$a$, $c(v)=b$ on~$H$, and $c(v)\ge h(v)$ on~$O$.
Then~$D$ covers~$R_0$, and
\begin{align}
 \sum_{v\in H}\bigl(h(v)-b\bigr)+(a+1)|H\setminus R|&\le a\ell,
 \label{simp:endpoint:excess}\\
 |U|&\le\sum_{v\in O}h(v)x_v,
 \label{simp:endpoint:leftover}\\
 \cost(x)-|D|&\ge\frac{w+\ell}{a+1}+P.
 \label{simp:endpoint:saving}
\end{align}
\end{lemma}

\begin{proof}[\proofname]
Put $Q=H\cap R$.
The sets $Q,L,U$ partition~$R$: they are the selected demands, the other demands covered by~$H$, and the demands that must still be selected.
Since $W,H,U$ are disjoint,
\[
 |D|=w+|R|-\ell+|H\setminus R|.
\]
The output covers~$R_0$, since~$W$ covers the demands removed first,~$H$ covers those removed next, and~$U$ contains every demand left.

For \eqref{simp:endpoint:excess}, count for each $v\in H$ its neighbors in~$R$, and also~$v$ itself when $v\in Q$.
Write $e(X)=|E(G[X])|$, and~$e(X,Y)$ for the number of edges between disjoint sets $X,Y$.
There is no edge from~$H$ to~$U$, so
\[
\begin{aligned}
 \sum_{v\in H}h(v)
 &=2e(Q)+e(Q,L)+e(H\setminus R,Q\cup L)+|Q|\\
 &\le e(H\cup L)+e(Q)+|Q|
  \le a(|H|+\ell)+(a+1)|Q|.
\end{aligned}
\]
Since $|H|=|Q|+|H\setminus R|$ and $b=2a+1$, this is \eqref{simp:endpoint:excess}.

A demand $u\in U$ has no neighbor in $W\cup H$, so its constraint $\sum_{v\in N_G[u]\cap C}x_v\ge1$ uses only candidates in~$O$.
Summing these constraints over~$U$ gives \eqref{simp:endpoint:leftover}, since each variable~$x_v$, $v\in O$, is counted $|N_G[v]\cap U|\le h(v)$ times.

For \eqref{simp:endpoint:saving}, no vertex in~$W$ covers a demand in~$R$, so summing the covering constraints over~$R$ gives $\sum_{v\in C\setminus W}h(v)x_v\ge|R|$.
A vertex of~$H$ has cost~$b$ and is not in~$W$, so $x_v\le t$, while $x_v>t$ for every $v\in W$.
Let $E=\sum_{v\in H}(h(v)-b)$ denote the excess. Since $h(v)\ge b$ on~$H$, the bounds on~$x_v$ give:
\[
\begin{aligned}
 \cost(x)
 &=b\sum_{v\in W}x_v+
  \sum_{v\in C\setminus W}h(v)x_v+P
    -\sum_{v\in H}(h(v)-b)x_v\\
 &\ge btw+|R|+P-tE.
\end{aligned}
\]
Subtracting $|D|$ and using \eqref{simp:endpoint:excess} in the form $E\le a\ell-(a+1)|H\setminus R|$ gives
\[
 \cost(x)-|D|
 \ge (bt-1)w+(1-at)\ell+\bigl((a+1)t-1\bigr)|H\setminus R|+P.
\]
This explains the threshold: with $t=1/(a+1)$, each vertex of $H\setminus R$ contributes zero, $1-at=1/(a+1)$, and $bt-1=a/(a+1)\ge1/(a+1)$.
\end{proof}

\paragraph{ Applying rounding with slack.}

Inequality \eqref{simp:endpoint:saving} gives a lower bound on the saving using $w+\ell$ and~$P$.
If these quantities also give an upper bound on $|D|$, the saving is enough to cover the solver's error.

\begin{lemma}
\label{simp:endpoint:paying}
In the setting of Lemma~\ref{simp:endpoint:rounding}, let~$\cost^*$ be the minimum cost of a fractional cover of~$R_0$ using only candidates in~$C$.
Suppose that $|D|\le M(w+\ell)+M'P$ for numbers $M,M'>0$, and put
\[
 \delta=\min\left\{\frac{1}{(a+1)M},\frac{1}{M'}\right\}.
\]
If $\cost(x)\le(1+\delta)\cost^*$, then $|D|\le\cost^*$.
\end{lemma}

\begin{proof}[\proofname]
\par\noindent
\par\noindent
Apply Proposition~\ref{sc:principle} with $S_1=w+\ell$, $S_2=P$, $\alpha_1=1/(a+1)$, $\alpha_2=1$, $\beta_1=M$, and $\beta_2=M'$, using \eqref{simp:endpoint:saving}.
\end{proof}

In the next two subsections, we prove this bound for every feasible vector~$x$, with constants $M,M'$ that depend only on~$a$.
Thus~$\delta$ is fixed before the solver runs.
Since $D\subseteq C$ covers~$R_0$ and has cost at most $b|D|$, we have $\cost^*\le b|D|$. Thus the saving is at least $\delta|D|\ge$ $\delta\cost^*/b$, a fixed fraction of the optimum.

\subsection{The integer case: selecting regular regions}

Assume first that $d=2a$ is an integer, which includes the planar case, and put $b=d+1$.
Let~$B$ be the set of defects, the vertices of degree different from~$d$.
Select
\[
 F=V(G)\setminus N_G[B].
\]
Thus a vertex belongs to~$F$ exactly when it and all its neighbors have degree~$d$, so every defect remains a demand.
The remaining demands are the vertices that are neither in~$F$ nor adjacent to it,
\[
 R_0=V(G)\setminus N_G[F],
\]
and every vertex outside~$F$ is a candidate.
We can think of~$b$ as a budget per unit of fractional weight.
For each candidate~$v$, summing the covering constraints over~$F$ uses one unit of this budget per neighbor in~$F$, so we use the remaining budget as its cost:
\[
 c(v)=b-|N_G(v)\cap F|\qquad(v\notin F).
\]
A demand has no neighbor in~$F$, so its constraint is the same in~$G$ and in $G-F$.
Let~$\cost^*$ be the fractional optimum of this remaining problem.

\begin{lemma}
\label{simp:endpoint:budget}
\leavevmode
\begin{enumerate}[label=(\roman*)]
\item $B\subseteq R_0$.
\item The costs are integers with $1\le c(v)\le b$, and $c(v)=b$ holds exactly when $v\in R_0$.
If $c(v)<b$, then $|N_G[v]\cap R_0|\le c(v)-1$.
\item $|F|+\cost^*\le b\gf(G)$.
\end{enumerate}
\end{lemma}

\begin{proof}
By the definition of~$F$, no defect belongs to~$F$ or has a neighbor in~$F$. Thus $B\subseteq R_0$, proving~(i).

For (ii), a candidate~$v$ lies in~$R_0$ exactly when it has no neighbor in~$F$, that is, when $c(v)=b$.
Otherwise~$v$ has a neighbor in~$F$, so it is not a defect and has degree~$d$.
Then $1\le b-c(v)\le d$.
The demands in~$N_G[v]$ exclude~$v$ itself and its $b-c(v)$ neighbors in~$F$, so there are at most $d-(b-c(v))=c(v)-1$ of them.

For (iii), let~$y$ be an optimal fractional dominating vector of~$G$.
Its restriction to $V(G)\setminus F$ covers~$R_0$, because no demand has a neighbor in~$F$, so $\cost^*\le\sum_{v\notin F}c(v)y_v$.
Summing the constraints of~$y$ over~$F$ gives
\[
 |F|\le\sum_{v\in V(G)}|N_G[v]\cap F|\,y_v.
\]
For $v\in F$ this coefficient is at most $\deg_G(v)+1\le b$, since~$v$ is not a defect, and for $v\notin F$ it equals $b-c(v)$.
Adding the two bounds gives $|F|+\cost^*\le b\sum_vy_v=b\gf(G)$.
\end{proof}

It remains to round an approximate solution of the remaining problem to at most~$\cost^*$ vertices.
The solver runs on $G-F$ with demands~$R_0$, costs~$c$ (we set $B_0=b$), and a fixed accuracy~$\delta$ chosen below. It returns a feasible vector~$x$ on $V(G)\setminus F$.
Since the constraints at the demands are the same in~$G$, we use the rounding rule in~$G$ with $C=V(G)\setminus F$ and verify the hypotheses of Lemma~\ref{simp:endpoint:rounding} below.

By Lemma~\ref{simp:endpoint:budget}(ii), a candidate with $c(v)<b$ has $h(v)\le c(v)-1<b$, so it lies neither in~$W$, which requires $c(v)=b$, nor in~$H$.
Hence $c=b$ on $W\cup H$.
A candidate of cost~$b$ outside $W\cup H$ has $h(v)\le b-1$ by the definition of~$H$.
Thus the hypotheses of Lemma~\ref{simp:endpoint:rounding} hold, and every $v\in O$ satisfies
\begin{equation}
 \label{simp:endpoint:slack}
 h(v)\le d\qquad\text{and}\qquad c(v)-h(v)\ge1.
\end{equation}

\begin{lemma}
\label{simp:endpoint:count}
For every feasible vector~$x$ on $V(G)\setminus F$, $|D|\le b^2(w+\ell)+db^2P$.
\end{lemma}

\begin{proof}[\proofname]
Put $Q=H\cap R$.
Here $|D|=w+|H\setminus R|+|Q|+|U|$.
The sum in \eqref{simp:endpoint:excess} is nonnegative, because $h\ge b$ on~$H$, so $|H\setminus R|\le\ell$.
By \eqref{simp:endpoint:leftover} and \eqref{simp:endpoint:slack}, $|U|\le\sum_{v\in O}h(v)x_v\le dP$.
It remains to show that $|Q|\le d(d+2)(w+\ell+|U|)$.
Indeed, this gives
\[
 |D|\le \bigl(1+d(d+2)\bigr)(w+\ell+|U|)
       =b^2(w+\ell+|U|).
\]

The bound on~$Q$ is the only place that needs the density bound for minors, rather than just subgraphs.
Contract every connected component of $G-R$ that contains a vertex of~$W$ to one vertex, and retain all other vertices.
Let~$A$ be the set of contracted vertices; then $|A|\le w$.
Removing loops and keeping one edge from each group of parallel edges gives a minor~$K$ in which~$G[R]$ remains unchanged.

The components of $G-R$ that are not contracted consist entirely of vertices of original degree~$d$.
Indeed, by Lemma~\ref{simp:endpoint:budget}(i), a defect outside~$R$ belongs to $R_0\setminus R\subseteq N_G[W]$.
Since $W\cap R=\varnothing$, it lies in a component of $G-R$ meeting~$W$, which is therefore contracted.
An uncontracted component has no edges to a contracted component, so every vertex in it still has degree~$d$ in~$K$.

Every $q\in Q$ has $h(q)\ge b$, so~$q$ has at least~$d$ neighbors in~$R$, and they lie in $Q\cup L$ because no vertex of~$U$ is adjacent to~$H$.
Put $B_Q=B\cap Q$.
Such a vertex has at least~$d$ neighbors in~$R$ and degree different from~$d$, hence degree at least $d+1$ in~$G$.
Each $q\in B_Q$ has degree at least $d+1$ in~$K$: it either already has more than~$d$ neighbors in~$R$, or it has an additional neighbor outside~$R$.
In the latter case at least one such neighbor survives, either as itself or as a contracted vertex of~$A$.
A vertex of $Q\setminus B_Q$ has original degree~$d$ and at least~$d$ neighbors in $Q\cup L$, so these are all its neighbors, also in~$K$.

Thus every vertex outside $S=A\cup L\cup U\cup B_Q$ has degree~$d$ in~$K$.
Since~$K$ is a minor of~$G$, $2|E(K)|\le d|V(K)|$.
Hence
\[
 (d+1)|B_Q|\le\sum_{v\in S}\deg_K(v)
 =2|E(K)|-d|V(K)\setminus S|\le d|S|.
\]
Consequently $|B_Q|\le d(|A|+\ell+|U|)$ and $|S|\le b(|A|+\ell+|U|)$.

Finally, every $q\in Q\setminus B_Q$ is outside~$F$ but not in~$B$ and hence has a neighbor in~$B$.
This neighbor lies in $Q\cup L$, and therefore in $B_Q\cup L\subseteq S$.
The same degree bound gives
\[
 |Q\setminus B_Q|\le\sum_{v\in S}\deg_K(v)
 \le d|S|\le db(|A|+\ell+|U|).
\]
Adding the bounds on~$B_Q$ and $Q\setminus B_Q$, and using $|A|\le w$, proves $|Q|\le d(d+2)(w+\ell+|U|)$.
\end{proof}

Lemma~\ref{simp:endpoint:paying} therefore applies with $M=b^2$ and $M'=db^2$.
Since $d=2a\ge a+1$, fix
\[
 \delta=\frac{1}{db^2}=\frac{1}{2a(2a+1)^2}.
\]
Running the solver with this accuracy and applying the rounding rule gives $|D|\le\cost^*$.
The set $F\cup D$ dominates~$G$, because~$F$ covers~$N_G[F]$ and~$D$ covers $R_0=V(G)\setminus N_G[F]$.
By Lemma~\ref{simp:endpoint:budget}(iii), $|F\cup D|\le b\gf(G)$, which proves the integer case.
For planar graphs, $a=3$, $d=6$, $b=7$, $M=49$, and $M'=294$, so the fixed accuracy $\delta=1/294$ suffices.

\subsection{The non-integer case: using the gap to integer degrees}

Now assume that~$2a$ is not an integer.
We prove part~\ref{sc:thmC:sub} of Theorem~\ref{sc:thmC}; the non-integer case of part~\ref{sc:thmC:minor} follows by taking $\rho=a$.
Thus let $G\in\mathcal C_\rho$ for a fixed rational $\rho\ge a$, and suppose that every subgraph of~$G$ has a ratio of edges to vertices of at most~$a$.
Put $k=\lfloor2a\rfloor$ and $\theta=2a-k\in(0,1)$.

The solver runs on~$G$ with all vertices as demands and cost one at every vertex. With the fixed accuracy~$\delta$ chosen below, it returns a feasible vector $x\in[0,1]^{V(G)}$ with $\sum_vx_v\le(1+\delta)\gf(G)$.
Apply Lemma~\ref{simp:endpoint:rounding} with $C=R_0=V(G)$ and the constant costs $c(v)=b$.
Then $\cost(x)=b\sum_vx_v$. The same scaling gives $\cost^*=b\gf(G)$, so $\cost(x)\le(1+\delta)\cost^*$.
The hypotheses of the lemma hold, because $c=b$ everywhere and $h(v)<b$ on~$O$.

Each~$h(v)$ is an integer.
On~$H$, $h(v)\ge b=k+1+\theta$, so $h(v)-b\ge1-\theta$, and \eqref{simp:endpoint:excess} gives $(1-\theta)|H|\le a\ell$.
On~$O$, $h(v)\le k+1$ and $c(v)-h(v)\ge\theta$, so \eqref{simp:endpoint:leftover} gives $|U|\le(k+1)P/\theta$.
Hence
\[
 |D|=w+|H|+|U|\le M(w+\ell)+M'P,\qquad
 M=\frac{a}{1-\theta},\quad M'=\frac{k+1}{\theta},
\]
using $M\ge1$.

Lemma~\ref{simp:endpoint:paying} applies with the fixed rational accuracy
\[
 \delta=\min\left\{\frac{1-\theta}{a(a+1)},\frac{\theta}{k+1}\right\},
\]
and gives $|D|\le\cost^*=b\gf(G)$, which proves part~\ref{sc:thmC:sub} of Theorem~\ref{sc:thmC}, and part~\ref{sc:thmC:minor} when~$2a$ is not an integer.

\paragraph{Proof of the second assertion of Theorem~\ref{sc:thmB}.}
In the integer case, we select~$F$ and assign the costs~$c$, then apply the rounding rule of Lemma~\ref{simp:endpoint:rounding} with $C=V(G)\setminus F$.
Lemma~\ref{simp:endpoint:budget}(iii) bounds the transformed optimum by $b\gf(G)-|F|$.
Lemmas~\ref{simp:endpoint:rounding} and~\ref{simp:endpoint:count} give slack $\delta=1/(db^2)$ by Proposition~\ref{sc:principle} with the quantities $w+\ell$ and~$P$.
In the non-integer case no initial selection is needed and the same quantities give the slack~$\delta$ chosen above.
In both cases the rounding needs only the three broadcasts described next.

\paragraph{Implementation of both cases.}
All parameters are fixed rational numbers.
In the integer case, a fixed number of local broadcasts identifies the defects, the set~$F$, and the costs. The solver runs on $G-F$ while ignoring edges leaving it; the original identifiers and message bound remain available in the induced subgraph.
In both cases three further broadcasts announce~$W$,~$R$, and~$H$, after which each vertex decides whether it belongs to~$U$.

Components of $G-R$ may have unbounded diameter.
Their contraction is used only in the proof: the algorithm does not find these components or perform any contractions.

\section{Approximation schemes}\label{sc:schemes}

We now prove Theorem~\ref{sc:thmD}.
In both cases only the partition step of the solver changes: instead of averaging over all partitions, one partition is used, which is either sampled or computed.
Each part then solves its task \emph{integrally} and exactly, and the same disjointness argument compares the result with the integral optimum.

The setting is that of \gcov{}, with integral optimum~$\OPT_c$; the solver is run up to and including the construction of the bounded-degree owner graph~$J_0$.
This gives the initial cover~$D_0$ with $N=|D_0|$, the heavy set~$P$, the high-degree owners~$Q$, and the partners.

\paragraph{Integral tasks.}
For a part~$C$ of any partition of~$J_0$, let~$\OPT_C$ be the minimum cost of an integral cover of its task, using the candidates of~$H_C$.

\begin{lemma}\label{sc:integral-task}
A deterministic broadcast CONGEST algorithm with a parameter-bounded number of rounds finds an integral cover of the task of~$C$ of cost~$\OPT_C$, simultaneously for all parts of one partition.
For every partition, $\sum_C\OPT_C\le\OPT_c$.
\end{lemma}

\begin{proof}
The search in the proof of Lemma~\ref{simp:local-solver}, with the changes for \gcov{}, is run with multiplicity $q=1$ and the copy budget $M=2B_0k$ in place of $\lceil(q+2p)K_0\rceil$; the candidates are the admissible vertices of~$H_C$.
The owners of~$C$ and their partners form an integral cover of cost at most $2B_0k$, and costs are at least one, so an optimal integral cover has at most~$M$ vertices.
The search finds a cheapest cover by at most~$M$ copies, because at every branching state one branch remains consistent with a fixed optimal completion, and the terminal step finds a cheapest completion directly.
A cover with a repeated copy is never cheaper than the same cover without the repetition, so the result is an optimal integral cover.

For the second assertion, restrict an optimal integral cover of the whole instance to the sets~$T_C$ of candidates that cover some demand of the task of~$C$.
The restriction to~$T_C$ covers the task, and the sets~$T_C$ are pairwise disjoint, by the proof of Lemma~\ref{simp:tasks}.
\end{proof}

For one partition with boundary owners~$B$, the output is
\[
 Y=P\cup\bigcup_{i\in Q\cup B}\{i,\pi(i)\}\cup\bigcup_{C\text{ a part}}Z^C,
\]
where~$Z^C$ is the cover of Lemma~\ref{sc:integral-task}.
As for the vector~$X$ in \gcovsec{},~$Y$ covers every demand, and its cost is at most
\begin{equation}\label{sc:scheme-cost}
 \OPT_c+B_0\bigl(|P|+2|Q|+2|B|\bigr)
 \le\OPT_c+B_0\Bigl(\frac{2a}L+\frac{4C_L}\lambda\Bigr)N+2B_0|B|.
\end{equation}
Here~$\gamma_R$ denotes the minimum cardinality of a cover.
The initializer, extended as for \gcov{}, gives $N\le K_a\gamma_R\le K_a\OPT_c$, since the costs are at least one, and~$J_0$ has at most~$N$ vertices.
Thus it suffices to find a partition of~$J_0$ into connected parts of size at most~$k$ with $\mathbb E|B|\le2\eta N$, or $|B|\le2\eta N$, for a sufficiently small fixed~$\eta$.

\paragraph{The randomized scheme.}
Every owner of~$J_0$ draws the fair bits of its slots, and the local rule of Lemma~\ref{simp:partition} runs once, each round being simulated on~$J_0$ by Lemma~\ref{simp:simulation}.
Every owner learns the identifier list of its part, and its owner of smallest identifier sends the task record to the physical vertices, as in Section~\ref{simp:averaging}.
A light vertex now belongs to at most~$L$ tasks, one for each part meeting its support, so the tasks can run in parallel as before.
By \eqref{simp:boundary-expectation} we have $\mathbb E|B|\le2\eta N$, and choosing
\[
 L\ge\max\Bigl\{2,4a,\frac{8B_0aK_a}{\varepsilon}\Bigr\},\qquad
 \lambda\ge\frac{16B_0C_LK_a}{\varepsilon},\qquad
 \eta\le\frac{\varepsilon}{16B_0K_a}
\]
makes the expected cost at most $(1+\varepsilon)\OPT_c$ by \eqref{sc:scheme-cost}.
The output is a feasible cover for every outcome of the bits, which proves the first part of Theorem~\ref{sc:thmD}.

\paragraph{A deterministic partition.}
For the second part, the random acceptance rule is replaced by a coloring.
The standard convention is that the vertices know a common polynomial upper bound on the identifiers, under which Linial's color reduction~\cite{Linial1992} properly colors a graph of maximum degree~$\Delta$ with~$\mathcal O(\Delta^2)$ colors in~$\mathcal O(\log^*n)$ rounds, sending only current colors.

\begin{lemma}\label{sc:det-partition}
Under the hypotheses of Lemma~\ref{simp:partition}, and with this convention, a deterministic algorithm partitions~$V(H)$ into connected sets of size at most~$k$ with at most $\eta|V(H)|$ crossing edges.
On a bounded-degree graph it uses~$\mathcal O(\log^*n)$ broadcast CONGEST rounds, with a constant depending only on $\lambda,C,\eta$.
\end{lemma}

\begin{proof}
The integer~$k$, the candidate sets, and the constants~$A$ and~$M$ are those of the proof of Lemma~\ref{simp:partition}.
First color the power~$H^{2k}$ properly with a palette of $\Phi=\mathcal O(\Delta_k^2)$ colors, where $\Delta_k=\sum_{j=1}^{2k}\lambda^j$ bounds the maximum degree of~$H^{2k}$; one round on~$H^{2k}$ is simulated by bounded flooding in~$H$.
A candidate set is colored by the pair formed by the color of its smallest-identifier vertex and its index among the at most~$A$ sets assigned to that vertex.
Two distinct intersecting sets receive different colors: if their smallest-identifier vertices are different, these are within distance~$2k-2$ of each other, and otherwise their indices differ.

In each stage, with remaining set~$U$, eligibility is defined as before and the~$\Phi A$ colors are processed in a fixed order.
A set of the current color is accepted if it is eligible and disjoint from all sets accepted earlier in the stage.
Sets of one color are disjoint, so they can be decided together.
The decision for a color depends on the decisions for the earlier colors within distance~$2k$, so all decisions of a stage depend only on the stage-start state within distance $2k\Phi A+k+1$.
At the end of the stage, every eligible set meets an accepted set.

The existence argument of Lemma~\ref{simp:partition} gives disjoint eligible sets covering at least $|U|/2$ vertices, hence at least $|U|/(2k)$ such sets.
Each meets an accepted set, and each accepted set meets at most~$M$ candidate sets, so at least $|U|/(2kM)$ sets are accepted and at least that many vertices are removed.
Thus $|U'|\le(1-1/(2kM))|U|$ deterministically.
Run~$T$ stages with $(1-1/(2kM))^T\le\eta/\lambda$ and make the remaining vertices singletons.
The count of crossing edges in the proof of Lemma~\ref{simp:partition} used only the eligibility test and the number of remaining vertices, so it gives at most $\eta|V(H)|$ crossing edges.
The coloring takes~$\mathcal O(\log^*n)$ rounds, and the~$T$ stages take a parameter-bounded number.
\end{proof}

\begin{proof}[Proof of Theorem~\ref{sc:thmD}]
The first part was proved above.
For the second, the partition of Lemma~\ref{sc:det-partition} is computed on~$J_0$, with each round simulated by Lemma~\ref{simp:simulation}; colors and identifiers have~$\mathcal O(\log n)$ bits.
The boundary owners then satisfy $|B|\le2\eta N$, and with the parameters chosen above, \eqref{sc:scheme-cost} gives a feasible cover of cost at most $(1+\varepsilon)\OPT_c$.
All other steps use a parameter-bounded number of rounds.
\end{proof}

The matching round lower bound is proved in Appendix~\ref{app:rounds-lower}.

\section{Exactly seven on every fixed surface}
\label{new:surface}

The density bound for a fixed surface has a constant extra term, and we now show how we can remove this term while keeping the same approximation factor.
For this, we solve exactly those components of the \emph{remaining covering problem} whose demands can be covered by a bounded number of candidates, so that we can use the budget that remains after our first selection.
Solving components of the original graph instead would not be enough, since this first selection may already spend most of the original budget.

To handle the extra term, we use a more accurate fractional solution and keep part of the rounding saving available.
The proof then uses two budgets for the remaining problem.
If every remaining incidence component is solved exactly, the integral budget bounds the size of the answer.
Otherwise, the rounding bound shows that the remaining fractional optimum is large enough for the unused saving to pay for the genus term, so the fractional budget gives the required bound.
These are two cases in the proof; the algorithm makes its replacements locally and does not need a global test to distinguish them.

\begin{theorem}
\label{new:surface-seven}\label{fv:imp:surface}
For every fixed integer $e\ge0$, there is a deterministic constant-round broadcast CONGEST algorithm that finds a dominating set~$D$ with
\[
 |D|\le7\gamma(G)
\]
on every graph~$G$ of Euler genus at most~$e$.
The algorithm uses~$\mathcal O(\log n)$ bits per message.
It needs no embedding, orientation, degree bound, initial cover, or knowledge of~$n$.
The statement includes both orientable and non-orientable surfaces.
\end{theorem}

We use the usual Euler genus, which is the least Euler genus of a closed surface in which the graph embeds.
An orientable surface of genus~$h$ has Euler genus~$2h$, while a non-orientable surface with~$h$ crosscaps has Euler genus~$h$.
For every minor~$J$ of~$G$, we have the surface density bound
\begin{equation}
 \label{new:surface-density}
 |E(J)|\le3|V(J)|+3e.
\end{equation}
We give the argument here so that the bound also covers disconnected graphs.
The empty graph satisfies the bound directly, and for a nonempty graph embedded in a connected surface of Euler genus at most~$e$, we first join its components by bridges without changing the surface.
Whenever more than one component remains, some region of the complement has two distinct graph components on its boundary, since otherwise the surface would be disconnected.
We can therefore draw a bridge through that region, and we repeat this step until we have a connected simple supergraph on the same vertices.

In this connected graph, we repeatedly delete vertices of degree at most one, and if nothing remains, the graph is a tree and the bound holds.
Otherwise, the remaining graph is connected and has minimum degree at least two, so we take a cellular embedding of this graph in a surface of its minimum Euler genus $e'\le e$.
Every facial boundary walk has length at least three, because an immediate reversal would require a vertex of degree one, and a simple graph has no loop or cycle of length two.
If the remaining graph has~$n'$ vertices,~$m'$ edges and~$f'$ faces, we use Euler's formula and count edge sides to obtain
\[
 n'-m'+f'=2-e',\qquad 3f'\le2m',
 \qquad m'\le3n'-6+3e'\le3n'+3e.
\]
When we restore a deleted vertex, we add at most one edge and therefore preserve the last bound, and we can then delete the added bridges without increasing the number of edges.
Finally, deleting and contracting edges in an embedding do not increase its Euler genus, so we can apply the same argument to every~$J$.

We therefore have $G\in\mathcal C_{3+3e}$, which allows us to use our fractional solver.
The coefficient seven in the theorem is optimal, since every surface class considered here contains all planar graphs, and their deterministic constant-round lower bound already holds in the LOCAL model.

\subsection{The remaining problem has two budgets}

We begin with the same first selection as in the planar proof, and we put
\[
 B=\{v:\deg_G(v)\ne6\},\qquad
 F=V(G)\setminus N_G[B],\qquad
 R_0=V(G)\setminus N_G[F].
\]
After we select~$F$, we use every vertex in $C=V(G)\setminus F$ as a candidate for covering~$R_0$, and we give it cost
\[
 c(v)=7-|N_G(v)\cap F|.
\]
We write~$\tau'$ for the fractional minimum cost of this problem,~$\OPT'_c$ for its integral minimum cost, and~$\OPT'_1$ for its integral minimum cardinality.
All costs lie in $\{1,\ldots,7\}$, so $\OPT'_1\le\OPT'_c$.

The budget calculation in Section~\ref{new:planar} gives the first inequality below, and its integral version gives the second:
\begin{equation}
 \label{new:surface-budgets}
 |F|+\tau'\le7\gamma_f(G),
 \qquad
 |F|+\OPT'_c\le7\gamma(G).
\end{equation}
To see the integral bound directly, we let~$Y$ be a minimum dominating set of~$G$.
We can use $Y\setminus F$ to cover~$R_0$, since a demand has no neighbor in~$F$, and hence
\[
 \OPT'_c\le\sum_{v\in Y\setminus F}c(v).
\]
We now count the vertices of~$F$ that are covered by~$Y$.
Each vertex $v\in Y\cap F$ covers at most seven vertices of~$F$, while each vertex $v\in Y\setminus F$ covers exactly $7-c(v)$ vertices of~$F$, so we have
\[
 |F|\le7|Y\cap F|+
             \sum_{v\in Y\setminus F}(7-c(v)).
\]
By adding these two inequalities, we obtain the integral bound.
We obtain the fractional bound by the same calculation, with the coordinates of a feasible fractional dominating vector in place of the indicator of~$Y$.

\subsection{The rounding loses only a constant}

We take any feasible fractional cover $x\in[0,1]^C$ of~$R_0$ and write $\cost(x)=\sum_{v\in C}c(v)x_v$ for its cost.
As in the planar proof, we apply our three-step rule with threshold $1/4$:
\[
\begin{aligned}
 W&=\{v\in C:c(v)=7,\ x_v>1/4\},
       &R&=R_0\setminus N_G[W],\\
 h(v)&=|N_G[v]\cap R|\quad(v\in C\setminus W),
       &H&=\{v\in C\setminus W:h(v)\ge7\},\\
 U&=R\setminus N_G[H],
       &D_0&=W\cup H\cup U.
\end{aligned}
\]
This gives us a set~$D_0$ that covers~$R_0$, and for our count we put
\[
 Q=H\cap R,\quad L=R\setminus(H\cup U),\quad
 O=C\setminus(W\cup H),\quad
 w=|W|,\quad\ell=|L|,
\]
and define
\[
 P=\sum_{v\in O}(c(v)-h(v))x_v.
\]
The degree properties in \eqref{new:planar-cost-properties} hold on every graph, so they also apply here.
In particular, we still have $c(v)=7$ on $W\cup H$, and every $v\in O$ satisfies $h(v)\le6$ and $c(v)-h(v)\ge1$, so $P\ge0$.

\begin{lemma}
\label{new:surface-rounding}
For every such vector~$x$,
\begin{align}
 \cost(x)-|D_0|&\ge\frac{w+\ell}{4}+P-\frac32e,
       \label{new:surface-saving}\\
 |D_0|&\le49(w+\ell)+294P+\frac{99}{2}e.
       \label{new:surface-count}
\end{align}
Consequently,
\begin{equation}
 \label{new:surface-slack}
 \left(1+\frac1{294}\right)|D_0|\le\cost(x)+2e.
\end{equation}
\end{lemma}

\begin{proof}
We use the proof of Lemma~\ref{prop:planar-rounding}, with $D_0$ in place of~$D$, and keep track of the extra terms in the surface density bound~\eqref{new:surface-density}.
Put $z=|H\setminus R|$ and $E=\sum_{v\in H}(h(v)-7)$.
The incidence count now gives
\[
 \sum_{v\in H}h(v)
 \le |E(G[H\cup L])|+|E(G[Q])|+|Q|
 \le3(|H|+\ell)+4|Q|+6e.
\]
Since $|H|=|Q|+z$, we obtain
\begin{equation}\label{new:surface-excess}
 E+4z\le3\ell+6e,
\end{equation}
and hence $z\le\ell+3e/2$.
The covering-constraint calculation is unchanged:
\[
 \cost(x)\ge\frac74w+|R|+P-\frac14E.
\]
Subtracting $|D_0|=w+|R|-\ell+z$ and using \eqref{new:surface-excess} gives
\[
 \cost(x)-|D_0|\ge\frac34w+\frac14\ell+P-\frac32e,
\]
which implies \eqref{new:surface-saving}.
The bound on~$U$ uses only the covering constraints and the costs, so it also remains unchanged:
\begin{equation}\label{new:surface-U}
 |U|\le\sum_{v\in O}h(v)x_v\le6P.
\end{equation}

Construct the minor~$K$ and define $A,B_Q,T,S$ as in the planar proof.
The degree properties and the existence of a defect neighbor for every vertex of $Q\setminus B_Q$ use only the first selection and the rounding rule, so they still hold.
The minor has Euler genus at most~$e$, and its degree sum gives
\[
 7|B_Q|\le\sum_{v\in S}\deg_K(v)
 =2|E(K)|-6|V(K)\setminus S|
 \le6|S|+6e.
\]
Since $|S|=T+|B_Q|$, we obtain
\begin{equation}\label{new:surface-defect-count}
 |B_Q|\le6T+6e,\qquad |S|\le7T+6e.
\end{equation}
Counting an edge to a defect neighbor for each vertex of $Q\setminus B_Q$ gives
\[
 |Q\setminus B_Q|\le\sum_{v\in S}\deg_K(v)
 \le6|S|+6e\le42T+42e.
\]
Thus $|Q|\le48T+48e\le48(w+\ell+|U|)+48e$.
Adding $w+z+|U|$ and using the bounds on~$z$ and~$U$ proves
\[
 |D_0|\le49(w+\ell+|U|)+\frac{99}{2}e
       \le49(w+\ell)+294P+\frac{99}{2}e.
\]
Finally, dividing by~$294$ and using \eqref{new:surface-saving} gives
\[
 \frac{|D_0|}{294}
 \le\cost(x)-|D_0|+
       \left(\frac32+\frac{99}{588}\right)e
 =\cost(x)-|D_0|+\frac{327}{196}e.
\]
Since $327/196<2$, this proves \eqref{new:surface-slack}.
\end{proof}

We use the contracted graph only for this count, so our algorithm does not need to find the components of $G-R$, whose diameter may be large.

\subsection{Solving small remaining components}

We now consider the incidence graph~$I$ of our residual problem, whose vertex set is~$C$ and which has an edge~$rv$ if $r\in R_0$, $v\in C\cap N_G(r)$.
This is a subgraph of $G-F$, and a candidate and each demand it covers lie in the same component of~$I$.
We mark a non-demand vertex isolated in~$I$ as solved with the empty cover, so every other component that we consider contains demands.

On these components, we use our exact small-component routine with \emph{unit} costs and budget
\begin{equation}
 \label{new:surface-K}
 K_0=\max\{1,1176e\}.
\end{equation}
With this budget, we solve every component~$Z$ for which the minimum number of candidates covering its demands is at most~$K_0$, and on every solved component we return a cover of minimum cardinality.
We can apply the routine because~$I$ has subgraph density at most $3+3e$ and its incidence relation is locally known.

This routine remains local even when a component has many vertices.
A connected incidence component with a cover of at most~$K_0$ vertices has diameter at most $5K_0-1$, since every demand is within distance one of that cover and every non-isolated candidate is within distance two.
To see this in detail, we assign each vertex to a cover vertex at distance at most two and join pairs of cover vertices whose distance is at most five.
These cover vertices form a connected graph, since otherwise an edge of~$I$ between two assignment classes would join the two supposed sides.
We then use a path through at most~$K_0$ cover vertices to obtain the diameter bound.

We put $d_0=5K_0-1$ and first flood the minimum identifier for~$d_0$ rounds in~$I$.
We then flag each vertex that has a neighbor with a different resulting label, and we flood these flags for~$2d_0$ rounds.
A vertex accepts if it sees no flag, and with this rule all vertices of an incidence component accept together or reject together.
To verify this, we take a vertex~$v$ in a component with more than one label and follow a path from~$v$ to a different label, writing~$w$ for the vertex just before the first change.
The vertex whose identifier is their common label lies within distance~$d_0$ of both~$v$ and~$w$, so~$v$ sees the flag at~$w$ within distance~$2d_0$ and rejects.
If a component has only one label, every vertex lies within distance~$d_0$ of its label vertex, so the component has diameter at most~$2d_0$.
In particular, every component of diameter at most~$d_0$ accepts.

On each accepted component, we run the bounded-diameter integral search with copy budget~$K_0$.
We use the same search as for integral owner tasks in Lemma~\ref{sc:integral-task}, with the component itself as our communication graph.
By the bounded branching rule, we find a smallest cover of size at most~$K_0$ if one exists, without gathering the graph.
Such a cover is also globally minimum in its component, since any better cover would have size at most~$K_0$ and would occur in the same search.
We mark precisely these components as solved, and their decisions and their output are known throughout the component.
Since the budgets, diameters and density bound are fixed, we use a constant number of rounds and records of~$\mathcal O(\log n)$ bits, even when the components have many vertices or some vertices have large degree.

\subsection{The final algorithm and its guarantee}

\begin{proof}[Proof of Theorem~\ref{new:surface-seven}]
We first select~$F$ and set up the residual problem, and we then run our weighted partial fractional solver on $G-F$ with density parameter $3+3e$, cost bound seven, and fixed accuracy
\[
 \delta'=\frac1{588}.
\]
This gives us a feasible $x\in[0,1]^C$ with $\cost(x)\le(1+\delta')\tau'$, which we round to~$D_0$ as above.
We also run our exact component routine with budget~$K_0$, and in every solved component~$Z$, we replace $D_0\cap Z$ by its minimum-cardinality cover~$D_Z$.
We keep the choices of~$D_0$ in all other components and write~$D_1$ for the resulting cover of~$R_0$.

Since no candidate covers demands in different incidence components, we know that $D_0\cap Z$ covers the demands of~$Z$, so replacing it by~$D_Z$ cannot increase cardinality.
We therefore have
\[
 |D_1|\le|D_0|,
 \qquad F\cup D_1\text{ dominates }G.
\]
We put $\delta=1/294$ and $q=2e$.
By \eqref{new:surface-slack},
\begin{equation}
 \label{new:surface-final-bound}
 (1+\delta)|D_0|\le(1+\delta')\tau'+q.
\end{equation}
We now consider two cases to prove the guarantee, although our algorithm uses the same steps in both cases and does not need to tell them apart.

If all incidence components are solved, then we have $|D_1|=\OPT'_1$, and the integral budget in~\eqref{new:surface-budgets} gives
\[
 |F\cup D_1|=|F|+\OPT'_1
       \le|F|+\OPT'_c\le7\gamma(G).
\]
Otherwise, some unsolved component has minimum cover size greater than~$K_0$, and since~$D_0$ covers that component, we have $|D_0|>K_0$.
By our choice of~$K_0$, we also have
\[
 K_0\ge\frac{q}{\delta-\delta'}=1176e.
\]
We combine this with \eqref{new:surface-final-bound} and obtain
\[
 \tau'>\frac{(1+\delta)K_0-q}{1+\delta'}
       \ge\frac{q}{\delta-\delta'}.
\]
For the last step, we substitute $K_0\ge q/(\delta-\delta')$, after which the numerator becomes $q(1+\delta')/(\delta-\delta')$.
We therefore have $q\le(\delta-\delta')\tau'$, and we can use \eqref{new:surface-final-bound} to conclude that $|D_1|\le|D_0|\le\tau'$.
The fractional budget now gives
\[
 |F\cup D_1|\le|F|+\tau'
       \le7\gamma_f(G)\le7\gamma(G).
\]

In all these steps, we use a number of rounds that depends only on~$e$.
We find degrees, defects,~$F$,~$R_0$, costs and the incidence flags with a fixed number of broadcasts.
When we work on $G-F$ or on~$I$, each vertex broadcasts one message to all its physical neighbors and ignores messages from edges outside the simulated graph.
Our fractional solver and exact component search use bounded lists and~$\mathcal O(\log n)$-bit records, while the three rounding steps use Boolean broadcasts.
Finally, each solved component makes its own replacement, so we carry out the whole algorithm without a global test, an embedding, or a component contraction.
\end{proof}

We need the integral benchmark in this theorem, since Proposition~\ref{sc:gap} gives a fixed-surface example with $\gamma(G)=15$ and $\gamma_f(G)\le2$, which rules out the bound~$7\gamma_f(G)$ on all fixed surfaces.

\label{new:core-end}

\clearpage
\begingroup
\phantomsection
\addcontentsline{toc}{section}{References}
\renewcommand{\bibfont}{\normalsize}
\setlength{\bibsep}{4pt}
\interlinepenalty=10000
\newcommand{\etalchar}[1]{$^{#1}$}

\endgroup

\clearpage
\appendix
\part*{Appendices}
\addtocontents{toc}{\protect\newpage}
\addcontentsline{toc}{part}{Appendices}

\begingroup
\section{Planar graphs of large girth}\label{sc:girth}
\endgroup

\noindent We prove part~\ref{sc:thmC:girth} of Theorem~\ref{sc:thmC}.
The example shows how the framework separates its hypotheses: the solver uses the planar minor-density bound, while the rounding rule uses the subgraph density bound.
Fix an integer $g\ge3$ and put
\[
 a=\frac g{g-2},\qquad 2a+1=3+\frac4{g-2}.
\]
For $g=3$ this is the planar case of Section~\ref{simp:endpoint:section}.
For $g\ge5$ with $g\ne6$, the number $2a$ is not an integer. To apply the non-integer case, it therefore suffices to show that every subgraph has density at most~$a$.
The two remaining girths, four and six, give an integer value of $d=2a$.
Here the one step of the integer case that uses the density of minors, the count of Lemma~\ref{simp:endpoint:count}, must be replaced, because contraction does not preserve girth.
The replacement is a count of short faces, which charges every face shorter than~$g$ to the contracted vertices that it meets.

\begingroup
\subsection{Short faces at specified vertices}
\endgroup

\begin{lemma}
\label{sc:face}
Let $g\ge3$ and suppose that~$K$ is a simple planar graph with an independent set $A\subseteq V(K)$ such that $K-A$ has girth at least~$g$.
If~$p$ edges of~$K$ are incident with~$A$, then
\[
 |E(K)|\le\frac g{g-2}|V(K)|+\frac{g-3}{g-2}\,p.
\]
In particular, with $A=\varnothing$, every planar graph of girth at least~$g$ has a ratio of edges to vertices of at most $a=g/(g-2)$, and so has each of its subgraphs.
\end{lemma}

\begin{proof}[\proofname]
The bound is immediate when~$K$ is empty. If~$K$ has a vertex~$v$ of degree at most one, the hypotheses still hold for $K-v$ with $A\setminus\{v\}$, and deleting~$v$ removes at most one edge while it decreases the right side by at least $a\ge1$.
By induction on the number of vertices, we may therefore assume that every vertex has degree at least two.

Fix a plane drawing and consider one component, with~$n$ vertices,~$m$ edges,~$f$ faces, and~$p_0$ edges incident with~$A$.
Every face has a closed boundary walk, and the lengths of these walks add up to~$2m$, since every edge has two sides.
A boundary walk never immediately reverses along the preceding edge, because every vertex has degree at least two.

A closed walk without immediate reversals in a simple graph contains a cycle of length at most the length of the walk: a shortest closed subwalk between two occurrences of the same vertex has length at least three, since there are no loops and no reversals, and its vertices are distinct.
Consequently every boundary walk has length at least three, and a walk of length $j<g$ contains a cycle shorter than~$g$, which must meet~$A$.
At such a vertex, the walk enters and leaves along two distinct edge sides incident with~$A$.

A face of length $j<g$ falls short of $g$ by $g-j$. To bound the sum of these differences, we will distribute them among the $p_0$ edges incident with~$A$ so that each edge receives at most $g-3$.
For each such face, we charge half of this difference, at most $(g-3)/2$, to each of these two edge sides.
Since~$A$ is independent, each edge side is charged at most once. Every edge incident with~$A$ has two sides, so the sum of these differences is at most $(g-3)p_0$, and
\[
 2m\ \ge\ gf-(g-3)p_0.
\]
Euler's formula $f=m-n+2$ now gives $(g-2)m\le g(n-2)+(g-3)p_0$, which is stronger than the claim for this component.
Summing over the components proves the lemma.
\end{proof}

\begingroup
\subsection{The count at girths four and six}
\endgroup

\noindent Let $g\in\{4,6\}$, so that $d=2a$ is an integer, and let~$G$ be planar of girth at least~$g$.
By Lemma~\ref{sc:face}, every subgraph of~$G$ has a ratio of edges to vertices of at most~$a$.
We use the construction from the integer case in Section~\ref{simp:endpoint:section} with the same notation: the defects~$B$, the selected set~$F$, the demands~$R_0$, the costs~$c$, and the rounding of Lemma~\ref{simp:endpoint:rounding} with the sets $W,R,H,U,D,Q,L$ and the quantities $w,\ell,P$.
All statements before Lemma~\ref{simp:endpoint:count} use only the degrees and the density of subgraphs, so they remain valid; the solver runs on the planar graph $G-F$, with $\rho=3$.
It remains to bound $|D|$.

\begin{lemma}
\label{sc:girth:count}
For every feasible vector~$x$ on $V(G)\setminus F$, $|D|\le b^2(w+\ell)+db^2P$ for $g=4$, and $|D|\le41(w+\ell)+63P$ for $g=6$.
\end{lemma}

\begin{proof}[\proofname]
Let $\kappa=(g-4)/(g-2)$, so $\kappa=0$ for $g=4$ and $\kappa=1/2$ for $g=6$.
Construct the minor~$K$ as in the proof of Lemma~\ref{simp:endpoint:count}: contract every component of $G-R$ that contains a vertex of~$W$, denote the set of contracted vertices by~$A$, so $|A|\le w$, and keep all other vertices.
Let~$p$ be the number of edges of~$K$ incident with~$A$.
We first show that
\[
 |Q|\le d(d+2)(w+\ell+|U|)+(d+2)\kappa p,\qquad
 p\le(g+4)w+(g+2)\ell+2|U|.
\]
To apply Lemma~\ref{sc:face}, observe first that~$K$ is planar, being a minor of~$G$.
The set~$A$ is independent, because different components of $G-R$ are not adjacent.
And $K-A$ is the subgraph of~$G$ induced by the vertices outside the contracted components, since the other vertices are unchanged; it therefore has girth at least~$g$.
Lemma~\ref{sc:face} gives
\[
 2|E(K)|\le d|V(K)|+\frac{2(g-3)}{g-2}\,p.
\]

As in the proof of Lemma~\ref{simp:endpoint:count}, put $B_Q=B\cap Q$ and $S=A\cup L\cup U\cup B_Q$.
Every vertex outside~$S$ has degree exactly~$d$ in~$K$, and every vertex of~$B_Q$ has degree at least $d+1$; both facts were proved there without using the density of~$K$.
Since~$A$ is independent, its vertices have degree sum~$p$.
Hence
\[
 \sum_{v\in S\setminus A}\deg_K(v)
 =2|E(K)|-d|V(K)\setminus S|-p
 \le d|S|+\Bigl(\frac{2(g-3)}{g-2}-1\Bigr)p
 =d|S|+\kappa p.
\]
This is the only change from the earlier proof, which had the bound $d|S|$.
The same degree count gives $(d+1)|B_Q|\le d|S|+\kappa p$, hence $|B_Q|\le d(|A|+\ell+|U|)+\kappa p$ and $|S|\le b(|A|+\ell+|U|)+\kappa p$.
Every vertex of $Q\setminus B_Q$ has a neighbor in $B_Q\cup L\subseteq S\setminus A$, so
\[
 |Q\setminus B_Q|\le\sum_{v\in S\setminus A}\deg_K(v)\le d|S|+\kappa p
 \le db(|A|+\ell+|U|)+(d+1)\kappa p.
\]
Adding the two bounds and using $|A|\le w$ proves the bound on $|Q|$.

It remains to bound~$p$, which matters only when $\kappa>0$.
The vertices of~$A$ are adjacent only to vertices of $R=Q\cup L\cup U$.
The edges between~$A$ and~$L$ form a simple bipartite planar graph, which has at most twice as many edges as vertices, so there are at most~$2(|A|+\ell)$ of them. Similarly, at most~$2(|A|+|U|)$ edges join~$A$ and~$U$. The bound of twice the number of vertices also holds for graphs with at most two vertices.
For the edges between~$A$ and~$Q$, whose number is denoted by~$p_Q$, we apply Lemma~\ref{sc:face} to the subgraph~$K'$ of~$K$ on $Q\cup L\cup A$ that keeps only the edges inside~$Q$, between~$Q$ and~$L$, and between~$Q$ and~$A$.
Its subgraph $K'-A$ is a subgraph of~$G$, so
\[
 |E(K')|\le a(|Q|+\ell+|A|)+\frac{g-3}{g-2}\,p_Q.
\]
On the other hand, every vertex of~$Q$ has at least~$d$ neighbors in $Q\cup L$, as $h\ge b$ on~$H$, so $2e(Q)+e(Q,L)\ge d|Q|$, while $e(Q)\le a|Q|$ by Lemma~\ref{sc:face}.
Therefore
\[
 |E(K')|=e(Q)+e(Q,L)+p_Q\ge d|Q|-e(Q)+p_Q\ge a|Q|+p_Q.
\]
Comparing the two bounds gives $p_Q\le(g-2)a(\ell+|A|)=g(\ell+|A|)$, and adding the three groups of edges proves the bound on~$p$.

Finally, $|D|=w+|H\setminus R|+|Q|+|U|$, with $|H\setminus R|\le\ell$ and $|U|\le dP$, as in Lemma~\ref{simp:endpoint:count}.
For $g=4$ we have $\kappa=0$, and the bound is that of Lemma~\ref{simp:endpoint:count}.
For $g=6$ we have $d=3$, $b=4$, and $\kappa=1/2$, so
\[
 |Q|\le15(w+\ell+|U|)+\tfrac52\bigl(10w+8\ell+2|U|\bigr)
 =40w+35\ell+20|U|,
\]
and $|D|\le41w+36\ell+21|U|\le41(w+\ell)+63P$.
\end{proof}

\begin{proof}[Proof of part~\ref{sc:thmC:girth} of Theorem~\ref{sc:thmC}]
For $g=3$, part~\ref{sc:thmC:minor} applies with $a=3$.
For $g\ge5$ with $g\ne6$, Lemma~\ref{sc:face} shows that every subgraph has a ratio of edges to vertices of at most $a=g/(g-2)\le3$, and~$2a$ is not an integer, so part~\ref{sc:thmC:sub} applies with $\rho=3$.
For $g\in\{4,6\}$, Lemma~\ref{sc:girth:count} gives the hypothesis of Lemma~\ref{simp:endpoint:paying} with $M=b^2$ and $M'=db^2$ for $g=4$, and with $M=41$ and $M'=63$ for $g=6$.
Run the solver with the fixed accuracies
\[
 \delta=\frac1{100}\quad(g=4),\qquad
 \delta=\min\Bigl\{\frac1{(a+1)41},\frac1{63}\Bigr\}=\frac2{205}\quad(g=6),
\]
which yields $|D|\le\cost^*$.
As in the integer case, Lemma~\ref{simp:endpoint:budget}(iii) then shows that $F\cup D$ is a dominating set of size at most $b\gf(G)$, giving approximation factors of five for $g=4$ and four for $g=6$.
The faces, the contractions, and the drawing are used only in the analysis.
\end{proof}

\section{Optimality of the approximation coefficients}\label{sc:optimality}

To prove Corollary~\ref{sc:optimal}, we need two facts for each class: an upper bound on the density of minors, which places it in some~$\mathcal C_a$, and a family of graphs in the class on which no deterministic constant-round algorithm beats $2a+1$.
We prove the lower bounds in the LOCAL model and against~$\gamma(G)$, so, because $\gf(G)\le\gamma(G)$, they also rule out smaller coefficients against~$\gf(G)$.

For $r\ge1$, the~$r$-th power~$P_n^r$ of a path has vertex set $\{1,\ldots,n\}$, with $i\ne j$ adjacent when $|i-j|\le r$.
Its pathwidth and treewidth equal~$r$ when $n\ge r+1$, so it has no~$K_{r+2}$ minor, and~$P_n^3$ is planar.
The lower bound $2r+1$ on these graphs is due to Amiri and Siebertz~\cite[Theorem~8.7]{AS2026}.
Proposition~\ref{new:ordinary-lower} gives its full proof, together with the grid bounds, using one common Ramsey argument.
The statement allows the vertices to know~$n$ and uses identifiers exactly $1,\ldots,n$.

\begin{proof}[Proof of Corollary~\ref{sc:optimal}]
\emph{Planar graphs and Euler genus at most two.}
A simple graph with $n\ge3$ vertices and Euler genus~$e$ has at most $3n-6+3e$ edges, and Euler genus is additive over components and does not increase under taking minors.
Hence all minors of a graph of Euler genus at most two have a ratio of edges to vertices of at most three, so the class lies in~$\mathcal C_3$, and Theorem~\ref{sc:thmC} gives coefficient seven.
The lower bound of Hilke, Lenzen, and Suomela~\cite{HLS2014} holds on planar graphs, which belong to both classes.

\emph{Treewidth and pathwidth at most~$r$.}
Graphs of treewidth at most~$r$ form a minor-closed class, and every such graph with~$n$ vertices has at most~$rn$ edges, so the class lies in~$\mathcal C_r$.
Graphs of pathwidth at most~$r$ form a subclass.
Both classes contain~$P_n^r$, so the lower bound of Amiri and Siebertz gives~$2r+1$.

\emph{$K_t$-minor-free graphs with $3\le t\le9$.}
Every~$K_t$-minor-free graph with~$n$ vertices has at most $(t-2)n$ edges in this range.
For $n<t$, the bound follows from $|E(G)|\le\binom n2\le(t-2)n$.
For $n\ge t$ and $t\le7$, Mader's theorem~\cite{Mader1968} gives the stronger bound $(t-2)n-\binom{t-1}2$.
For $t=8$ the required density bound follows from J{\o}rgensen's theorem~\cite{Jorgensen1994}, and for $t=9$ from the theorem of Song and Thomas~\cite{SongThomas2006}.\footnote{
For $n\ge8$, J{\o}rgensen's bound is $6n-21$ except for the $(K_{2,2,2,2,2},5)$-cockades, which have $5m+5$ vertices and $30m+10=6n-20$ edges for some $m\ge1$.
For $n\ge9$, the bound of Song and Thomas is $7n-28$ except for~$K_{2,2,2,3,3}$, with~$12$ vertices and~$57$ edges, and the $(K_{1,2,2,2,2,2},6)$-cockades, with $5m+6$ vertices and $35m+15$ edges for some $m\ge1$.
Both kinds of exceptions have exactly $7n-27$ edges.
All exceptions have at most $(t-2)n$ edges.}
The class is minor-closed, so it lies in~$\mathcal C_{t-2}$, and Theorem~\ref{sc:thmC} gives $2t-3$.
The graphs~$P_n^{t-2}$ have no~$K_t$ minor, so the lower bound gives $2(t-2)+1=2t-3$.
\end{proof}

The same argument gives optimality whenever a class lies in~$\mathcal C_r$ and contains all powers~$P_n^r$.
For triangle-free planar graphs, the grid bound noted by Bonamy, Cook, Groenland, and Wesolek~\cite[Section~4]{BonamyCook2021} gives coefficient five.
Proposition~\ref{new:ordinary-lower} proves it in full.
Hence part~\ref{sc:thmC:girth} of Theorem~\ref{sc:thmC} is optimal at girth four as well.
Paths have every girth, so the case $r=1$ of the theorem of Amiri and Siebertz shows that the girth coefficients $3+4/(g-2)$ converge to the optimal limit three.
For the Euler coefficients of Theorem~\ref{sc:thmC}, no optimality is claimed at the other girths or for minor density~$a$ in general.
Appendix~\ref{fv:hg} gives a separate argument that achieves the optimal coefficient three at planar girth at least seventeen.

\subsection{Full proofs of the local lower bounds}
\label{new:ordinary-lower-section}

We use the Ramsey argument of Hilke, Lenzen, and Suomela~\cite{HLS2014} for both grids and powers of paths.
By choosing the identifiers suitably, we make the algorithm give the same output at many vertices in a large regular region, and this output must be one because a vertex and all its neighbors cannot all remain unselected.
We now give the details, including the small dominating sets that we use to compare the output with the optimum.

\paragraph{Block graphs.}
We let $q,s\ge1$ and let $S\subseteq\mathbb Z^q\setminus\{0\}$ be a finite set with $S=-S$ and $\max_{\sigma\in S}\|\sigma\|_\infty=s$.
We put $d=|S|$.
The \emph{lattice graph} on~$\mathbb Z^q$ joins~$x$ and $x+\sigma$ for all $\sigma\in S$.
For an integer~$m$, the set $\{0,\ldots,m-1\}^q$ is a \emph{block}.
A vertex of a block is \emph{full} if all its coordinates lie in $[s,m-1-s]$.
An \emph{$(m,S)$-block graph with~$b$ blocks} is a graph whose vertex set is a disjoint union of~$b$ blocks such that
\begin{enumerate}[label=(\alph*)]
\item two vertices of the same block are adjacent if and only if their difference lies in~$S$, and
\item two vertices of different blocks are adjacent only if neither of them is full.
\end{enumerate}
Thus a full vertex has exactly the~$d$ neighbors $x+\sigma$ with $\sigma\in S$, all in its block.
There are three families.
\begin{itemize}
\item \emph{Triangular grid:} $q=2$, $S=\{\pm(1,0),\pm(0,1),\pm(1,1)\}$, $s=1$, $d=6$.
The vertex $(m-1,0)$ of block~$j$ is joined to the vertex $(0,0)$ of block $j+1$ for $j<b$, and no other edges are added.
\item \emph{Square grid:} $q=2$, $S=\{\pm(1,0),\pm(0,1)\}$, $s=1$, $d=4$, with the same joining edges.
\item \emph{Path powers:} $q=1$, $S=\{\pm1,\ldots,\pm t\}$, $s=t$, $d=2t$.
The graph~$P_n^t$ has vertex set $\{1,\ldots,n\}$ with $n=bm$, and $i\sim j$ if $1\le|i-j|\le t$.
Block~$j$ consists of the vertices $(j-1)m+1+x$ with position $x\in\{0,\ldots,m-1\}$.
\end{itemize}
In all three cases, property (a) follows directly from the construction.
For the grids, a joining edge has endpoints with a coordinate equal to~$0$ or $m-1$, so neither endpoint is full.
For~$P_n^t$, an edge between consecutive blocks joins a position at least $m-t$ to a position at most $t-1$, and again neither endpoint is full, which proves (b).

\begin{lemma}\label{new:forced-selection}
Let $T\ge0$, let $m\ge2(T+3)s+1$ and let~$\mathcal A$ be a deterministic~$T$-round \LOCAL{} algorithm.
There is an integer~$\ell$, depending only on~$T$,~$m$ and~$S$, such that the following holds for every $b>\ell$ and every $(m,S)$-block graph~$G$ with~$b$ blocks.
Put $n=bm^q$.
There is an assignment of the identifiers $1,\ldots,n$ to~$V(G)$ such that, if~$\mathcal A$ returns a dominating set~$D$ on~$G$ with these identifiers and knowledge of~$n$, then
\[
 |D|\ge(b-\ell)\,(m-2(T+2)s)^q .
\]
\end{lemma}

\begin{proof}
At every full vertex, we number its~$d$ lattice edges by the position of the corresponding~$\sigma$ in a fixed enumeration of~$S$.
At every other vertex, we use any valid port numbering.
Since all balls used below contain only full vertices, their port numbers agree with those of the lattice.
We use the convention that after~$T$ rounds a vertex knows its labeled ball of radius $T+1$, including identifiers, degrees, and port numbers, as well as the number~$n$.
Its output is a function of this information, which also covers the usual conventions of the model.

\emph{Deep vertices.}
A vertex of a block is \emph{deep} if all its coordinates lie in $[(T+2)s,m-1-(T+2)s]$.
We let $B\subseteq\mathbb Z^q$ be the set of points at lattice distance at most $T+1$ from~$0$.
We claim that for every deep vertex~$v$ the ball of radius $T+1$ around~$v$ in~$G$ is $v+B$ with the lattice edges, the lattice port numbers, and all degrees equal to~$d$.
Induction on $j\le T+1$ shows that the vertices at distance at most~$j$ from~$v$ are the vertices $v+y$ in the block of~$v$, where $y\in B$ has lattice distance at most~$j$.
A vertex at distance $j\le T+1$ has each coordinate within~$js$ of that of~$v$, hence in $[s,m-1-s]$, so it is full.
A full vertex has exactly its~$d$ lattice neighbors, so the vertices at distance $j+1$ are the lattice neighbors of those at distance~$j$.
By (a) the edges among the vertices of the ball are exactly the lattice edges, and every vertex of the ball is full, so it has degree~$d$ and lattice ports.
This proves the claim.
Hence the labeled ball of a deep vertex is determined by the identifiers on~$v+B$, read in the order of~$B$.

\emph{Ramsey argument.}
We let $p=|B|$.
The lexicographic order on~$\mathbb Z^q$ is preserved by translations.
We fix~$n$ and color each~$p$-element set $X\subseteq\{1,\ldots,n\}$ by the bit that~$\mathcal A$ outputs at the center of~$B$ when the elements of~$X$ are assigned in increasing order to the points of~$B$ in lexicographic order.
By the claim, this is well defined, and it does not depend on how the identifiers are assigned outside the ball.
We let $M=m^q$ and let~$R$ be a Ramsey number such that every two-coloring of the~$p$-element subsets of an~$R$-element set has a homogeneous subset of size~$M$.
We put $\ell=\lceil R/M\rceil$.
Both numbers depend only on~$p$ and~$M$, hence on~$T$,~$m$ and~$S$, and not on~$b$.

We proceed through $j=1,\ldots,b-\ell$ in order.
At the start of step~$j$, at least $n-(j-1)M=(b-j+1)M\ge\ell M\ge R$ identifiers are still unused, so they contain a homogeneous set~$H_j$ of size~$M$.
We assign this set to block~$j$ in increasing order along the lexicographic order of the block, then remove these identifiers from the unused set.
After these steps, we assign the remaining identifiers arbitrarily.

We let~$v$ be a deep vertex of a block $j\le b-\ell$.
The identifiers on $v+B$ form a~$p$-subset of~$H_j$, and they increase along the lexicographic order of~$B$, since this is the restriction of the order of the block.
So~$\mathcal A$ outputs at~$v$ the color~$c_j$ of~$H_j$, the same bit for all deep vertices of the block.

\emph{The color is one.}
Since $m\ge2(T+3)s+1$, the point~$u$ of the block with all coordinates equal to $(T+3)s$ exists and is full.
All its neighbors $u+\sigma$ have coordinates in $[(T+2)s,(T+4)s]\subseteq [(T+2)s,m-1-(T+2)s]$, so they are deep.
If $c_j=0$, none of these neighbors is selected, and the vertex~$u$ is also deep and has bit zero.
Its closed neighborhood would therefore contain no selected vertex, which contradicts domination.
Hence $c_j=1$, and~$D$ contains all $(m-2(T+2)s)^q$ deep vertices of each of the $b-\ell$ blocks.
\end{proof}

\begin{proposition}
\label{new:ordinary-lower}
For every $\varepsilon>0$ and every fixed number of rounds, no deterministic LOCAL algorithm achieves coefficient $7-\varepsilon$ on planar graphs, $5-\varepsilon$ on triangle-free planar graphs, or $2r+1-\varepsilon$ on the~$r$-th powers of paths, for any fixed integer $r\ge1$.
All three bounds are against the integral domination number.
They hold with the order known and identifiers exactly $1,\ldots,n$.
\end{proposition}

\begin{proof}
We fix a~$T$-round algorithm and first consider the triangular and square block grids above, whose interior vertices have degree $d=6$ and $d=4$, respectively.
We put $k=d+1$, so $k=7$ or~$k=5$.
In each~$m$ by~$m$ block, we select every boundary vertex and every vertex with coordinates~$(i,j)$ satisfying $i+2j\equiv0\pmod k$.
This gives a dominating set, because every boundary vertex is selected, while the closed neighborhood of an interior vertex contains all residues modulo~$k$: the changes in the square grid are $0,\pm1,\pm2$, and those in the triangular grid are $0,\pm1,\pm2,\pm3$.
Thus we select a vertex in every such closed neighborhood, and adding bridges between blocks preserves this coverage.

For each fixed row there are at most $m/k+1$ selected vertices of the residue class, and there are at most~$4m$ boundary vertices per block.
Consequently
\[
 \gamma(G)\le b\left(\frac{m^2}{k}+5m\right).
\]
By Lemma~\ref{new:forced-selection}, there is a labeling on which any feasible output has at least $(b-\ell)(m-2T-4)^2$ vertices.
The ratio of this lower bound to the displayed upper bound tends to~$k$ as first~$b$ and then~$m$ tend to infinity.
We choose~$m$ and then~$b$ large enough to make the ratio greater than $k-\varepsilon$.
The grids and their bridges are planar; the square case has no triangle.
This proves the first two bounds.

For the last bound, we use the path power~$P_{bm}^r$, with $q=1$ and $s=r$ in Lemma~\ref{new:forced-selection}.
We partition the underlying path into consecutive intervals of at most $2r+1$ vertices and select a middle vertex of each interval.
These vertices dominate the path power, so
\[
 \gamma(P_{bm}^r)\le\left\lceil\frac{bm}{2r+1}\right\rceil.
\]
The forced-selection bound is $(b-\ell)(m-2(T+2)r)$.
Its ratio to this upper bound tends to $2r+1$ as first~$b$ and then~$m$ tend to infinity.
We therefore again choose~$m$ and then~$b$ large enough.
The lemma uses identifiers exactly $1,\ldots,bm$ and allows the algorithm to know~$bm$, as required.
\end{proof}

\section{A round lower bound for approximation schemes}\label{app:rounds-lower}

On the classes of Corollary~\ref{sc:optimal}, the deterministic constant-round coefficients of Theorem~\ref{sc:thmC} are optimal, so an algorithm that approaches the integral optimum there must use either random bits or more than a constant number of rounds.

\paragraph{The deterministic round bound is optimal.}
The factor $\log^*n$ in the second part of Theorem~\ref{sc:thmD} cannot be removed, already for dominating set on cycles.
Every minor of a cycle has at most as many edges as vertices, so cycles lie in~$\mathcal C_1$ and hence in every class~$\mathcal C_\rho$ of the theorem.
Lenzen and Wattenhofer~\cite{LW2008} derived $\log^*n$ lower bounds from Linial's lower bound for coloring cycles: for maximum independent set on cycles, and for dominating set on unit disk graphs.
Czygrinow, Han\'ckowiak, and Wawrzyniak~\cite{CHW2008} independently proved the bound for maximum independent set, and by repeated applications of Ramsey's theorem a lower bound of $5-\varepsilon$ for dominating set on planar graphs, which Hilke, Lenzen, and Suomela~\cite{HLS2014} raised to $7-\varepsilon$ for constant-round algorithms.
The graphs in the bound of Lenzen and Wattenhofer for dominating set are powers of cycles whose exponent grows with the running time, so they lie in no class~$\mathcal C_a$.
The following proof runs their argument on cycles themselves, so it applies to every class~$\mathcal C_\rho$, including those with $\rho<3$, such as graphs of treewidth two.

\begin{proposition}
\label{sc:rounds-lower}
Let $0<\varepsilon<1/4$, and let~$T(n)$ be the number of rounds that a deterministic LOCAL algorithm uses on cycles with~$n$ vertices.
If, on every cycle, the algorithm returns a dominating set of size at most $(1+\varepsilon)\gamma$, then $T(n)=\Omega(\log^*n)$.
This holds even when the vertices know~$n$ and the identifiers are $1,\ldots,n$.
\end{proposition}

Since $\gf(C_n)=n/3\le\gamma(C_n)$, the same bound holds for approximations of~$\gf$, and, being a LOCAL bound, it holds in broadcast CONGEST.
Together with the first part of Theorem~\ref{sc:thmD}, it shows that random bits are exactly what removes the factor $\log^*n$.

\begin{proof}
Two facts are needed about a cycle~$C_n$ whose vertices know~$n$ and have identifiers $1,\ldots,n$.
By Linial's theorem~\cite{Linial1992}, in the form of~\cite[Theorem~7]{LW2008}, a deterministic algorithm that properly colors every such cycle with three colors uses at least $\frac12(\log^*n-1)$ rounds.
Call an assignment of bits $b(v)\in\{0,1\}$ to the vertices \emph{$\ell$-alternating} if every run of consecutive vertices with equal bits has fewer than~$\ell$ vertices.
By~\cite[Lemma~9]{LW2008}, an~$\ell$-alternating assignment computed in~$\mathcal O(\ell)$ rounds yields a proper~$3$-coloring in~$\mathcal O(\ell)$ rounds.

Fix the algorithm and~$N$, let $T=T(N)$, and let $n=\lfloor N/2\rfloor$.
We may assume $T\ge1$, by adding an idle round, and $n\ge4T+3$, since otherwise $T\ge(n-3)/4$.
A labeling of~$C_N$ uses each of the identifiers $1,\ldots,N$ exactly once.
In~$T$ rounds, the output of a vertex is determined by~$N$ and by the sequence of identifiers within distance~$T$ of it.
A sequence $w=(w_1,\ldots,w_{4T})$ of distinct identifiers is \emph{bad} if, in every labeling of~$C_N$ in which~$w$ occupies consecutive vertices, the~$2T$ vertices labeled $w_{T+1},\ldots,w_{3T}$ all join the output.
Whether~$w$ is bad depends only on~$w$, because these vertices see no identifier outside~$w$.

\emph{Bad sequences cover less than half of~$C_N$.}
Suppose that, in some labeling,~$s$ pairwise disjoint segments of~$4T$ consecutive vertices carry bad sequences, with $4Ts\ge N/2$, and let~$D$ be the output.
The set~$F$ of the~$2Ts$ middle vertices of the segments lies in~$D$, and $C_N-F$ consists of at most~$s$ paths.
The inner vertices of a path with~$\ell$ vertices are dominated from inside the path, and every vertex dominates at most three vertices, so the path contains at least $(\ell-2)/3$ vertices of~$D$.
Summing over the paths, and using $T\ge1$,
\[
 |D|\ge|F|+\frac{N-|F|-2s}3=\frac N3+\frac{(4T-2)s}3
 \ge\frac N3+\frac{4Ts}6\ge\frac{5N}{12}.
\]
Since $\gamma(C_N)=\lceil N/3\rceil\le(N+2)/3$, this exceeds $(1+\varepsilon)\gamma(C_N)$ when $N>8(1+\varepsilon)/(1-4\varepsilon)$, which we assume.

\emph{Many identifiers avoid bad sequences.}
Choose the largest possible number~$s$ of pairwise disjoint bad segments over all labelings of~$C_N$.
Fix a labeling with~$s$ pairwise disjoint bad segments, and let~$S$ be the identifiers outside them.
Then $|S|>N/2\ge n$, and no bad sequence can consist only of identifiers in~$S$.
Otherwise, place the~$s$ segments consecutively, followed by this sequence and then by the remaining identifiers, to obtain $s+1$ disjoint segments, a contradiction.

\emph{Simulation on~$C_n$.}
Fix an injection $\varphi\colon\{1,\ldots,n\}\to S$.
On a cycle~$C_n$ with identifiers $1,\ldots,n$, every vertex~$v$ simulates the algorithm for~$T$ rounds as if the number of vertices were~$N$ and every identifier~$x$ were~$\varphi(x)$, and sets $b(v)=1$ if it joins the output.
The identifiers within distance $T+1$ of~$v$ form a sequence of $2T+3$ distinct elements of~$S$, which can be placed on consecutive vertices of some labeling of~$C_N$.
There,~$v$ and its two neighbors compute the same bits as in the simulation, and the output is a dominating set, so one of them has bit~$1$.
Hence no three consecutive vertices of~$C_n$ have bit~$0$.
If~$2T$ consecutive vertices had bit~$1$, then, with the~$T$ vertices on either side, they would carry a sequence of~$4T$ distinct elements of~$S$ that is bad, since the middle bits depend only on these identifiers.
Thus the assignment is $(2T+1)$-alternating and is computed in~$T$ rounds, so~$C_n$ is properly~$3$-colored in~$\mathcal O(T)$ rounds.
By Linial's theorem, $T(N)\ge c\log^*\lfloor N/2\rfloor-c'$ for constants $c,c'>0$, which proves $T(N)=\Omega(\log^*N)$.
\end{proof}

\section{Applications to other covering problems}\label{sc:applications}

Theorems~\ref{sc:thmA} and~\ref{sc:thmD} hold for every local covering problem with bounded costs, but a deterministic constant-round integral cover also needs a rounding rule.
This section gives one rounding rule that works for all covering sets $C(r)\subseteq N_G[r]$, and lists the problems that it covers.
Its coefficient includes an error term: the first assertion of Theorem~\ref{sc:thmB} removes the error only for a rule with slack.
Such savings are proved for ordinary domination in Section~\ref{simp:endpoint:section} and for total domination in Appendix~\ref{fv:total}.

Here the setting is that of \gcov{}.
The \emph{incidence graph}~$I$ of the instance has vertex set~$V(G)$, and an edge~$rv$ for every demand~$r$ and every $v\in C(r)\setminus\{r\}$.
It is a subgraph of~$G$, and its density may be much smaller than that of~$G$; for red-blue domination, for instance, it contains only the edges between the two colors.

\begin{corollary}
\label{sc:rounding-all}
Fix a rational $\rho\ge1$, integers $a,b\ge1$, and $\varepsilon>0$.
There is a deterministic broadcast CONGEST algorithm with a constant number of rounds that, on every covering instance in the setting of \gcov{} with $G\in\mathcal C_\rho$, costs in $\{1,\ldots,b\}$, and an incidence graph whose subgraphs~$F$ all satisfy $|E(F)|\le a|V(F)|$, returns a feasible integral cover of cost at most $(2a+1+\varepsilon)\tau_c$.
If no demand belongs to its own covering set, the cost is at most $(2a+\varepsilon)\tau_c$.
\end{corollary}

\begin{proof}
We run Theorem~\ref{sc:thmA} with accuracy $\varepsilon/(2a+1)$, obtaining~$x$ with cost at most $(1+\varepsilon/(2a+1))\tau_c$.
Each demand~$r$ broadcasts whether $r\in C(r)$.
A candidate~$v$ sets
\[
 s_v=\begin{cases}
 2a+1,&\text{if }v\in C(r)\text{ for some demand }r\in C(r),\\
 2a,&\text{otherwise}.
 \end{cases}
\]
Since both ends know the covering incidences, one broadcast determines~$s_v$ locally.
We select every admissible~$v$ with $x_v\ge1/s_v$ and broadcast its selection bit and cost, after which every demand still uncovered requests a cheapest candidate in~$C(r)$, with ties broken by identifiers.
We add the requested candidates to the output, and each request consists of one identifier broadcast to all neighbors.
In this way, we cover every demand in a constant number of further rounds.

Let~$W$ be the threshold-selected set.
Its cost is at most $(2a+1)\sum_{v\in W}c_vx_v$.
Fix an orientation of the incidence graph~$I$ with outdegree at most~$a$, using Lemma~\ref{simp:orientation}.
We use this orientation only in the proof of the bound, and for a requesting demand~$r$ we put $\mu_r=\min_{v\in C(r)}c_v$.
If $r\in C(r)$, all its candidates have scale $2a+1$, and hence coordinate below $1/(2a+1)$.
Its self-coordinate and its at most~$a$ outgoing candidates have total mass less than $(a+1)/(2a+1)$.
Its incoming candidates therefore have mass greater than $a/(2a+1)$.
If $r\notin C(r)$, every candidate has coordinate below~$1/(2a)$, and there are at most~$a$ outgoing candidates.
The incoming candidates then have mass greater than $1/2$, which is at least $a/(2a+1)$.
In either case,
\[
 \mu_r\le\frac{2a+1}{a}
       \sum_{v\in C(r),\ v\to r}\mu_rx_v
 \le\frac{2a+1}{a}
       \sum_{v\in C(r),\ v\to r}c_vx_v.
\]
Each candidate occurs in at most~$a$ of these sums, and none lies in~$W$, so summing bounds the cost of the requests by $(2a+1)\sum_{v\notin W}c_vx_v$.
Adding the threshold selections gives total cost at most $(2a+1)\sum_vc_vx_v\le(2a+1+\varepsilon)\tau_c$.

If no demand covers itself, no candidate receives a flag, so our algorithm uses scale~$2a$ everywhere and the incoming mass is greater than $1/2$ for every requesting demand.
We can therefore repeat the count with~$2a$ in place of $2a+1$ to obtain cost at most $2a\sum_vc_vx_v$, which is at most $(2a+\varepsilon)\tau_c$, since $2a(1+\varepsilon/(2a+1))\le2a+\varepsilon$.
\end{proof}

For dominating set, this gives $(2a+1+\varepsilon)\gf(G)$ for the density~$a$ of subgraphs; Theorem~\ref{sc:thmC} removes the error term, and the density of minors may then be rational.

\paragraph{Instances.}
Each problem below is a choice of demands~$R$, of a set~$B$ of permitted candidates, and of covering sets.
Feasibility means that every covering set is nonempty, and costs lie in $\{1,\ldots,b\}$.
For each problem, Theorem~\ref{sc:thmA} gives a deterministic fractional $(1+\varepsilon)$-approximation in a constant number of rounds, Theorem~\ref{sc:thmD} gives a randomized constant-round and a deterministic~$\mathcal O(\log^*n)$-round integral $(1+\varepsilon)$-approximation, and Corollary~\ref{sc:rounding-all} gives the deterministic constant-round coefficient listed in Table~\ref{sc:app-table}.
The planar entries use that every planar graph has at most~$3n$ edges, and that every bipartite or triangle-free planar graph has at most~$2n$ edges, so that $a=3$ or $a=2$ in the corollary; the same bounds hold for all subgraphs.
\begin{itemize}
\item \emph{Weighted, partial, and restricted dominating set}: $C(r)=N_G[r]\cap B$ for $r\in R$.
The demands are the vertices that must be dominated, and only vertices of~$B$ may be chosen; see~\cite{DGI2024} for weighted domination.
Here partial domination means covering the prescribed set~$R$.
The partial-domination problem of Kneis, M\"olle, and Rossmanith~\cite{KneisMolleRossmanith2007} instead asks to dominate at least~$t$ vertices, whose identities are not fixed in advance.
\item \emph{Directed domination}~\cite{CaroHenning2012}: the edges of~$G$ carry directions known to both endpoints, and~$C(r)$ consists of~$r$ and its in-neighbors.
A vertex must be chosen or have a chosen in-neighbor.
\item \emph{Total domination}~\cite{CockayneDawesHedetniemi1980,HenningYeo2013}, also weighted, partial, and restricted: $C(r)=N_G(r)\cap B$.
Every demand, chosen or not, needs a chosen neighbor.
\item \emph{Red-blue domination}~\cite{GarneroSauThilikos2017}: the vertices are colored red and blue, the demands are the red vertices, and~$C(r)$ is the set of blue neighbors of~$r$.
The incidence graph is the bipartite graph of red-blue edges.
\item \emph{Set cover and hitting set}~\cite{Johnson1974,Chvatal1979,MustafaRay2010}: the network is the incidence graph of a set system, with one vertex for every element and every set.
For set cover, the demands are the elements and~$C(e)$ is the set of sets containing~$e$; for hitting set, the roles are exchanged.
Both are red-blue domination on a bipartite graph, so the only hypothesis is that the incidence graph of the set system lies in~$\mathcal C_\rho$; no bound on the sizes or frequencies of the sets is needed.
\end{itemize}

\begin{table}[ht]
\centering\normalsize

\begin{tabularx}{\linewidth}{@{}>{\raggedright\arraybackslash}Xl>{\raggedright\arraybackslash}X@{}}
\toprule
Problem & Coefficient & Planar\\
\midrule
weighted, partial, restricted domination & $2a+1+\varepsilon$ & $7+\varepsilon$\\
directed domination & $2a+1+\varepsilon$ & $7+\varepsilon$\\
total domination (all variants) & $2a+\varepsilon$ & $6+\varepsilon$; triangle-free $4+\varepsilon$\\
red-blue domination & $2a+\varepsilon$ & $4+\varepsilon$\\
set cover, hitting set & $2a+\varepsilon$ & $4+\varepsilon$ (planar incidence graph)\\
\bottomrule
\end{tabularx}
\caption{Deterministic constant-round coefficients of Corollary~\ref{sc:rounding-all}, against the fractional optimum~$\tau_c$, where every subgraph of the incidence graph has a ratio of edges to vertices of at most~$a$.
All problems also have the approximation schemes of Theorem~\ref{sc:thmD}.}
\label{sc:app-table}
\end{table}

The coefficients for total domination in Table~\ref{sc:app-table} also cover bounded costs and restricted instances.
For unweighted total domination on all vertices, Appendix~\ref{fv:total} gives the stronger exact coefficients, with a separate rounding argument.

\Needspace{16\baselineskip}

\section{Improved coefficients and additive density bounds}\label{fv:imp}

The same fractional solver also gives better coefficients when the rounding has more room than the basic density bound suggests.
When~$2a$ is not an integer, a different threshold in the rule of Section~\ref{simp:endpoint:section} gives a smaller coefficient, and this improves the planar girth bounds of Theorem~\ref{sc:thmC}.
A second application concerns a fixed additive excess in the density.
The general small-optimum routine below removes such a term against the integral benchmark, and it will also be used for red-blue covering in Appendix~\ref{fv:rb}.
For ordinary domination on a fixed surface, the stronger bound seven itself has already been proved in Section~\ref{new:surface}, using the residual instance and its two budgets.
Throughout this section,~$e(F)$ denotes the number of edges of a graph~$F$.

\subsection{Threshold rounding with a density excess}

An additive excess in the density costs only an additive term in the plain threshold rounding.
The proof uses a fractional orientation, whose existence follows from a max-flow argument; the orientation is used only in the analysis.
The rounding-with-excess argument of Amiri and Siebertz~\cite[Lemmas~7.1 and~7.2]{AS2026} is the model for this proof.

\begin{lemma}
\label{fv:imp:additive}
Let $a>0$ and $\eta\ge0$, and suppose $e(F)\le a|V(F)|+\eta$ for every subgraph~$F$ of~$G$.
Let $b=2a+1$, let $x\in[0,1]^{V}$ be a feasible fractional dominating set, and put $W=\{v:x_v\ge1/b\}$ and $U=V\setminus N[W]$.
Then $W\cup U$ is a dominating set and
\[
 |W\cup U|\le b\sum_vx_v+\frac{2\eta}{a}.
\]
It is computed with one broadcast after~$x$.
\end{lemma}

\begin{proof}
\emph{A fractional orientation.}
Consider the network with a source, a node for every edge of~$G$, a node for every vertex, and a sink.
The source has an arc of capacity one to every edge node, every edge node has arcs of unlimited capacity to its two endpoints, and every vertex has an arc of capacity~$a$ to the sink.
In a finite cut, let~$S$ be the set of vertices and~$Y$ the set of edge nodes on the source side.
Every edge in~$Y$ has both ends in~$S$, so $Y\subseteq E(G[S])$, and the capacity of the cut is at least
\[
 |E(G)|-|Y|+a|S|\ge|E(G)|-e(G[S])+a|S|\ge|E(G)|-\eta.
\]
Hence a maximum flow sends at least $|E(G)|-\eta$ units through the edge nodes.
For an edge~$uv$, let the flow assign the parts~$f_{uv}(u)$ and~$f_{uv}(v)$, with sum at most one, to its ends.
Give the unassigned rest of the edge to~$u$, say, and let $o_{uv}(u),o_{uv}(v)\ge0$ be the resulting shares, so that $o_{uv}(u)+o_{uv}(v)=1$.
Let~$\eta_v$ be the total unassigned part given to~$v$; then $\sum_v\eta_v\le\eta$.
The flow into~$v$ is at most~$a$, so $\operatorname{load}(v)=\sum_{u\in N(v)}o_{uv}(v)$ satisfies $\operatorname{load}(v)\le a+\eta_v$.

\emph{The count.}
Let $u\in U$.
No vertex of~$N[u]$ lies in~$W$, so every coordinate in~$N[u]$ is less than $1/b$.
Feasibility at~$u$ gives
\[
 \sum_{v\in N(u)}o_{uv}(v)x_v
 =\sum_{v\in N(u)}x_v-\sum_{v\in N(u)}o_{uv}(u)x_v
 \ge\Bigl(1-\frac1b\Bigr)-\frac{a+\eta_u}b=\frac{a-\eta_u}b,
\]
because $b-1-a=a$.
Now sum over $u\in U$.
On the right, the sum is at least $(a|U|-\eta)/b$.
On the left, only vertices $v\notin W$ occur, because no vertex of~$U$ has a neighbor in~$W$, and~$v$ occurs with total share at most $\operatorname{load}(v)\le a+\eta_v$.
Since $x_v<1/b$ for $v\notin W$, the left side is at most
\[
 \sum_{v\notin W}(a+\eta_v)x_v\le a\sum_{v\notin W}x_v+\frac\eta b.
\]
Therefore $|U|\le b\sum_{v\notin W}x_v+2\eta/a$.
Moreover $|W|\le b\sum_{v\in W}x_v$, which proves the bound.
The set dominates~$G$, because~$W$ dominates~$N[W]$ and~$U$ contains every remaining vertex.
\end{proof}

\subsection{An improved rounding when \texorpdfstring{$2a$}{2a} is not an integer}

Let us return to the rule of Section~\ref{simp:endpoint:section} in the case that~$2a$ is not an integer, with a threshold that depends on a parameter~$\mu$ in place of $1/(a+1)$.
Put $b=2a+1$, $k=\lfloor2a\rfloor$, and $\theta=2a-k\in(0,1)$, as in the non-integer case there.
The rule is that of Section~\ref{simp:endpoint:section} with candidates and demands $C=R_0=V$ and constant cost $c\equiv b$.
For a feasible $x\in[0,1]^V$ and a threshold $t>0$:
\begin{enumerate}[label=\textbf{Step \arabic*.},leftmargin=*,itemsep=2pt,parsep=0pt,topsep=4pt]
\item $W=\{v:x_v>t\}$ and $R=V\setminus N[W]$.
\item $h(v)=|N[v]\cap R|$ for $v\notin W$, and $H=\{v\notin W:h(v)\ge b\}$.
\item $U=R\setminus N[H]$, and the output is $D=W\cup H\cup U$.
\end{enumerate}
The set~$D$ is dominating:~$W$ dominates~$N[W]$,~$H$ dominates the demands of~$R$ in~$N[H]$, and~$U$ contains the others.
The rule needs three Boolean broadcasts after~$x$.

\begin{lemma}
\label{fv:imp:lemma}
Let $a\ge1$ be rational with $2a\notin\mathbb Z$, and let $\eta\ge0$.
Suppose $e(F)\le a|V(F)|+\eta$ for every subgraph~$F$ of~$G$.
Let~$\mu$ be a rational number with $0\le\mu\le\theta$ and
\begin{equation}\label{fv:imp:mu}
 (\mu+1)(\mu+1-\theta)\le\frac{b(1-\theta)}a,
\end{equation}
and put
\[
 t=\max\Bigl\{\frac{\mu+1}b,\ \frac{1-\theta}{(a+1)(\mu+1-\theta)}\Bigr\}.
\]
Then for every feasible $x\in[0,1]^V$, the set~$D$ of the rule with threshold~$t$ satisfies
\[
 |D|\le(b-\mu)\sum_vx_v+q_\eta,\qquad
 q_\eta=2\eta t\Bigl(1+\frac\mu{1-\theta}\Bigr).
\]
In particular $|D|\le(b-\mu)\sum_vx_v$ if $\eta=0$.
\end{lemma}

The value $\mu=0$ satisfies the hypotheses, since $b>a$, and then $t=1/(a+1)$ is the threshold of Section~\ref{simp:endpoint:section}.
The left side of \eqref{fv:imp:mu} increases with~$\mu$ on $\mu\ge0$, so the admissible~$\mu$ are the rational numbers in $[0,\mu_{\max}]$, where
\[
 \mu_{\max}=\min\{\theta,\mu^*\},\qquad
 \mu^*=\frac{\theta-2+\sqrt{(2-\theta)^2+4(1-\theta)(1+1/a)}}2.
\]
The number~$\mu_{\max}$ may be irrational.
Indeed $\mu^*>0$ is the positive root of the quadratic obtained from equality in \eqref{fv:imp:mu}, using $b/a-1=1+1/a$.

\begin{proof}
The notation is that of Lemma~\ref{simp:endpoint:rounding}: $Q=H\cap R$, $L=R\setminus(H\cup U)$, $O=V\setminus(W\cup H)$, $w=|W|$, $\ell=|L|$, $z=|H\setminus R|$, and $P=\sum_{v\in O}(b-h(v))x_v$.
Since $h(v)<b$ on~$O$, $P\ge0$.
The sets $Q,L,U$ partition~$R$, and $W,H,U$ are disjoint, so
\[
 |D|=w+|R|-\ell+z.
\]

\emph{Excess.}
For every $v\in H$, count its neighbors in~$R$, and also~$v$ itself if $v\in Q$.
There is no edge from~$H$ to~$U$, so as in Lemma~\ref{simp:endpoint:rounding},
\[
 \sum_{v\in H}h(v)\le e(H\cup L)+e(Q)+|Q|
 \le a(|H|+\ell)+(a+1)|Q|+2\eta .
\]
Since $b|H|=(2a+1)(|Q|+z)$, the excess $E=\sum_{v\in H}(h(v)-b)$ satisfies
\begin{equation}\label{fv:imp:E}
 E\le a\ell-(a+1)z+2\eta.
\end{equation}
Every~$h(v)$ is an integer, and $h(v)\ge b=k+1+\theta$ on~$H$, so $h(v)-b\ge1-\theta$ on~$H$.
Hence $E\ge(1-\theta)|H|$ and
\begin{equation}\label{fv:imp:Hbound}
 (1-\theta)|H|\le a\ell-(a+1)z+2\eta.
\end{equation}

\emph{Cost.}
No vertex of~$W$ lies in a constraint of~$R$, so the constraints at the demands of~$R$ give $\sum_{v\notin W}h(v)x_v\ge|R|$.
A vertex of~$H$ is not in~$W$, so $x_v\le t$ there, and $h(v)-b\ge0$.
Therefore
\[
\begin{aligned}
 b\sum_vx_v&=b\sum_{v\in W}x_v+\sum_{v\notin W}h(v)x_v+P
 -\sum_{v\in H}(h(v)-b)x_v\\
 &\ge btw+|R|+P-tE .
\end{aligned}
\]
Subtracting $|D|$ and using \eqref{fv:imp:E} gives
\begin{equation}\label{fv:imp:S}
 b\sum_vx_v-|D|\ge(bt-1)w+(1-at)\ell+\bigl((a+1)t-1\bigr)z+P-2\eta t.
\end{equation}

\emph{Mass.}
Every vertex of~$W$ has $x_v\le1$, every vertex of~$H$ has $x_v\le t$, and every $v\in O$ has $h(v)\le k+1$, so $b-h(v)\ge\theta$.
Hence $P\ge\theta\sum_{v\in O}x_v$, and with \eqref{fv:imp:Hbound} and $\mu\le\theta$,
\[
 \mu\sum_vx_v\le\mu w+\mu t|H|+\frac\mu\theta P
 \le\mu w+\frac{\mu t}{1-\theta}\bigl(a\ell-(a+1)z+2\eta\bigr)+P.
\]

\emph{Comparison.}
By \eqref{fv:imp:S}, the claim $|D|\le(b-\mu)\sum_vx_v+q_\eta$ follows once $\mu\sum_vx_v$ is at most the right side of \eqref{fv:imp:S} plus~$q_\eta$.
By the mass bound, and since~$P$ cancels, it suffices to compare the coefficients of $w,\ell,z$ and the constants.
We need
\begin{align*}
 \mu&\le bt-1,\\
 \frac{\mu ta}{1-\theta}&\le1-at,\\
 -\frac{\mu t(a+1)}{1-\theta}&\le(a+1)t-1,
\end{align*}
and the constant terms satisfy $2\eta\mu t/(1-\theta)=q_\eta-2\eta t$.
The first inequality is $t\ge(\mu+1)/b$.
The second is equivalent to $at(\mu+1-\theta)\le1-\theta$, that is, $t\le(1-\theta)/(a(\mu+1-\theta))$.
The third is equivalent to $(a+1)t(\mu+1-\theta)\ge1-\theta$.
Both lower bounds on~$t$ hold by definition.
The upper bound is at least $(\mu+1)/b$ by \eqref{fv:imp:mu}, and it is larger than the other lower bound since $a<a+1$.
Hence it holds for the maximum.
\end{proof}

\subsection{Planar graphs of large girth}

For planar graphs of girth~$g$, Appendix~\ref{sc:girth} gives $3+4/(g-2)$.
When~$2a$ is not an integer, the improved rounding is better, and every rational coefficient above the improved one is achieved with no error term.
For $g\ge5$ with $g\ne6$, let $a=g/(g-2)$, so that $2a\notin\mathbb Z$, define $b,\theta,\mu^*,\mu_{\max}$ as above, and put
\[
 c_g=2a+1-\mu_{\max}=2a+1-\min\{\theta,\mu^*\}.
\]

\begin{corollary}
\label{fv:imp:girth}
Fix $g\ge5$ with $g\ne6$ and a rational $c>c_g$.
There is a deterministic constant-round broadcast CONGEST algorithm that returns on every planar graph~$G$ of girth at least~$g$ a dominating set~$D$ with $|D|\le c\,\gf(G)$.
\end{corollary}

The numbers are $c_5=4$, $c_7\approx3.562$, $c_8\approx3.320$, $c_9\approx3.158$, $c_{10}\approx3.041$, and $c_g=3$ for all $g\ge11$.
For $g\ge7$, $\theta=2a-2$, and $\theta\le\mu^*$ exactly when $6a^2-5a-3\le0$, that is, when $g\ge11$.
In words, the rounding gives $4+\varepsilon$ at girth five and $3+\varepsilon$ at girth at least eleven, which improves $13/3$ and $3+4/(g-2)$ in Theorem~\ref{sc:thmC}.
The values~$c_g$ themselves, that is,~$4$ and~$3$, are not reached by this rounding, because for $\mu=\theta$ the coefficient of~$P$ in the proof of Lemma~\ref{fv:imp:lemma} has no slack.
Appendix~\ref{fv:hg} reaches~$3$ exactly at girth at least~$17$ by another rounding.

\begin{proof}
Here $\mu_{\max}>0$, so $c_g<b$.
A bound with a smaller coefficient also holds with a larger coefficient, so $c<b$ is no loss of generality.
Put $\mu'=b-c\in (0,\mu_{\max})$ and fix a rational~$\mu$ with $\mu'<\mu<\mu_{\max}$, so that~$\mu$ satisfies the hypotheses of Lemma~\ref{fv:imp:lemma}.
Put
\[
 \delta=\frac{\mu-\mu'}{b-\mu}>0.
\]
By Lemma~\ref{sc:face} with $A=\varnothing$, every subgraph of~$G$ has a ratio of edges to vertices of at most~$a$.
Planar graphs lie in~$\mathcal C_3$, so the solver of Theorem~\ref{sc:thmA} applies with $\rho=3$, unit costs, and the rational accuracy~$\delta$.
It gives a feasible $x\ge0$ with $\sum_vx_v\le(1+\delta)\gf(G)$.
Capping the output at one keeps it feasible and does not increase its cost, so $x\in[0,1]^V$.
The rule of Lemma~\ref{fv:imp:lemma} with $\eta=0$ then satisfies
\[
 |D(x)|\le(b-\mu)\sum_vx_v\le(b-\mu)(1+\delta)\gf(G)=(b-\mu')\gf(G)
 =c\,\gf(G),
\]
because $(b-\mu)(1+\delta)=b-\mu+\mu-\mu'$.
\end{proof}

\subsection{Absorbing a fixed additive term}

An additive term can be removed by solving small components exactly.
Both lemmas of this subsection hold for any covering problem with covering sets $C(r)\subseteq N_G[r]$ and integer costs, so that they can be used for other problems (see Corollary~\ref{fv:rb:surface-opt}).
The input is that of \gcov{}: demands~$R$, covering sets $C(r)\subseteq N_G[r]$ such that for every edge~$rv$ both endpoints know whether $v\in C(r)$, and costs $c_v\in\{1,\ldots,B\}$ on the admissible vertices.
Let~$\OPT_c$ and~$\tau_c$ denote the integral and the fractional optimum.
The \emph{incidence graph}~$I$ has the vertex set~$V(G)$ and an edge~$rv$ whenever $r\in R$, $v\in C(r)$, and $v\ne r$.
It is a spanning subgraph of~$G$, and every vertex knows its edges in~$I$ after one round.
For domination, $I=G$.
Every vertex of~$C(r)$ lies in the component of~$I$ that contains~$r$, so both optima are the sums of the optima over the components of~$I$, called \emph{instance components}.

\begin{lemma}
\label{fv:imp:exact}
Fix a rational $a\ge1$ and integers $B,K\ge1$.
There is a deterministic broadcast CONGEST algorithm with messages of~$\mathcal O(\log n)$ bits and a number of rounds that depends only on $a,B,K$, with the following property.
On every covering instance with costs in $\{1,\ldots,B\}$ on a graph~$G$ all of whose subgraphs have a ratio of edges to vertices of at most~$a$, it marks some instance components as solved and outputs in every solved component~$Z$ a cover~$D_Z$ of the demands of~$Z$ of cost~$\OPT_c(Z)$.
Every instance component~$Z$ with $\OPT_c(Z)\le K$ is solved.
Moreover, such a component has diameter at most $3K-1$ in~$I$ if every vertex is a demand, and at most $5K-1$ otherwise.
\end{lemma}

\begin{proof}
\emph{Diameter.}
Let~$Z$ be an instance component with $\OPT_c(Z)\le K$.
Costs are at least one, so some cover $D\subseteq Z$ of the demands of~$Z$ has at most~$K$ vertices.
Let $s=1$ if every vertex is a demand, and $s=2$ otherwise.
Every vertex of~$Z$ is within distance~$s$ in~$I$ of a vertex of~$D$: a demand~$r$ has a vertex of~$D$ in~$C(r)$, which is~$r$ or an~$I$-neighbor of~$r$, and a vertex that is not a demand and not isolated in~$I$ has an~$I$-neighbor that is a demand.
A vertex that is isolated in~$I$ and not a demand has no demand to cover and forms~$Z$ alone, with diameter zero.
Assign to every vertex of~$Z$ such a vertex of~$D$, and consider the graph on~$D$ in which two vertices are adjacent if their distance in~$I$ is at most $2s+1$.
If it were disconnected, the vertices assigned to the two sides would give two nonempty parts of~$Z$.
Since~$Z$ is connected, an edge~$uv$ of~$I$ joins them, and the assigned vertices of~$u$ and~$v$ are at distance at most $2s+1$, a contradiction.
So any two vertices of~$D$ are at distance at most $(2s+1)(K-1)$, and the diameter of~$Z$ is at most $(2s+1)(K-1)+2s=(2s+1)K-1$.

\emph{Acceptance.}
Put $d=5K-1$, which is at least $3K-1$.
Every vertex computes the minimum identifier~$\lambda(v)$ among the vertices within distance~$d$ in~$I$ by~$d$ rounds of minimum flooding.
It exchanges labels with its~$I$-neighbors and is \emph{flagged} if some~$I$-neighbor has a different label.
The flags are flooded for~$2d$ rounds, and a vertex \emph{accepts} if it sees no flag within distance~$2d$.
Every message is an identifier or one bit.

Let~$Z$ be an instance component and $v\in Z$.
If all vertices of~$Z$ have one label, nobody in~$Z$ is flagged, so all accept.
Otherwise a path in~$I$ leads from~$v$ to a vertex with a label different from~$\lambda(v)$.
Let~$w$ be the vertex before the first such vertex on this path.
Then~$w$ has label~$\lambda(v)$ and is flagged.
The vertex~$z$ that carries the identifier~$\lambda(v)$ is within distance~$d$ of both~$v$ and~$w$, so~$v$ sees the flag of~$w$ within distance~$2d$, and~$v$ does not accept.
Thus acceptance is the same for all vertices of a component.
An accepted component has one label, so every vertex lies within distance~$d$ of the vertex carrying it, and the diameter is at most~$2d$.
Conversely, a component of diameter at most~$d$ has one label, so by the diameter bound every component with $\OPT_c\le K$ is accepted.

\emph{Search.}
On every accepted component~$Z$ separately, the search in the proof of Lemma~\ref{simp:local-solver} runs on the graph~$I[Z]$, with density bound~$a$, diameter bound~$2d$, and costs in $\{1,\ldots,B\}$.
As in the proof of Lemma~\ref{sc:integral-task}, the multiplicity is one, the copy budget is~$K$ in place of $\lceil(q+2p)K_0\rceil$, and a candidate~$v$ covers a demand~$r$ when $v\in C(r)$.
Every subgraph of~$I$ has a ratio of edges to vertices of at most~$a$, and a candidate covers at most the demands of its closed neighborhood in~$I$, so the branching bound of that proof applies unchanged.
The search returns a cheapest cover of the demands of~$Z$ with at most~$K$ vertices, or reports that there is none; a repeated copy never makes a cover cheaper.
If the cover found has cost at most~$K$,~$Z$ is marked solved and the cover is output as~$D_Z$.
It is optimal: an optimal cover has cost~$\OPT_c(Z)$ at most the cost of the one found, hence at most~$K$, so it has at most~$K$ vertices and was a candidate for the search.
If $\OPT_c(Z)\le K$, the cover found is optimal and has cost at most~$K$, so~$Z$ is solved.
Different components are disjoint and do not interact.
The rounds are those of Lemma~\ref{simp:local-solver}, and the records are bounded lists and counts at most~$n$, so messages have~$\mathcal O(\log n)$ bits.
\end{proof}

\begin{lemma}
\label{fv:imp:absorb}
Fix a rational $\rho\ge1$, integers $B\ge1$ and $q\ge0$, a real $r\ge1$, and a class of covering instances with costs in $\{1,\ldots,B\}$ whose graphs have a ratio of edges to vertices of at most~$\rho$ in every subgraph.
Suppose that for every $\delta>0$ there is a deterministic constant-round broadcast CONGEST algorithm~$\mathcal A_\delta$ that returns on every instance of the class a feasible cover~$D_0$ with
\[
 c(D_0)\le r(1+\delta)\tau_c+q.
\]
Then for every $\varepsilon>0$ there is a deterministic constant-round broadcast CONGEST algorithm that returns on every instance of the class a feasible cover of cost at most $(r+\varepsilon)\OPT_c$.
\end{lemma}

\begin{proof}
Choose a rational $\delta\le\varepsilon/(2r)$ and put $K=\max\{1,\lceil2q/\varepsilon\rceil\}$.
Run~$\mathcal A_\delta$ and the algorithm of Lemma~\ref{fv:imp:exact} with $a=\rho$ and budget~$K$.
Every vertex of a solved component~$Z$ takes the output~$D_Z$, and every other vertex takes its output in~$D_0$.
This is a local decision, so the number of rounds is constant.

\emph{Feasibility.}
A vertex of~$D_0$ covers only demands in its own instance component.
For every solved~$Z$, the set $D_0\cap Z$ covers the demands of~$Z$, so $c(D_0\cap Z)\ge\OPT_c(Z)=c(D_Z)$.
The output~$D$ is a cover of the demands of every instance component, and $c(D)\le c(D_0)$.

\emph{Cost.}
If every instance component is solved, then $c(D)=\sum_Z\OPT_c(Z)=\OPT_c\le(r+\varepsilon)\OPT_c$, since $r\ge1$.
Otherwise some component~$Z$ is not solved, so $\OPT_c(Z)>K$ by Lemma~\ref{fv:imp:exact}, and $\OPT_c\ge\OPT_c(Z)>K\ge2q/\varepsilon$.
Since $\tau_c\le\OPT_c$,
\[
 c(D)\le c(D_0)\le r(1+\delta)\OPT_c+q
 \le\Bigl(r+\frac\varepsilon2\Bigr)\OPT_c+\frac\varepsilon2\OPT_c
 =(r+\varepsilon)\OPT_c .
\]
The additive term~$q$ is global, but it is charged only in the second case, where the whole instance has a large optimum.
Instances with many small components fall under the first case or are absorbed in~$\OPT_c$.
\end{proof}

\subsection{Fixed surfaces}

The Euler genus of a connected graph is the least Euler genus of a surface in which it embeds, and the Euler genus of a graph is the sum over its components.
Three standard facts are used~\cite[Propositions~3.4.1 and~3.4.2, Theorem~4.4.3]{MoharThomassen2001}.
Euler genus does not increase under taking subgraphs and minors of a connected graph; it is additive over the blocks of a graph, so deleting a bridge does not increase it either, and the first fact extends to disconnected graphs; and a connected graph has a cellular embedding in a surface of its Euler genus, so that every face is an open disc.
Thus the graphs of Euler genus at most~$e$ form a minor-closed class.

\begin{lemma}
\label{fv:imp:euler}
Let~$G$ be a simple graph of Euler genus at most~$e$ and girth at least $g\ge3$, and put $a=g/(g-2)$.
Every subgraph~$F$ of~$G$ satisfies
\[
 e(F)\le a|V(F)|+ae.
\]
In particular, every graph of Euler genus at most~$e$ has $e(F)\le3|V(F)|+3e$ for all its subgraphs~$F$, and lies in~$\mathcal C_{3+3e}$.
\end{lemma}

\begin{proof}
Let~$F$ be a subgraph of~$G$.
Its Euler genus is at most~$e$ and is the sum of the Euler genera~$e_j$ of its components~$F_j$, so $\sum_je_j\le e$.
It suffices to show $e(C)\le a|V(C)|+ae_j$ for every component $C=F_j$, and then add these bounds.

Delete vertices of degree at most one repeatedly.
Every deletion removes at most one edge and decreases $a|V|$ by $a\ge1$, so it suffices to prove the bound for the graph~$C'$ that remains.
This graph is connected, and it is empty or a single vertex, where the bound is trivial, or it has minimum degree at least two.
In the last case~$C'$ has Euler genus $e'\le e_j$.
Take a cellular embedding in a surface of Euler genus~$e'$, with~$n$ vertices,~$m$ edges, and~$f$ faces, so that $n-m+f=2-e'$.
The boundary walk of a face has no immediate reversal, as every degree is at least two.
As in the proof of Lemma~\ref{sc:face}, a closed walk without immediate reversals in a simple graph contains a cycle of length at most its length, so every facial walk has length at least~$g$.
Every edge has two sides, so~$2m\ge gf$, and
\[
 2-e'=n-m+f\le n-m+\frac{2m}g,
\]
that is, $m\le\frac g{g-2}(n-2+e')\le an+ae_j$.

For $g=3$, every simple graph qualifies and $a=3$.
A nonempty minor~$J$ of~$G$ has Euler genus at most~$e$, so $e(J)\le3|V(J)|+3e\le(3+3e)|V(J)|$, which shows $G\in\mathcal C_{3+3e}$.
\end{proof}

Theorem~\ref{new:surface-seven} gives the exact bound~$7\gamma(G)$ on every fixed surface, without an embedding.
The bound cannot in general be strengthened to~$7\gf(G)$: Proposition~\ref{sc:gap} gives a finite graph with $\gamma(G)=15$ and $\gf(G)\le2$, so even a sequential algorithm with unrestricted computation cannot guarantee~$7\gf(G)$ on every fixed surface.
The following bounds use additional girth information and the general absorption lemma.

\begin{corollary}
\label{fv:imp:surface-girth}
Fix $e\ge0$, $\varepsilon>0$, and $g\ge4$, and let $a=g/(g-2)$.
There is a deterministic constant-round broadcast CONGEST algorithm that returns on every graph of Euler genus at most~$e$ and girth at least~$g$ a dominating set~$D$ with
\begin{enumerate}[label=(\roman*)]
\item $|D|\le\bigl(3+\tfrac4{g-2}+\varepsilon\bigr)\gamma(G)$, and
\item $|D|\le(c_g+\varepsilon)\gamma(G)$ if $g=5$ or $g\ge7$.
\end{enumerate}
\end{corollary}

\begin{proof}
By Lemma~\ref{fv:imp:euler}, $G\in\mathcal C_{3+3e}$ and every subgraph satisfies $e(F)\le a|V(F)|+ae$.
The solver of Theorem~\ref{sc:thmA} is run with $\rho=3+3e$ and a rational accuracy~$\delta$, and its output is capped at one.
For~(i), round with Lemma~\ref{fv:imp:additive}, which gives $|D_0|\le(2a+1)(1+\delta)\gf(G)+2e$ with $2a+1=3+4/(g-2)$.
For~(ii), $2a\notin\mathbb Z$; choose a rational $\mu<\mu_{\max}$ with $b-\mu\le c_g+\varepsilon/2$ and round with Lemma~\ref{fv:imp:lemma} and $\eta=ae$, which gives $|D_0|\le(b-\mu)(1+\delta)\gf(G)+q_\eta$.
In both cases the hypothesis of Lemma~\ref{fv:imp:absorb} holds, with $r=2a+1$ and $q=2e$ in~(i), and with $r=b-\mu$ and $q=\lceil q_\eta\rceil$ in~(ii), and the lemma is applied with~$\varepsilon$ in~(i) and $\varepsilon/2$ in~(ii).
\end{proof}

\paragraph{Remarks.}
Graphs of Euler genus at most two, which include the projective plane, the torus, and the Klein bottle, have no excess beyond~$3n$: they lie in~$\mathcal C_3$, since $m\le3n-6+3e\le3n$ for $e\le2$.
The upper bound seven is therefore part~\ref{sc:thmC:minor} of Theorem~\ref{sc:thmC}, and Corollary~\ref{sc:optimal} shows that it is optimal.
The exact coefficient three on pseudoforests is Theorem~\ref{sc:thmC} with $a=1$.
The randomized constant-round scheme and the deterministic~$\mathcal O(\log^*n)$-round scheme of Theorem~\ref{sc:thmD} apply to every class of Euler genus at most~$e$, since these classes lie in~$\mathcal C_{3+3e}$, so on every fixed surface the optimum itself can be approached with these resources.

\section{Exact coefficients for total domination}\label{fv:total}

A set~$D$ is a \emph{total dominating set} of a graph~$G$ if every vertex of~$G$, whether it belongs to~$D$ or not, has a neighbor in~$D$.
Such a set exists exactly when~$G$ has no isolated vertex, and~$G$ is then called \emph{isolate-free}.
The \emph{fractional total domination number}~$\gamma_{t,f}(G)$ is the minimum of $\sum_vy_v$ over all $y\ge0$ with $\sum_{u\in N(v)}y_u\ge1$ for every vertex~$v$.
Corollary~\ref{sc:rounding-all} gives total domination the coefficient $2a+\varepsilon$, for integer~$a$, on graphs in~$\mathcal C_\rho$ whose subgraphs have a ratio of edges to vertices of at most~$a$.
We remove the error and prove the exact coefficient~$2a$, using the same two tools as for domination: the solver of Theorem~\ref{sc:thmA} and the principle of Proposition~\ref{sc:principle}.

\begin{theorem}
\label{fv:total:exact}
There are deterministic constant-round~$\BC$ algorithms that return, on an isolate-free graph~$G$, a total dominating set~$D$ with $|D|\le2a\,\gamma_{t,f}(G)$ in each of the following cases.
\begin{enumerate}[label=(\roman*)]
\item\label{fv:total:exact:minor} $G\in\mathcal C_a$, for every fixed rational $a\ge1$.
\item\label{fv:total:exact:sub} $G\in\mathcal C_\rho$ and every subgraph of~$G$ has a ratio of edges to vertices of at most~$a$, for fixed rationals $1\le a\le\rho$ such that~$2a$ is not an integer.
\end{enumerate}
\end{theorem}

The coefficient is one smaller than for domination, because a demand is covered only by a neighbor and never by itself.
We must account for this difference in the proof.
A selected vertex may itself be undominated, and then many later candidates can be adjacent to the same selected vertex.
We pay for this family of candidates with the fractional weight of the selected vertex, which is the new step compared with Section~\ref{simp:endpoint:section}.
As there, two cases arise.
When~$2a$ is not an integer, the gap between~$2a$ and the integer degrees gives a saving directly (Section~\ref{fv:total:nonint}).
When~$2a$ is an integer, the regions in which all degrees equal~$2a$ are selected first, and their budget is transferred to the remaining candidates (Section~\ref{fv:total:int}).
The planar girth bounds and the treewidth bound follow from these two cases (Sections~\ref{fv:total:girth:sec} and~\ref{fv:total:tw:sec}).

\subsection{The rounding rule}
\label{fv:total:rule}

Throughout this section, $a\ge1$ is a fixed rational, $d=2a$, and $n_0=\lceil d\rceil$.
Assume that~$G$ is isolate-free and that every subgraph~$J$ of~$G$ satisfies $|E(J)|\le a|V(J)|$.
Write $N(v)=N_G(v)$, and~$e(X)$ for the number of edges of~$G[X]$ and~$e(X,Y)$ for the number of edges between disjoint sets~$X$ and~$Y$.

\paragraph{The rule.}
The input is a set $K\subseteq V(G)$ of candidates, a set $R_0\subseteq V(G)$ of demands, costs $c(v)\ge0$ for $v\in K$, a set $\Pi\subseteq K$ of vertices that may be selected in the first step, a threshold $t>0$, and a vector $x\in[0,1]^K$, extended by zero to all of~$V(G)$, such that $\sum_{v\in N(r)}x_v\ge1$ for every $r\in R_0$.
\begin{enumerate}[label=\textbf{Step \arabic*.},leftmargin=*,itemsep=2pt,parsep=0pt,topsep=4pt]
\item Select the vertices of~$\Pi$ with fractional weight greater than~$t$, and let~$R$ be the demands that have no selected neighbor:
\[
 W=\{v\in\Pi:x_v>t\},\qquad R=R_0\setminus N(W).
\]
\item For each remaining candidate, count the demands of~$R$ that it covers, and select those that cover at least~$n_0$:
\[
 h(v)=|N(v)\cap R|\quad(v\in K\setminus W),\qquad
 H=\{v\in K\setminus W:h(v)\ge n_0\}.
\]
\item Let $Y=R\cap N(H)$ and $U=R\setminus Y$.
Every demand of~$U$ has a neighbor, because~$G$ is isolate-free, and requests the neighbor of smallest identifier.
Let~$D'$ be the union of $W\cup H$ and the requested vertices.
\end{enumerate}
A vertex of~$W$ can belong to~$R$: it is removed from~$R_0$ only if it has a neighbor in~$W$.
Every demand of $R_0\setminus R$ has a neighbor in~$W$, every demand of~$Y$ has a neighbor in~$H$, and every demand of~$U$ has a requested neighbor.
Hence every demand of~$R_0$ has a neighbor in~$D'$.

\paragraph{Notation for the analysis.}
Let $O=K\setminus(W\cup H)$, and let $\cost(x)=\sum_{v\in K}c(v)x_v$.
Write
\[
 w=|W|,\qquad j=|Y|-|H|,\qquad E=\sum_{v\in H}\bigl(h(v)-d\bigr),
\]
\[
 P=\sum_{v\in O}\bigl(c(v)-h(v)\bigr)x_v,\qquad
 \Omega=\sum_{r\in R}\Bigl(\sum_{v\in N(r)}x_v-1\Bigr).
\]
The quantity $\Omega\ge0$ is the total excess coverage at the remaining demands.
The quantity~$E$ is nonnegative, because $h\ge n_0\ge d$ on~$H$.
Every vertex of~$K$ lies in~$W$, in~$H$ or in~$O$.

\begin{lemma}
\label{fv:total:rounding}
Assume the following three conditions.
\begin{description}[leftmargin=2.2em,itemsep=1pt,parsep=0pt,topsep=2pt,labelwidth=2em]
\item[(R1)] $c(v)=d$ for every $v\in W\cup H$.
\item[(R2)] $c(v)\ge h(v)$ for every $v\in O$.
\item[(R3)] $x_v\le t$ for every $v\in H$ with $h(v)>d$.
\end{description}
Then $P\ge0$, and
\begin{align}
 E&\le a\,j,\qquad\text{in particular } j\ge0,\label{fv:total:E}\\
 |D'|&\le w+|R|-j,\label{fv:total:size}\\
 \cost(x)-|D'|&\ge(dt-1)w+(1-at)j+P+\Omega.\label{fv:total:saving}
\end{align}
\end{lemma}

\begin{proof}
The inequality $P\ge0$ is (R2).
The sets $W,H$ are disjoint, and $|U|=|R|-|Y|=|R|-|H|-j$, so at most $|W|+|H|+|U|=w+|R|-j$ vertices are selected, which is \eqref{fv:total:size}.

For \eqref{fv:total:E}, count the pairs $(v,r)$ with $v\in H$, $r\in R$ and $vr\in E(G)$.
There are $\sum_{v\in H}h(v)=d|H|+E$ of them.
Every such~$r$ lies in~$Y$, so the pair is an edge of~$G[H\cup Y]$ counted once or twice, and it is counted twice only if both ends lie in $H\cap Y$.
Hence
\[
 d|H|+E\le e(H\cup Y)+e(H\cap Y)\le a\bigl(|H\cup Y|+|H\cap Y|\bigr)
 =a\bigl(|H|+|Y|\bigr).
\]
Since $d=2a$, this reads $E\le a(|Y|-|H|)=aj$.
As $E\ge0$, also $j\ge0$.

For \eqref{fv:total:saving}, sum the constraints at the demands of~$R$.
No vertex of~$W$ is adjacent to a demand of~$R$, so $\sum_{v\in K\setminus W}h(v)x_v=|R|+\Omega$.
By (R1),
\[
 \cost(x)=d\sum_{v\in W}x_v+d\sum_{v\in H}x_v+\sum_{v\in O}h(v)x_v+P
 =d\sum_{v\in W}x_v+|R|+\Omega+P-\sum_{v\in H}\bigl(h(v)-d\bigr)x_v.
\]
Here $x_v>t$ on~$W$.
On~$H$, a vertex with $h(v)=d$ contributes zero to the last sum, and a vertex with $h(v)>d$ has $x_v\le t$ by (R3), so the last sum is at most~$tE$.
Therefore $\cost(x)\ge dtw+|R|+\Omega+P-tE$.
Subtract \eqref{fv:total:size} and use $E\le aj$ and $t>0$.
\end{proof}

\subsection{When \texorpdfstring{$2a$}{2a} is not an integer}
\label{fv:total:nonint}

Put $k=\lfloor d\rfloor$ and $\theta=d-k\in(0,1)$.
All vertices are candidates and demands, all costs equal~$d$, and every vertex may be selected in the first step.
The next lemma rounds any feasible fractional total dominating vector without error.

\begin{lemma}
\label{fv:total:noncrit}
Let $d=2a$ be not an integer, and let~$\mu$ be a positive rational with
\[
 \mu\le\theta,\qquad(1+\mu)(1+\mu-\theta)\le2(1-\theta).
\]
There is a deterministic constant-round~$\BC$ rule that turns every feasible fractional total dominating vector~$x$ of an isolate-free graph~$G$, all of whose subgraphs have a ratio of edges to vertices of at most~$a$, into a total dominating set~$D$ with
\[
 |D|\le(d-\mu)\sum_vx_v.
\]
\end{lemma}

\begin{proof}
Every coordinate that exceeds one may be replaced by one, which preserves feasibility and does not increase $\sum_vx_v$.
Apply the rule of Section~\ref{fv:total:rule} with $K=R_0=\Pi=V(G)$, $c\equiv d$, $t=(1+\mu)/d$, and $n_0=k+1$, and take $D=D'$.
Then~$D$ is a total dominating set.
Condition (R1) holds, (R2) holds since $h(v)\le k<d$ on~$O$, and (R3) holds since $\Pi=V(G)$, so~$W$ contains every vertex of weight above~$t$.
Also $\cost(x)=d\sum_vx_v$.

On~$H$ we have $h(v)\ge k+1$, so every term of~$E$ is at least $1-\theta$, and $|H|\le E/(1-\theta)\le aj/(1-\theta)$ by \eqref{fv:total:E}.
On~$O$ we have $d-h(v)\ge\theta$, so $\sum_{v\in O}x_v\le P/\theta$.
Since $x_v\le1$ on~$W$ and $x_v\le t$ on~$H$,
\[
 \sum_vx_v\le w+t|H|+\sum_{v\in O}x_v\le w+\frac{ta}{1-\theta}\,j+\frac P\theta.
\]
By \eqref{fv:total:saving} with $dt-1=\mu$ and $\Omega\ge0$, $\cost(x)-|D|\ge\mu w+(1-at)j+P$.
Compare coefficients with $\mu\sum_vx_v$.
For~$w$ they are equal, and for~$P$ we have $1\ge\mu/\theta$.
For~$j$ we need $1-at\ge\mu ta/(1-\theta)$.
Since $at=(1+\mu)/2$, this is $(1+\mu)(1-\theta+\mu)\le2(1-\theta)$, which is our second hypothesis.
As $w,j,P\ge0$, it follows that $d\sum_vx_v-|D|\ge\mu\sum_vx_v$.

The rule needs one broadcast of the bit for membership in~$W$, one for~$R$, one for~$H$, and one broadcast of the requested identifier.
This takes a constant number of rounds after the identifiers of the neighbors are known.
\end{proof}

The second condition is $f(\mu):=\mu^2+(2-\theta)\mu+\theta-1\le0$, which is equivalent to $\mu(1+\mu)\le(1-\mu)(1-\theta)$.
The function~$f$ is increasing for $\mu>0$ and $f(0)<0$, so the condition holds exactly for $\mu\le(\theta-2+\sqrt{\theta^2-8\theta+8})/2$.
Write
\[
 \mu_{\max}(\theta)=\min\Bigl\{\theta,\
 \frac{\theta-2+\sqrt{\theta^2-8\theta+8}}2\Bigr\}>0 .
\]
A rational~$\mu$ satisfies the hypotheses of Lemma~\ref{fv:total:noncrit} exactly when $0<\mu\le\mu_{\max}(\theta)$.

\begin{corollary}
\label{fv:total:noncrit-cor}
Fix a rational $\rho\ge1$.
Let~$G$ be isolate-free, $G\in\mathcal C_\rho$, and let every subgraph of~$G$ have a ratio of edges to vertices of at most~$a$, where $d=2a$ is not an integer and $\theta=d-\lfloor d\rfloor$.
There are deterministic constant-round~$\BC$ algorithms that return a total dominating set~$D$ with
\begin{enumerate}[label=(\roman*)]
\item $|D|\le(d-\mu_0)\,\gamma_{t,f}(G)$, for every fixed rational~$\mu_0$ with $0<\mu_0<\mu_{\max}(\theta)$;
\item $|D|\le(d-\mu_{\max}(\theta)+\varepsilon)\,\gamma_{t,f}(G)$, for every fixed $\varepsilon>0$.
\end{enumerate}
In particular, $|D|\le d\,\gamma_{t,f}(G)$, which is part~\ref{fv:total:exact:sub} of Theorem~\ref{fv:total:exact}.
\end{corollary}

\begin{proof}
Let~$\mu$ be a rational with $\mu_0<\mu\le\mu_{\max}(\theta)$ in case~(i), and with $0<\mu\le\mu_{\max}(\theta)$ and $\mu\ge\mu_{\max}(\theta)-\varepsilon/2$ in case~(ii).
The solver of Theorem~\ref{sc:thmA} runs on~$G$ with all vertices as demands, covering sets $C(r)=N(r)$, unit costs, and an accuracy~$\delta$ fixed below.
The covering sets are nonempty since~$G$ is isolate-free.
Capping the output~$x$ at one does not increase its cost, and Lemma~\ref{fv:total:noncrit} then applies.
Since $\sum_vx_v\le(1+\delta)\gamma_{t,f}(G)$, this gives $|D|\le(d-\mu)(1+\delta)\gamma_{t,f}(G)$.
In case~(i) take $\delta=(\mu-\mu_0)/(d-\mu)$, so that $(d-\mu)(1+\delta)=d-\mu_0$.
In case~(ii) take a rational $\delta\le\varepsilon/(2d)$; then $(d-\mu)(1+\delta)\le d-\mu_{\max}+\varepsilon/2+d\delta=d-\mu_{\max}+\varepsilon$.
The last statement follows from~(i), since $d-\mu_0\le d$.
\end{proof}

Part~\ref{fv:total:exact:sub} of Theorem~\ref{fv:total:exact} is the last statement of the corollary.
Part~\ref{fv:total:exact:minor} for $2a\notin\mathbb Z$ is the special case $\rho=a$.
Only the integer case remains.

\subsection{When \texorpdfstring{$2a$}{2a} is an integer}
\label{fv:total:int}

Now $d=2a$ is an integer, so $d\ge2$ and $n_0=d$.
We first apply the budget transfer of Section~\ref{simp:endpoint:section}.
Let~$G$ be isolate-free, and let $B=\{v:\deg(v)\ne d\}$ be the set of \emph{defects}.
Put
\[
 F_0=V(G)\setminus N^2[B],
\]
remove from~$F_0$ every vertex that forms a singleton component of~$G[F_0]$; the remaining set is~$F$.
All of~$F$ is selected, and $Z=V(G)\setminus F$.
Every vertex of~$F$ has a neighbor in~$F$, so~$F$ totally dominates itself.
The vertices of~$F$ have distance at least three from~$B$, which is why all vertices within distance two of~$F$ have degree~$d$.
The demands that remain are the vertices of~$Z$, and the candidate costs are
\[
 c(v)=d-|N(v)\cap F|,\qquad K=\{v:c(v)>0\}.
\]
Let~$\cost^*$ be the minimum of $\sum_{v\in K}c(v)x_v$ over all $x\in[0,1]^K$ with $\sum_{v\in N(r)}x_v\ge1$ for every $r\in Z$.
Capping at one does not change the optimum, so~$\cost^*$ is also the optimum of the covering problem of Theorem~\ref{sc:thmA} with demands~$Z$ and covering sets~$N(r)$.

\begin{lemma}
\label{fv:total:budget}
\leavevmode
\begin{enumerate}[label=(\roman*)]
\item Every vertex within distance two of~$F$ has degree~$d$.
No vertex of~$N[B]$ belongs to~$F$ or has a neighbor in~$F$, so $N[B]\subseteq Z$.
If $Z\ne\varnothing$, then $B\ne\varnothing$, and every vertex of~$Z$ has a neighbor in~$Z$ and a path of length at most three inside~$Z$ to~$B$.
\item The costs satisfy $c(v)\ge0$ for all~$v$, and $c(v)\in\{1,\ldots,d\}$ on~$K$.
We have $N(Z)\subseteq K$ and $Z\subseteq K$.
\item $|F|+\cost^*\le d\,\gamma_{t,f}(G)$.
\end{enumerate}
\end{lemma}

\begin{proof}
(i) Let~$v$ be within distance two of $f\in F$.
Since $f\notin N^2[B]$, the distance from~$v$ to~$B$ is at least one, so $v\notin B$ and $\deg(v)=d$.
A vertex of~$N[B]$ is within distance one of~$B$, hence it is not in~$F$ and has no neighbor in~$F$.
If $B=\varnothing$, then $F_0=V(G)$, and since~$G$ is isolate-free no vertex is removed, so $F=V(G)$ and $Z=\varnothing$.
So let $B\ne\varnothing$, and let $z\in Z$.
If $z\in N^2[B]$, a shortest path from~$z$ to~$B$ has length at most two and lies in $N^2[B]\subseteq Z$.
Otherwise $z\in F_0\setminus F$ is a singleton component of~$G[F_0]$.
Since~$G$ is isolate-free,~$z$ has a neighbor, which lies outside~$F_0$, hence in~$N^2[B]$, and from there a path of length at most two leads to~$B$.
This gives a path inside~$Z$ of length at most three.
Every vertex of~$Z$ also has a neighbor in~$Z$: the second vertex of such a path, or, if $z\in B$, any neighbor of~$z$, which lies in $N[B]\subseteq Z$.

(ii) A vertex with a neighbor in~$F$ has degree~$d$ by (i), so $c(v)\ge0$ for all~$v$.
Let $r\in Z$ and $u\in N(r)$.
If $c(u)=0$, then~$u$ has~$d$ neighbors in~$F$, so all its neighbors lie in~$F$, which contradicts $r\in N(u)\setminus F$.
So $N(Z)\subseteq K$.
Every $z\in Z$ has a neighbor in~$Z$, so $z\in N(Z)\subseteq K$.

(iii) Let~$y$ be an optimal fractional total dominating vector of~$G$.
Its restriction to $N(Z)\subseteq K$ covers all demands of~$Z$, so $\cost^*\le\sum_{v\in K}c(v)y_v=\sum_vc(v)y_v$, where the sum may run over all~$v$ because $c(v)=0$ outside~$K$.
Summing the constraints of~$y$ over~$F$ gives $|F|\le\sum_v|N(v)\cap F|\,y_v$.
Since $c(v)+|N(v)\cap F|=d$ for all~$v$, this gives $|F|+\cost^*\le d\sum_vy_v=d\gamma_{t,f}(G)$.
\end{proof}

\paragraph{The rounding.}
Let $L_F=Z\cap N(F)$ be the demands of~$Z$ that~$F$ already dominates, and put
\[
 R_0=Z\setminus L_F,\qquad\Pi=\{v\in K:\deg(v)>d\},\qquad t=\frac{3}{2d}.
\]
The solver of Theorem~\ref{sc:thmA} runs on~$G$, which lies in~$\mathcal C_\rho$ for a~$\rho$ fixed below, with the demands~$Z$, the covering sets~$N(r)$, the costs $c\in\{1,\ldots,d\}$ on~$K$, and a fixed accuracy~$\delta$ chosen below.
The solver returns a feasible vector, which we cap at one and extend by zero to~$K$; call this vector~$x$.
The rule of Section~\ref{fv:total:rule} is then applied with $K,R_0,c,\Pi,t$, and the output is $F\cup D'$.
This is a total dominating set:~$F$ dominates $F\cup L_F$, and~$D'$ dominates $R_0=Z\setminus L_F$.
The sets $W,R,H,Y,U,O$ and the quantities $w,j,E,P,\Omega$ are those of the rule.

\paragraph{Notation for the analysis.}
Beyond the notation of the rule, put
\[
 Q=H\cap R,\qquad A=H\setminus R,\qquad L=Y\setminus H,\qquad I=Q\setminus Y,
 \qquad\ell=|L|.
\]
Then $R=Q\cup L\cup U$ with $Q\cap L=\varnothing$ and $Q\cap U=I$: a vertex of~$R$ in~$H$ lies in~$Q$, a vertex of~$Y$ outside~$H$ lies in~$L$, and $Q\setminus Y\subseteq R\setminus Y=U$.
Moreover $Y=(Q\setminus I)\cup L$, so
\[
 j=\ell-|A|-|I|.
\]
No vertex of~$U$ has a neighbor in~$H$, so $N(v)\cap R\subseteq Q\cup L$ for $v\in H$.

\begin{lemma}
\label{fv:total:setup}
For every feasible $x\in[0,1]^K$:
\begin{enumerate}[label=(\roman*)]
\item Conditions (R1)-(R3) hold, $N[B]\subseteq R_0$, and $H\subseteq R_0$.
\item $|L_F|\le P$.
\item $\sum_{v\in O}h(v)x_v\le2(d-1)P$, and hence $|U|\le2(d-1)P$.
\end{enumerate}
\end{lemma}

\begin{proof}
(i) The vertices of~$N[B]$ lie in~$Z$ and have no neighbor in~$F$, so $N[B]\subseteq Z\setminus L_F=R_0$.
For~(R1), a vertex $v\in W$ has degree above~$d$, so it has no neighbor in~$F$ by Lemma~\ref{fv:total:budget}(i), and $c(v)=d$.
Let $v\in H$.
If~$v$ had a neighbor in~$F$, then $\deg(v)=d$ and this neighbor is not in $R\subseteq Z$, so $h(v)\le d-1<n_0$, contradicting $v\in H$.
Hence $c(v)=d$ on~$H$.
Moreover~$v$ is not in~$F$ and not in~$L_F$, because every vertex of $F\cup L_F$ has a neighbor in~$F$.
Hence $v\in Z\setminus L_F=R_0$.
For (R2), let $v\in O$.
If~$v$ has no neighbor in~$F$, then $c(v)=d>h(v)$ because $v\notin H$.
Otherwise $\deg(v)=d$ and $c(v)=|N(v)\cap Z|\ge|N(v)\cap R|=h(v)$.
For (R3), if $h(v)>d$ for $v\in H$, then $\deg(v)>d$, so $v\in\Pi$ and $x_v\le t$, since $v\notin W$.

(ii) Let $\lambda\in L_F$ and $v\in N(\lambda)$.
Then~$v$ is within distance two of~$F$, so $\deg(v)=d$, $v\notin\Pi$, and $c(v)=|N(v)\cap Z|$.
The sets $N(v)\cap R$ and $N(v)\cap L_F$ are disjoint subsets of $N(v)\cap Z$, and the second one is nonempty.
So $c(v)\ge h(v)+|N(v)\cap L_F|$ and $h(v)<c(v)\le d$; also $v\in K$ since $N(Z)\subseteq K$.
Thus $v\in O$.
Summing the constraints at~$L_F$ gives
\[
 |L_F|\le\sum_{v\in O}|N(v)\cap L_F|\,x_v\le\sum_{v\in O}\bigl(c(v)-h(v)\bigr)x_v=P.
\]

(iii) A demand $u\in U$ has no neighbor in $W\cup H$, and all its neighbors are in~$K$, so its constraint involves only vertices of~$O$.
Summing over~$U$ gives $|U|\le\sum_{v\in O}|N(v)\cap U|\,x_v\le\sum_{v\in O}h(v)x_v$.
Let $O^+=\{v\in O:c(v)>h(v)\}$.
For $v\in O^+$ we have $c(v)-h(v)\ge1$, since costs and~$h$ are integers, and $h(v)\le c(v)-1\le d-1$.
So $\sum_{v\in O^+}h(v)x_v\le(d-1)P$.
Now let $v\in O\setminus O^+$ with $h(v)\ge1$.
Then $c(v)=h(v)\le d-1$, so~$v$ has a neighbor in~$F$.
If $v\in F$, then all its neighbors in~$Z$ lie in~$L_F$, so $h(v)=0$, which is excluded.
Hence $v\in Z\cap N(F)=L_F$.
As $x_v\le1$ and $h(v)\le d-1$, these vertices contribute at most $(d-1)|L_F|\le(d-1)P$, by (ii).
Altogether the sum is at most $2(d-1)P$.
\end{proof}

\subsection{The count of selected vertices}

The remaining analysis bounds $|D'|$ by a combination of $w,j,P,\Omega$ with constant coefficients.
The first step is to control $\ell=|L|$, the demands that are covered by~$H$ but not selected.
This is where the excess coverage is needed, because a vertex of $A=H\setminus R$ is selected but is not a remaining demand.

\begin{lemma}
\label{fv:total:oversupply}
For every feasible $x\in[0,1]^K$,
\[
 \ell\le\frac{d(d+2)}2\,w+\frac{d^2}2\,j+(d-1)(d^2+2)P+d\,\Omega.
\]
\end{lemma}

\begin{proof}
Every vertex of~$A$ lies in $H\subseteq R_0$ but not in~$R$, so it has a neighbor in~$W$.
Let~$A_0$ consist of the vertices of~$A$ with a neighbor in $W\cap R$, let $A_1=A\setminus A_0$, and put $W_{\rm out}=W\setminus R$.
Each vertex of~$A_1$ has a neighbor in~$W_{\rm out}$.
Every vertex of~$A_0$ is a neighbor of some demand of $W\cap R$, so by the constraints at these demands
\[
 \sum_{v\in A_0}x_v\le\sum_{r\in W\cap R}\sum_{v\in N(r)}x_v\le w+\Omega.
\]

The sets $Q,A,L,W_{\rm out}$ are pairwise disjoint.
Consider the edges among them.
By the notation above,
\[
 d|H|+E=\sum_{v\in H}h(v)=2e(Q)+e(Q,L)+e(A,Q)+e(A,L),
 \qquad
 \sum_{v\in A}h(v)=e(A,Q)+e(A,L)\ge d|A|.
\]
Every vertex of~$A_1$ also gives an edge to~$W_{\rm out}$, and these $|A_1|$ edges are distinct and not among the edges above.
Add the two displays, then~$e(Q,L)$ and $2|A_1|$, and count every edge at most twice:
\[
 d|H|+E+d|A|+e(Q,L)+2|A_1|\le2e\bigl(Q\cup A\cup L\cup W_{\rm out}\bigr)
 \le d\bigl(|Q|+|A|+\ell+|W_{\rm out}|\bigr).
\]
The last inequality is the density bound.
With $|H|=|Q|+|A|$, $\ell-|A|=j+|I|$ and $|W_{\rm out}|\le w$, this gives
\begin{equation}\label{fv:total:A-edge}
 E+e(Q,L)+2|A_1|\le d\,(j+|I|+w).
\end{equation}

Now sum the constraints at the vertices of $L\subseteq R$.
A vertex of~$W$ is adjacent to no vertex of~$R$.
A vertex $v\in Q$ contributes at most $|N(v)\cap L|$, since $x_v\le1$, so~$Q$ contributes at most~$e(Q,L)$.
A vertex $v\in A$ contributes at most $h(v)x_v$, and $h(v)x_v\le dx_v+t(h(v)-d)$ by (R3), so~$A$ contributes at most $d\sum_{v\in A_0}x_v+d|A_1|+tE$.
A vertex of~$O$ contributes at most $h(v)x_v$, so~$O$ contributes at most $2(d-1)P$ by Lemma~\ref{fv:total:setup}(iii).
Therefore
\[
 \ell\le e(Q,L)+d|A_1|+tE+d(w+\Omega)+2(d-1)P.
\]
Since $d\ge2$ and $t=3/(2d)\le d/2$, $e(Q,L)+d|A_1|+tE\le\frac d2\bigl(e(Q,L)+2|A_1|+E\bigr)\le\frac{d^2}2(j+|I|+w)$ by \eqref{fv:total:A-edge}.
Finally $|I|\le|U|\le2(d-1)P$, and collecting terms gives the claim: the coefficient of~$w$ is $d^2/2+d$, and the coefficient of~$P$ is $d^2(d-1)+2(d-1)$.
\end{proof}

The next lemma controls the degree defects among the demands that remain.
It is the only place where the density of minors, rather than of subgraphs, is used.
It uses a minor that contracts the vertices of $N(W)\setminus W$ into stars centered at the vertices of~$W$.

\begin{lemma}
\label{fv:total:defects}
For every feasible $x\in[0,1]^K$,
\begin{enumerate}[label=(\roman*)]
\item if $G\in\mathcal C_a$, then $|B\cap Q|\le a(\ell+w)$;
\item if $a=g/(g-2)$ for an integer $g\ge3$ and~$G$ is planar of girth at least~$g$, then $|B\cap Q|\le g(\ell+w)$.
\end{enumerate}
\end{lemma}

\begin{proof}
No edge of~$G$ joins~$Q$ to~$W$, since $Q\subseteq R$ and $R\cap N(W)=\varnothing$, and $Q\cap W=\varnothing$ because $Q\subseteq H$.
Assign every vertex of $N(W)\setminus W$ to a neighbor in~$W$ and contract each vertex of~$W$ together with its assigned vertices, called its leaves.
The leaves are not in~$R$, since $R\cap N(W)=\varnothing$.
Let~$J$ be the graph on $Q\cup L\cup W$ with three kinds of edges: the edges of~$G[Q]$, the edges of~$G$ between~$Q$ and~$L$, and the~$p$ edges between~$Q$ and~$W$ of the minor after the contraction (so $q\in Q$ and $w'\in W$ are adjacent when~$q$ is adjacent to a leaf of~$w'$).
Since no edge of~$G$ joins~$Q$ to~$W$, the three kinds are disjoint even if $W\cap L\ne\varnothing$, and~$J$ is a subgraph of a minor of~$G$.
Put $E_Q=\sum_{q\in Q}(h(q)-d)\ge0$.
For $q\in Q$,~$h(q)$ is the number of neighbors of~$q$ in $Q\cup L$, so $2e(Q)+e(Q,L)=d|Q|+E_Q$ and
\[
 e(J)+e(Q)=d|Q|+E_Q+p.
\]
Moreover $|V(J)|\le|Q|+\ell+w$.

First, $|B\cap Q|\le E_Q+p$.
Let $q\in B\cap Q$.
Then $h(q)\ge d$ and $\deg(q)\ne d$, so $\deg(q)>d$.
If $h(q)>d$, then~$q$ contributes at least one to~$E_Q$.
If $h(q)=d$, then~$q$ has a neighbor $u\notin R$.
Since $u\in N[B]\subseteq R_0$, we have $u\in N(W)$.
Also $u\notin W$, since otherwise $q\in N(W)$.
So~$u$ is a leaf of some $w'\in W$, and~$J$ has the edge~$qw'$.
Different defects give different edges, because their endpoints in~$Q$ differ.

(i) Since~$J$ is a subgraph of a minor of $G\in\mathcal C_a$, $e(J)\le a|V(J)|$.
As $e(Q)\le a|Q|$, it follows that $d|Q|+E_Q+p\le a(|Q|+\ell+w)+a|Q|=d|Q|+a(\ell+w)$, and $|B\cap Q|\le E_Q+p\le a(\ell+w)$.

(ii) The graph~$J$ is planar, as a subgraph of a minor of a planar graph.
The set~$W$ is independent in~$J$, since all edges of~$J$ at~$W$ go to~$Q$.
The graph $J-W$ is a subgraph of~$G$, so it has girth at least~$g$, and exactly~$p$ edges of~$J$ are incident with~$W$.
By Lemma~\ref{sc:face}, $e(J)\le a|V(J)|+\frac{g-3}{g-2}p$, and by the same lemma for~$G[Q]$, $e(Q)\le a|Q|$.
Therefore $d|Q|+E_Q+p\le d|Q|+a(\ell+w)+\frac{g-3}{g-2}p$, that is, $E_Q+\frac p{g-2}\le a(\ell+w)$.
Multiplying by $g-2\ge1$ and using $E_Q\ge0$ gives $|B\cap Q|\le E_Q+p\le(g-2)a(\ell+w)=g(\ell+w)$.
\end{proof}

\begin{lemma}
\label{fv:total:count}
Let $\kappa\ge0$ be a constant with $|B\cap Q|\le\kappa(\ell+w)$ for every feasible $x\in[0,1]^K$, and put $M=d^3-d^2+d+1$ and $\Gamma=M(\kappa+1)$.
Then for every feasible $x\in[0,1]^K$,
\[
 |D'|\le\beta_ww+\beta_jj+\beta_PP+\beta_\Omega\Omega,
\]
where
\[
 \beta_w=1+M\kappa+\Gamma\,\frac{d(d+2)}2,\quad
 \beta_j=\Gamma\,\frac{d^2}2,\quad
 \beta_P=2(d-1)M+\Gamma(d-1)(d^2+2),\quad
 \beta_\Omega=\Gamma d.
\]
\end{lemma}

\begin{proof}
Let $T=\{r\in R:\deg_{G[R]}(r)\ne d\}\cup(B\cap R)$.
A vertex $q\in Q\setminus B$ has degree~$d$ and $h(q)\ge d$, so $\deg_{G[R]}(q)=h(q)=d$ and $q\notin T$.
Since $R=Q\cup L\cup U$, we obtain
\[
 |T|\le|B\cap Q|+\ell+|U|\le\kappa w+(\kappa+1)\ell+2(d-1)P
\]
by Lemma~\ref{fv:total:setup}(iii).

Every vertex $r\in R$ has distance at most three from~$T$ in~$G[R]$.
Indeed, by Lemma~\ref{fv:total:budget}(i) there is a path $r=z_0,\ldots,z_s$ in~$Z$ with $s\le3$ and $z_s\in B$.
Let~$i$ be the largest index with $z_0,\ldots,z_i\in R$.
If $i=s$, then $z_s\in B\cap R\subseteq T$.
Otherwise~$z_i$ has the neighbor $z_{i+1}\notin R$.
If $z_i\in B$ then $z_i\in T$, and if $z_i\notin B$ then $\deg(z_i)=d$ and $\deg_{G[R]}(z_i)<d$, so $z_i\in T$ as well.
In both cases the path $z_0,\ldots,z_i$ lies in~$G[R]$.

Vertices of $R\setminus T$ have degree exactly~$d$ in~$G[R]$.
Since~$G[R]$ has at most $a|R|$ edges, $\sum_{v\in T}\deg_{G[R]}(v)\le d|T|$.
So at most $d|T|$ vertices of~$R$ are at distance one from~$T$ in~$G[R]$, and a vertex outside~$T$ has at most $d-1$ neighbors besides the one from which it is reached.
Hence at most $d(d-1)|T|$ vertices are at distance two and at most $d(d-1)^2|T|$ at distance three, and $|R|\le M|T|$.

By \eqref{fv:total:size} and $j\ge0$, $|D'|\le w+|R|\le w+M|T|$.
Insert the bound on $|T|$ and then Lemma~\ref{fv:total:oversupply} for~$\ell$.
\end{proof}

\begin{proof}[Proof of Theorem~\ref{fv:total:exact} for integer $2a$]
Let $G\in\mathcal C_a$ be isolate-free with $d=2a$ an integer.
The construction applies with $\rho=a$ and $\kappa=a$, which Lemma~\ref{fv:total:defects}(i) allows.
If $Z=\varnothing$, then $F=V(G)$ is a total dominating set with $|F|\le d\gamma_{t,f}(G)$ by Lemma~\ref{fv:total:budget}(iii).
Otherwise let
\[
 \delta=\min\Bigl\{\frac1{2\beta_w},\ \frac1{4\beta_j},\
 \frac1{\beta_P},\ \frac1{\beta_\Omega}\Bigr\},
\]
a rational number that depends only on~$a$.
The quantities $w,j,P,\Omega$ are nonnegative.
For every feasible $x\in[0,1]^K$, Lemma~\ref{fv:total:rounding} applies by Lemma~\ref{fv:total:setup}(i), and $t=3/(2d)$ gives $dt-1=\frac12$ and $1-at=\frac14$, so
\[
 \cost(x)-|D'|\ge\tfrac12w+\tfrac14j+P+\Omega.
\]
Lemma~\ref{fv:total:count} gives $|D'|\le\beta_ww+\beta_jj+\beta_PP+\beta_\Omega\Omega$.
By Proposition~\ref{sc:principle} with $\alpha=(\frac12,\frac14,1,1)$, the rule has slack~$\delta$ on the program of the costs~$c$ over $[0,1]^K$, and every feasible~$x$ with $\cost(x)\le(1+\delta)\cost^*$ satisfies $|D'|\le\cost^*$.
The solver of Theorem~\ref{sc:thmA}, followed by capping, returns such an~$x$.
Hence $F\cup D'$ is a total dominating set with
\[
 |F\cup D'|\le|F|+\cost^*\le d\,\gamma_{t,f}(G)
\]
by Lemma~\ref{fv:total:budget}(iii).
\end{proof}

\paragraph{Implementation.}
All parameters are fixed rationals.
The vertices learn~$B$,~$N[B]$,~$N^2[B]$,~$F_0$,~$F$, the costs, and membership in~$Z$ and in~$L_F$ by a constant number of broadcasts of one bit each, after degrees and identifiers are exchanged.
The solver runs on~$G$ with demands~$Z$, and the rule is then carried out by broadcasts of the bits for membership in~$W$,~$R$ and~$H$, followed by the requests, as in Lemma~\ref{fv:total:noncrit}.
The contractions, the edge counts and the drawings are used only in the analysis.
This completes the proof of Theorem~\ref{fv:total:exact}.

\subsection{Planar graphs of large girth}
\label{fv:total:girth:sec}

For planar graphs with a girth bound, the solver runs on the planar input, and the rounding uses the smaller density $a=g/(g-2)$.
Here $d=2a=2+\frac4{g-2}$.

\begin{corollary}
\label{fv:total:girth}
For every fixed integer $g\ge3$, there is a deterministic constant-round~$\BC$ algorithm that returns, on every isolate-free planar graph~$G$ of girth at least~$g$, a total dominating set of size at most
\[
 \frac{2g}{g-2}\,\gamma_{t,f}(G).
\]
In particular the coefficients are~$6$,~$4$ and~$3$ for $g=3,4,6$.
For $g=5$ and $g\ge7$, Table~\ref{fv:total:table} gives smaller coefficients.
The coefficients~$6$ and~$4$ are optimal by \cref{fv:total:lower}.
\end{corollary}

\begin{proof}
By Lemma~\ref{sc:face} with $A=\varnothing$, every subgraph of~$G$ has a ratio of edges to vertices of at most $a=g/(g-2)$, and $G\in\mathcal C_3$.
If~$2a$ is not an integer, Corollary~\ref{fv:total:noncrit-cor} applies with $\rho=3$.
Otherwise $g\in\{3,4,6\}$ and $d\in\{6,4,3\}$.
For $g=3$ the graph lies in $\mathcal C_3=\mathcal C_a$, and Theorem~\ref{fv:total:exact}\ref{fv:total:exact:minor} applies.
For $g\in\{4,6\}$ the proof of the integer case is repeated with $\rho=3$.
Apart from Lemma~\ref{fv:total:defects}, it used only the density of subgraphs, and Lemma~\ref{fv:total:defects}(ii) provides the hypothesis of Lemma~\ref{fv:total:count} with $\kappa=g$.
\end{proof}

It remains to explain Table~\ref{fv:total:table}.
For $g\ge7$, $d=2+\frac4{g-2}<3$, so $k=2$ and $\theta=\frac4{g-2}$.
For $g=5$, $d=\frac{10}3$, $k=3$ and $\theta=\frac13$.
A row for $7\le g\le13$ lists a rational~$\mu_0$ with $\mu_0<\theta$ and $(1+\mu_0)(1+\mu_0-\theta)<2(1-\theta)$, which means $\mu_0<\mu_{\max}(\theta)$.
Then Corollary~\ref{fv:total:noncrit-cor}(i) gives the coefficient $d-\mu_0$ without error.
Here is the arithmetic:
\[
\begin{array}{c|ccccc}
g&\theta&\mu_0&(1+\mu_0)(1+\mu_0-\theta)&2(1-\theta)&d-\mu_0\\\hline
7&4/5&2/15&17/45&2/5&8/3\\
8&2/3&1/6&7/12&2/3&5/2\\
9&4/7&5/21&52/63&6/7&7/3\\
10&1/2&1/4&15/16&1&9/4\\
11&4/9&5/18&115/108&10/9&13/6\\
12&2/5&3/10&117/100&6/5&21/10\\
13&4/11&69/220&5491/4400&14/11&41/20
\end{array}
\]
In every row $\mu_0<\theta$, and the fourth entry is smaller than the fifth.
If $g=5$, then $\theta=\frac13$, and $\mu=\theta$ satisfies the quadratic condition with equality, because $(1+\frac13)\cdot1=2(1-\frac13)$.
So $\mu_{\max}=\frac13$, and Corollary~\ref{fv:total:noncrit-cor}(ii) gives $d-\mu_{\max}+\varepsilon=3+\varepsilon$.
When $g\ge14$, $\theta\le\frac13$ and $\mu=\theta$ satisfies the condition, because $1+\theta\le2(1-\theta)$.
So $\mu_{\max}=\theta$ and the coefficient is $d-\theta=2+\varepsilon$.
At girth~$19$, the coefficient two is achieved exactly by \cref{fv:hg:total-two}.

\begin{table}[ht]
\centering\small
\begin{tabular}{@{}rll@{}}
\toprule
Girth~$g$ & Coefficient against~$\gamma_{t,f}$ & Source\\
\midrule
3 &~$6$ (optimal) & integer case, $\kappa=3$\\
4 &~$4$ (optimal) & integer case, $\kappa=4$\\
5 & $3+\varepsilon$ & $\mu_{\max}=\theta=1/3$\\
6 &~$3$ & integer case, $\kappa=6$\\
7 & $8/3$ & $\mu_0=2/15$\\
8 & $5/2$ & $\mu_0=1/6$\\
9 & $7/3$ & $\mu_0=5/21$\\
10 & $9/4$ & $\mu_0=1/4$\\
11 & $13/6$ & $\mu_0=5/18$\\
12 & $21/10$ & $\mu_0=3/10$\\
13 & $41/20$ & $\mu_0=69/220$\\
$\ge14$ & $2+\varepsilon$ & $\mu_{\max}=\theta\le1/3$\\
$\ge19$ &~$2$ & \cref{fv:hg:total-two}\\
\bottomrule
\end{tabular}
\caption{Coefficients for isolate-free planar graphs of girth at least~$g$ against~$\gamma_{t,f}$.
The rows with~$\varepsilon$ hold for every fixed $\varepsilon>0$.
We claim optimality only for $g=3,4$ (\cref{fv:total:lower}).}
\label{fv:total:table}
\end{table}

The coefficient two is also reached exactly in every class that excludes a fixed minor once the girth is large enough; see \cref{fv:hg:minor-total-two}.

\subsection{Bounded treewidth}
\label{fv:total:tw:sec}

\begin{corollary}
\label{fv:total:treewidth}
For every fixed integer $t\ge1$, there is a deterministic constant-round~$\BC$ algorithm that returns, on every isolate-free graph of treewidth at most~$t$, a total dominating set of size at most $2t\,\gamma_{t,f}(G)$.
The coefficient~$2t$ is optimal by \cref{fv:total:lower-tw}.
\end{corollary}

\begin{proof}
Every minor of a graph of treewidth at most~$t$ has treewidth at most~$t$, and a graph of treewidth at most~$t$ is~$t$-degenerate, so it has a ratio of edges to vertices of at most~$t$.
Thus $G\in\mathcal C_t$, and Theorem~\ref{fv:total:exact}\ref{fv:total:exact:minor} with $a=t$ gives the bound.
\end{proof}

\subsection{Lower bounds for total domination}\label{fv:total:lowersec}

The coefficients of the total-domination results are optimal against the fractional optimum, already in the \LOCAL{} model.
Here~$\gamma_{t,f}(G)$ is the minimum of $\sum_vx_v$ over all $x\ge0$ with $\sum_{u\in N(v)}x_u\ge1$ for every vertex~$v$, where~$N(v)$ is the open neighborhood, and~$\gamma_t(G)$ is the same minimum over $\{0,1\}$-vectors.
Both are defined on graphs without isolated vertices, and $\gamma_{t,f}\le\gamma_t$.
A lower bound against~$\gamma_{t,f}$ is therefore weaker than a lower bound against~$\gamma_t$.

\begin{theorem}\label{fv:total:lower}
Let $\varepsilon>0$ and $T\ge0$.
No deterministic~$T$-round \LOCAL{} algorithm returns, on every connected planar graph~$G$, a total dominating set of size at most $(6-\varepsilon)\gamma_{t,f}(G)$.
No such algorithm achieves $(4-\varepsilon)\gamma_{t,f}(G)$ on every connected triangle-free planar graph.
This holds even if the vertices know~$n$ and the identifiers are exactly $1,\ldots,n$.
\end{theorem}

\begin{proposition}\label{fv:total:lower-tw}
Let $t\ge1$, $\varepsilon>0$ and $T\ge0$.
No deterministic~$T$-round \LOCAL{} algorithm returns, on every connected graph~$G$ of treewidth at most~$t$, a total dominating set of size at most $(2t-\varepsilon)\gamma_{t,f}(G)$.
For $t=1$ the same holds with~$\gamma_t(G)$ in place of~$\gamma_{t,f}(G)$, on paths.
This holds even if the vertices know~$n$ and the identifiers are exactly $1,\ldots,n$.
\end{proposition}

For $t\ge2$, only the fractional benchmark is claimed.
The integral optimum of~$P_n^t$ is at most about $2n/(3t+1)$, since two vertices~$i$ and $i+t$ totally dominate the $3t+1$ vertices from $i-t$ to $i+2t$, so against it the argument below forces a ratio of at most about $(3t+1)/2<2t$.

The argument follows Hilke, Lenzen, and Suomela~\cite{HLS2014}, who obtain their lower bounds for dominating set from repeated applications of Ramsey's theorem.
The only property of the problem that the argument needs is that the common output on a large regular region cannot be zero.
For total domination this holds because every vertex needs a selected neighbor.
A first lemma covers the triangular grid, the square grid, and the powers of paths.

\paragraph{Block graphs.}
Let $q,s\ge1$ and let $S\subseteq\mathbb Z^q\setminus\{0\}$ be a finite set with $S=-S$ and $\max_{\sigma\in S}\|\sigma\|_\infty=s$.
Put $d=|S|$.
The \emph{lattice graph} on~$\mathbb Z^q$ joins~$x$ and $x+\sigma$ for all $\sigma\in S$.
For an integer~$m$, the set $\{0,\ldots,m-1\}^q$ is a \emph{block}.
A vertex of a block is \emph{full} if all its coordinates lie in $[s,m-1-s]$.
An \emph{$(m,S)$-block graph with~$b$ blocks} is a graph whose vertex set is a disjoint union of~$b$ blocks such that
\begin{enumerate}[label=(\alph*)]
\item two vertices of the same block are adjacent if and only if their difference lies in~$S$, and
\item two vertices of different blocks are adjacent only if neither of them is full.
\end{enumerate}
Thus a full vertex has exactly the~$d$ neighbors $x+\sigma$ with $\sigma\in S$, all in its block.
There are three families.
\begin{itemize}
\item \emph{Triangular grid:} $q=2$, $S=\{\pm(1,0),\pm(0,1),\pm(1,1)\}$, $s=1$, $d=6$.
The vertex $(m-1,0)$ of block~$j$ is joined to the vertex $(0,0)$ of block $j+1$ for $j<b$, and no other edges are added.
\item \emph{Square grid:} $q=2$, $S=\{\pm(1,0),\pm(0,1)\}$, $s=1$, $d=4$, with the same joining edges.
\item \emph{Path powers:} $q=1$, $S=\{\pm1,\ldots,\pm t\}$, $s=t$, $d=2t$.
The graph~$P_n^t$ has vertex set $\{1,\ldots,n\}$ with $n=bm$, and $i\sim j$ if $1\le|i-j|\le t$.
Block~$j$ consists of the vertices $(j-1)m+1+x$ with position $x\in\{0,\ldots,m-1\}$.
\end{itemize}
In all three cases (a) is clear.
In the grids a joining edge has endpoints with a coordinate equal to~$0$ or $m-1$, so they are not full.
In~$P_n^t$, an edge between consecutive blocks joins a position at least $m-t$ to a position at most $t-1$, and neither is full.
So (b) holds.

\begin{lemma}\label{fv:total:forced}
Let $T\ge0$, let $m\ge2(T+3)s+1$ and let~$\mathcal A$ be a deterministic~$T$-round \LOCAL{} algorithm.
There is an integer~$\ell$, depending only on~$T$,~$m$ and~$S$, such that the following holds for every $b>\ell$ and every $(m,S)$-block graph~$G$ with~$b$ blocks.
Put $n=bm^q$.
There is an assignment of the identifiers $1,\ldots,n$ to~$V(G)$ such that, if~$\mathcal A$ returns a total dominating set~$D$ on~$G$ with these identifiers and knowledge of~$n$, then
\[
 |D|\ge(b-\ell)\,(m-2(T+2)s)^q .
\]
\end{lemma}

\begin{proof}
At each vertex~$x$, the edge to $x+\sigma$ is numbered by the position of~$\sigma$ in a fixed enumeration of~$S$, and all other edges receive larger numbers.
This is possible by (a).
The convention is that after~$T$ rounds a vertex knows the labeled ball of radius $T+1$ around it, with identifiers, degrees and port numbers, and the number~$n$.
Its output is a function of this information.
This covers the usual conventions of the model.

\emph{Deep vertices.}
A vertex of a block is \emph{deep} if all its coordinates lie in $[(T+2)s,m-1-(T+2)s]$.
Let $B\subseteq\mathbb Z^q$ be the set of points at lattice distance at most $T+1$ from~$0$.
We claim that for every deep vertex~$v$ the ball of radius $T+1$ around~$v$ in~$G$ is $v+B$ with the lattice edges, the lattice port numbers, and all degrees equal to~$d$.
Induction on $j\le T+1$ shows that the vertices at distance at most~$j$ from~$v$ are the vertices $v+y$ in the block of~$v$, where $y\in B$ has lattice distance at most~$j$.
A vertex at distance $j\le T+1$ has each coordinate within~$js$ of that of~$v$, hence in $[s,m-1-s]$, so it is full.
A full vertex has exactly its~$d$ lattice neighbors, so the vertices at distance $j+1$ are the lattice neighbors of those at distance~$j$.
By (a) the edges among the vertices of the ball are exactly the lattice edges, and every vertex of the ball is full, so it has degree~$d$ and lattice ports.
This proves the claim.
Hence the labeled ball of a deep vertex is determined by the identifiers on $v+B$, read in the order of~$B$.

\emph{Ramsey argument.}
Let $p=|B|$.
The lexicographic order on~$\mathbb Z^q$ is preserved by translations.
Fix~$n$ and color each~$p$-element set $X\subseteq\{1,\ldots,n\}$ by the bit that~$\mathcal A$ outputs at the center of~$B$ when the elements of~$X$ are assigned in increasing order to the points of~$B$ in lexicographic order.
By the claim, this is well defined, and it does not depend on how the identifiers are assigned outside the ball.
Let $M=m^q$ and let~$R$ be a Ramsey number such that every two-coloring of the~$p$-element subsets of an~$R$-element set has a homogeneous subset of size~$M$.
Put $\ell=\lceil R/M\rceil$.
Both numbers depend only on~$p$ and~$M$, hence on~$T$,~$m$ and~$S$, and not on~$b$.

For $j=1,\ldots,b-\ell$, proceed as follows.
At the start of step~$j$ at least $n-(j-1)M=(b-j+1)M\ge\ell M\ge R$ identifiers are unused, so they contain a homogeneous set~$H_j$ of size~$M$.
Assign it to block~$j$ in increasing order along the lexicographic order of the block, and remove it from the unused identifiers.
The remaining identifiers are assigned arbitrarily.

Let~$v$ be a deep vertex of a block $j\le b-\ell$.
The identifiers on $v+B$ form a~$p$-subset of~$H_j$, and they increase along the lexicographic order of~$B$, since this is the restriction of the order of the block.
So~$\mathcal A$ outputs at~$v$ the color~$c_j$ of~$H_j$, the same bit for all deep vertices of the block.

\emph{The color is one.}
Since $m\ge2(T+3)s+1$, the point~$u$ of the block with all coordinates equal to $(T+3)s$ exists and is full.
All its neighbors $u+\sigma$ have coordinates in $[(T+2)s,(T+4)s]\subseteq [(T+2)s,m-1-(T+2)s]$, so they are deep.
If $c_j=0$, none of them is selected, and~$u$ has no selected neighbor, which contradicts that~$D$ is a total dominating set.
Hence $c_j=1$, and~$D$ contains all $(m-2(T+2)s)^q$ deep vertices of each of the $b-\ell$ blocks.
\end{proof}

\begin{proof}[Proof of Theorem~\ref{fv:total:lower}]
Let $d\in\{6,4\}$ correspond to the triangular and the square grid, and let~$G$ be the block graph with~$b$ blocks and side $m\ge2(T+3)+1$ defined above, with $s=1$, $q=2$.
It is connected.
It is planar: each block is a plane grid, with the diagonals drawn inside the unit squares in the triangular case, and the joining edges are bridges, so the~$2$-connected components of~$G$ are planar.
For the square grid, each block is bipartite and the bridges lie on no cycle, so~$G$ has no triangle.

\emph{The fractional solution.}
Call a vertex of a block a boundary vertex if some coordinate is~$0$ or $m-1$.
Put $x_v=1$ on boundary vertices and $x_v=1/d$ on all other vertices.
A non-boundary vertex is full and has~$d$ neighbors, each of weight at least $1/d$.
A boundary vertex~$v$ has a coordinate~$i$ equal to~$0$ or $m-1$.
Changing the other coordinate by $+1$ or $-1$, whichever stays in $[0,m-1]$, gives a neighbor of~$v$ with the same coordinate~$i$, which is again a boundary vertex of weight one.
Joining edges only add neighbors.
So~$x$ is a fractional total dominating set of~$G$, and
\[
 \gamma_{t,f}(G)\le b\Bigl(\frac{(m-2)^2}{d}+4m-4\Bigr).
\]

\emph{The ratio.}
Suppose that~$\mathcal A$ returns a total dominating set of size at most $(d-\varepsilon)\gamma_{t,f}$ on every such graph.
Lemma~\ref{fv:total:forced} gives identifiers for which the output on~$G$ has size at least $(b-\ell)(m-2T-4)^2$, so
\[
 \frac{(1-\ell/b)(m-2T-4)^2}{(m-2)^2/d+4m-4}\le d-\varepsilon .
\]
The left side tends to~$d$ as first $b\to\infty$ and then $m\to\infty$, because~$\ell$ depends on~$m$ but not on~$b$.
This is a contradiction for~$m$ and then~$b$ large.
The identifiers are $1,\ldots,n$ and $n=bm^2$ is known to the algorithm, as the lemma allows.
\end{proof}

\begin{proof}[Proof of Proposition~\ref{fv:total:lower-tw}]
The graph~$P_n^t$ is connected, and the sets $\{i,\ldots,i+t\}$ form a path decomposition of width~$t$, so its treewidth is at most~$t$.
Take $G=P_n^t$ with $n=bm$ and the blocks defined above.

\emph{The fractional solution.}
Put $x_i=1$ for $i\le t+1$ and $i\ge n-t$, and $x_i=1/(2t)$ otherwise.
A vertex~$i$ with $t+2\le i\le n-t-1$ has the~$2t$ neighbors $i\pm1,\ldots,i\pm t$, each of weight at least $1/(2t)$.
A vertex $i\le t+1$ is adjacent to the~$t$ other vertices of $\{1,\ldots,t+1\}$, of total weight $t\ge1$, and symmetrically for $i\ge n-t$.
So~$x$ is feasible, and
\[
 \gamma_{t,f}(G)\le\frac{n}{2t}+2(t+1)\le\frac n{2t}+2t+2 .
\]
(Here $n\ge2t+2$.)

\emph{The ratio.}
Suppose that~$\mathcal A$ always returns a total dominating set of size at most $(2t-\varepsilon)\gamma_{t,f}$.
By Lemma~\ref{fv:total:forced} with $q=1$ and $s=t$, its output on~$G$ has, for suitable identifiers, size at least $(b-\ell)(m-2(T+2)t)$.
Hence
\[
 \frac{(b-\ell)(m-2(T+2)t)}{bm/(2t)+2t+2}\le2t-\varepsilon .
\]
As first $b\to\infty$ and then $m\to\infty$, the left side tends to~$2t$, a contradiction.

\emph{Integral benchmark for $t=1$.}
Let~$4\mid n$ and select the vertices~$i$ with $i\equiv2$ or $3\pmod4$, which are $n/2$ vertices.
Each vertex $i\equiv1,3\pmod4$ has the selected neighbor $i+1$ or $i-1$, respectively, namely the vertex $4k+2$ in both cases, and each vertex $i\equiv2,0\pmod4$ has the selected neighbor $i+1$ or $i-1$, namely $4k+3$.
Hence $\gamma_t(P_n)\le n/2$.
Taking~$4\mid m$, so that~$4\mid n$, Lemma~\ref{fv:total:forced} gives output size at least $(b-\ell)(m-2(T+2))$ against $\gamma_t(P_n)\le bm/2$, and the ratio tends to~$2$.
So no algorithm achieves $(2-\varepsilon)\gamma_t$ on paths.
\end{proof}

\section{Exact coefficients at large girth}\label{fv:hg}

On paths, no deterministic constant-round algorithm achieves a coefficient below three for dominating set, and part~\ref{sc:thmC:girth} of Theorem~\ref{sc:thmC} gives $3+4/(g-2)$ on planar graphs of girth at least~$g$.
This section shows that the limit itself is achieved once the girth is large.
A further budget transfer gives the saving that is needed to achieve this coefficient.
The rounding needs only that every subgraph has a ratio of edges to vertices of at most~$a$ for a fixed $a<8/7$.
In general, this bound on subgraph density does not imply the bounded minor density required by Theorem~\ref{sc:thmA}.
We therefore state the main result for every approximate fractional solution of a fixed accuracy.
It gives unconditional algorithms on planar graphs of girth at least~$17$, and at sufficiently large girth in every class that excludes a fixed minor, where Theorem~\ref{sc:thmA} gives the fractional solution.

\subsection{Coefficient three at a fixed density below eight sevenths}

\paragraph{The transformation.}
Let~$G$ be a graph, and call a vertex of degree at least three a \emph{branch vertex}.
Put
\[
 B=\{v:\deg_G(v)\ge3\},\qquad Z=N_G^2[B],\qquad
 F=V(G)\setminus Z,\qquad R_0=Z\setminus N_G[F],
\]
where~$N_G^2[B]$ is the set of vertices at distance at most two from~$B$.
The algorithm selects~$F$ and keeps~$R_0$ as demands.
Every vertex of~$Z$ is a candidate, with the cost
\[
 c(v)=3-|N_G(v)\cap F|\qquad(v\in Z).
\]
A demand has no neighbor in~$F$, so its closed neighborhood lies in~$Z$, and its constraint is the same in~$G$ and in~$G[Z]$.
Let~$\cost^*$ be the minimum cost of a fractional cover of~$R_0$ in~$G[Z]$ with these costs.
Compared with Section~\ref{simp:endpoint:section}, the branch vertices play the role of the defects, and only vertices at distance at least three from them are selected.
This distance is what makes the residual demands short paths below.

\begin{lemma}
\label{fv:hg:budget}
\leavevmode
\begin{enumerate}[label=(\roman*)]
\item Every vertex of~$F$ has degree at most two, and $N_G[B]\subseteq R_0$.
\item The costs are integers with $1\le c(v)\le3$, and $c(v)=3$ holds exactly when $v\in R_0$.
If $c(v)<3$, then $\deg_G(v)\le2$ and $|N_G[v]\cap R_0|\le c(v)-1$.
\item $|F|+\cost^*\le3\gf(G)$.
\end{enumerate}
\end{lemma}

\begin{proof}
(i) No vertex of~$F$ lies in~$B$.
Every neighbor of a vertex $u\in N_G[B]$ lies in~$Z$, so~$u$ is neither in~$F$ nor adjacent to it.

(ii) A candidate~$v$ lies in~$R_0$ exactly when it has no neighbor in~$F$, that is, when $c(v)=3$.
Otherwise~$v$ has a neighbor in~$F$, which lies outside~$Z$, so $v\notin N_G[B]$ and $\deg_G(v)\le2$.
Hence $1\le3-c(v)\le2$.
The demands in~$N_G[v]$ exclude~$v$ itself and its $3-c(v)$ neighbors in~$F$, so there are at most $2-(3-c(v))=c(v)-1$ of them.

(iii) Let~$y$ be an optimal fractional dominating vector of~$G$.
Its restriction to~$Z$ covers~$R_0$, so $\cost^*\le\sum_{v\in Z}c(v)y_v$.
Summing the constraints of~$y$ over~$F$ gives $|F|\le\sum_v|N_G[v]\cap F|\,y_v$.
For $v\in F$ this coefficient is at most $\deg_G(v)+1\le3$, and for $v\in Z$ it equals $3-c(v)$.
Adding the two bounds gives $|F|+\cost^*\le3\sum_vy_v=3\gf(G)$.
\end{proof}

\Needspace{16\baselineskip}
\paragraph{The rounding.}
Let $x\in[0,1]^Z$ be a feasible fractional cover of~$R_0$.
The rounding consists of the three steps of Section~\ref{simp:endpoint:section} with $C=Z$, the cost $b=3$, and the threshold $t=1/2$, with one change: the second step selects only candidates that cover at least four demands.
Thus
\[
\begin{gathered}
 W=\{v\in Z:c(v)=3,\ x_v>1/2\},\qquad R=R_0\setminus N_G[W],
 \qquad h(v)=|N_G[v]\cap R|,\\
 H=\{v\in Z\setminus W:h(v)\ge4\},\qquad U=R\setminus N_G[H],
 \qquad D=W\cup H\cup U.
\end{gathered}
\]
As there, $O=Z\setminus(W\cup H)$, $L=R\setminus(H\cup U)$, $w=|W|$, $\ell=|L|$, and $P=\sum_{v\in O}(c(v)-h(v))x_v$.
Put also $Q=H\cap R$ and $z=|H\setminus R|$, so that~$R$ is the disjoint union of~$Q$,~$L$, and~$U$.
For $r\in R$, the \emph{excess} $\omega_r=\sum_{v\in N_G[r]}x_v-1$ is nonnegative; let $\Omega=\sum_{r\in R}\omega_r$.
Finally,~$O^+$ denotes the set of $v\in O$ with $c(v)>h(v)$.

We use the following facts throughout.
By definition, $W\subseteq R_0$ and $c=3$ on~$W$.
A vertex of~$H$ has at least three neighbors, so $H\subseteq B\subseteq R_0$ by Lemma~\ref{fv:hg:budget}(i), and $c=3$ on~$H$.
Every $v\in R_0\setminus W$ has $c(v)=3$ and therefore $x_v\le1/2$.
A vertex of~$O$ has $h(v)\le3$, so $c(v)\ge h(v)$ when $c(v)=3$, and $c(v)-h(v)\ge1$ when $c(v)<3$ by Lemma~\ref{fv:hg:budget}(ii).
Thus $P\ge0$, and every $v\in O^+$ satisfies $c(v)-h(v)\ge1$ and $h(v)\le2$.

\begin{lemma}
\label{fv:hg:saving}
Suppose that every subgraph of~$G$ has a ratio of edges to vertices of at most~$a$, where $1\le a<6/5$.
Then~$D$ covers~$R_0$, $|H|\le a\ell/(3-2a)$, and
\[
 \cost(x)-|D|\ge\frac w2+\beta\ell+P+\Omega,
 \qquad\beta=\frac{6-5a}{2(3-2a)}>0.
\]
\end{lemma}

\begin{proof}
The set~$W$ covers $R_0\setminus R$, the set~$H$ covers $R\setminus U$, and $U\subseteq D$.
The sets $W,H,U$ are disjoint, so $|D|=w+|H|+|U|=w+z+|R|-\ell$.

Put $E_0=\sum_{v\in H}(h(v)-3)$.
As in Lemma~\ref{simp:endpoint:rounding}, the neighbors in~$R$ of a vertex of~$H$ lie in $Q\cup L$.
Write $e(X)=|E(G[X])|$, and~$e(X,Y)$ for the number of edges between disjoint sets $X,Y$.
Then
\[
\begin{aligned}
 \sum_{v\in H}h(v)
 &=2e(Q)+e(Q,L)+e(H\setminus R,Q\cup L)+|Q|\\
 &\le e(H\cup L)+e(Q)+|Q|
 \le a(|H|+\ell)+(a+1)(|H|-z).
\end{aligned}
\]
Hence $E_0\le(2a-2)|H|+a\ell-(a+1)z$.
Since $h\ge4$ on~$H$, we have $|H|\le E_0$, and since $2a-2\ge0$, this gives $(3-2a)E_0\le a\ell-(a+1)z$.
In particular $|H|\le E_0\le a\ell/(3-2a)$.

No vertex of~$W$ lies in~$R$ or is adjacent to it.
Summing the constraints over~$R$ therefore gives $\sum_{v\in Z\setminus W}h(v)x_v=|R|+\Omega$.
Since $x_v>1/2$ on~$W$ and $x_v\le1/2$ on~$H$,
\[
\begin{aligned}
 \cost(x)&=3\sum_{v\in W}x_v+\sum_{v\in Z\setminus W}h(v)x_v
   -\sum_{v\in H}(h(v)-3)x_v+P\\
 &\ge\tfrac32w+|R|+\Omega-\tfrac12E_0+P.
\end{aligned}
\]
Subtracting $|D|$ and inserting the bound on~$E_0$ yields
\[
 \cost(x)-|D|\ge\frac w2+\ell-z-\frac{a\ell-(a+1)z}{2(3-2a)}+P+\Omega
 =\frac w2+\beta\ell+\frac{5a-5}{2(3-2a)}\,z+P+\Omega.
\]
The coefficient of~$z$ is nonnegative because $a\ge1$.
\end{proof}

\paragraph{Attachments.}
The saving does not involve $|U|$, so it remains to bound the residual demands~$U$ by~$w$,~$\ell$,~$P$, and~$\Omega$.
Let~$T$ be the set of vertices outside~$W$ that have a neighbor in~$W$ and a neighbor in~$U$, and put $s=|T|$.
An \emph{attachment} is an edge between~$U$ and~$T$.
Let~$q$ be the number of attachments, and call a vertex of~$U$ an \emph{attachment vertex} if it is incident with one.
Thus $s\le q$.

\begin{lemma}
\label{fv:hg:residual}
\leavevmode
\begin{enumerate}[label=(\roman*)]
\item Every vertex of~$U$, and every vertex of~$R$ adjacent to~$U$, has at most two neighbors in~$R$.
Hence every component of~$G[U]$ is a path or a cycle, where a single vertex counts as a path.
\item Every neighbor of a vertex $u\in U$ in $R_0\setminus R$ lies in~$T$.
Every vertex of $B\cap U$ is an attachment vertex.
\item Let $u\in U$, and let~$K$ be its component in~$G[U]$.
Then~$u$ has distance at most two in~$K$ from a vertex of $B\cap K$, or distance at most one in~$K$ from an attachment vertex, or there is a path of length at most two from~$u$ to a vertex of~$L$ whose other vertices lie in~$K$.
\item $|U|\le5q+4\ell$.
\end{enumerate}
\end{lemma}

\begin{proof}
(i) A vertex $v\in R$ with $h(v)\ge4$ lies in~$H$, because~$R$ is disjoint from~$W$.
Neither a vertex of~$U$ nor a neighbor of it lies in~$H$, so these vertices have $h\le3$, that is, at most two neighbors in~$R$.

(ii) The set $R_0\setminus R$ lies in~$N_G[W]$, and no neighbor of~$u$ lies in~$W$, since $u\notin N_G[W]$.
This proves the first claim.
A vertex of~$B$ has all its neighbors in $N_G[B]\subseteq R_0$.
A vertex of $B\cap U$ has at least three neighbors and at most two of them in~$R$, so it has a neighbor in $R_0\setminus R$, which lies in~$T$.

(iii) Since $u\in Z$, there is a path $u=p_0,\ldots,p_k$ in~$G$ with $k\le2$ and $p_k\in B$.
All its vertices lie in~$R_0$: the vertex~$p_0$ lies in~$R$, and the others lie in~$N_G[B]$.
If all of them lie in~$U$, then the path lies in~$K$, and $p_k\in B\cap K$.
Otherwise, let $j\ge1$ be the least index with $p_j\notin U$, so that $p_0,\ldots,p_{j-1}$ lie in~$K$.
If $p_j\in R$, then $p_j\in Q\cup L$, and $p_j\notin H$ because $p_{j-1}\in U$; thus $p_j\in L$.
If $p_j\notin R$, then $p_j\in T$ by (ii), and~$p_{j-1}$ is an attachment vertex at distance at most $j-1\le1$ from~$u$ in~$K$.

(iv) Assign each $u\in U$ to a vertex given by (iii): an attachment vertex at distance at most two in~$K$, using (ii) in the first case, or a vertex of~$L$.
By (i), a vertex of~$U$ has at most two neighbors in~$R$, so an attachment vertex has at most two neighbors in~$U$, and each of these has at most one further neighbor in~$U$.
This gives at most $1+2+2=5$ vertices assigned to each of the at most~$q$ attachment vertices.
A vertex of~$L$ is adjacent to at most two vertices of~$U$, and each of them has at most one further neighbor in~$U$, so at most four vertices are assigned to it.
\end{proof}

The bound in (iv) is not enough on its own.
A path of~$G[U]$ has one edge fewer than vertices, and if each attachment vertex ends its own path, the density bound gives nothing.
Call a path component~$K$ of~$G[U]$ \emph{free} if no vertex of~$K$ is adjacent to~$L$, and put
\[
 Q_K=\sum_{v\in O^+\cap N_G[K]}x_v,\qquad
 \Omega_K=\sum_{r\in K}\omega_r.
\]
Free paths with few attachments give fractional slack.

\begin{lemma}
\label{fv:hg:free}
Let~$K$ be a free path.
\begin{enumerate}[label=(\roman*)]
\item Every neighbor $y\notin K$ of a vertex of~$K$ lies in~$T$ or in $Z\setminus R_0$; in the second case $y\in O^+$ and $h(y)\le1$.
\item A vertex of~$K$ of degree at least three lies in~$B$, and all its neighbors outside~$K$ lie in~$T$.
An inner vertex of~$K$ that is not an attachment vertex has no neighbor outside~$K$.
\item Every vertex of~$K$ has distance at most two in~$K$ from a vertex of $B\cap K$, or distance at most one from an attachment vertex.
All vertices of $B\cap K$ are attachment vertices; in particular,~$K$ has an attachment vertex.
\item Every vertex of~$K$ has $x_v\le1/2$, and the ends of~$K$ lie in~$O^+$.
Every vertex of~$T$ adjacent to~$K$ lies in~$O^+$ or has coordinate at most $1/2$.
\end{enumerate}
\end{lemma}

\begin{proof}
(i) The vertex~$y$ is not in another component of~$G[U]$, not in~$L$ because~$K$ is free, and not in $Q\subseteq H$ because $K\subseteq U$.
Hence $y\notin R$.
Since $K\subseteq R_0$ has no neighbor in~$F$, the vertex~$y$ lies in~$Z$.
If $y\in R_0$, then $y\in T$ by Lemma~\ref{fv:hg:residual}(ii).
Otherwise $c(y)<3$, so $y\notin W\cup H$, and Lemma~\ref{fv:hg:budget}(ii) gives $h(y)\le c(y)-1\le1$; thus $y\in O^+$.

(ii) A vertex of degree at least three lies in~$B$, so all its neighbors lie in~$R_0$, and by (i) those outside~$K$ lie in~$T$.
An inner vertex that is not an attachment vertex has no neighbor in~$T$, so by (i) its neighbors outside~$K$ lie outside~$R_0$.
Hence it is not in~$B$, and its two neighbors in~$K$ are all its neighbors.

(iii) This follows from Lemma~\ref{fv:hg:residual}(iii) and (ii), since~$K$ is free.

(iv) The path lies in $R_0\setminus W$, so $x_v\le1/2$ on~$K$.
By (i), the neighbors in~$R$ of a vertex $v\in K$ lie in~$K$, so an end of~$K$ has $h(v)\le2<3=c(v)$, and it lies in~$O$ because $U\cap(W\cup H)=\varnothing$.
A vertex $p\in T$ is not in~$W$ by definition, and not in~$H$ since it has a neighbor in~$U$; so $p\in O$.
If $p\notin O^+$, then $c(p)\le h(p)$, which forces $c(p)=3$, and hence $p\in R_0\setminus W$ and $x_p\le1/2$.
\end{proof}

\begin{lemma}
\label{fv:hg:half}
Let~$K$ be a free path that has exactly one attachment, or exactly two attachments at two vertices of distance exactly five in~$K$.
Then $Q_K+\Omega_K\ge1/2$.
Consequently, if~$c_1$ and~$b_2$ count the free paths of these two kinds, then $c_1+b_2\le4P+2\Omega$.
\end{lemma}

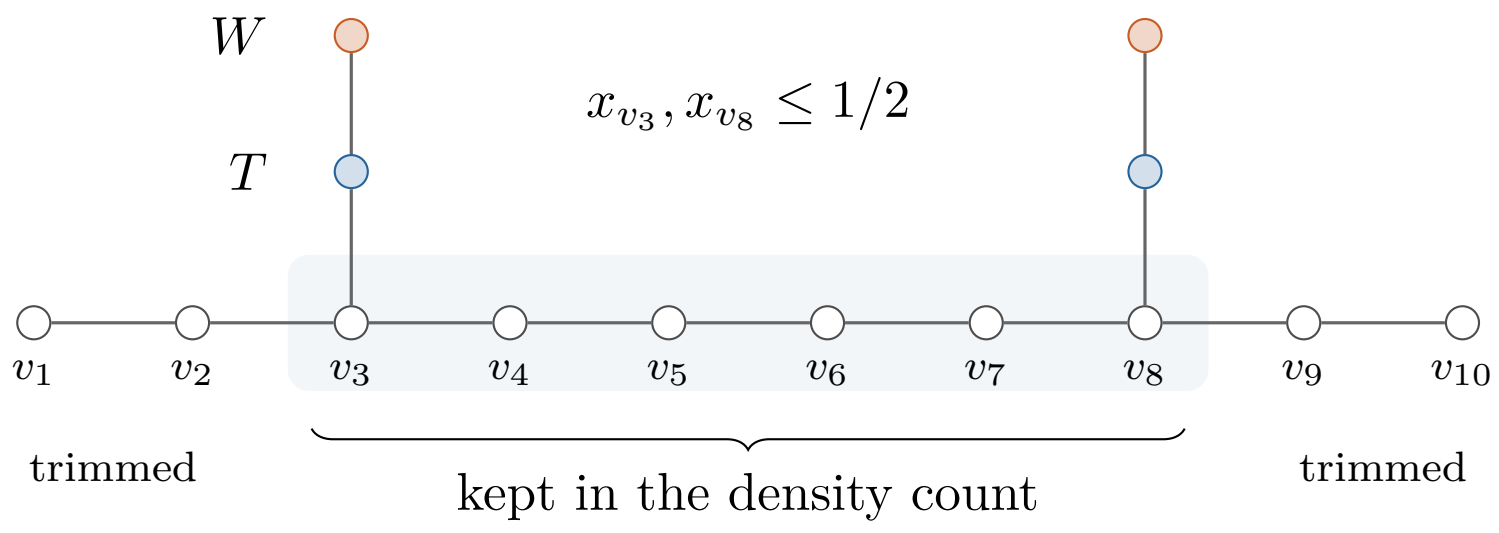
\begin{figure}[tb]
\centering
\begin{tikzpicture}[x=1.05cm,y=1cm]
 \fill[simpblue!6,rounded corners] (1.6,-.45) rectangle (7.4,.45);
 \foreach \i in {1,...,10} {
   \node[sv,label=below:{\scriptsize $v_{\i}$}] (v\i) at (\i-1,0) {};
 }
 \foreach \i/\j in {1/2,2/3,3/4,4/5,5/6,6/7,7/8,8/9,9/10} {
   \draw[se] (v\i)--(v\j);
 }
 \foreach \i in {3,8} {
   \node[so] (t\i) at (\i-1,1) {};
   \node[sb] (w\i) at (\i-1,1.9) {};
   \draw[se] (v\i)--(t\i);
   \draw[se] (t\i)--(w\i);
 }
 \node[font=\small,anchor=east] at (1.6,1) {$T$};
 \node[font=\small,anchor=east] at (1.6,1.9) {$W$};
 \node[font=\small,align=center] at (4.5,1.45)
   {$x_{v_3},x_{v_8}\le1/2$};
 \draw[decorate,decoration={brace,mirror,amplitude=4pt}]
   (1.75,-.7)--(7.25,-.7);
 \node[font=\small,align=center] at (4.5,-1.15)
   {kept in the density count};
 \node[font=\scriptsize,align=center] at (.5,-.95) {trimmed};
 \node[font=\scriptsize,align=center] at (8.5,-.95) {trimmed};
\end{tikzpicture}
\caption{A free path with two attachments at distance five, the longest case of Lemma~\ref{fv:hg:half}.
Its ends are removed only from the graph used in the density count of Lemma~\ref{fv:hg:count}; all ten constraints are kept, and they force half a unit of unused cost or excess coverage.
The two vertices of~$T$ may coincide, and so may the two vertices of~$W$.}
\label{fv:hg:fig}
\end{figure}

\begin{proof}
The argument uses the constraints $\sum_{v\in N_G[r]}x_v=1+\omega_r$ at vertices $r\in K$.
By Lemma~\ref{fv:hg:free}, if an end of~$K$ is not an attachment vertex, then every coordinate in its constraint belongs to $O^+\cap N_G[K]$, except possibly the coordinate of its neighbor in~$K$.
For $v_i\in K$, let~$e_i$ be the sum of the coordinates of the neighbors of~$v_i$ outside~$K$.

\emph{One attachment, at~$b$.}
By Lemma~\ref{fv:hg:free}(iii), $B\cap K\subseteq\{b\}$.
Suppose first that $b\notin B$.
Then every vertex of~$K$ has distance at most one from~$b$, and~$b$ has degree at most two, so~$K$ has at most two vertices.
If $K=\{b\}$, the constraint at~$b$ contains~$x_b$ with $b\in O^+$, the coordinate of the attachment neighbor, which is at most $1/2$ or belongs to~$O^+$, and otherwise coordinates of~$O^+$; hence $Q_K\ge1/2$.
If $K=\{b,v\}$, both ends lie in~$O^+$, and the constraint at~$v$ gives $Q_K\ge1$.

Suppose next that $b\in B$.
Then~$b$ has only one neighbor outside~$K$, so it has two neighbors in~$K$, and every vertex of~$K$ has distance at most two from~$b$.
If an end~$v$ of~$K$ is adjacent to~$b$, the constraint at~$v$ and $x_b\le1/2$ give $Q_K\ge1/2$.
Otherwise $K=v_1\cdots v_5$ with $b=v_3$, and the constraints at~$v_1$ and~$v_5$ give
\[
 x_{v_2}+x_{v_4}\ge2-x_{v_1}-x_{v_5}-e_1-e_5.
\]
The neighbors counted in~$e_1$ and~$e_5$ lie in~$O^+$ and have only one neighbor in~$R$, so no coordinate occurs twice.
The excess at~$v_3$ is at least $x_{v_2}+x_{v_4}-1$, and hence $Q_K+\Omega_K\ge1$.

\emph{Two attachments, at~$a$ and~$a'$ of distance five.}
The vertex of~$K$ between them at distance two from~$a$ has distance three from~$a'$, so $a\in B$ by Lemma~\ref{fv:hg:free}(iii); similarly $a'\in B$.
Each of them has only one neighbor outside~$K$, so both are inner vertices of~$K$.
A vertex beyond~$a$, on the side away from~$a'$, has distance at most two from~$a$, so~$K$ extends at most two vertices beyond~$a$, and similarly beyond~$a'$.
If~$K$ extends by only one vertex beyond~$a$ or~$a'$, the constraint at that end gives $Q_K\ge1/2$ as before.
Otherwise $K=v_1\cdots v_{10}$ with $a=v_3$ and $a'=v_8$, as in Figure~\ref{fv:hg:fig}.
The vertices $v_2,v_4,\ldots,v_7,v_9$ have no neighbor outside~$K$, by Lemma~\ref{fv:hg:free}(ii), and~$v_3$ has only its attachment neighbor~$p$.
The constraints at $v_1,v_3,v_4,v_6,v_7$ give, in turn,
\[
\begin{gathered}
 x_{v_2}\ge1-x_{v_1}-e_1,\qquad
 x_{v_3}+x_{v_4}\le1+\omega_{v_3}-x_{v_2}-x_p\le x_{v_1}+e_1+\omega_{v_3},\\
 x_{v_5}\ge1-x_{v_1}-e_1-\omega_{v_3},\qquad
 x_{v_6}+x_{v_7}\le x_{v_1}+e_1+\omega_{v_3}+\omega_{v_6},\\
 x_{v_8}\ge1-x_{v_1}-e_1-\omega_{v_3}-\omega_{v_6}.
\end{gathered}
\]
Since $x_{v_8}\le1/2$, this gives $Q_K+\Omega_K\ge x_{v_1}+e_1+\omega_{v_3}+\omega_{v_6}\ge1/2$.

For the second assertion, the sets~$K$ are disjoint, so their excesses sum to at most~$\Omega$.
A vertex $v\in O^+$ lies in~$N_G[K]$ for at most $|N_G[v]\cap U|\le h(v)\le2$ of these paths, and $x_v\le(c(v)-h(v))x_v$.
Hence $\sum_KQ_K\le2P$, and $(c_1+b_2)/2\le2P+\Omega$.
\end{proof}

\begin{lemma}
\label{fv:hg:count}
Suppose that every subgraph of~$G$ has a ratio of edges to vertices of at most~$a$, where $1\le a<8/7$.
Then for every feasible~$x$,
\[
 \frac{8-7a}2\,q\le aw+(4a-2)\ell+(4-3a)(2P+\Omega),
 \qquad
 |D|\le w+\frac{a}{3-2a}\,\ell+5q+4\ell.
\]
\end{lemma}

\begin{proof}
The second bound follows from $|D|=w+|H|+|U|$, Lemma~\ref{fv:hg:saving}, and Lemma~\ref{fv:hg:residual}(iv).

For the first, shorten the free paths, but only in the graph to which the density bound is applied; all constraints stay as they are.
Let~$c$ be the number of free paths, and let~$c_1$ and~$c_2$ count those with exactly one and exactly two attachments.
Replace every free path~$K$ by its shortest subpath~$K'$ that contains all attachment vertices of~$K$, keep all other components, and let~$U'$ be the resulting set.
By Lemma~\ref{fv:hg:free}(iii), every vertex of~$K$ has distance at most two from an attachment vertex, so consecutive attachment vertices have distance at most five.
If~$K$ has~$q_K$ attachments, then $|K'|\le5q_K-4$, and if $q_K=2$ and~$K$ is not counted by~$b_2$, then $|K'|\le5$.
The other components are bounded by the assignment in the proof of Lemma~\ref{fv:hg:residual}(iv).
Altogether,
\[
 |U'|\le5q+4\ell-4c-c_2+b_2.
\]

Now apply the density bound to the subgraph formed by~$U'$,~$T$, and~$W$, with the edges of~$G[U']$, all~$q$ attachments, whose ends in~$U$ lie in~$U'$, and one edge from each vertex of~$T$ to~$W$.
Every component of~$G[U]$ keeps a nonempty connected part, so~$G[U']$ has $|U'|-c_{\mathrm{p}}$ edges, where~$c_{\mathrm{p}}$ is the number of path components of~$G[U]$.
Each path component that is not free has a neighbor in~$L$, and by Lemma~\ref{fv:hg:residual}(i) a vertex of~$L$ is adjacent to at most two components, so $c_{\mathrm{p}}\le c+2\ell$.
The density bound and $s\le q$ give
\[
 |U'|-c_{\mathrm{p}}+q+s\le a(|U'|+s+w),\quad\text{hence}\quad
 (2-a)q\le(a-1)|U'|+c+2\ell+aw.
\]
With the bound on $|U'|$, this becomes
\[
 (7-6a)q\le aw+(4a-2)\ell+(5-4a)c-(a-1)(c_2-b_2).
\]
The free paths not counted by~$c_1$ or~$c_2$ have at least three attachments, so $c\le(q+2c_1+c_2)/3$, and $c_1+2c_2\le q$.
Substituting the first bound, using $5-4a>0$, and then the second, using $8-7a>0$, gives
\[
 \tfrac{16-14a}3\,q\le aw+(4a-2)\ell+\tfrac{10-8a}3\,c_1
 +\tfrac{8-7a}3\,c_2+(a-1)b_2,
 \qquad
 \tfrac{8-7a}2\,q\le aw+(4a-2)\ell+\tfrac{4-3a}2\,c_1+(a-1)b_2.
\]
Since $a-1\le(4-3a)/2$, Lemma~\ref{fv:hg:half} completes the proof.
\end{proof}

\paragraph{The theorem.}
Lemmas~\ref{fv:hg:saving} and~\ref{fv:hg:count} provide the two inequalities of Proposition~\ref{sc:principle}, with the quantities $w,\ell,P,\Omega$.

\begin{theorem}
\label{fv:hg:three}
Let $1\le a<8/7$ be rational.
There is a rational $\delta_a>0$ with the following property.
Let~$G$ be a graph every subgraph of which has a ratio of edges to vertices of at most~$a$, let~$F$,~$Z$,~$R_0$, and~$c$ be as above, and let $x\in[0,1]^Z$ be any feasible fractional cover of~$R_0$ in~$G[Z]$ with $\cost(x)\le(1+\delta_a)\cost^*$.
Then $F\cup D$ is a dominating set of~$G$ of size at most~$3\gf(G)$.
The sets~$F$ and~$R_0$ and the costs~$c$ are computed in a constant number of broadcast rounds with messages of constant size, and so is $F\cup D$ once~$x$ is given.
For $a=17/15$, the value $\delta_a=5/8482$ suffices.
\end{theorem}

\begin{proof}
By Lemma~\ref{fv:hg:count},~$5q$ is at most $\frac{10}{8-7a}\bigl(aw+(4a-2)\ell+(4-3a)(2P+\Omega)\bigr)$, and hence
\[
 |D|\le\beta_ww+\beta_\ell\ell+\beta_PP+\beta_\Omega\Omega
\]
with the positive constants $\beta_w=1+\frac{10a}{8-7a}$, $\beta_\ell=4+\frac a{3-2a}+\frac{10(4a-2)}{8-7a}$, $\beta_P=\frac{20(4-3a)}{8-7a}$, and $\beta_\Omega=\frac{10(4-3a)}{8-7a}$.
Lemma~\ref{fv:hg:saving} gives the saving with the coefficients $1/2,\beta,1,1$.
By Proposition~\ref{sc:principle}, the rounding has slack
\[
 \delta_a=\min\Bigl\{\frac1{2\beta_w},\frac\beta{\beta_\ell},
 \frac1{\beta_P},\frac1{\beta_\Omega}\Bigr\},
\]
and so $|D|\le\cost^*$.
The set~$F$ dominates~$N_G[F]$, and~$D$ covers $R_0=V(G)\setminus N_G[F]$, so $F\cup D$ dominates~$G$, and Lemma~\ref{fv:hg:budget}(iii) gives $|F\cup D|\le3\gf(G)$.
For $a=17/15$, the constants are $\beta=5/22$, $\beta_w=171$, $\beta_\ell=384+17/11$, $\beta_P=180$, and $\beta_\Omega=90$, so $\delta_a=5/8482$.

Each vertex learns its degree, its membership in~$B$,~$Z$,~$F$, and~$R_0$, and its cost from a constant number of broadcasts.
Given~$x$, the vertices announce membership in~$W$,~$R$, and~$H$ in three Boolean broadcasts, after which each vertex decides whether it belongs to~$U$.
The attachments and free paths appear only in the proof.
\end{proof}

\begin{corollary}
\label{fv:hg:planar17}
There is a deterministic constant-round broadcast CONGEST algorithm that returns a dominating set of size at most~$3\gf(G)$ on every planar graph~$G$ of girth at least~$17$.
No deterministic constant-round LOCAL algorithm achieves a coefficient $3-\varepsilon$ against~$\gamma(G)$ on this class, for any $\varepsilon>0$.
\end{corollary}

\begin{proof}
By Lemma~\ref{sc:face}, every subgraph of~$G$ has a ratio of edges to vertices of at most $17/15$, and $17/15<8/7$ because $119<120$.
The induced subgraph~$G[Z]$ is planar and hence lies in~$\mathcal C_3$.
After computing~$F$,~$R_0$, and~$c$, run the algorithm of Theorem~\ref{simp:weighted-solver} for the class~$\mathcal C_3$ on~$G[Z]$, with costs at most three and accuracy $5/8482$; the vertices outside~$Z$ stay silent, and edges leaving~$Z$ are ignored.
Capping the output at one keeps it feasible and does not increase its cost.
Theorem~\ref{fv:hg:three} applies to the result.
For the lower bound, paths are planar and have no cycles, and the case $r=1$ of the theorem of Amiri and Siebertz~\cite[Theorem~8.7]{AS2026}, quoted in Appendix~\ref{sc:optimality}, excludes every coefficient below three.
\end{proof}

At girth~$16$ the planar density bound is exactly $8/7$, and the final count in Lemma~\ref{fv:hg:count} degenerates.
Whether coefficient three can be achieved there is open.

\subsection{Large girth in classes excluding a fixed minor}

A large girth makes every class~$\mathcal C_\rho$ almost a forest for the purpose of an edge count.
The idea is to group the vertices into connected clusters of radius at most~$2r$ that each contain at least $r+1$ vertices.
By the girth bound, the clusters induce trees and are joined by at most one edge each, so contracting them yields a minor in which every edge between different clusters survives.
The clusters appear only in the proof; the algorithm never computes them.

\begin{lemma}
\label{fv:hg:dilute}
Let $\rho\ge1$, let $r\ge1$ be an integer, and let $G\in\mathcal C_\rho$ have girth greater than $8r+2$, where a forest has infinite girth.
Then every subgraph~$G'$ of~$G$ satisfies
\[
 |E(G')|\le\Bigl(1+\frac{\rho-1}{r+1}\Bigr)|V(G')|.
\]
\end{lemma}

\begin{proof}
It suffices to prove the bound for every component~$K$ of~$G'$ and then add these bounds.
Let~$n$ and~$m$ be the numbers of vertices and edges of~$K$, and measure distances in~$K$.
If $n\le r$, then~$K$ is a tree, since a cycle in~$K$ would have length at most~$r$; so $m=n-1<n$.

Let $n\ge r+1$.
Choose an inclusion-maximal set of centers with pairwise distances greater than~$2r$, and order the centers.
Assign every vertex to a nearest center, breaking ties by the order.
By maximality, every vertex has distance at most~$2r$ from its center.
If~$v$ is assigned to~$c$, then the vertex after~$v$ on a shortest path from~$v$ to~$c$ is also assigned to~$c$: a closer center, or an equally close earlier one, would also be preferred to~$c$ at~$v$.
Hence each cluster is connected and contains a shortest path from each of its vertices to its center, so it has a spanning tree in which every vertex has distance at most~$2r$ from the center.

Every vertex~$v$ at distance at most~$r$ from a center~$c$ is assigned to~$c$, because every other center has distance at least $2r+1$ from~$c$ and hence at least $r+1$ from~$v$.
Since~$K$ is connected and $n\ge r+1$, this ball contains at least $r+1$ vertices: either some vertex has distance at least~$r$ from~$c$, and a shortest path to it gives $r+1$ vertices of the ball, or the ball is all of~$K$.
So the number~$N$ of clusters satisfies $N\le n/(r+1)$.

An edge inside a cluster that is not in its tree closes a cycle of length at most $4r+1$ with the tree path between its ends.
Two distinct edges $x_1y_1$ and $x_2y_2$ between the same two clusters, with $x_1,x_2$ in one and $y_1,y_2$ in the other, close a cycle of length at most $8r+2$ with the tree paths from~$x_1$ to~$x_2$ and from~$y_1$ to~$y_2$.
This is a cycle because the two tree paths are disjoint and, as the edges are distinct, at most one of them is trivial.
Both cases contradict the girth bound.
So each cluster induces its tree, and contracting the clusters gives a minor~$J$ of~$G$ with~$N$ vertices, in which every edge between clusters survives.
Since $G\in\mathcal C_\rho$,
\[
 m=(n-N)+|E(J)|\le n+(\rho-1)N\le\Bigl(1+\frac{\rho-1}{r+1}\Bigr)n.
 \qedhere
\]
\end{proof}

\begin{corollary}
\label{fv:hg:minor-three}
Let $\rho\ge1$ be rational, and put
\[
 r_\rho=\max\{1,\lfloor7(\rho-1)\rfloor\},\qquad g_\rho=8r_\rho+3.
\]
There is a deterministic constant-round broadcast CONGEST algorithm that returns a dominating set of size at most~$3\gf(G)$ on every graph $G\in\mathcal C_\rho$ of girth at least~$g_\rho$.
In particular, every class that excludes a fixed minor has such a girth threshold.
The coefficient three is optimal on the class of graphs in~$\mathcal C_\rho$ of girth at least~$g_\rho$, since it contains all paths.
More generally, it is optimal on the graphs of sufficiently large girth in every class that excludes a fixed minor and contains all paths.
\end{corollary}

\begin{proof}
By Lemma~\ref{fv:hg:dilute}, every subgraph of~$G$ has a ratio of edges to vertices of at most $a=1+(\rho-1)/(r_\rho+1)$.
This number is rational, at least one, and smaller than $8/7$, because $r_\rho+1>7(\rho-1)$.
The induced subgraph~$G[Z]$ is a minor of~$G$ and so lies in~$\mathcal C_\rho$.
Run the algorithm of Theorem~\ref{simp:weighted-solver} for~$\mathcal C_\rho$ on~$G[Z]$, with costs at most three and accuracy~$\delta_a$.
Capping the output at one keeps it feasible and does not increase its cost.
Then apply Theorem~\ref{fv:hg:three}.
Every class that excludes a fixed minor lies in some~$\mathcal C_\rho$~\cite{Mader1967}.
Every minor of a path is a disjoint union of paths, so paths lie in~$\mathcal C_1$ and have every girth; the theorem of Amiri and Siebertz~\cite[Theorem~8.7]{AS2026} with $r=1$ then gives optimality.
\end{proof}

For planar graphs, $\rho=3$ gives $r_3=14$ and the threshold~$115$, much larger than the threshold~$17$ of Corollary~\ref{fv:hg:planar17}, which uses the sharper density bound of Lemma~\ref{sc:face}.
Neither threshold is claimed to be best possible.

\subsection{Coefficient two for total domination at a fixed density below nine eighths}

Recall from Appendix~\ref{fv:total} that a set~$D$ is a total dominating set of an isolate-free graph~$G$ if every vertex has a neighbor in~$D$, and that~$\gamma_{t,f}(G)$ is the optimum of the relaxation with the constraints $\sum_{u\in N(v)}y_u\ge1$.
Throughout this subsection $N(v)=N_G(v)$ is the open neighborhood.
On paths, no deterministic constant-round algorithm beats the coefficient two (Proposition~\ref{fv:total:lower-tw} with $t=1$).
The construction follows the one for domination, with two changes.
A selected vertex does not cover itself, so the selected exterior must have no isolated vertex.
And the demands now include the vertices that the exterior already covers; their constraints give an additional saving.

\paragraph{The transformation.}
Let~$G$ be isolate-free, and put
\[
 B=\{v:\deg_G(v)\ge3\},\qquad Z_0=N_G^2[B],\qquad F_0=V(G)\setminus Z_0.
\]
Let~$F$ be~$F_0$ without the vertices that are isolated in~$G[F_0]$, and put $Z=V(G)\setminus F$.
The algorithm selects~$F$.
Every vertex is given the cost $c(v)=2-|N(v)\cap F|$, the candidates are the vertices of $K=\{v:c(v)>0\}$, and every vertex of~$Z$ is a demand, to be covered by a candidate in its open neighborhood.
Let~$\cost^*$ be the optimum of this program and $\cost(x)=\sum_{v\in K}c(v)x_v$.

\begin{lemma}
\label{fv:hg:t-budget}
\leavevmode
\begin{enumerate}[label=(\roman*)]
\item The graph~$G[F]$ has no isolated vertex.
Every vertex of~$Z$ is joined to~$B$ by a path of length at most three whose vertices other than the first lie in~$Z_0$.
No vertex of~$N_G[B]$ has a neighbor in~$F$.
\item A vertex with a neighbor in~$F$ has degree at most two.
Hence $c(v)\in\{0,1,2\}$ for every~$v$, and $c=2$ on~$N_G[B]$.
A vertex of cost zero has no neighbor in~$Z$.
Moreover $Z\subseteq K$, and every demand has all its neighbors in~$K$.
\item $|F|+\cost^*\le2\gamma_{t,f}(G)$.
\end{enumerate}
\end{lemma}

\begin{proof}
(i) The first claim holds by the definition of~$F$.
A vertex of~$Z_0$ has distance at most two from~$B$, and a shortest path to~$B$ stays in~$Z_0$.
A vertex of $Z\setminus Z_0$ is isolated in~$G[F_0]$, and it has a neighbor because~$G$ is isolate-free; this neighbor lies in~$Z_0$.
Every neighbor of a vertex of~$N_G[B]$ lies in~$Z_0$, which is disjoint from~$F$.

(ii) A neighbor of a vertex in~$F$ has distance at least two from~$B$, so it is not in~$B$.
A vertex of cost zero therefore has exactly two neighbors, both in~$F$.
A vertex of~$Z_0$ at distance two from~$B$ has a neighbor in~$Z_0$, so it has at most one neighbor in~$F$; vertices of~$N_G[B]$ have none; and a vertex of $Z\setminus Z_0$ has no neighbor in~$F_0$.
Thus $c\ge1$ on~$Z$.
A neighbor of a demand has a neighbor in~$Z$, so its cost is positive.

(iii) Let~$y$ be an optimal fractional total dominating vector of~$G$.
By (ii), its restriction to~$K$ covers every demand, so $\cost^*\le\sum_{v\in K}c(v)y_v$.
Summing the constraints of~$y$ over~$F$ gives $|F|\le\sum_v|N(v)\cap F|\,y_v$.
Since $|N(v)\cap F|+c(v)=2$ for every~$v$, adding the two bounds gives $|F|+\cost^*\le2\gamma_{t,f}(G)$.
\end{proof}

\Needspace{16\baselineskip}
\paragraph{The rounding.}
Let $x\in[0,1]^K$ be a feasible fractional solution, extended by zero outside~$K$.
The rounding is the rule of Section~\ref{fv:total:rule} with $\Pi=B$, threshold $t=2/3$, and three in place of~$n_0$ in Step~2, except that a demand with a neighbor in~$F$ requests nothing in Step~3:
\[
\begin{gathered}
 W=\{v\in B:x_v>2/3\},\qquad R=Z\setminus N(W),\qquad
 h(v)=|N(v)\cap R|,\\
 H=\{v\in K\setminus W:h(v)\ge3\},\qquad Y=R\cap N(H),\qquad
 U=R\setminus Y.
\end{gathered}
\]
Every vertex of~$U$ without a neighbor in~$F$ requests its neighbor of smallest identifier, and~$D$ is the union of $W\cup H$ and the requested vertices.
The output is $F\cup D$.
Put $O=K\setminus(W\cup H)$ and
\[
 w=|W|,\quad j=|Y|-|H|,\quad f=|U\cap N(F)|,\quad
 E=\sum_{v\in H}(h(v)-2),\quad P=\sum_{v\in O}(c(v)-h(v))x_v,
\]
and, for $r\in R$, let $\omega_r=\sum_{v\in N(r)}x_v-1\ge0$ and $\Omega=\sum_{r\in R}\omega_r$.

A vertex of~$H$ has at least three neighbors, so $W\cup H\subseteq B$, and $c=2$ on $W\cup H$ by Lemma~\ref{fv:hg:t-budget}(ii).
Every $v\in B\setminus W$ has $x_v\le2/3$.
On~$O$ we have $c\ge h$: if $c(v)=2$, then $h(v)\le2$, and if $c(v)=1$, then~$v$ has a neighbor in~$F$ and at most one neighbor in~$Z$.
Hence $P\ge0$.
Note that a vertex of~$W$ may lie in~$R$.

\begin{lemma}
\label{fv:hg:t-saving}
Suppose that every subgraph of~$G$ has a ratio of edges to vertices of at most~$a$, where $1\le a<9/8$.
Then $F\cup D$ is a total dominating set, $j\ge0$, $|H|\le aj/(3-2a)$, $|Y|\le(3-a)j/(3-2a)$, and
\[
 \cost(x)-|D|\ge\frac w3+\beta'j+f+P+\Omega,\qquad
 \beta'=\frac{9-8a}{3(3-2a)}>0.
\]
\end{lemma}

\begin{proof}
The set~$F$ covers~$F$ by Lemma~\ref{fv:hg:t-budget}(i),~$W$ covers $Z\setminus R$,~$H$ covers~$Y$, and every vertex of~$U$ has a neighbor in~$F$ or a requested neighbor.
Since $|U|=|R|-|Y|=|R|-|H|-j$, the size of~$D$ is at most $w+|H|+|U|-f=w+|R|-j-f$.

Count the pairs $(v,r)$ with $v\in H$, $r\in R$, and $vr\in E(G)$.
There are $2|H|+E$ of them, and each is an edge of~$G[H\cup Y]$, counted twice if both ends lie in $H\cap Y$ and once otherwise.
Hence the number of pairs is $e(G[H\cup Y])+e(G[H\cap Y])$.
Each term is at most~$a$ times its number of vertices, and $|H\cup Y|+|H\cap Y|=|H|+|Y|$.
So $2|H|+E\le a(|H|+|Y|)=a(2|H|+j)$, that is, $E\le(2a-2)|H|+aj$.
Since $h\ge3$ on~$H$, we have $|H|\le E$, and therefore $|H|\le E\le aj/(3-2a)$; in particular $j\ge0$.
The bound on $|Y|$ follows from $|Y|=|H|+j$.

No vertex of~$W$ is adjacent to~$R$, and every neighbor of a demand lies in~$K$.
Summing the constraints over~$R$ gives $\sum_{v\in K\setminus W}h(v)x_v=|R|+\Omega$.
With $x_v>2/3$ on~$W$ and $x_v\le2/3$ on~$H$,
\[
 \cost(x)=2\sum_{v\in W}x_v+\sum_{v\in K\setminus W}h(v)x_v
 -\sum_{v\in H}(h(v)-2)x_v+P
 \ge\tfrac43w+|R|+\Omega-\tfrac23E+P.
\]
Subtracting the bound on $|D|$ and using $E\le aj/(3-2a)$ gives the saving.
\end{proof}

\paragraph{Attachments.}
Let $U_0=U\setminus(W\cup H)$.
Let~$T$ be the set of vertices outside~$W$ with a neighbor in~$W$ and a neighbor in~$U_0$, and put $s=|T|$.
An \emph{attachment} is an edge between~$U_0$ and~$T$; let~$q$ be their number, so $s\le q$, and call a vertex of~$U_0$ incident with an attachment an \emph{attachment vertex}.

\begin{lemma}
\label{fv:hg:t-residual}
\leavevmode
\begin{enumerate}[label=(\roman*)]
\item Every vertex of~$U_0$ has at most two neighbors in~$R$, and they lie in $U_0\cup(Y\setminus H)$.
Every vertex of~$Y$ adjacent to~$U_0$ has at most two neighbors in~$R$.
Hence every component of~$G[U_0]$ is a path or a cycle.
\item Every neighbor of a vertex $u\in U_0$ that lies in $Z\setminus R$ belongs to~$T$.
Every vertex of $B\cap U_0$ is an attachment vertex.
\item Let $u\in U_0$, and let~$C$ be its component in~$G[U_0]$.
Then~$u$ has distance at most three in~$C$ from a vertex of $B\cap C$, or distance at most two in~$C$ from an attachment vertex, or there is a path of length at most three from~$u$ to a vertex of~$Y$ whose other vertices lie in~$C$.
\item $|U_0|\le7q+6|Y|$.
\end{enumerate}
\end{lemma}

\begin{proof}
(i) A vertex of~$U_0$ lies in $K\setminus(W\cup H)$ by Lemma~\ref{fv:hg:t-budget}(ii), so it has $h\le2$.
Its neighbors in~$R$ are not in~$W$, since $R\cap N(W)=\varnothing$, and not in~$H$, since $U\cap N(H)=\varnothing$; so they lie in~$U_0$ or in $Y\setminus H$.
A vertex of~$Y$ adjacent to~$U_0$ is thus in $K\setminus(W\cup H)$, and has $h\le2$.

(ii) The set $Z\setminus R$ lies in~$N(W)$, and no neighbor of~$u$ lies in~$W$.
A vertex of $B\cap U_0$ has at least three neighbors, all in~$Z_0$, and at most two of them in~$R$.

(iii) By Lemma~\ref{fv:hg:t-budget}(i), there is a path $u=p_0,\ldots,p_k$ with $k\le3$, $p_k\in B$, and $p_1,\ldots,p_k\in Z$.
If it lies in~$U_0$, then $p_k\in B\cap C$.
Otherwise let $j\ge1$ be least with $p_j\notin U_0$.
If $p_j\in R$, then $p_j\in Y$ by (i).
If $p_j\notin R$, then $p_j\in T$ by (ii), and~$p_{j-1}$ is an attachment vertex at distance at most two from~$u$.

(iv) Assign each vertex of~$U_0$ to a vertex given by (iii), which by (ii) is an attachment vertex at distance at most three in~$C$, or a vertex of~$Y$.
An attachment vertex has at most seven vertices of its component within distance three.
A vertex of~$Y$ has at most two neighbors in~$U_0$, and the routes from them inside~$U_0$ reach at most six vertices.
\end{proof}

Call a path component~$C$ of~$G[U_0]$ \emph{free} if it has no neighbor in~$Y$ and no vertex in~$N(F)$, and put
\[
 Q_C=\sum_{v\in O,\ N(v)\cap C\ne\varnothing}(c(v)-h(v))x_v,\qquad
 \Omega_C=\sum_{r\in C}\omega_r.
\]

\begin{lemma}
\label{fv:hg:t-free}
Let~$C$ be a free path.
\begin{enumerate}[label=(\roman*)]
\item Every neighbor of~$C$ outside~$C$ lies in~$T$.
All vertices of~$C$ and of~$T$ have cost two.
\item A vertex of~$C$ of degree at least three is an attachment vertex in~$B$.
A vertex of~$C$ that is not an attachment vertex has all its neighbors in~$C$.
\item Every vertex of~$C$ has distance at most three in~$C$ from an attachment vertex in~$B$, or at most two from an attachment vertex.
\item Every vertex of $C\cap B$ has $x_v\le2/3$.
Every end~$v$ of a path~$C$ with at least two vertices has $h(v)\le1$, so~$x_v$ is a summand of~$Q_C$; if $C=\{b\}$, the proof of Lemma~\ref{fv:hg:t-bad} uses~$x_p$ instead.
Every positive summand of~$Q_C$ comes from a vertex with exactly one neighbor in~$R$.
\end{enumerate}
\end{lemma}

\begin{proof}
(i) A neighbor $y\notin C$ of $v\in C$ is not in~$F$, since $v\notin N(F)$.
It is not in~$R$, because by Lemma~\ref{fv:hg:t-residual}(i) it would lie in~$Y$.
Hence $y\in Z\setminus R$, and $y\in T$ by Lemma~\ref{fv:hg:t-residual}(ii).
Vertices of~$T$ lie in $N(W)\subseteq N_G[B]$, and vertices of~$C$ have no neighbor in~$F$, so both have cost two.

(ii) A vertex of degree at least three lies in~$B$ and, by (i), has a neighbor in~$T$.
A vertex with no neighbor in~$T$ has, by (i), all its neighbors in~$C$.

(iii) This follows from Lemma~\ref{fv:hg:t-residual}(ii) and (iii), since~$C$ has no neighbor in~$Y$.

(iv) We have $C\cap W=\varnothing$.
The neighbors of $v\in C$ in~$R$ lie in~$C$, so an end has $h\le1<2=c$.
A vertex $v\in O$ with $c(v)>h(v)$ and a neighbor in $C\subseteq R$ has $h(v)\ge1$, hence $c(v)=2$ and $h(v)=1$.
\end{proof}

A free path is \emph{bad} if it has exactly one attachment, or if its attachments are at distinct vertices and, counted from some end, its first two attachment vertices have distance six or seven.
A bad path is \emph{exceptional} if it is a single vertex~$b$ whose attachment neighbor~$p$ has degree two and a neighbor in $W\cap R$.

\begin{lemma}
\label{fv:hg:t-bad}
Every bad path~$C$ that is not exceptional satisfies $Q_C+\Omega_C\ge1/3$.
The number~$b$ of bad paths satisfies $b\le w+3P+3\Omega$.
\end{lemma}

\begin{proof}
By Lemma~\ref{fv:hg:t-free}(ii), the constraint at a vertex of~$C$ that is not an attachment vertex involves only coordinates of vertices of~$C$.

\emph{One attachment, at~$b$, with neighbor $p\in T$.}
If $b\in B$, then~$b$ has two neighbors in~$C$, and every vertex of~$C$ has distance at most three from~$b$.
Otherwise~$b$ has degree at most two, so it is an end of~$C$, and~$C$ has at most three vertices, all within distance two of~$b$.

If $C=\{b\}$, the constraint at~$b$ gives $x_p=1$.
The vertex~$p$ is not in~$W$, and not in~$H$ since $U\cap N(H)=\varnothing$, so $p\in O$ and $1\le h(p)\le2$.
If $h(p)=1$, then $Q_C\ge x_p=1$.
If $h(p)=2$, then $p\notin B$, because $x_p>2/3$; so~$p$ has degree two, and its neighbor in~$W$ lies in~$R$, which makes~$C$ exceptional.
If $C=\{b,v\}$, the constraint at~$v$ gives $x_b=1$, and $h(b)=1$.
If~$C$ has three vertices and~$b$ is an end, the constraint at the middle vertex shows that the two ends have total coordinate at least one.

If~$b$ is an inner vertex, then $b\in B$ and $x_b\le2/3$.
An end adjacent to~$b$ would force $x_b=1$.
If an end~$v$ has distance two from~$b$, the constraint at the vertex between them gives $x_v\ge1-x_b\ge1/3$.
Otherwise $C=v_1\cdots v_7$ with $b=v_4$.
The constraints at~$v_2$ and~$v_6$ give $x_{v_3}\ge1-x_{v_1}$ and $x_{v_5}\ge1-x_{v_7}$, so $\omega_{v_4}\ge1-x_{v_1}-x_{v_7}$, and $Q_C+\Omega_C\ge1$.

\emph{At least two attachments.}
Each attachment vertex has exactly one attachment, so an inner one lies in~$B$ and has coordinate at most $2/3$, and the coordinate of an end is a summand of~$Q_C$.
Number the vertices of~$C$ from the chosen end so that the first two attachment vertices are~$v_0$ and~$v_d$, $d\in\{6,7\}$, and the vertices before~$v_0$ are $v_{-1},\ldots,v_{-\ell}$; by Lemma~\ref{fv:hg:t-free}(iii), $\ell\le3$.
If $\ell=1$, the constraint at~$v_{-1}$ gives $x_{v_0}=1$, which is impossible.
If $\ell=2$, the constraint at~$v_{-1}$ gives $x_{v_{-2}}\ge1-x_{v_0}\ge1/3$.

If $\ell=0$, then~$v_0$ has degree two and is not in~$B$.
If $d=7$, the vertex~$v_3$ has distance three from~$v_0$ and four from~$v_7$, which contradicts Lemma~\ref{fv:hg:t-free}(iii).
So $d=6$, and the constraints at $v_1,v_3,v_5$ give
\[
 x_{v_2}\ge1-x_{v_0},\qquad x_{v_4}\le x_{v_0}+\omega_{v_3},\qquad
 x_{v_6}\ge1-x_{v_0}-\omega_{v_3}.
\]

If $\ell=3$ and $d=6$, the constraints at~$v_{-3}$ and~$v_{-1}$ give $x_{v_{-2}}=1$ and $x_{v_0}=\omega_{v_{-1}}$, and then the constraints at $v_1,v_3,v_5$ give
\[
 x_{v_2}\ge1-\omega_{v_{-1}},\qquad
 x_{v_4}\le\omega_{v_{-1}}+\omega_{v_3},\qquad
 x_{v_6}\ge1-\omega_{v_{-1}}-\omega_{v_3}.
\]
If $\ell=3$ and $d=7$, put $e=x_{v_{-3}}$, a summand of~$Q_C$.
The constraints at $v_{-2},v_0,v_2,v_4,v_6$ give, in turn,
\[
 x_{v_{-1}}\ge1-e,\quad x_{v_1}\le e+\omega_{v_0},\quad
 x_{v_3}\ge1-e-\omega_{v_0},\quad
 x_{v_5}\le e+\omega_{v_0}+\omega_{v_4},\quad
 x_{v_7}\ge1-e-\omega_{v_0}-\omega_{v_4}.
\]
In each of the three cases, compare the bound on~$x_{v_d}$ with $x_{v_d}\le2/3$ if~$v_d$ is inner, or use that~$x_{v_d}$ is a summand of~$Q_C$ if~$v_d$ is an end.
Either way $Q_C+\Omega_C\ge1/3$.

For the count, the sets~$C$ are disjoint, and by Lemma~\ref{fv:hg:t-free}(iv) every positive summand of~$Q_C$ belongs to one path only.
So the number of bad paths that are not exceptional is at most $3P+3\Omega_U$, where~$\Omega_U$ is the excess over~$U_0$.
For an exceptional path, $x_p=1$, and~$b$ is the only neighbor of~$p$ in~$U_0$, so different exceptional paths have different vertices~$p$.
Summing the constraints at $W\cap R$ shows that the number of exceptional paths is at most $|W\cap R|+\Omega_W$, where~$\Omega_W$ is the excess over $W\cap R$.
Since $W\cap U_0=\varnothing$, we have $\Omega_U+\Omega_W\le\Omega$.
\end{proof}

\begin{lemma}
\label{fv:hg:t-count}
Suppose that every subgraph of~$G$ has a ratio of edges to vertices of at most~$a$, where $1\le a<9/8$.
Then for every feasible~$x$,
\[
 \frac{9-8a}2\,q\le\frac{5-2a}2\,w+(6a-4)|Y|+f+\frac{15-12a}2\,(P+\Omega),
 \qquad |D|\le2w+2|H|+6|Y|+7q.
\]
\end{lemma}

\begin{proof}
The second bound follows from $|D|\le w+|H|+|U|$, from $|U|\le|U_0|+w+|H|$, and from Lemma~\ref{fv:hg:t-residual}(iv).

For the first, let~$c$ be the number of free paths and~$c_1$ the number of those with one attachment.
Replace every free path~$C$ by its shortest subpath~$C'$ containing all its attachment vertices, keep the other components of~$G[U_0]$, and let~$U'$ be the resulting set.
By Lemma~\ref{fv:hg:t-free}(iii), every vertex of~$C$ has distance at most three from an attachment vertex, so consecutive attachment vertices have distance at most seven.
If~$C$ has~$q_C$ attachments, then $|C'|\le7q_C-8+2[C\text{ is bad}]$.
Indeed, $|C'|=1$ if $q_C=1$; if there are $k<q_C$ distinct attachment vertices, then $|C'|\le7k-6\le7q_C-13$; and otherwise $|C'|\le1+5+7(q_C-2)$ unless~$C$ is bad, in which case $|C'|\le7q_C-6$.
The other components are bounded as in the proof of Lemma~\ref{fv:hg:t-residual}(iv).
Hence
\[
 |U'|\le7q+6|Y|-8c+2b.
\]

Apply the density bound to the subgraph formed by~$U'$,~$T$, and~$W$, with the edges of~$G[U']$, all~$q$ attachments, and one edge from each vertex of~$T$ to~$W$.
If~$c_{\mathrm p}$ is the number of path components of~$G[U_0]$, then~$G[U']$ has $|U'|-c_{\mathrm p}$ edges.
A path component that is not free has a neighbor in~$Y$ or a vertex in $U\cap N(F)$, and a vertex of~$Y$ is adjacent to at most two components by Lemma~\ref{fv:hg:t-residual}(i), so $c_{\mathrm p}\le c+2|Y|+f$.
With $s\le q$ this gives
\[
 |U'|-c_{\mathrm p}+q+s\le a(|U'|+s+w),\quad\text{hence}\quad
 (2-a)q\le(a-1)|U'|+c+2|Y|+f+aw.
\]
With the bound on $|U'|$,
\[
 (9-8a)q\le aw+(6a-4)|Y|+f+(9-8a)c+2(a-1)b.
\]
Every free path not counted by~$c_1$ has at least two attachments, and every path counted by~$c_1$ is bad, so $c\le(q+c_1)/2\le(q+b)/2$.
Since $9-8a>0$, this gives $\frac{9-8a}2q\le aw+(6a-4)|Y|+f+\frac{5-4a}2b$, and Lemma~\ref{fv:hg:t-bad} completes the proof.
\end{proof}

\begin{theorem}
\label{fv:hg:total-two}
\leavevmode
\begin{enumerate}[label=(\roman*)]
\item Let $1\le a<9/8$ be rational.
There is a rational $\delta'_a>0$ with the following property.
Let~$G$ be an isolate-free graph every subgraph of which has a ratio of edges to vertices of at most~$a$, let~$F$,~$Z$,~$K$, and~$c$ be as above, and let $x\in[0,1]^K$ be any feasible fractional solution with $\cost(x)\le(1+\delta'_a)\cost^*$.
Then $F\cup D$ is a total dominating set of~$G$ of size at most~$2\gamma_{t,f}(G)$.
The sets~$F$,~$Z$,~$K$ and the costs are computed in a constant number of broadcast rounds, and so is $F\cup D$ once~$x$ is given.
\item There is a deterministic constant-round broadcast CONGEST algorithm that returns a total dominating set of size at most~$2\gamma_{t,f}(G)$ on every isolate-free planar graph~$G$ of girth at least~$19$.
The coefficient two is optimal on this class, also against~$\gamma_t(G)$.
\end{enumerate}
For $a=19/17$, the value $\delta'_a=1/62514$ suffices.
\end{theorem}

\begin{proof}
(i) By Lemmas~\ref{fv:hg:t-count} and~\ref{fv:hg:t-saving}, the size of~$D$ is at most $\beta_ww+\beta_jj+\beta_ff+\beta_P(P+\Omega)$, where
\[
 \beta_w=2+\tfrac{7(5-2a)}{9-8a},\quad
 \beta_j=\tfrac{2a+6(3-a)}{3-2a}+\tfrac{14(6a-4)(3-a)}{(9-8a)(3-2a)},\quad
 \beta_f=\tfrac{14}{9-8a},\quad
 \beta_P=\tfrac{7(15-12a)}{9-8a}
\]
are positive, and the saving of Lemma~\ref{fv:hg:t-saving} has the coefficients $1/3,\beta',1,1,1$.
By Proposition~\ref{sc:principle} the rounding has slack
\[
 \delta'_a=\min\Bigl\{\frac1{3\beta_w},\frac{\beta'}{\beta_j},
 \frac1{\beta_f},\frac1{\beta_P}\Bigr\},
\]
so $|D|\le\cost^*$, and Lemma~\ref{fv:hg:t-budget}(iii) gives $|F\cup D|\le2\gamma_{t,f}(G)$.
For $a=19/17$ the constants are $\beta'=1/39$, $\beta_w=331$, $\beta_j=20838/13$, $\beta_f=238$, and $\beta_P=189$, so $\delta'_a=1/62514$.

Each vertex learns membership in~$B$,~$Z_0$,~$F$, and~$K$, and its cost, from a constant number of broadcasts.
Given~$x$, the vertices announce membership in~$W$,~$R$, and~$H$ in three Boolean broadcasts, and every requesting vertex then broadcasts one identifier.

(ii) By Lemma~\ref{sc:face}, every subgraph of~$G$ has a ratio of edges to vertices of at most $19/17$, and $19/17<9/8$ because $152<153$.
The graph~$G[K]$ is planar, so it lies in~$\mathcal C_3$.
Run the algorithm of Theorem~\ref{sc:thmA} on~$G[K]$ with $\rho=3$, the demands~$Z$, the covering sets~$N(r)$, costs at most two, and accuracy $1/62514$.
Capping the output at one keeps it feasible and does not increase its cost.
Then apply (i).
Paths are planar and have no cycles, so Proposition~\ref{fv:total:lower-tw} with $t=1$ gives optimality.
\end{proof}

\begin{corollary}
\label{fv:hg:minor-total-two}
Let $\rho\ge1$ be rational, and put $r'_\rho=\max\{1,\lfloor8(\rho-1)\rfloor\}$ and $g'_\rho=8r'_\rho+3$.
There is a deterministic constant-round broadcast CONGEST algorithm that returns a total dominating set of size at most~$2\gamma_{t,f}(G)$ on every isolate-free graph $G\in\mathcal C_\rho$ of girth at least~$g'_\rho$.
The coefficient two is optimal on this class, since it contains all paths with at least two vertices.
More generally, it is optimal on the isolate-free graphs of sufficiently large girth in every class that excludes a fixed minor and contains all paths with at least two vertices.
\end{corollary}

\begin{proof}
By Lemma~\ref{fv:hg:dilute}, every subgraph of~$G$ has a ratio of edges to vertices of at most $a=1+(\rho-1)/(r'_\rho+1)$, and $a<9/8$ because $r'_\rho+1>8(\rho-1)$.
The graph~$G[K]$ lies in~$\mathcal C_\rho$.
Run the algorithm of Theorem~\ref{sc:thmA} on~$G[K]$ as in the proof of Theorem~\ref{fv:hg:total-two}(ii), with accuracy~$\delta'_a$.
Capping the output at one keeps it feasible and does not increase its cost.
Then apply Theorem~\ref{fv:hg:total-two}(i).
Paths lie in~$\mathcal C_1$ and have every girth, so Proposition~\ref{fv:total:lower-tw} with $t=1$ gives optimality.
\end{proof}

\section{Red-blue domination}\label{fv:rb}

In red-blue domination, the vertices of the network~$G$ are colored red and blue, and we seek a cheapest set of blue vertices such that every red vertex has a chosen blue neighbor~\cite{GarneroSauThilikos2017}.
This is the general covering instance with~$R$ the set of red vertices and~$C(r)$ the set of blue neighbors of~$r$.
Equivalently, this is set cover on the bipartite incidence graph, with blue vertices representing sets and red vertices representing elements.
Exchanging the two sides gives hitting set.
There is no bound on set sizes or element frequencies.
Let~$\tau_c$ be its fractional optimum and~$\OPT_c$ its integral optimum.
No red vertex belongs to its own covering set, so Corollary~\ref{sc:rounding-all} gives the coefficient $2a+\varepsilon$ for an integer~$a$ whenever all subgraphs of the red-blue incidence graph have a ratio of edges to vertices of at most~$a$; it is used here with the integer $a=2$.
In this section we state the planar case, prove that its coefficient four cannot be improved by any deterministic constant-round algorithm, and show what the same rounding gives on a fixed surface.

\paragraph{Planar networks.}
Every minor of a planar graph is planar, so planar graphs lie in~$\mathcal C_3$.
The red-blue incidence graph is bipartite, and every bipartite planar graph with~$n$ vertices has at most~$2n$ edges, so $a=2$ for all its subgraphs.
Corollary~\ref{sc:rounding-all} therefore applies with the integer $a=2$.
Nothing else is needed: no initial cover, no embedding, and no bound on the degrees.

\begin{corollary}\label{fv:rb:upper}
Fix $\varepsilon>0$ and an integer $b\ge1$.
There are
\begin{enumerate}[label=(\roman*)]
\item a deterministic constant-round broadcast CONGEST algorithm that returns, on every feasible red-blue instance whose network is planar and whose costs lie in $\{1,\ldots,b\}$, a fractional cover of cost at most $(1+\varepsilon)\tau_c$;
\item a deterministic constant-round broadcast CONGEST algorithm that returns, on every such instance, an integral cover of cost at most $(4+\varepsilon)\tau_c$;
\item a randomized constant-round algorithm whose output, on every such instance, is always a cover and has expected cost at most $(1+\varepsilon)\OPT_c$, and, if the vertices know a polynomial bound on the identifiers, a deterministic~$\mathcal O(\log^*n)$-round algorithm that returns, on every such instance, a cover of cost at most $(1+\varepsilon)\OPT_c$.
\end{enumerate}
The same holds for set cover and hitting set whose incidence graph is planar.
\end{corollary}

\begin{proof}
Parts~(i) and~(iii) are Theorems~\ref{sc:thmA} and~\ref{sc:thmD} with $\rho=3$; in~(i) the accuracy is a rational $\delta\le\varepsilon$.
Part~(ii) follows from Corollary~\ref{sc:rounding-all} with the integer $a=2$; no red vertex lies in its own covering set, so the coefficient is $2a+\varepsilon=4+\varepsilon$.
For set cover and hitting set, the incidence graph of the set system is the network.
\end{proof}

Thus the framework also gives the optimal limiting factor four for planar set cover and hitting set, without a bound on set sizes or element frequencies.
Here the positive~$\varepsilon$ remains because the rounding gives a cover of cost at most four times the cost of the fractional solution, while the solver guarantees only a cost of at most $(1+\delta)\tau_c$.
Together, these bounds give $4(1+\delta)\tau_c$, so choosing $\delta\le\varepsilon/4$ gives the factor $4+\varepsilon$.
For ordinary and total domination, the saving in the rounding compensates for this fractional error, but we have not proved a corresponding saving for red-blue domination.
In fact, the present rounding rule can be tight even when the fractional solution is optimal.
Consider one red vertex with four blue neighbors of unit cost, each with fractional value~$1/4$.
The fractional cost is one, but the threshold rule selects all four blue vertices.
Thus this rounding rule has no positive saving that holds for every feasible fractional input.
This does not rule out an exact factor four obtained by another argument.
The lower bound in Theorem~\ref{fv:rb:lower} excludes every coefficient below four.

Weighted, partial, and restricted dominating set on planar graphs are handled in the same way, with coefficient $7+\varepsilon$, as in Table~\ref{sc:app-table}.

\paragraph{The coefficient four is optimal.}
The coefficient in Corollary~\ref{fv:rb:upper}(ii) is not an artifact of the rounding.
No deterministic constant-round algorithm beats it, even in the LOCAL model and even on instances where the fractional and the integral optimum agree.
The proof follows Hilke, Lenzen, and Suomela~\cite{HLS2014}, who obtain the coefficient seven for dominating set on planar graphs from repeated applications of Ramsey's theorem, and applies the argument to a grid of red squares whose corners are blue.

\begin{theorem}\label{fv:rb:lower} Let $\varepsilon>0$ and $t\ge0$.
No deterministic~$t$-round LOCAL algorithm returns, on every feasible planar red-blue instance, a set of blue vertices that covers all red vertices and has size at most $(4-\varepsilon)\tau_c$.
This holds even if the network is connected, has maximum degree four, and has $\tau_c=\OPT_c$, if the vertices know~$n$, and if the identifiers are exactly $1,\ldots,n$.
\end{theorem}

Since $\tau_c=\OPT_c$ on the instances of the proof, the bound holds against both benchmarks, and together with Corollary~\ref{fv:rb:upper}(ii) it shows that four is the infimum of the coefficients of deterministic constant-round algorithms for planar red-blue domination.
It also holds for set cover and hitting set on planar incidence graphs.
It does not apply to the fractional solution of Corollary~\ref{fv:rb:upper}(i): on a patch below, the vector $1/4$ on the blue vertices is an optimal fractional cover, so the indistinguishability argument does not apply to fractional outputs.
Nor does it apply to randomized algorithms, for which Theorem~\ref{sc:thmD} gives ratio $1+\varepsilon$.

\begin{proof}
Fix a deterministic algorithm with~$t$ rounds.
An even integer~$L$ and an integer~$m$ are chosen later.
The \emph{patch} has blue vertices~$b_{i,j}$ for $0\le i,j<L$ and red vertices~$r_{i,j}$ for $0\le i,j<L-1$, where~$r_{i,j}$ is adjacent to the four blue vertices at the corners of the unit square with lower-left corner~$(i,j)$.
Place~$r_{i,j}$ inside its square and draw its four edges inside the square, so the patch is planar.
It has $s=L^2+(L-1)^2$ vertices and maximum degree four.
The port numbers are fixed by direction: a blue vertex numbers its red neighbors by the four quadrants, and a red vertex numbers its blue neighbors by the four corners.

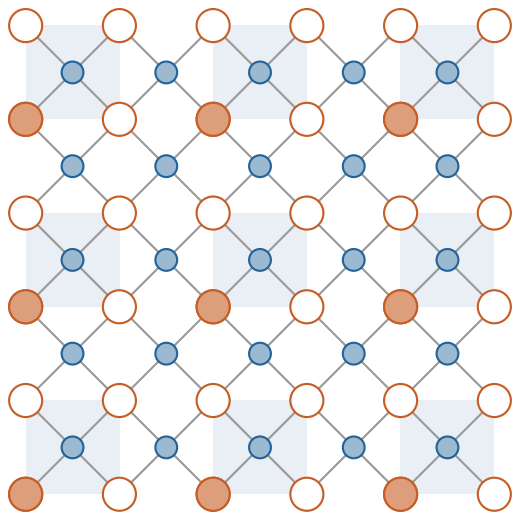
\begin{figure}[t]
\centering
\begin{tikzpicture}[scale=.62,
 rbblue/.style={circle,draw=simporange,fill=white,inner sep=2.2pt},
 rbred/.style={circle,draw=simpblue,fill=simpblue!45,inner sep=1.45pt}]
\foreach \i in {0,2,4} {
  \foreach \j in {0,2,4} {
    \fill[simpblue!10] (\i,\j) rectangle ({\i+1},{\j+1});
  }
}
\foreach \i in {0,...,4} {
  \foreach \j in {0,...,4} {
    \foreach \a/\b in {0/0,0/1,1/0,1/1} {
      \draw[black!40,line width=.4pt]
        ({\i+.5},{\j+.5}) -- ({\i+\a},{\j+\b});
    }
    \node[rbred] at ({\i+.5},{\j+.5}) {};
  }
}
\foreach \i in {0,...,5} {
  \foreach \j in {0,...,5} {
    \node[rbblue] at (\i,\j) {};
  }
}
\foreach \i in {0,2,4} {
  \foreach \j in {0,2,4} {
    \node[rbblue,fill=simporange!60] at (\i,\j) {};
  }
}
\end{tikzpicture}
\caption{The patch for $L=6$.
The nine filled blue vertices cover all red vertices.
The nine red vertices in the shaded squares have pairwise disjoint neighborhoods, so $\tau_c=\OPT_c=9$.
Illustration: an algorithm that cannot distinguish the blue vertices chooses all~$36$.}
\label{fv:rb:fig}
\end{figure}

The blue vertices with both coordinates even cover every red vertex, and there are $L^2/4$ of them.
Conversely, the red vertices~$r_{2i,2j}$ with $0\le i,j<L/2$ have pairwise disjoint blue neighborhoods, which together contain all blue vertices.
Summing the covering constraints of these $L^2/4$ red vertices shows that every fractional cover has cost at least $L^2/4$.
Hence a patch has $\tau_c=\OPT_c=L^2/4$.

The instance~$G_{L,m}$ consists of~$m$ patches and $m-1$ further red vertices.
The~$k$th further red vertex is adjacent to~$b_{0,0}$ in patches~$k$ and $k+1$, and these edges are drawn in the outer face.
The graph~$G_{L,m}$ is connected, planar, and bipartite, and has maximum degree four, since~$b_{0,0}$ has one neighbor in its patch and at most two further neighbors.
The even blue vertices also cover the further red vertices, and the disjoint constraints of the patches remain available, so
\begin{equation}\label{fv:rb:opt}
 n=m(s+1)-1,\qquad \tau_c=\OPT_c=\frac{mL^2}4 .
\end{equation}

\emph{Interior vertices.}
Put $h=t+2$, choose $L>2h+2$, and call a blue vertex \emph{interior} if both coordinates lie in $\{h,\ldots,L-1-h\}$.
The red vertices sit at the centers of their squares, so every edge changes each coordinate by $1/2$.
Hence all vertices within distance $t+1$ of an interior vertex have both coordinates in $[1,L-2]$, so they lie in the patch and have full degree.
Therefore the ball of radius $t+1$ around an interior vertex, with colors and port numbers, is the same for all interior vertices of all patches: it is a ball in the infinite grid.
Radius $t+1$ is used so that the argument also covers conventions in which a vertex starts by knowing the identifiers of its neighbors.
Let~$k$ be the number of vertices of this ball, which is ordered lexicographically by their coordinates, where~$r_{i,j}$ has coordinates $(i+\frac12,j+\frac12)$.
This order is preserved by translations.

\emph{Ramsey argument.}
Fix~$n$ as in~\eqref{fv:rb:opt}.
Color every~$k$-element set $X\subseteq\{1,\ldots,n\}$ by a bit: assign the elements of~$X$ in increasing order to the ordered ball, and take the bit that the algorithm computes at its center.
This bit depends only on the ball, its identifiers, and~$n$, so it does not depend on how the labeling is completed elsewhere, and the color is well defined.
Let~$R$ be a Ramsey number such that every two-coloring of the~$k$-element subsets of an~$R$-element set has a homogeneous subset of size~$s$, and put $\ell=\lceil R/s\rceil$.
Both numbers depend on~$t$ and~$L$ but not on~$m$.

We reserve $m-1$ identifiers for the further red vertices, which leaves exactly~$ms$ for the patches.
For $j=1,\ldots,m-\ell$, at the start of step~$j$ at least $(m-j+1)s\ge\ell s\ge R$ identifiers are unused, so they contain a homogeneous set of size~$s$.
Assign it to patch~$j$ in increasing order along the lexicographic order of the patch coordinates, and remove it from the unused identifiers.
The remaining identifiers are assigned arbitrarily.

Consider an interior blue vertex of a patch $j\le m-\ell$.
The identifiers of its ball form a~$k$-subset of the homogeneous set of the patch, and they increase along the lexicographic order of the ball, since this order is the restriction of the order of the patch.
Hence its output bit is the color of the homogeneous set, which is the same for all interior vertices of the patch.
The bit must be one: since $L-2h\ge2$, the patch contains a red vertex all of whose four blue neighbors are interior, and this vertex has a chosen blue neighbor only if the bit is one.

The algorithm therefore chooses at least $(m-\ell)(L-2h)^2$ blue vertices.
By~\eqref{fv:rb:opt}, its output has size at least
\[
 4\Bigl(1-\frac\ell m\Bigr)\Bigl(1-\frac{2h}L\Bigr)^2\tau_c .
\]
Choose~$L$ large enough, and then~$m$ large enough, so that this exceeds $(4-\varepsilon)\tau_c$.
\end{proof}

\paragraph{Fixed surfaces.}
Graphs of Euler genus at most~$g$ lie in~$\mathcal C_{3+3g}$ by Lemma~\ref{fv:imp:euler}, so Theorem~\ref{sc:thmA} applies.
Their red-blue incidence graphs have at most $2n+2g$ edges, not~$2n$, and this additive excess prevents us from using Corollary~\ref{sc:rounding-all} with the integer $a=2$.
Using it with the smallest integer~$a$ that covers the excess gives a coefficient that grows with~$g$.
The same rounding, with an orientation that allows for the excess, gives coefficient four up to an additive term.

\begin{proposition}
\label{fv:rb:surface}
Fix integers $g\ge0$ and $b\ge1$, and $\varepsilon>0$.
There is a deterministic constant-round broadcast CONGEST algorithm that, on every feasible red-blue instance whose network has Euler genus at most~$g$ and whose costs lie in $\{1,\ldots,b\}$, returns a cover of cost at most
\[
 (4+\varepsilon)\tau_c+bg .
\]
In particular, the cost is at most $(4+2\varepsilon)\tau_c$ whenever $\tau_c\ge bg/\varepsilon$.
\end{proposition}

\begin{proof}
Every minor of a graph of Euler genus at most~$g$ has Euler genus at most~$g$, and a simple graph with $n\ge3$ vertices and Euler genus~$e$ has at most $3n-6+3e$ edges.
Hence the network lies in~$\mathcal C_\rho$ with $\rho=3g+3$.
Let~$I$ be the red-blue incidence graph.
Euler genus is additive over components and does not increase for subgraphs.
A connected bipartite simple graph with $n\ge3$ vertices and Euler genus~$e$ has all faces of length at least four in a cellular embedding in a surface of Euler genus~$e$, so Euler's formula gives at most $2n-4+2e$ edges, and components with at most two vertices have at most~$2n$ edges.
Thus every subgraph~$F$ of~$I$ satisfies $|E(F)|\le2|V(F)|+2g$.

\emph{Orientation.}
We claim that~$I$ has an orientation in which every vertex~$v$ has outdegree at most~$2+e_v$, for integers $e_v\ge0$ with $\sum_ve_v\le2g$.
Consider the bipartite graph whose left side is~$E(I)$ and whose right side consists of two slots at every vertex of~$I$, where an edge is adjacent to the slots of its two ends.
An edge set~$E'$ with end set~$V'$ has $|E'|\le2|V'|+2g$ by the bound above, and~$2|V'|$ is the number of slots adjacent to~$E'$.
By the deficiency version of Hall's theorem, there is a matching that leaves at most~$2g$ edges unmatched.
Every matched edge is oriented away from the vertex that owns its slot, every unmatched edge arbitrarily, and~$e_v$ is the number of unmatched edges that leave~$v$.
The orientation is used only in the analysis.

\emph{Algorithm.}
Theorem~\ref{sc:thmA} computes, with a rational accuracy $\delta\le\varepsilon/4$, a fractional cover~$x$ of cost at most $(1+\delta)\tau_c$.
Every blue vertex with $x_v\ge1/4$ joins the output.
Every red vertex without a chosen neighbor then requests a cheapest blue neighbor, breaking ties by identifiers, and the requested vertices join.
This takes two rounds after the solver, and every red vertex is covered.
The threshold vertices cost at most $4\sum_{x_v\ge1/4}c_vx_v$.

\emph{The requests.}
Let~$r$ request a vertex, of cost $\mu_r=\min_{v\in C(r)}c_v$.
Then $x_v<1/4$ for all $v\in C(r)$.
Every member of~$C(r)$ is a neighbor of~$r$ in~$I$, and at most $2+e_r$ of them are out-neighbors of~$r$, with total weight less than $(2+e_r)/4$.
The members $v\in C(r)$ with $v\to r$ therefore carry weight more than $(2-e_r)/4$, and so
\[
 \mu_r\le2\sum_{v\in C(r),\,v\to r}\mu_rx_v+\frac{e_r\mu_r}2
 \le2\sum_{v\in C(r),\,v\to r}c_vx_v+\frac{be_r}2 .
\]
A blue vertex~$v$ occurs on the right only for the at most $2+e_v$ demands~$r$ with $v\to r$, and only if $x_v<1/4$, in which case $e_vc_vx_v\le be_v/4$.
Summing over all requesting demands, the requests cost at most
\[
 4\sum_{x_v<1/4}c_vx_v+\frac b2\sum_{v}e_v
 \le4\sum_{x_v<1/4}c_vx_v+bg .
\]
Altogether the output costs at most $4\sum_vc_vx_v+bg\le(4+\varepsilon)\tau_c+bg$.
\end{proof}

The additive term is removed by solving the small instance components exactly, as in Appendix~\ref{fv:imp}.

\begin{corollary}
\label{fv:rb:surface-opt}
Fix integers $g\ge0$ and $b\ge1$, and $\varepsilon>0$.
There is a deterministic constant-round broadcast CONGEST algorithm that, on every feasible red-blue instance whose network has Euler genus at most~$g$ and whose costs lie in $\{1,\ldots,b\}$, returns a cover of cost at most $(4+\varepsilon)\OPT_c$.
\end{corollary}

\begin{proof}
Every subgraph of the network has at most $(3g+3)|V(F)|$ edges.
For every $\delta>0$, Proposition~\ref{fv:rb:surface} with~$4\delta$ in place of~$\varepsilon$ gives a cover of cost at most $4(1+\delta)\tau_c+bg$.
This is the hypothesis of Lemma~\ref{fv:imp:absorb} with $\rho=3g+3$, $B=b$, $r=4$, and $q=bg$, and the lemma gives the claim.
\end{proof}

On the plane, Theorem~\ref{fv:rb:lower} shows that four cannot be replaced by any smaller constant, so four is the limiting coefficient on every fixed surface.

\paragraph{Other variants.}
Refinements of a given cover, partial domination relative to a given dominating set, and initializations on classes with separators are not needed here, because Theorems~\ref{sc:thmA} and~\ref{sc:thmD} take arbitrary demands, candidates, and costs, as in the setting of \gcov{}, and need no initial solution.
Geometric set cover and hitting set rely on incidence graphs of polynomial expansion, which need not exclude a minor, so they are outside this framework.

\section{Benchmarks and limits of the method}\label{simp:closing}

Our approach needs a fractional solver with fixed accuracy, which needs sparsity and short local descriptions, and a rounding rule with slack, which sets the coefficient. We now discuss the limits of this split and the questions it leaves open.

\paragraph{The surface benchmark.}
On a surface of Euler genus~$e$, a graph with $n\ge3$ vertices has at most $3n-6+3e$ edges.
The solver can handle this additive excess, but the excess changes what we can achieve by rounding.
The following example shows that, on some fixed surface, no algorithm, distributed or otherwise, can return a dominating set of size at most $c\,\gf(G)$ with $c<15/2$.

\begin{proposition}
\label{sc:gap}
Let~$G$ consist of a clique~$Z$ on~$28$ vertices and, for every~$14$-element subset $S\subseteq Z$, one further vertex adjacent exactly to~$S$.
Then $\gf(G)\le2$ and $\gamma(G)=15$.
\end{proposition}

\begin{proof}
We give weight $1/14$ to every vertex of~$Z$, so that each clique vertex has coverage two and each further vertex has coverage one, while the total cost is two.
On the other hand, any~$15$ vertices of~$Z$ dominate~$G$, because they meet every~$14$-subset.
If a dominating set contains $j\le14$ vertices of~$Z$, then each of the $\binom{28-j}{14}\ge1$ subsets avoiding them contributes its own vertex, so the set has at least $j+\binom{28-j}{14}\ge15$ vertices.
\end{proof}

Since the graph is finite, it embeds in a surface of some fixed genus, and therefore a factor seven against~$\gf$ cannot hold on every fixed surface.
This is why our exact surface theorem uses the integral optimum~$\gamma$: the change of benchmark is necessary.
For Euler genus at most two, the stronger bound~$7\gf$ does hold, since every minor has a ratio of edges to vertices of at most three.

\paragraph{Where minors are really needed.}
The solver uses deep minors in two places: the residual paths of the initializer, whose contraction has unbounded depth, and the separator theorem behind the partitions.
The ordinary-domination rounding in Section~\ref{simp:endpoint:section} uses one deep minor, the count of Lemma~\ref{simp:endpoint:count}, and only when~$2a$ is an integer.
The other minor constructions in the core solver and that rounding proof have depth at most two.
The large-girth dilution argument in Lemma~\ref{fv:hg:dilute} also contracts clusters of bounded radius, with the radius depending on the class and girth parameters.

\paragraph{Open problems.}
\begin{enumerate}
\item For~$K_t$-minor-free graphs with $t\ge10$, our coefficient is $2a_t+1$, where $a_t=\Theta(t\sqrt{\log t})$ is the extremal density, while the lower bound from powers of paths is $2t-3$.
Which coefficient is optimal?
\item The coefficient $2a+1$ is set by the density of minors, while part~\ref{sc:thmC:sub} of Theorem~\ref{sc:thmC} shows that the density of subgraphs suffices when~$2a$ is not an integer.
For integral~$2a$, does a subgraph-density bound of~$a$, together with an excluded minor, suffice to obtain the coefficient~$2a+1$?
We prove this for planar graphs of girth at least four and at least six, with~$a=2$ and~$a=3/2$, respectively, by counting faces (part~\ref{sc:thmC:girth} of Theorem~\ref{sc:thmC}), but a general argument remains open.
\item The constants in our round bounds are large, because the search tries many short lists and the averaging enumerates many local choices of bits.
Can the exact coefficients be achieved with a moderate number of rounds, say polynomial in the class parameters?
\item Other covering problems, such as distance-$r$~domination or connected domination, need a different budget transfer.
Which local covering problems admit a rounding rule with slack at their critical coefficient?
\end{enumerate}

\section{Formal verification and acknowledgments}\label{new:verification}

We have formalized and checked all 81 named results of this paper in Lean~4.
The formalization consists of 1612 modules and is available on GitHub.\footnote{\url{https://github.com/wwawrzy/paying-for-fractional-error-lean}}
Of these results, 73 use only Lean's standard axioms, while the remaining eight also use additional axioms based on the following classical results and constructions:
\begin{itemize}
\item Rad\'o's triangulation theorem;
\item the Euler--Poincar\'e formula;
\item the regular-neighborhood construction and the signed Heffter--Edmonds--Ringel rotation principle;
\item Mader's density bound for $K_7$-minor-free graphs~\cite{Mader1968};
\item J{\o}rgensen's density bound for $K_8$-minor-free graphs~\cite{Jorgensen1994};
\item the Song--Thomas density bound for $K_9$-minor-free graphs~\cite{SongThomas2006};
\item the Kuratowski--Wagner planarity theorem.
\end{itemize}

\section*{AI methodology}
\phantomsection\label{stoc:ai-methodology}
\addcontentsline{toc}{section}{AI methodology}
This work was developed jointly by the authors and generative AI models.
The authors provided the initial starting points: a constant-round CONGEST algorithm with a near-optimal constant factor for planar graphs, along with an approach based on set cover with bounded frequency and a sunflower argument.
Claude Opus~5.5 (Anthropic) and GPT Astra~6 (OpenAI) then extended this work and developed a complete initial proof.
The initial paper containing the proof was 199 pages long.
The authors revised it, producing the present, much shorter version, which is accessible to human readers.
The sunflower idea was replaced completely: the final proof uses essential subsets of short lists and a direct count of pairs instead.
AI models were also used for independent reviews and for a formalization of the proofs in Lean.
The authors take full responsibility for the contents of this paper.

\end{document}